\documentclass[useAMS,usenatbib]{mn2e}
\usepackage{graphicx}
\usepackage{txfonts}
\usepackage{longtable}
\usepackage{rotating}
\usepackage{multirow} 
\usepackage{appendix}
\usepackage{latexsym}
\usepackage{lscape}
\usepackage{afterpage}
\usepackage{caption}
\usepackage{hyperref}

\title[Optical spectroscopy of local type-1 AGN LINERs]{Optical spectroscopy of local type-1 AGN LINERs}
\author[S.~Cazzoli et al.]{S.~Cazzoli$^{1}$\thanks{E-mail: sara@iaa.es}, I.~M{\'a}rquez$^{1}$, J.~Masegosa$^{1}$, A.~del Olmo$^{1}$, M.~Povi{\'c}$^{2,1}$, O.~Gonz{\'a}lez-Mart{\'i}n$^{3}$,\newauthor   B.~Balmaverde$^{4}$, L.~Hern{\'a}ndez-Garc{\'i}a$^{5}$,  S.~Garc{\'i}a-Burillo$^{6}$\\
$^{1}$ IAA - Instituto de Astrof{\'i}sica de Andaluc{\'i}a (CSIC), Apdo. 3004, 18080, Granada, Spain\\
$^{2}$ ESSTI Ethiopian Space Science and Technology Institute (ESSTI), Entoto Observatory and Research Center (EORC), \\
 Astronomy and Astrophysics Research Division, P.O. Box 33679, Addis Ababa, Ethiopia\\
$^{3}$ Instituto de Radioastronom{\'i}a y Astrof{\'i}sica (IRyA-UNAM), 3-72 (Xangari), 8701, Morelia, Mexico\\
$^{4}$ INAF - Osservatorio Astronomico di Brera, INAF, via Brera 28, I-20121 Milano, Italy\\
$^{5}$ Instituto de F{\'i}sica y Astronom{\'i}a, Facultad de Ciencias, Universidad de Valpara{\'i}so, Gran Bretana 1111, Playa Ancha, Valpara{\'i}so, Chile\\
$^{6}$ Observatorio Astron{\'o}mico Nacional (OAN-IGN)-Observatorio de Madrid, Alfonso XII, 3, 28014, Madrid, Spain}

\begin{document}
\date{Accepted 2018 July 4. Received 2018 June 29; in original form 2018 April 18.}
\pagerange{\pageref{firstpage}--\pageref{lastpage}} \pubyear{2017}
\maketitle
\label{firstpage}

\begin{abstract}
The Balmer emission originated in  the broad line region (BLR) of  active galactic nuclei (AGNs) could be  either weak and difficult to detect, or even absent, for low luminosity AGNs,  as LINERs.  Our goals in this paper are threefold. First, we want to explore the AGN-nature of nearby type-1 LINERs. Second, we aim at deriving a reliable interpretation for the different components of emission lines by studying their kinematics and ionization mechanism. Third, we intend to probe the neutral gas in the nuclei of these LINERs. We study  the 22 local (z\,$<$\,0.025) type-1 LINERs  from the Palomar Survey, on the basis of optical ground- and space-based long-slit spectroscopic observations taken with TWIN/CAHA and ALFOSC/NOT. Kinematics and fluxes of a set of emission lines, from H$\beta$$\lambda$4861 to [S\,II]$\lambda$$\lambda$6716,6731, and the NaD$\lambda$$\lambda$5890,5896 doublet in absorption  have been modelled and measured, after the subtraction of the underlying starlight. We also use ancillary spectroscopic data from \textit{HST}/STIS. We found that the broad H$\alpha$ component is sometimes elusive in our ground-based spectroscopy whereas it is ubiquitous for space-based data. By combining optical  diagnostic diagrams, theoretical models (for AGNs, pAGB-stars and shocks) and the weak/strong-[O\,I] classification, we exclude the pAGBs-stars scenario in favor of the AGN as the dominant mechanism of ionisation in these LINERs, being shocks however relevant. The kinematical properties of the emission lines may indicate the presence of ionized outflows, preferentially seen in [O\,I]. However, the neutral gas outflows, diagnosed by NaD, would appear to be less frequent.

\end{abstract}


\section{Introduction}
It is nowadays accepted that all kinds of active galactic nuclei (AGNs) could be fit in the so-called \lq unified AGN model\rq \ (see \citealt{Padovani2017} for a review). However, Low ionization nuclear emission-line regions (LINERs) remain challenging to be accommodated within such unification scheme \citep{Netzer2015}. \\
\noindent  LINERs were considered from the beginning to be a distinct class of low-luminosity AGNs showing strong low-ionisation and faint high-ionisation emission lines \citep{Heckman1980}.  LINERs are  interesting objects  as they might represent the most numerous local AGN population, and as they may bridge the gap between normal and active galaxies, as suggested e.g. by their low X-ray luminosities \citep{Ho2008}. \\ 
Over the past 20 years, the ionising source in LINERs has been studied through a multi-wavelength approach via different tracers (see \citealt{Ho2008} for a review). Nevertheless, a long standing issue  is the origin and excitation mechanism of the ionised gas studied via optical emission lines. 
In addition to the AGN scenario, two more alternatives have been proposed to  explain the  optical properties of these ambiguous low-luminosity AGNs. On the one hand,  models  of  post-asymptotic giant branch stars (pAGBs, e.g. \citealt{Binette1994}) seem to be successful in reproducing the observed H$\alpha$$\lambda$6563 equivalent widths and the observed LINER-like emission ratios (e.g. \citealt{Stasinska2008}). This would classify LINERs as retired galaxies instead of genuine AGNs  (\citealt{Sarzi2010, CidFernandes2011, Singh2013} and references therein).  On the other hand, shocks might play a significant role in the ionization of the gas with the optical emission line ratios well fitted by shock-heating models only (e.g. \citealt{Dopita1996}) excluding any AGN contribution.   Shocks related to outflows and jets powered by the accretion onto a central supermassive black  hole (SMBH) do  reach those high velocities (300-500\,km\,s$^{-1}$, \citealt{Annibali2010}) required by shock models (e.g. \citealt{Groves2004}). However, arguments based on the velocity width  of emission lines seem to disfavor shock-heating as the dominant ionization mechanism \citep{Yan2012}. \\
Despite all these scenarios invoked to explain the observed LINER-like ratios, type-1 LINERs (analogue of  Seyfert-1s, i.e. the BLR is visible in the line of sight) seems to be genuine AGNs as they are observed as single compact hard X-ray sources \citep{GM2009} showing in some cases time-variability (e.g. \citealt{Younes2011, HG2014, HG2016}) and nearly the 80\,$\%$ of them are IR-bright \citep{Satyapal2004}. \\  
Type-1 LINERs are ideal targets to explore the true AGN-nature of the LINER-family as  they are  viewed face-on to the opening of the possible AGN-torus allowing the direct and unambiguous detection of broad Balmer emission lines indicative of the BLR existence. \\
In the debate about the AGN nature of LINERs, it remains also  to be disentangled if the  break-down of the AGN unification is related to the disappearance of the torus or the broad-line region (BLR) in these low-luminosity AGNs \citep{Elitzur2006,Elitzur2014,OGM2017}.\\ 
In LINERs, outflows  are  common as suggested by their \textit{HST}-H$\alpha$ morphologies \citep{Masegosa2011}. To open a new window to explore the AGN-nature  and the excitation mechanism of the LINER emission,  we  propose to infer the role of outflows (generally identified as intermediate to broad kinematic component in spectral lines) in the broadening of emission lines. This broadening effect may limit the spectroscopic classification, as the contribution of outflows may overcome the determination of whether the possibly faint and  broad (BLR-originated) H$\alpha$ component is present.  In this context, tasks as the starlight subtraction and the strategy for the line modelling have a crucial role in the detection of the BLR-component.\\
Outflows are made up by a number of gas-phases and are observed in starbursts and AGNs via long slit (e.g. \citealt{Rupke2005b,  Heckman2000, Harrison2012}) and integral field spectroscopy  (IFS, e.g. \citealt{Harrison2014,Maiolino2017}) of emission and absorption lines. In the optical regime,  H$\alpha$ and [O\,III] have been widely used to trace warm outflow signatures in AGN-hosts locally both at high (e.g. \citealt{Villar2014, Maiolino2017}) and low AGN-luminosities  (e.g.\citealt{Walsh2008, Dopita2015}), and at high redshift (e.g. \citealt{Harrison2014, Carniani2015}). Neutral gas outflows have been studied in detail only in starbursts and luminous and ultra-luminous infrared galaxies (U/LIRGs) via the NaD$\lambda$$\lambda$5890,5896 absorption with only few cases for the most IR-bright LINER-nuclei   (e.g. \citealt{Rupke2005a, Rupke2005b, Rupke2005c, Cazzoli2014, Cazzoli2016}).\\

\noindent In this paper, we used ground- and space-based optical slit-spectroscopic data taken with TWIN/CAHA, ALFOSC/NOT and \textit{HST}/STIS to investigate the presence of the broad H$\alpha$ emission in previously classified type-1 LINERs. Our main goals are to investigate the AGN nature of these LINERs and to characterize all the components by studying their kinematics and dominant ionization mechanism. Furthermore, we are able to probe for the first time the neutral gas content in some of the LINER-nuclei. \\
This paper is organized as follows. In  Section\,\ref{Sample_data}  the sample of local type-1 LINERs is presented as well as the observations and the data reduction. In Section\,\ref{Analysis} and Section\,\ref{Obs_res} are presented the spectroscopic analysis (including stellar subtraction and line modelling) and the main observational results, respectively. In Section\,\ref{discussion_results} we  discuss present and previous BLR measurements. Moreover, we explore, classify and discuss the kinematics of the different components used to model emission and absorption lines. Finally,  the main conclusions are presented in Section\,\ref{summary_conclusions}. \\
Throughout the paper we will assume H$_{\rm 0}$\,=\,70 km\,s$^{-1}$Mpc$^{-1}$ and the standard $\Omega_{\rm m}$\,=\,0.3,  $\Omega_{\rm \Lambda}$\,=\,0.7 cosmology.

\section{Sample, observations and data reduction}
\label{Sample_data}

\begin{table}
  \caption{General properties for the type-1 LINERs sample discussed in this paper.} 
    \begin{tabular}{l c c c c c c}
  \hline
ID    & z & Scale & Morphology   & \textit{i} &$\Delta$V$^{c}$  & A$_{i}$   \\
       &    &  (pc/$^{\prime}$$^{\prime}$) & & ($^{\circ}$) &  (km\,s$^{-1}$) & \\
   \hline
NGC\,0266      & 0.0155 & 326 &         SB(rs)ab$^{(a)}$ & 12 & 1004 & 0.01 \\
NGC\,0315      & 0.0165 & 344 &        E$^{(b)}$ & 52 &  -  & 0.00 \\
NGC\,0841      & 0.0151 & 315 &   Sab $^{(c)}$ & 57 & 507 & 0.31 \\        
NGC\,1052      & 0.0050 & 102 &     E4$^{(d)}$  & - & 368 & 0.00 \\
NGC\,2681      & 0.0023 &  49  & S0-a(s)$^{(b)}$  & 24  & -& 0.03  \\
NGC\,2787      & 0.0023 &  56 &   S0-a(sr)$^{(b)}$ & 51 & 476 & 0.00 \\                 
NGC\,3226      & 0.0044 &  79 &   E$^{(b)}$ & - & 220 & 0.00 \\
NGC\,3642      & 0.0053 & 112 &  SA(r)bc$^{(e)}$ & 34 & 114& 0.12 \\
NGC\,3718      & 0.0033 &  71  &  SB(s)a$^{(f)}$  & 62 & 528 & 0.32 \\
NGC\,3884      & 0.0234 & 461 & SA0/a$^{(g)}$  & 51 & 678 &0.14\\                     
NGC\,3998      & 0.0035 &  76 &   S0(r)$^{(b)}$  &34 & 1117 &0.00 \\
NGC\,4036      & 0.0046 & 100 &    E-S0$^{(b)}$ &69 & - & 0.00 \\
NGC\,4143      & 0.0032 &  65 & E-S0$^{(h)}$  & 52 &- &0.00\\    
NGC\,4203      & 0.0036 &  71 &  SAB0$^{(i)}$ & 21 &524 &0.00\\
NGC\,4278      & 0.0021 & 60 &   E$^{(b)}$ &  - & 417 &0.00\\
NGC\,4438      & 0.0002 & 27 &  Sa$^{(b)}$  & 71 &- & 0.32 \\
NGC\,4450      & 0.0065 & 124 &  Sab$^{(i)}$ & 43 &449 &0.17 \\
NGC\,4636      & 0.0031 & 53 & E$^{(b)}$   & -&767 &0.00 \\    
NGC\,4750      & 0.0054 & 120 &  S(r)$^{(l)}$ &24 & 584 & 0.05 \\
NGC\,4772      & 0.0035 & 60 & Sa$^{(m)}$  & 62 & 515 & 0.30 \\
NGC\,5005      & 0.0032 & 65 &   Sbc$^{(b)}$ & 63& 474&0.47\\
NGC\,5077      & 0.0094 & 175 & E3-E4$^{(n)}$  &-&-& 0.00 \\               
\hline
\end{tabular}
\label{T_sample}

\textit{Notes.}  \lq z\rq \ and \lq scale\rq: redshift and  scale distance from the Local Group, respectively. Both are from the NED. \lq Morphology\rq: Hubble classification. \lq \textit{i}\rq, \lq$\Delta$V$^{c}$\rq and \lq A$_{i}$\rq: respectively, inclination angle, rotational amplitude corrected for inclination and internal extinction from \citet{Ho1997b}.\\
\textit{References:}  $^{(a)}$\citet{Font2017}; $^{(b)}$\citet{GM2009}; $^{(c)}$\citet{Kuo2008}; $^{(d)}$\citet{Pogge2000}; $^{(e)}$\citet{Hughes2003}; $^{(f)}$\citet{HG2016}; $^{(g)}$\citet{Dudik2009}; $^{(h)}$\citet{Cappellari2011}; $^{(i)}$\citet{Ho2000}; $^{(l)}$\citet{Riffel2015}; $^{(m)}$\citet{Haynes2000}; $^{(n)}$\citet{DeFrancesco2008}.
\end{table}

\begin{table*}
  \caption{Optical observations details for the type-1 LINERs sample discussed in this paper.}
  \begin{tabular}{l c c c c c c c c c}
  \hline
ID&RA& DEC & Night             & EXP\,(s) & Air\,mass  &  Seeing &Slit\,PA &   Nuclear & Nuclear\ \\
&&  &              & (s) &   &  ($^{\prime}$$^{\prime}$) &($^{\circ}$) &   Aperture\,($^{\prime}$$^{\prime}$) & Aperture\,(kpc) \\
   \hline
NGC\,0266 & 00 49 47.80 & +32 16 39.8 & 23 Dec 2014 & 3\,$\times$\,1800 &  1.23 & 1.4 &  70 &  1.2\,$\times$\,2.8  & 0.39\,$\times$\,0.91 \\
NGC\,0315$^{ *}$  & 00 57 48.88 & +30 21 08.8 & 30 Sep 2013 & 3\,$\times$\,1800 & 1.23 & 0.6 &  99 & 1.0\,$\times$\,1.9 & 0.34\,$\times$\,0.65 \\
NGC\,0841 & 02 11 17.36 & +37 29 49.8 & 03 Dec 2012 & 4\,$\times$\,1800 &  1.01 &  1.0 & 270 &  1.2\,$\times$\,2.8 & 0.38\,$\times$\,0.88 \\	 
NGC\,1052$^{ *}$ & 02 41 04.80 & -08 15 20.8 & 30 Sep 2013 & 3\,$\times$\,1800 & 1.30 & 0.7 & 155 & 1.0\,$\times$\,2.3 & 0.10\,$\times$\,0.23 \\
NGC\,2681 & 08 53 32.74 & +51 18 49.2 & 04 Dec 2012 & 4\,$\times$\,1800 &  1.09 & 1.0  & 240 &  1.2\,$\times$\,3.9 & 0.06\,$\times$\,0.19 \\
NGC\,2787 & 09 19 18.59 & +69 12 11.7 & 22 Dec 2014 & 3\,$\times$\,1800 &  1.43 & 0.8 & 261 &  1.2\,$\times$\,2.2 & 0.07\,$\times$\,0.13 \\
NGC\,3226 & 10 23 27.01 & +19 53 54.7 & 23 Dec 2014 & 3\,$\times$\,1800 &  1.37 & 1.4 & 33  &  1.2\,$\times$\,2.2 & 0.10\,$\times$\,0.18 \\
NGC\,3642 & 11 22 17.89 & +59 04 28.3 & 23 Dec 2014 & 3\,$\times$\,1800 &  1.88 & 1.4 & 21  &  1.2\,$\times$\,2.8 & 0.13\,$\times$\,0.31 \\
NGC\,3718 & 11 32 34.85 & +53 04 04.5 & 24 Dec 2014 & 3\,$\times$\,1800 &  1.34 & 1.2 & 2   &  1.2\,$\times$\,2.8 & 0.09\,$\times$\,0.20 \\
NGC\,3884 & 11 46 12.18 & +20 23 29.9 & 23 Dec 2014 & 3\,$\times$\,2400 &  1.24 &  1.4   & 35  &  1.2\,$\times$\,2.2 & 0.55\,$\times$\,1.03 \\   
NGC\,3998 & 11 57 56.13 & +55 27 12.9 & 24 Dec 2014 & 3\,$\times$\,1800 &  1.16 & 1.0 & 250 &  1.2\,$\times$\,2.8 & 0.09\,$\times$\,0.21 \\
NGC\,4036 & 12 01 26.75 & +61 53 44.8 & 14 Apr 2013 & 4\,$\times$\,1800 &  1.52 & 1.2 & 102 &  1.2\,$\times$\,2.8 & 0.12\,$\times$\,0.28 \\
NGC\,4143 & 12 09 36.06 & +42 32 03.0 & 12 Apr 2013 & 4\,$\times$\,1800 &  1.12 & 2.0 & 90  &  1.5\,$\times$\,3.4 & 0.10\,$\times$\,218  \\	    
NGC\,4203 & 12 15 05.05 & +33 11 50.4 & 12 Apr 2013 & 4\,$\times$\,1800 &  1.27 & 1.0 & 20  &  1.2\,$\times$\,1.7 & 0.09\,$\times$\,0.12 \\
NGC\,4278 & 12 20 06.82 & +29 16 50.7 & 13 Apr 2013 & 4\,$\times$\,1800 &  1.07 & 1.0 & 61  &  1.2\,$\times$\,1.7 & 0.07\,$\times$\,0.10 \\
NGC\,4438 & 12 27 45.59 & +13 00 31.8 & 13 Apr 2013 & 4\,$\times$\,1800 &  1.53 & 1.0 & 39  &  1.2\,$\times$\,2.2 & 0.03\,$\times$\,0.06 \\
NGC\,4450 & 12 28 29.63 & +17 05 05.8 & 14 Apr 2013 & 4\,$\times$\,1800 &  1.28 & 1.5 & 60  &  1.2\,$\times$\,2.2 & 0.15\,$\times$\,0.28 \\
NGC\,4636 & 12 42 49.83 & +02 41 16.0 & 09 Jun 2015 & 3\,$\times$\,1800 &  1.29 & 1.2 & 38  &  1.2\,$\times$\,2.8 & 0.06\,$\times$\,0.15 \\
NGC\,4750 & 12 50 07.27 & +72 52 28.7 & 25 Dec 2014 & 3\,$\times$\,1800 &  1.31 & 1.0 & 231 &  1.2\,$\times$\,2.8 & 0.14\,$\times$\,0.34 \\
NGC\,4772 & 12 53 29.16 & +02 10 06.2 & 14 Apr 2013 & 3\,$\times$\,1800 &  1.23 &   1.2  & 21  &  1.2\,$\times$\,2.2 & 0.07\,$\times$\,0.13 \\
NGC\,5005 & 13 10 56.23 & +37 03 33.1 & 11 Apr 2013 & 3\,$\times$\,1800  &  1.23 &   1.0  & 76  &  1.2\,$\times$\,2.2 & 0.08\,$\times$\,0.15 \\ 
NGC\,5077 & 13 19 31.67 & -12 39 25.1 & 14 Apr 2013 & 4\,$\times$\,1800 &  1.62 & 1.2 & 13  &  1.2\,$\times$\,2.2 & 0.21\,$\times$\,0.39 \\ 
\hline
\end{tabular}
\label{T_Obs_sample}
\begin{flushleft}
\textit{Notes.} \lq ID\rq: object designation as in  Table\,\ref{T_sample}. \lq RA\rq \ and  \lq DEC\rq \  are the coordinates. \lq Night\rq: date the object was observed.  \lq EXP\rq: exposure time for the observations.  \lq Air mass\rq: full air mass range of the observations.  \lq Slit PA\rq : slit position angle of the observations (as measured from North and eastwards on the sky). \lq Nuclear Aperture\rq : columns indicate the nuclear aperture which is the nuclear region corresponding to the spectra presented in this work, expressed as angular size and spatial extent. These sizes are indicated as:  slit width $\times$ selected region during the extraction of the final spectrum (see Sect.\,\ref{Sample_data}).  $^{*}$\,marks those LINERs observed with ALFOSC/NOT instead of TWIN/CAHA.
\end{flushleft}
\end{table*}

The sample contains the nearby 22  type-1 LINERs (i.e. those with detected broad permitted emission lines, the analogous to Seyfert-1s) from the Palomar Survey. Except for individual discoveries of type-1 LINERs (\citealt{StorchiBergmann1993,Ho1997,Eracleous2001,Martinez2008}), the only systematic work is by \citet{Ho2003}. They used an homogeneous detection method on a magnitude limited spectroscopic catalogue, so theirs is the best defined sample of type-1 LINERs. \\
These LINERs-1 nuclei live mainly in elliptical and early type spirals. The average redshift is $\sim$\,0.0064; the average distance is 29.8\,Mpc (we consider average redshift-independent distance estimates, when available, or the redshift-estimated distance otherwise, both from the NASA Extragalactic Database, NED\footnote{\url{https://ned.ipac.caltech.edu/}}). Table\,\ref{T_sample} summarises the most important properties.  \\
Most observations were carried out with the Cassegrain TWIN Spectrograph (TWIN) mounted on the 3.5m telescope of the Calar Alto Observatory (CAHA) in different semesters from 2012 to 2015. The spectrograph was equipped with a \textsc{Site$\#$22b} (blue) and \textsc{Site$\#$20b} (red) independent CCDs allowing a resolution of about 0.5 \AA/pixel. We used the  T05 grating in the blue arm to cover the  range 4150\,-\,5450 \AA \ and the T06 grating in the red arm, covering the  5900\,-\,7100 \AA . The slit was generally set to be 1\arcsec.2 width,  or 1\arcsec.5 in case of poor seeing.  For two LINERs (NGC\,0226 and NGC\,3884)  the wavelength coverage does not include [S\,II] lines.\\
For two LINERs in the sample (NGC\,0315 and NGC\,1052) the spectroscopic data were acquired with the Andalucia Faint Object Spectrograph and Camera (ALFOSC) attached to  the 2.6m North Optical Telescope (NOT) at the Roque de los Muchachos Obsevatory, in 2013. We used two gratings: \textsc{gr$\#$08} and \textsc{gr$\#$14}, providing a resolution of 1.5 \AA/pixel in two spectral windows: 3200\,-\,6380 \AA \ and 5680\,-\,8580 \AA ; the slit width was set to be 1$\farcs$0. Table\,\ref{T_Obs_sample} summarises the details of the observations. \\
Both sets of data were achieved with the slit oriented at the parallactic angle (see Table\,\ref{T_Obs_sample}) thus without any orientational prescription. \\
For both data sets, several target exposures were taken for cosmic rays and bad pixel removal. Arc lamp exposures were obtained before and after each target observation. At least two standard stars (up to four) were observed at the beginning and at the end of each night through a 10$^{\prime\prime}$ width slit.  We reduced  raw data in the \textsc{IRAF}\footnote{\url{http://iraf.noao.edu}} environment. The reduction process included bias subtraction, flat-fielding, bad pixel removal, detector distortion, wavelength calibrations and sky subtraction.\\
For the final flux calibration, we only considered the combination of those stars where the difference of their computed instrumental sensitivity function was lower than 10$\%$. The sky background level was determined by taking median averages over two strips on both sides of the galaxy signal, and subtracting it from the final combined galaxy spectra.  The nuclear aperture for the extraction of the nuclear spectra has been done considering the seeing of the observations (Table\,\ref{T_Obs_sample}).\\
We checked the width of the instrumental profile and wavelength calibration using the [O\,I]$\lambda$$\lambda$6300.3,6363.7  sky lines. 
For TWIN/CAHA (ALFOSC/NOT) observations, the values for the central wavelengths were 6300.22\,$\pm$\,0.02  (6300.35\,$\pm$\,0.08) \AA \ and 6363.63\,$\pm$\,0.02 (6363.80\,$\pm$ 0.09)  \AA, respectively, and the full width at half maximum (FWHM) was 1.2\,$\pm$\,0.05 (5.3\,$\pm$\,0.2) \AA. \\
The line width of the sky lines  represents the instrumental dispersion ($\sigma_{\rm INS}$) that will be used in Sect.\,\ref{Analysis_LF_ground} to calculate the velocity dispersion values.  \\ 

\noindent Our data set includes observations of ten normal galaxies  selected to be of the same morphological types as the LINER-hosts (Table\,\ref{T_sample}) from the list of 48 templates from the initial catalogue of 79 galaxies in \citet{Ho1997b}. The optical observations details for these ten non-active galaxies and their optical spectra are presented in Appendix\,\ref{App_templates}, specifically we refer to Table\,\ref{T_Obs_templ}  and Fig.\,\ref{Panel_templates}  respectively. These normal galaxies  served as template to test stellar light subtraction (see Sect.\,\ref{Analysis_St_Sub}).

\subsection{Ancillary data}
\label{aux_data} 

For 12 LINERs, archival  spectroscopic data obtained with the Space Telescope Imaging Spectrograph (STIS) on board the Hubble Space Telescope (\textit{HST}) were analysed.  These data are part of a larger data-set of 24 nearby galaxies (16 LINERs) in which the presence of a BLR has been reported from their Palomar spectra \citep{Ho1997b} previously analysed by \citet{Balmaverde2014}. The spectra were obtained with a slit-width of $\leq$0\arcsec.2 using the medium-resolution \textsc{G750M} grism. The instrumental dispersion of these space-based data is $\sigma_{\rm INS}$\,$\sim$\,1.34\,\AA \ (\textit{HST}/STIS handbook). \\
We refer to \citet{Balmaverde2014} (hereafter \textit{BC14}) and references therein for the details about these \textit{HST}/STIS observations.

\begin{figure*}
\centering
\includegraphics[trim = 2.3cm 12.6cm 3.2cm 6.9cm, clip=true, width=1.\textwidth]{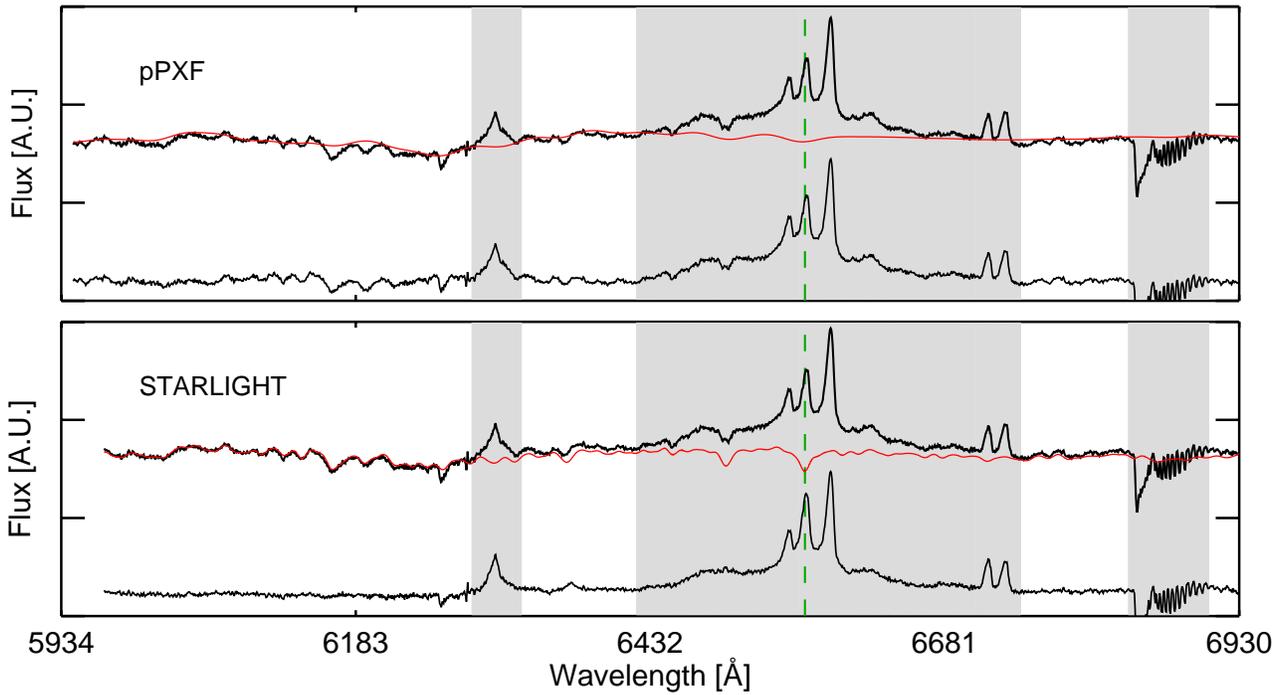}\\
\caption{Example of stellar continuum modelling and its subtraction for NGC\,4203 performed with different methods. The red line indicates the modelled stellar spectrum that matches the observed continuum, obtained applying the \textsc{pPXF} (top) and \textsc{STARLIGHT} (bottom) methods (Sect.\,\ref{Analysis_St_Sub}).   The most relevant spectral features blocked for modelling a line-free continuum are shown in gray. The rest frame H$\alpha$ wavelength is marked in green with dashed lines as reference. In this case, the output of \textsc{STARLIGHT}  better reproduce the stellar continuum with respect to \textsc{pPXF}. This LINER shows complex H$\alpha$-[N\,II] emission-line profiles which are not studied in detail in this paper (see text).}

 \label{Panel_stsub} 			 
\end{figure*}

\begin{figure*}
\centering
\includegraphics[trim = 2.3cm 12.4cm 3.2cm 6.7cm, clip=true, width=1.\textwidth]{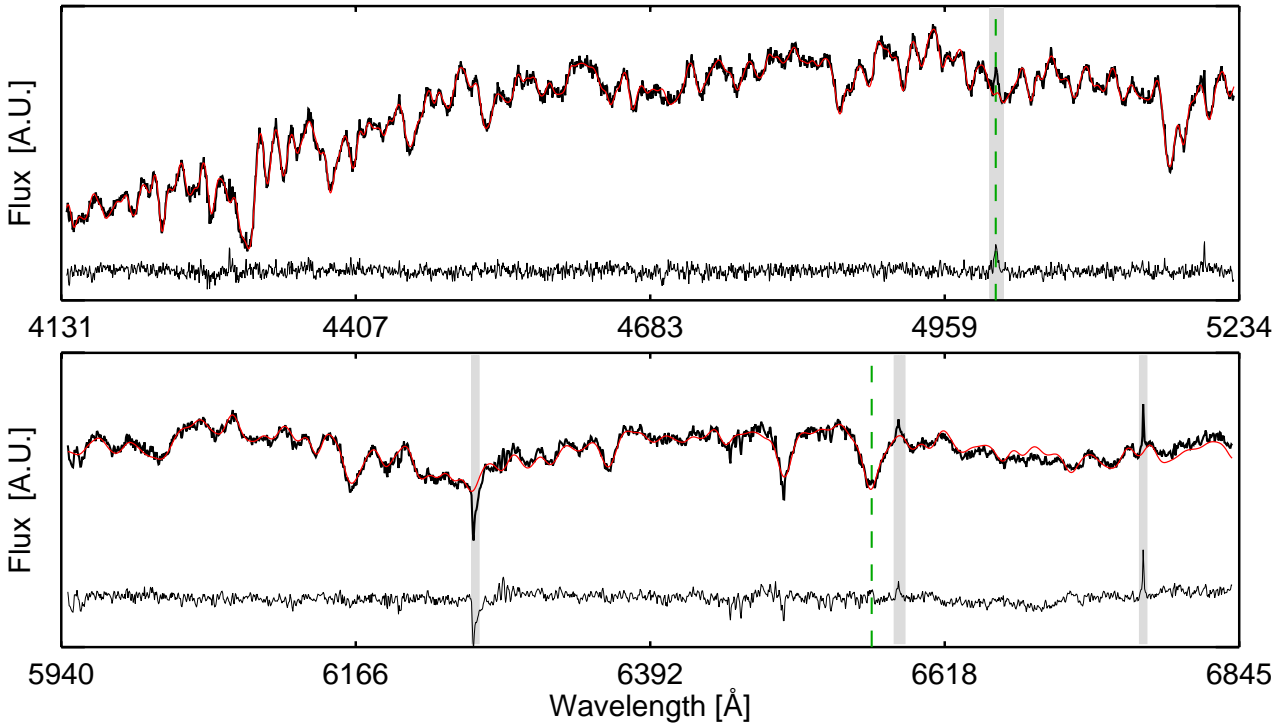}\\
\caption{Stellar continuum modelling  obtained with \textsc{pPXF} (in red) for the template galaxy  NGC\,4026 (Sections \ref{Sample_data} and  \ref{Analysis_St_Sub}) in both blue (top) and red (bottom)  bandpasses. The spectral window excluded are shown in gray as in Fig.\,\ref{Panel_stsub}.   The rest frame [O\,III]$\lambda$5007 and H$\alpha$ wavelength are marked in green with dashed lines as reference.}
 \label{Panel_template} 
 \end{figure*}

\section{Analysis of the nuclear spectra}
\label{Analysis}

All the spectra have been shifted to rest-frame wavelengths using the values of the redshift provided by NED (Table\,\ref{T_sample}).

\subsection{Stellar Subtraction}
\label{Analysis_St_Sub}

Reliable measurements of emission and absorption lines require a proper account of the starlight contamination.  We applied a penalized PiXel fitting analysis (\textsc{pPXF}  version 4.71; \citealt{Cappellari2004,Cappellari2017}) for the recovery of the shape of the stellar continuum.  This underlying continuum spectrum is then subtracted to the one observed to obtain a purely interstellar medium (ISM) spectrum. \\
To produce a model of the stellar spectrum that matches the observed line-free continuum (any ISM features, atmospheric absorption and residuals from sky lines subtraction from data reduction were masked out), we used the Indo-U.S. stellar library \citep{Valdes2004} as in \citet{Cazzoli2014,Cazzoli2016}. Briefly, this library is constituted of 1273 stars selected to provide a broad coverage of the atmospheric parameters  (effective temperature T$_{\rm eff}$, surface gravity log(\textit{g}) and metallicity  [Fe/H]) as well as spectral types. Nearly all the stellar spectra (885) have a spectral coverage from 3460 to 9464 \AA, at a resolution of $\sim$\,1\,\AA \ FWHM \citep{Valdes2004}. 
Additionally, we set the default value for the bias parameter (see pag.144 in \citealt{Cappellari2004}), and the keyword for linear regularization was fixed to zero. We also assumed a constant noise (i.e. the standard deviation calculated from a line-free continuum considering a wavelength range of either 20 or 60 \AA) per pixel.\\ 
The  stellar model subtraction leaves in few cases some residuals in the region blueward of [O\,I] (e.g. NGC\,2681, Fig.\,\ref{Panel_NGC2681}). Despite this, the result of this approach is a model that in general reproduces the continuum shape well (the residuals are typically $<$\,10$\%$). \\
In order to check the robustness of the \textsc{pPXF} modelling for all the LINERs in the  sample,  we also modelled the stellar continuum with \textsc{STARLIGHT}  V.\,04 \citep{CidFernandes2005,CidFernandes2009},  using single stellar populations from \citet{BC2003}. The spectra in this library have a spectral resolution of 3\,\AA \ guaranteeing enough spectral coverage to fit our data ($\sim$\,3200-9500\,\AA).  Our templates include simple stellar populations of 25 different stellar ages (from 0.001 to 18 Gyr) and solar metallicity. We used the  extinction law of \citet{Cardelli1989}, as in  \citet{Povic2016}. During the fits, any region with either ISM lines, or atmospheric absorption  were ignored, as for the \textsc{pPXF}-modelling. \\ 
The reliability of the stellar model is difficult to evaluate considering the large wavelength coverage of our data ($\sim$\,1000-2000\,\AA)
and the absence of unambiguous stellar absorption features (e.g. in the near-IR the CO$\lambda$$\lambda$2.293,2.322 bands). For example,  the model might  fit well the stellar continuum in the blue part of the spectra but not in the red one (e.g. NGC\,2681, Fig.\,\ref{Panel_comparison_stars}). Therefore, we visually inspected all the results from both \textsc{pPXF}  and \textsc{STARLIGHT} fitting procedures noting that generally the \textsc{pPXF} model better reproduces the global continuum shape with lower residuals. In those few cases for which the \textsc{pPXF} modelling was not satisfactory, we selected the \textsc{STARLIGHT} output for the following analysis. Specifically, we adopted the \textsc{STARLIGHT} continuum model for blue and red spectra in one and six cases, respectively.  In only one case (NGC\,4036) we assumed the  continuum model from \textsc{STARLIGHT} for both spectra.  An example of the \textsc{pPXF}-\textsc{STARLIGHT} comparison is shown in Fig.\,\ref{Panel_stsub}; for all the other cases we refer to  Fig.\,\ref{Panel_comparison_stars} of  Appendix\,\ref{App_templates}. \\
Applying the  \textsc{pPXF}  fitting analysis also to our template galaxies (Sect.\,\ref{Sample_data}), in  five cases (NGC\,2950, NGC\,3838, NGC\,4026, NGC\,4371, NGC\,4382 and NGC\,7332, Table\,\ref{T_Obs_templ})  we found the presence of weak emission lines (mostly [O\,III]$\lambda$5007 and [N\,II]$\lambda$6584) in the residual spectra. An example is shown in Fig.\,\ref{Panel_template}, for the galaxy NGC\,4026, for all the nine cases we refer to  Fig.\,\ref{Panel_templates}. The possible presence of ISM emission lines in the spectra of template galaxies  may compromise the attainment of a final purely ISM-spectrum of LINERs. Hence, to account for this issue, we prefer the stellar subtraction obtained with either \textsc{pPXF} or \textsc{STARLIGHT} rather than using template galaxies. \\
For a more detailed discussion of  different techniques for starlight subtraction we refer to \citet{Cappellari2017}. \\
The final selected procedure is listed for all the LINERs in Tab.\,\ref{T_kin} (column 3). The observed spectrum, its stellar continuum model and the final emission-line spectrum, are shown for each LINER in Appendix\,\ref{App_comments_panels}.\\
We did not apply any procedure for the subtraction of the underlying stellar light for \textit{HST}/STIS spectra as their limited wavelength coverage prevents an optimal stellar-continuum modelling. However,  the contamination by the host-galaxy stellar continuum is expected to be small as the \textit{HST}/STIS aperture is much smaller than those  for ground-based observations (see Sect.\,\ref{Sample_data} and Table\,\ref{T_Obs_sample}). The results of the stellar continuum modelling of \textit{HST}/STIS spectra by \citet{Shields2007}, for  four LINERs  in common with our sample (NGC\,2787, NGC\,4143, NGC\,4203 and NGC\,4450),  support that the stellar light contamination is small at \textit{HST}/STIS scales.

\subsection{Emission line fitting}
\label{Analysis_LF}

After the subtraction of the stellar contribution (Sect.\,\ref{Analysis_St_Sub}), each of the emission lines [O\,I]$\lambda$$\lambda$6300,6363, H$\alpha$$\lambda$6563,  [N\,II]$\lambda$$\lambda$6548,6584 and  [S\,II]$\lambda$$\lambda$6716,6731 in the spectra  were modelled with  single or  multiple Gaussian-profiles with a Levenberg-Marquardt least-squares fitting routine (\textsc{MPFITEXPR}, implemented by \citealt{Markwardt2009}) within the Interactive Data Analysis\footnote{\url{http://www.harrisgeospatial.com/SoftwareTechnology/IDL.aspx} (\textsc{IDL})} environment.  We have imposed that the intensity ratios between the [N\,II]$\lambda$6548 and the [N\,II]$\lambda$6583 lines, and  the [O\,I]$\lambda$6363 and the [O\,I]$\lambda$6300 lines satisfied the 1:3 and 1:2.96 relations, respectively \citep{Osterbrock2006}.  The fit was performed simultaneously for all lines using Gaussians and we consider one (or two) template(s) as reference for central wavelength(s) and line width(s) of the Gaussian curves. We did not use H$\alpha$ or [N\,II] as template since these lines are generally blended, and this may compromise  the results of the fitting. Thus, each spectrum is fitted with three distinct models using [S\,II], [O\,I] or both, as reference. \\

\noindent\textbf{\textit{S}-model.} The first model consists on modelling the [S\,II] lines and then tie all narrow lines to follow the same shifts and widths. \\
\textbf{\textit{O}-model.} The second model is similar to the  \textit{S}-model but considering the [O\,I] lines as reference. This is particularly helpful as Oxygen lines are typically not-blended contrary to [S\,II]. \\
\textbf{\textit{M}-model.} The last model we tested is a \lq mixed\rq \ line-modelling. It encompasses two possibilities with different assumptions. On the one hand, we considered that [S\,II] probes [N\,II], and  [O\,I]  traces the narrow H$\alpha$ (\textit{M1}). On the other hand, we assumed that narrow H$\alpha$ and [N\,II] follow [O\,I]  with [S\,II] line behaving otherwise (\textit{M2}).\\

\noindent The use of [S\,II] or [O\,I] lines as template for modelling the H$\alpha$-[N\,II] is motivated by the stratification density in the narrow line region (NLR). Since a high difference in critical densities do exist for forbidden lines ions, $\sim$\,10$^{3}$\,cm$^{-3}$ for [S\,II], $\sim$\,10$^{4}$\,cm$^{-3}$ for [N\,II] and $\sim$\,10$^{6}$\,cm$^{-3}$ for [O\,I], the line profiles are expected not to be the same (see \citealt{Balmaverde2014}). Moreover, in the case of strong shocks being present in the ionised regions an enhanced  [O\,I] emission is expected in relation to the other ions.\\
The procedure is aimed at obtaining the best-fitting to all emission lines and is organized in four steps as follows. \\
First, we tested all the models considering one Gaussian per each forbidden line and narrow H$\alpha$ (hereafter, narrow component). Second, after visual inspecting all the spectra and the results from the precedent step, for strongly asymmetric forbidden lines profiles, we tested a two-components line fitting. In these cases, the procedure adds a second Gaussian of intermediate width  (hereafter, second component) following the three models listed above.  \\
To prevent overfit models and allow for the appropriate number of Gaussians,  we first calculated the standard deviation of a portion of the continuum free of both emission and absorption lines ($\varepsilon_{\rm c}$). Then we compare this value with the standard deviation estimated from the residuals under [S\,II] and [O\,I] obtained once the underlying stellar population has been substracted  ($\varepsilon^{\rm line}$).  We are quite conservative and have considered a reliable fit when $\varepsilon^{\rm line}$\,$<$\,3\,$\times$\,$\varepsilon_{\rm c}$. Table\,\ref{T_rms} in Appendix\,\ref{App_templates} summarises all the $\varepsilon_{\rm c}$ and $\varepsilon^{\rm line}$ measurements.\\ 
Third, a  broad H$\alpha$  component  is added if needed to reduce significantly the residuals and the standard deviation when compared to single and double component(s) modelling. To assess whether the addition of a broad component is significant we estimated  the $\varepsilon^{\rm line}$-level of confidence of our fitting as in the previous step (but in correspondence with  H$\alpha$-[N\,II]) and adopting the same criterion to avoid overfitting. \\ 
Finally, we considered the best-fitting model for  H$\alpha$ as  template for  H$\beta$$\lambda$4861, as these lines are expected to arise in the same region. Similarly, the [O\,III]$\lambda$$\lambda$4959,5007 emission lines are tied to follow the [S\,II] emission lines as  they both originate in the NLR. Considering this the general behaviour, exceptions could be explained in terms of a more complex stratification in the ionization state and/or density of the gas in the NLR (being the critical density of  [O\,III]  of $\sim$\,10$^{5}$\,cm$^{-3}$). For this step of the procedure, we also imposed the intensity ratios between the [O\,III]$\lambda$4959 and the [O\,III]$\lambda$5007 satisfied the 1:2.96 relation \citep{Osterbrock2006}. This last step of the procedure has been applied only to ground spectroscopic data as only the red \textit{HST}/STIS bandpass is available (Sect.\,\ref{Sample_data}).\\
Overall, our fitting procedure includes three different physical models either with (\textit{M}-models) or without (\textit{S}- and \textit{O}-models) stratification of the NLR. These models span  four possible combinations of the components to reproduce the emission line profiles. Specifically, one or two components per forbidden line and narrow H$\alpha$, and each of these alternatives with an additional broad component in H$\alpha$. \\
A summary of the different models and combinations considering all the LINERs and data sets is given in Table\,\ref{T_summary}. The adopted line modelling of the observed line profiles for both ground- and space-based spectroscopic data are shown in Figures from \ref{Panel_NGC0266} to \ref{Panel_NGC5077} of Appendix\,\ref{App_comments_panels};  we also briefly describe the line profiles seen in ground-based data and their modelling (in the captions of the figures in Appendix\,\ref{App_comments_panels}).\\
The selected best-fitting models are listed for each LINER in Tab.\,\ref{T_kin}, along with the velocity and velocity dispersion for each component. We corrected the observed line width $\sigma_{\rm line}$ for the effect of instrumental dispersion  (Sections \ref{Sample_data} and \ref{aux_data}) by subtracting it in quadrature from the observed line dispersion ($\sigma_{\rm obs}$):  $\sigma_{\rm line}$=\,$\sqrt{\sigma_{\rm obs}^{2}\,-\,\sigma_{\rm INS}^{2}}$.

\subsubsection{Ground-based spectroscopy}
\label{Analysis_LF_ground}
We tested the three models listed in Sect.\,\ref{Analysis_LF} for our ground-based spectroscopic data. In  two cases (NGC\,0226 and NGC\,3884) it was not possible to apply either  \textit{S}- or  \textit{M}-models as the wavelength coverage does not include [S\,II] lines (Sect.\,\ref{Sample_data}, Figures \ref{Panel_NGC0266} and \ref{Panel_NGC3884}).\\
NGC\,4203 represents an extreme case as three line components (narrow, second and broad) are not sufficient to reproduce well the H$\alpha$ line profile (Fig.\,\ref{Panel_NGC4203}). Its very broad and double-peaked emission line profile is likely originated in the outer parts of the accretion disc surrounding the SMBH (e.g. for a discussion of such model see \citealt{StorchiBergmann2017}).  Thus, a more detailed modelling is needed (e.g.  with a skewed Gaussian component as in \citealt{Balmaverde2014}) but this is beyond the aim of this work. Therefore, we excluded NGC\,4203 from the analysis of ground-based data.\\
The use of a single  Gaussian component for the forbidden lines and narrow H$\alpha$ provides already a good fit only in two cases: NGC\,0841 and NGC\,4772 (Table\,\ref{T_summary}, Figures \ref{Panel_NGC0841} and  \ref{Panel_NGC4772}). \\
From our visual inspection, we note that [S\,II] and [O\,I] have strongly asymmetric profiles and their modelling requires two Gaussians (i.e. a single component fitting is an oversimplification) in many cases (i.e. 15/21 objects, Table\,\ref{T_summary}). Therefore, in these cases, we applied a two-components line fitting to forbidden lines and narrow H$\alpha$.  A two-Gaussians fit  reproduces well the profiles of forbidden lines  in  all the 15 cases producing  a final modelling accurate at $<$\,3 $\varepsilon^{\rm line}$ confidence (Table\,\ref{T_rms}). \\  
\noindent In four cases (NGC\,1052, NGC\,3998, NGC\,4438 and NGC\,5005) the level of confidence calculated for H$\alpha$-[N\,II] is between 3.5 and 8  ($\varepsilon^{\rm line}$ in Table\,\ref{T_rms}). However, even if we tried to add the broad H$\alpha$ component in the modelling (step 3 in our procedure, Sect.\ref{Analysis_LF}),  the residuals were not lowered. Therefore, to avoid overfitting with a   questionable Gaussian component we did not  add the broad H$\alpha$  \noindent component. 

\begin{landscape}
\begin{table}  
\caption[Results from the analysis of the spectra.]{Results from the analysis of  the optical spectra.}
\begin{tabular}{l l c c  c| c c c c | c c c c | c c}
\hline 						
ID &  Obs.    &  SF & Mod.  & Comp. & V$_{\rm N}$$^{\rm [S\,II]}$ & $\sigma$$_{\rm N}$$^{\rm [S\,II]}$  &    V$_{\rm N}$$^{\rm[O\,I]}$ & $\sigma$$_{\rm N}$$^{\rm[O\,I]}$ &  V$_{\rm S}$$^{\rm[S\,II]}$ & $\sigma$$_{\rm S}$$^{\rm[S\,II]}$  &   V$_{\rm S}$$^{\rm[O\,I]}$ & $\sigma$$_{\rm S}$$^{\rm[O\,I]}$ &   V$_{\rm B}$$^{\rm H\alpha}$ & $\sigma$$_{\rm B}$$^{\rm H\alpha}$  \\
    &              &       &            &             & km\,s$^{-1}$& km\,s$^{-1}$& km\,s$^{-1}$& km\,s$^{-1}$& km\,s$^{-1}$& km\,s$^{-1}$& km\,s$^{-1}$& km\,s$^{-1}$& km\,s$^{-1}$& km\,s$^{-1}$\\
 \hline  
  
NGC\,0226 &   CAHA$^{\dagger}$ &p/p& O & N\,+\,S & - & -& 13\,$\pm$\,2 & 165\,$\pm$\,17 & - & - & -296\,$\pm$\,19 & 475\,$\pm$\,50 &- & - \\  
NGC\,0315 & 	   [NOT] &p/p& M1 & N\,+\,S\,+\,B &  -4\,$\pm$\,6 & 88\,$\pm$\,5 & -24\,$\pm$\,5 &   118\,$\pm$\,31 &      -209\,$\pm$\,148 & 485\,$\pm$\,50 & -288\,$\pm$\,57 & 711\,$\pm$\,144 &374\,$\pm$\,74 & 1051\,$\pm$\,210  \\
& HST $^{\dagger}$ &   - &  S  & N\,+\,S\,+\,B &  54\,$\pm$\,6 & 168\,$\pm$\,8  &- & - & -19\,$\pm$\,4    & 397\,$\pm$\,60   & - & - & 465\,$\pm$\,70 &  1370\,$\pm$\,165  \\	    
NGC\,0841 & CAHA	  &p/p& S & N &	     31\,$\pm$\,5 & 142\,$\pm$\,10 & (31\,$\pm$\,5) & (142\,$\pm$\,10) &  - &	  - &  - &  - &- &  -  \\
NGC\,1052 &  [NOT] 	  &p/p& M2  &N\,+\,S &     -43 \,$\pm$\,2 & 121\,$\pm$\,36 & -123\,$\pm$\,25 & 224\,$\pm$\,49 &    11\,$\pm$\,3 &342\,$\pm$\,71 &-303\,$\pm$\,16 &761\,$\pm$\,38 & - & - \\
&&   	 		  &   &&  &  &  -111\,$\pm$\,10 &   265\,$\pm$\,12 	   & &&-340\,$\pm$\,19   &   522\,$\pm$\,78   & &   \\
 & [HST] &-& M1  & N\,+\,S\,+\,B & -104\,$\pm$\,3 & 174\,$\pm$\,9 & -91\,$\pm$\,2 &186\,$\pm$\,14 &83\,$\pm$\,17 & 331\,$\pm$\,7 & -66\,$\pm$\,5 &535\,$\pm$\,5 &12\,$\pm$\,12 & 1238\,$\pm$\,22  \\
NGC\,2681 & CAHA	  &p/SL& O & N\,+\,S &  (-41\,$\pm$\,2) & (75\,$\pm$\,26 ) & -41\,$\pm$\,2 & 75\,$\pm$\,26 & (-180\,$\pm$\,8) & (209\,$\pm$\,12) & -180\,$\pm$\,8 & 209\,$\pm$\,12 & - & -  \\ 
NGC\,2787 & 	CAHA   &p/p& S & N\,+\,B &  -5\,$\pm$\,4 & 157\,$\pm$\,13 &  (-5\,$\pm$\,4) & (157\,$\pm$\,13)  & -&-&-&-& 214\,$\pm$\,67 & 542\,$\pm$\,83  \\
 &   HST $^{\ddagger}$ &-& S &  N\,+\,B & 110\,$\pm$\,10 &204\,$\pm$\,6 &  (110\,$\pm$\,10) & (204\,$\pm$\,6) & -&-&-&-& 228\,$\pm$\,46 & 968\,$\pm$\,32 \\
NGC\,3226 & CAHA 	  &p/p& O & N\,+\,S & (-58\,$\pm$\,6) & (185\,$\pm$\,20) & -58\,$\pm$\,6 & 185\,$\pm$\,20 & (-155\,$\pm$\,21) & (618\,$\pm$\,45) & -155\,$\pm$\,21 &  618\,$\pm$\,45 & - & -  \\ 
NGC\,3642 & 	[CAHA]  &p/p& M2  &  N\,+\,S\,+\,B &   -33\,$\pm$\,2 & 49\,$\pm$\,25 & -32\,$\pm$\,6 & 70\,$\pm$\,14 & -49\,$\pm$\,6 & 174\,$\pm$\,7 & -335\,$\pm$\,67 & 300\,$\pm$\,60 & -228\,$\pm$\,45 & 1341\,$\pm$\,240  \\ 
& [HST] $^{\dagger}$ &   	-  &  S   & N\,+\,B  &   71\,$\pm$\,13 & 120\,$\pm$\,21  &  - &  - &	 -    &  -  & - &  - & 191\,$\pm$\,33 &  959\,$\pm$\,81 \\	    
NGC\,3718 & 	CAHA  &p/p& M1  & N\,+\,B &   -110\,$\pm$\,5 & 202\,$\pm$\,13 & -91\,$\pm$\,7 & 261\,$\pm$\, 7     & - & - & - & - & -49\,$\pm$\,10 & 1096\,$\pm$\,219  \\
NGC\,3884&  CAHA $^{\dagger}$  &p/p& O & N\,+\,S & - & -  & -64\,$\pm$\,6 & 194\,$\pm$\,10 & - & - & -336\,$\pm$\,32  & 456\,$\pm$\,91 & - & -  \\
NGC\,3998 & 	[CAHA]  &p/p& O & N\,+\,S & (-26\,$\pm$\,2) &  (203\,$\pm$\,20) & -26\,$\pm$\,2  & 203\,$\pm$\,20 &  (-73\,$\pm$\,15) & (713\,$\pm$\,75) & -73\,$\pm$\,15 & 713\,$\pm$\,75 & - & -  \\       
&&   	 		  &   &&     & &  -105\,$\pm$\,7 &   196\,$\pm$\,6      &  & & -228\,$\pm$\,17   &   531\,$\pm$\,21  &   \\ 
  & HST $^{\ddagger}$ &-& M1  & N\,+\,B & 194\,$\pm$\,18 & 272\,$\pm$\,4 & 202\,$\pm$\,10 &474\,$\pm$\,5 & -&-& -&-& 0\,$\pm$\,1 &1870\,$\pm$\,36  \\
NGC\,4036 & CAHA 	  & SL/SL&M1  & N\,+\,S & 6\,$\pm$\,3 & 165\,$\pm$\,10 & -5\,$\pm$\,3 & 85\,$\pm$\,10 & 48\,$\pm$\,5 & 379\,$\pm$\,7 & 35\,$\pm$\,12 & 332\,$\pm$\,32   & - & -  \\ 
 & HST   &-& M1  & N\,+\,B &  240\,$\pm$\,5 & 201\,$\pm$\,13 & 242\,$\pm$\,12 &180\,$\pm$\,4 & -&-& -& -& 191\,$\pm$\,38 &1051\,$\pm$\,210 \\
NGC\,4143& CAHA 	  &p/SL & M2   & N\,+\,S & 46\,$\pm$\,4 & 122\,$\pm$\,9 & 32\,$\pm$\,3 & 100\,$\pm$\,10 & -30\,$\pm$\,6 & 222\,$\pm$\,44 & -15\,$\pm$\,3 &  570\,$\pm$\,22 & - & -  \\
&   HST $^{\ddagger}$ &-& S & N\,+\,S\,+\,B & 145\,$\pm$\,9 & 168\,$\pm$\,36 & (145\,$\pm$\,9) & (168\,$\pm$\,36) & 6\,$\pm$\,2 &320\,$\pm$\,65 &(6\,$\pm$\,2 ) & (320\,$\pm$\,65)& 540\,$\pm$\,42 & 1492\,$\pm$\,59  \\
  NGC\,4203& 	CAHA  &p/SL& M1   & N\,+\,S\,+\,B &  37: & 110:  & -4:  & 210:  & -197: & 356:  & 545:    & 846:  & -251:  &  3494:  \\ 
&&   	 		  &   & &    53: &	141:   & &  &   104: 	&   368:   &  &  &  &  \\
 &  [HST] &-& M1  & N\,+\,S\,+\,B &  102:  & 68:  & 82:  & 111:  & 20:  & 369:  & 163:  & 474:  & -285:  & 3191:   \\
NGC\,4278 & 	 CAHA &p/p& M2   & N\,+\,S & 6\,$\pm$\,7 & 177\,$\pm$\,35 & 21\,$\pm$\,2 & 180\,$\pm$\,14 &  77\,$\pm$\,34 & 240\,$\pm$\,4 & 14\,$\pm$\,46 & 669\,$\pm$\,67 & - &- \\ 	     
 &  HST $^{\ddagger}$  &-& M1  & N\,+\,S\,+\,B & 105\,$\pm$\,12 & 172\,$\pm$\,3 &96\,$\pm$\,27 & 189\,$\pm$\,2 & 122\,$\pm$\,106 & 536\,$\pm$\,2 & 69\,$\pm$\,28 & 737\,$\pm$\,2 &165\,$\pm$\,32 & 1142\,$\pm$\,228  \\
 NGC\,4438& 	[CAHA]  &p/p& M1   & N\,+\,S & -50\,$\pm$\,2 & 87\,$\pm$\,22 & -62\,$\pm$\,2 & 68\,$\pm$\,30 & -5\,$\pm$\,5 & 203\,$\pm$\,10 & 35\,$\pm$\,10 & 213\,$\pm$\,42 & - &-  \\     
NGC\,4450 & CAHA	  &p/p& M2   & N\,+\,S & -17\,$\pm$\,8 & 101\,$\pm$\,6  & -6\,$\pm$\,4 & 133\,$\pm$\,10 & -15\,$\pm$\,2 & 220\,$\pm$\,30 & -94\,$\pm$\,15 & 451\,$\pm$\,45  & - &-  \\
&  HST &-& M1  & N\,+\,S\,+\,B & 108\,$\pm$\,2 & 117\,$\pm$\,13 &109\,$\pm$\,10 &113\,$\pm$\,3  &-75\,$\pm$\,43 & 442\,$\pm$\,89 & 88\,$\pm$\,19 &442\,$\pm$\,2 & -89\,$\pm$\,28 & 3125\,$\pm$\,34  \\
 NGC\,4636 & CAHA	  &p/SL& S & N\,+\,B & 165\,$\pm$\,7 & 154\,$\pm$\,8 &	 (165\,$\pm$\,7) & (154\,$\pm$\,8) &   -&-& - &  -& -44\,$\pm$\,97 & 914\,$\pm$\,182 \\
NGC\,4750  & CAHA 	  &SL/p&  O & N\,+\,S\,+\,B & (-32\,$\pm$\,5) & (172\,$\pm$\,28) & -32\,$\pm$\,5 & 172\,$\pm$\,28 & (-296\,$\pm$\,21) & (380\,$\pm$\,42) &-296\,$\pm$\,21 & 380\,$\pm$\,42 & 328\,$\pm$\,42 & 1005\,$\pm$\,45   \\ 
 NGC\,4772 & 	CAHA  &p/p& S & N & -21\,$\pm$\,5 & 249\,$\pm$\,12 &   (-21\,$\pm$\,5) & (249\,$\pm$\,12) &  - &     - &  - &  - &- &  -\\      
 NGC\,5005   & [CAHA]	  &p/SL& S  & N\,+\,S& 92\,$\pm$\,18 & 237\,$\pm$\,47 & (92\,$\pm$\,18) & (237\,$\pm$\,47) & -108\,$\pm$\,22 & 446\,$\pm$\,90 & (-108\,$\pm$\,22) & (446\,$\pm$\,90) & - & - \\ 

& HST $^{\dagger}$& -  &  S & N\,+\,S\,+\,B &  -96\,$\pm$\,12  &  98\,$\pm$\,20 & - &  - & -111\,$\pm$\,22 &  302\,$\pm$\,62  & - & - & 145\,$\pm$\,29 & 914\,$\pm$\,135  \\	    
 
 NGC\,5077 & 	CAHA  &p/SL& S& N\,+\,B & -14\,$\pm$\,4 &195\,$\pm$\,18 & (-14\,$\pm$\,4) & (195\,$\pm$\,18) &  - &	    - &  - &  - & 168\,$\pm$\,34  &  1188\,$\pm$\,238   \\	     

& HST $^{\dagger}$& - & S  & N\,+\,S\,+\,B &   93\,$\pm$\,14 & 282\,$\pm$\,16  & - & - & -184\,$\pm$\,37	     & 397\,$\pm$\,80   &	 -    & -  & 419\,$\pm$\,80 & 1142\,$\pm$\,170   \\	    

\hline 
 \label{T_kin}
\end{tabular}

\textit{Notes.}  \lq ID\rq: object designation as in Table\,\ref{T_sample}. \lq OBS\rq:  origin of the optical data.   \lq Stellar fit\rq: assumed model for the stellar continuum in the blue and red spectra,  \lq p\rq \ and  \lq SL\rq \ stand for \textsc{pPXF} and \textsc{STARLIGHT} method, respectively. \lq Mod.\rq: best-fitting model for emission lines.  \lq S\rq,   \lq O\rq \ and   \lq M\rq \ stand for models based on [S\,II] or [O\,I] or the two mixed-types (see text).  \lq Comp.\rq: components used to achieve the best-fitting model. \lq N\rq , \lq S\rq \ and \lq B\rq \ stand for narrow,  second and broad components, respectively. The  velocity (\lq V\rq ) and velocity dispersion (\lq$\sigma$\rq),  for each component listed in column 5 and  for the set of emission lines: [S\,II], [O\,I] and H$\alpha$.  For three LINERs we cannot constrain H$\beta$ and [O\,III] using the results obtained for the red spectrum. Hence, we report V and $\sigma$ of these emission lines in a different line.  Symbols indicate: \\
$( \ )$ the values obtained for those cases for which the shift and width of emission lines has been constrained using a different forbidden template (e.g. [S\,II] values when the \textit{O}-model is adopted); \\
$[ \ ]$ the data for which  the fit of  the H$\alpha$-[N\,II] emission is not constrained well;\\
$^{\dagger}$ the observed spectrum lacks  either [S\,II] or [O\,I] preventing to test all the proposed fitting model;\\
$^{\ddagger}$  the  [O\,I]  lines are at the edge of \textit{HST}/STIS spectra;\\
: the measurements which are only indicative owed the need of a more elaborate physical modelling (NGC\,4203). 
\end{table}
\end{landscape}

\begin{table*}
\caption{Summary of the models and components used to model emission lines.}
\begin{tabular}{l c c c c c c c c c }
\hline
      & \hspace{.5cm} & \multicolumn{4}{c}{GROUND}  & \hspace{1.2cm} & \multicolumn{2}{c}{SPACE} \\
         \hline
       & &     [S\,II]     &     [O\,I]     &     M1    & M2    &  &  [S\,II]      & M1    \\
   \hline
      
 Narrow     && NGC\,0841 	&           	     		&           &              	 &  &\\
          && NGC\,4772 	&           	     			&           &              	 &  &\\ 
   \hline	
 Narrow\,+\,Second && [NGC\,5005] 	&           	     			&           &              	 &  &\\
	&&           		& NGC\,0266 $^{\dagger}$   & &                       	 &  &\\
	&&           		& NGC\,2681 	     &           &&              	   &\\
	& &          		& NGC\,3226 	     &           &&              	   &\\
	& &          		& NGC\,3884 $^{\dagger}$    &&                         	 &  &\\ 
	& &         		 & [NGC\,3998] 	     &           &&              	  &\\
        & &          		&           	     &   &[NGC\,1052]&       &                  &\\	       
        & &        		  &           	     & NGC\,4036 &&                         &\\
        & &          		&           	     &  & NGC\,4143&                         &\\
        & &          		&           	     & &  NGC\,4278&                         &\\  
        & &          &           	     & [NGC\,4438] &&                        &\\
         & &          &           	     &  & NGC\,4450&                         &\\												 
   \hline
 Narrow\,+\,Second\,+\,Broad$_{\rm H\alpha}$          &&                     & NGC\,4750 &	     &                                  &  &\\
        &&           &           	     & [NGC\,0315] &&                         &\\
        &&           &            	     &  & [NGC\,3642]&                         &\\
        &&           &           	     &           &&                         &&  [NGC\,1052] \\
        &&           &           	     &           && &NGC\,0315 $^{\dagger}$    &\\
        &&           &           	     &           && &NGC\,4143 $^{\ddagger}$     &\\        
        &&           &           	     &           && &NGC\,5005 $^{\dagger}$    &\\     
       &&           &           	     &           && &NGC\,5077 $^{\dagger}$    &\\                	
        &&           &          	   	     &           &&  		           &&  NGC\,4278 $^{\ddagger}$\\
        &&           &           	     &           & & 		         &   &NGC\,4450 \\						    
    \hline
 Narrow\,+\,Broad$_{\rm H\alpha}$   && NGC\,2787 &           	     &           &&  		           &\\
        && NGC\,4636 &           	     &           &&  		           &\\  
        && NGC\,5077 &         	             &           && 		           &\\ 
        &&           &          	             & NGC\,3718 &  		         &  &\\
        &&           &          	             &           & &&NGC\,2787 $^{\ddagger}$    &\\
        &&           &          	             &           & && [NGC\,3642] $^{\dagger}$    &\\        
	&&           &          	             &           &&                       &  & NGC\,3998 $^{\ddagger}$\\
	&&           &          	             &           &&                       &  & NGC\,4036\\						     
\hline
\end{tabular}
\label{T_summary}
\begin{flushleft}%
\textit{Notes.} Rows display the four possible combinations of the components to  reproduce the emission line profiles (Sect.\,\ref{Analysis_LF}). Columns indicate the different physical models (Sect.\,\ref{Analysis_LF}) considered for both ground- and space-based data as indicated on the top.  The  object designation is as in Table\,\ref{T_sample}. $^{\dagger}$ and $^{\ddagger}$ symbols and square-brackets are the same of Table\,\ref{T_kin}.\\ 
\end{flushleft}
\end{table*}

\noindent The addition of a broad H$\alpha$ component reduces significantly the residuals from the two-components fitting in four cases (3/15, NGC\,0315, NGC\,3642 and NGC\,5750). Among these, in  two cases (NGC\,0315 and NGC\,3642), the modelling is accurate to $\sim$\,4-5 ($\varepsilon^{\rm line}$  in  Table\,\ref{T_rms}; Figures \ref{Panel_NGC0315} and \ref{Panel_NGC3642}).\\
In four cases, namely  NGC\,2787, NGC\,3718, NGC\,4636 and NGC\,5077, the forbidden lines do not require any second component,  but H$\alpha$ is clearly broad and the BLR component is essential  for an adequate fitting (Figures \ref{Panel_NGC2787}, \ref{Panel_NGC3718}, \ref{Panel_NGC4636} and \ref{Panel_NGC5077}). \\
 In the  peculiar case of  NGC\,1052  (Fig.\,\ref{Panel_NGC1052})  to obtain a good fit we displace the second Gaussian needed to model [O\,III] lines with a shift larger then the uncertainties estimated when  modelling [S\,II]. In the exceptional case of  NGC\,3998 (Fig.\,\ref{Panel_NGC3998}), we also shifted the narrow component of [O\,III].  

\subsubsection{Space spectroscopy}
\label{Analysis_LF_space}

We applied the same procedure described in Sect.\,\ref{Analysis_LF} for  modelling the observed emission line profiles in \textit{HST}/STIS spectroscopic data (Sect.\,\ref{aux_data}) for the 12 LINERs in common with \citet{Balmaverde2014}. As for ground-based spectroscopy, we exclude NGC\,4203. This LINER  is an extreme case (Sect.\,\ref{Analysis_LF_ground}) whose detailed modelling is beyond the aim of this work.\\
We attempted to test the three models for the remaining 11 LINERs  except for four cases, namely NGC\,0315, NGC\,3642, NGC\,5005 and NGC\,5077. For these objects it was not possible to apply either  \textit{O}- or  \textit{M}-models since the wavelength coverage of our \textit{HST}/STIS does not include the [O\,I] lines. This is a strong limitation in the case of  NGC\,3642, as we were not able to obtain a satisfactory fit using only  [S\,II] as reference, since these lines are rather weak (Fig.\,\ref{Panel_NGC3642}).\\
For four LINERs (namely, NGC\,2787, NGC\,3998, NGC\,4143 and NGC\,4278), the line modelling of  [O\,I] is complicated by the fact that these forbidden lines are observed at the edge of the \textit{HST}/STIS spectra (Figures \ref{Panel_NGC2787}, \ref{Panel_NGC3998}, \ref{Panel_NGC4143} and \ref{Panel_NGC4278}). This hinders a robust estimate of the line shifts and widths of any second component (generally identified as a wing of the line profiles).\\
\noindent Overall,  in four (seven) cases is adequate one (two) Gaussian(s) per reference forbidden line and narrow H$\alpha$. In all 11 cases, a broad H$\alpha$ BLR-originated component is required (Table\,\ref{T_kin}). In all 11 cases, the modelling of forbidden lines is accurate at $<$\,3\,$\varepsilon^{\rm line}$ level. However, this is not the case for the H$\alpha$-[N\,II] complex since in two cases (2/11, NGC\,1052 and  NGC\,N3642) the level of confidence is $\varepsilon^{\rm line}$\,$\sim$\,5-6.

\subsection{Absorption line fitting in ground-based data}
\label{Analysis_Abs_LF}

\begin{figure*}
\centering
\includegraphics[trim = 1.5cm 20.85cm 8.20cm 2.5cm, clip=true, width=1.\textwidth]{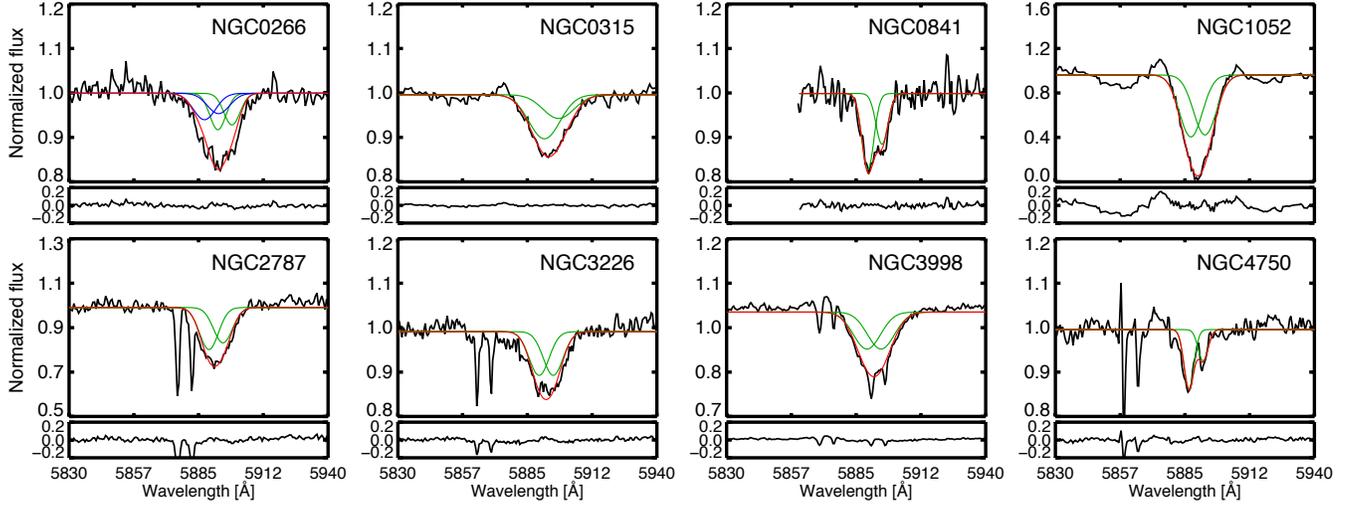}\\
\caption{Normalized spectra of the NaD absorption-line profile after the subtraction of the stellar contribution, for those LINERs having robust detection of the doublet (see Sect.\,\ref{Analysis_Abs_LF}). Each displayed spectrum covers a rest-frame range of $\sim$\,110\,\AA \, (i.e. $\sim$\,5500 km\,s$^{-1}$). The different kinematic components are shown in green and blue, i.e. narrow and second component, according to our classification (see Sect.\,\ref{Analysis_LF}). The red curve shows the total contribution coming from the NaD Gaussian fit. In the lower panels, the residuals (i.e. data - model) are presented.}
 \label{Panel_NaD} 		 
\end{figure*}

In 11 out of 22 of the LINERs in the sample, the wavelength coverage of our ground spectroscopic data allows to probe the NaD$\lambda$$\lambda$5890,5896 absorption lines (see figures in Appendix\,\ref{App_comments_panels}). Such a spectral feature originates both in the cold-neutral ISM of galaxies and in the atmospheres of old stars (e.g. K-type giants, \citealt{Jacoby1984}). The residual spectrum (after the stellar subtraction, Sect.\,\ref{Analysis_St_Sub}) grants the study of the cold neutral gas in LINERs based on a purely-ISM NaD  feature.  Hence, the study of the NaD ISM-absorption allows us to infer  whether the  cold neutral gas is either  participating to the ordinary disc-rotation or entraining in a multiphase outflow. These two possible scenarios have different impacts on the host-galaxy evolution (e.g. \citealt{Cazzoli2016}).\\
In the case of NGC\,3718 (Fig.\,\ref{Panel_NGC3718}), the spectral region from 5860 to 5910 \AA \, is dominated by telluric absorption, preventing any putative detection of the NaD doublet. In two cases (NGC\,3642 and NGC\,3884, Figures \ref{Panel_NGC3642} and  \ref{Panel_NGC3884})  the ISM-NaD seems to be present as a weak resonant NaD  emission that may indicate the presence of dusty outflows (see \citealt{Rupke2015}). However, in these two cases, data at higher S/N are needed to confirm such an unfrequent detection. We then excluded  for the line modelling and the following analysis these three cases as the NaD detection is highly uncertain. \\
In  the remaining eight cases (namely NGC\,0266, NGC\,0315, NGC\,0841, NGC\,1052, NGC\,2787, NGC\,3226, NGC\,3998 and NGC\,4750) the subtraction of a purely-stellar NaD feature leaves a strong residual, suggesting that its origin is mainly interstellar. In  three cases (NGC\,2787, NGC\,3226 and NGC\,3998), the purely ISM-originated NaD is either weak and/or  affected by sky/telluric lines especially evident blueward of the doublet (see Figures \ref{Panel_NGC2787}, \ref{Panel_NGC3226} and \ref{Panel_NGC3998}). \\
As for emission lines (Sect.\,\ref{Analysis_LF}), the ISM-NaD absorption was modelled with Gaussians. We first considered two Gaussian profiles i.e. a single kinematic component (as in \citealt{Davis2012}, \citealt{Cazzoli2014} and  \citealt{Cazzoli2016}). Specifically,  the central wavelength of the NaD$\lambda$5890 is a free parameter, while the widths are constrained to be equal for the two lines. In addition, the ratio between the equivalent widths (EW) of the two lines, R$_{\rm NaD}$\,=\,EW$_{5890}$/EW$_{5896}$, is restricted to vary from 1 (i.e. optically thick limit) to 2 (i.e. optically thin absorbing gas) according to \citet{Spitzer1978}.  This fitting method allows to infer global neutral gas kinematics as with our data we are not able to map and constrain individual gas clouds motions (i.e. many subcomponents) eventually present along the line of sight. For a more detailed discussion of the limitations of the method we refer to \citet{Rupke2005a} and \citet{Cazzoli2016}.\\
In the procedure for modelling the NaD absorption, we did not take into account the He\,I$\lambda$5876 line (as done in some previous works e.g. \citealt{Cazzoli2016}) since it is generally  not observed in these objects. The only two exceptions are the rather weak He\,I detections in NGC\,0315 and NGC\,1052 (Figures \ref{Panel_NGC0315} and \ref{Panel_NGC1052}).  \\
In general, the modelling of the NaD line profile is rather complicated and sometimes leads to unphysical or non-unique solutions. Therefore, in order to preserve against spurious results, as an initial guess, the Gaussian components were constrained to have the same velocity shift and line width of H$\alpha$, testing both narrow and second (if present) components (similarly to \citealt{Cazzoli2014}). From this initial guess, we tried to obtain the  best-fitting to the NaD absorption preventing overfitting.\\
For all cases, we tested a two-component modelling by adding a second broader kinematic component to the one Gaussian-pair fit (similarly to the procedure described in  Sect.\,\ref{Analysis_LF}), but  the majority of the spectra have not strongly asymmetric profiles and their modelling requires only one  component. Only in the case of NGC\,0266,  we find that a two-Gaussian components model per doublet led to a remarkably good fit of the NaD absorption, also reducing the residuals (especially in the wings of the absorption profile) with respect to one-Gaussian fits (from  $\sim$2\,$\varepsilon^{\rm line}$ to 1.3\,$\varepsilon^{\rm line}$ confidence level).  In this case, the broadest component is called \lq second component\rq \ (similarly to emission lines).\\
The line modelling of the NaD profiles in ground-based spectroscopic data are shown in Fig.\,\ref{Panel_NaD}. The  line properties (shift and width) and line ratios (R$_{\rm NaD}$) for each component are reported in  Table\,\ref{T_NaD}.

\section{Main observational results}
\label{Obs_res}

\begin{figure*}
\centering
\includegraphics[trim = 0.75cm .85cm 2.275cm 11.0cm, clip=true, width=1.\textwidth]{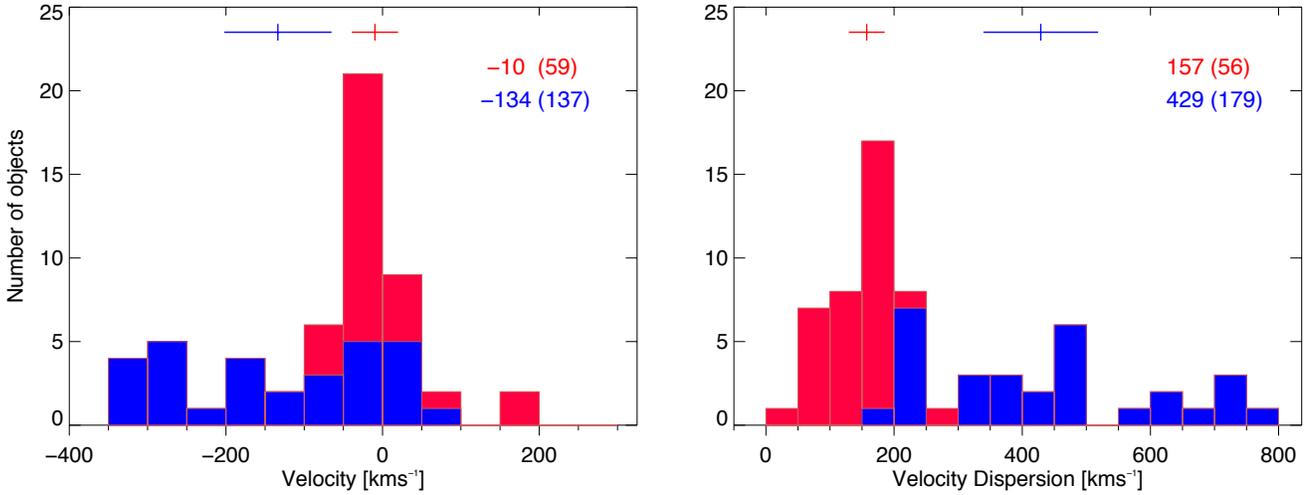}\\
\caption{Distribution of velocity (left) and velocity dispersion values (right). In both panels, narrow and second components are indicated in red and blue, respectively. The crosses represent the average values reported on top right, in parenthesis the standard deviation. The horizontal size of the crosses is equal to the standard deviation of the measurements.}  \label{Panel_kin_models} 		 
\end{figure*}

\begin{figure*}
\centering
\includegraphics[trim = 0.75cm .85cm 2.275cm 11.0cm, clip=true, width=1.\textwidth]{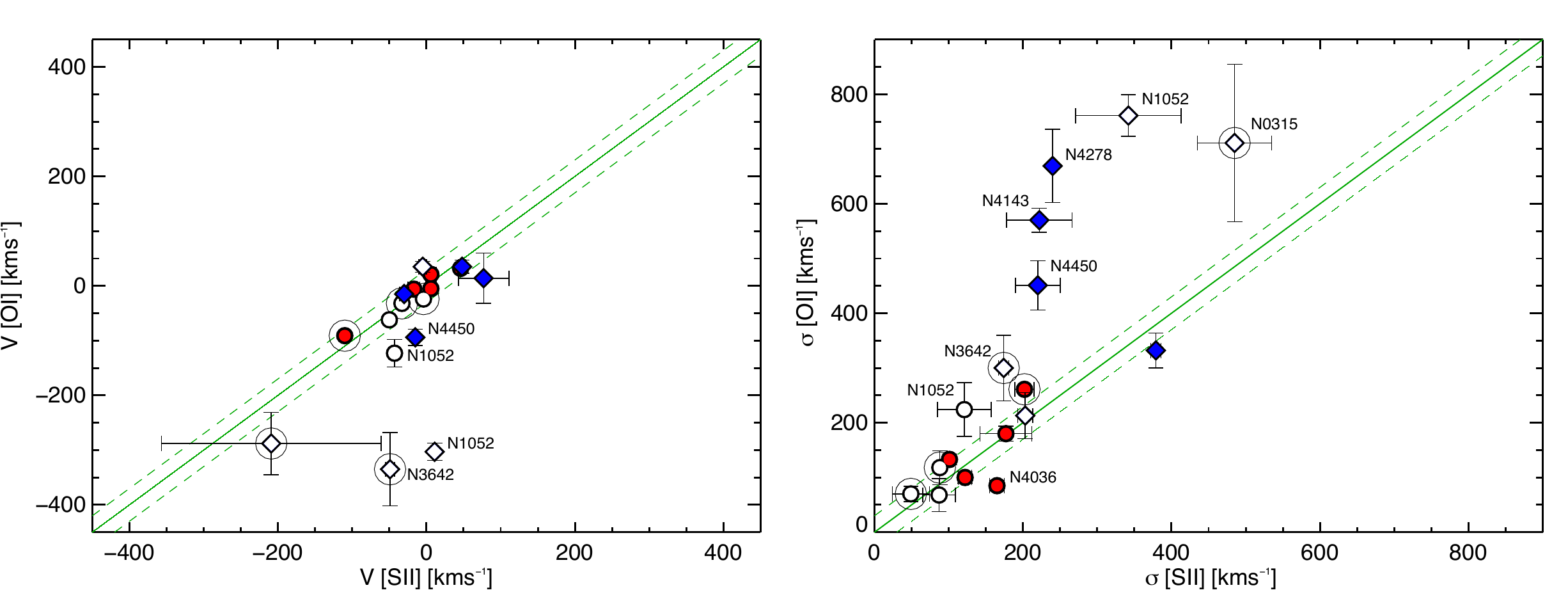}\\
\caption{Comparison between the velocity (left) and velocity dispersion (right) values for the different components in \textit{M}-models (both \textit{M1} and \textit{M2}). Red circles and blue diamonds indicate narrow and second components. Empty symbols indicate that the corresponding H$\alpha$ measurement is less reliable (Table\ref{T_rms} and Sect.\,\ref{Analysis_LF_ground}). An additional circle marks those cases  for which a broad component is needed to improve the fit. The green continuos line marks the 1:1 relation. The green dashed lines represents the positions  for which the velocity (velocity dispersion) of [S\,II] is equal to $\pm$\,30\,km\,s$^{-1}$  that of the [O\,I].  The total range of 60\,km\,s$^{-1}$ roughly corresponds to the instrumental dispersion of our ground-based TWIN/CAHA data (Sect.\,\ref{Sample_data}). The IDs of outliers are also indicated (see text for details).} 
 \label{Panel_kin_mix} 		 
\end{figure*}

For ground-based data, the \textit{S}-model and \textit{O}-model  reproduce well  the line profiles in six of the cases each, while a larger fraction of cases (i.e. 9/21) require \textit{M}-models (4 and 5 cases for \textit{M1} and \textit{M2}, respectively) for a satisfactory fit (Table\,\ref{T_kin}). \\
Of the four  possible combinations of components listed in Sect.\,\ref{Analysis_LF} (see also rows in Table\,\ref{T_summary}), one Gaussian per forbidden line  (with width lower than 250\,km\,s$^{-1}$) is adequate in  6 out of 21 cases.  More specifically,  we found that only in two cases (NGC\,0841 and NGC\,4772, Figures \ref{Panel_NGC0841} and  \ref{Panel_NGC4772}), one Gaussian reproduce well all emission lines including H$\alpha$ (the broad component is not required). A broad Gaussian for H$\alpha$ is required in the remaining four cases (see Table\,\ref{T_summary}).\\  
In the large majority of the cases (15/21), two Gaussians per forbidden line are required for a satisfactory modelling (Tables \ref{T_kin} and Table\,\ref{T_summary}). Among this 15 cases, only in three cases (NGC\,0315, NGC\,3642 and NGC\,4750, Figures \ref{Panel_NGC0315}, \ref{Panel_NGC3642} and  \ref{Panel_NGC4750}) an additional broad H$\alpha$ component is required to  reproduce well the observed line profiles (i.e. for a total of three kinematic components). For the remaining 12 cases the broad component is not required (Table\,\ref{T_summary}). For this subsample of 15 LINERs, there are only six cases  (Table\,\ref{T_summary}) for which the presence/absence of the BLR-component is less reliable in terms of $\varepsilon^{\rm line}$-uncertainty (Sect.\,\ref{Analysis_LF_ground} and Table\,\ref{T_rms}).\\
Overall we required the broad H$\alpha$ component in 7 out of 21 cases.\\
Table\,\ref{T_sum_kin} summarises the mean(median) value(s) for the velocity (V), velocity dispersion ($\sigma$) and FWHM of the different components for each data set, as well as the standard deviation for the distribution of our measurements\footnote{The inclusion of less reliable measurements (Table\,\ref{T_rms}) does not strongly affect the values listed in Table\,\ref{T_sum_kin}. Specifically, considering average values,  the variation is less than  5\,$\%$ ($\leq$\,30\,km\,s$^{-1}$) and 7.5\,$\%$ ($\leq$\,170\,km\,s$^{-1}$) for narrow/second and broad components, respectively.}. \\
In Fig.\,\ref{Panel_kin_models}, we show the comparison between the  velocity and velocity dispersion of the narrow and second  components used to fit emission lines for each of the proposed models. Velocities of the narrow components  are close to rest frame (on average V\,=\,-10\,km\,s$^{-1}$, Table\,\ref{T_sum_kin}), varying between $\pm$\,110 km\,s$^{-1}$ (Table\,\ref{T_kin}). There are two exceptions: NGC\,1052 (whose fitting is less reliable, Table\,\ref{T_rms}, Fig.\,\ref{Panel_NGC1052}) and NGC\,4636 (Fig.\,\ref{Panel_NGC4636}) having [O\,I] velocities of   $\sim$\,-125 km\,s$^{-1}$ and $\sim$\,165\,km\,s$^{-1}$, respectively. The velocity dispersion value for the narrow components is $\sigma$\,=\,157\,km\,s$^{-1}$, on average (Table\,\ref{T_sum_kin}). For the second components the velocity range is larger varying from -350 km\,s$^{-1}$ to 100 km\,s$^{-1}$. The velocity dispersion varies  between 150 and 800 km\,s$^{-1}$ being generally broader (on average $\sigma$\,=\,429\,km\,s$^{-1}$, Table\,\ref{T_sum_kin}) than those of \noindent narrow components. \\ 
Fig.\,\ref{Panel_kin_mix} shows the comparison between the  velocity and velocity dispersion of the narrow and second components used to  fit  [O\,I] and [S\,II]  emission lines for the nine LINERs for which we adopt \textit{M}-models to fit emission lines (Table\,\ref{T_summary}).\\
\begin{table}\caption{Table that summarises all the results from  NaD absorption.}
\begin{tabular}{l c c c }
\hline
 ID              & V$_{\rm NaD}$ & $\sigma$$_{\rm NaD}$ & R$_{\rm NaD}$ \\
                  & km\,s$^{-1}$        & km\,s$^{-1}$   \\
    \hline
NGC\,0266 &  165\,$\pm$\,33 & 171\,$\pm$\,34 & 1.2	\\
                   & -125\,$\pm$\,25 & 238\,$\pm$\,48 & 1.3	\\	
NGC\,0315 &  117\,$\pm$\,48 & 335\,$\pm$\,67 & 1.9	\\
NGC\,0841 &      2\,$\pm$\,2   & 126\,$\pm$\,23 & 1.5	\\
NGC\,1052 & -122\,$\pm$\,24 & 245\,$\pm$\,49 & 1.0	\\
NGC\,2787 &   -28\,$\pm$\,6   & 202\,$\pm$\,40 & 1.2	\\
NGC\,3226 &      2\,$\pm$\,2   & 212\,$\pm$\,42 & 1.0	\\
NGC\,3998 &   -18\,$\pm$\,4   & 292\,$\pm$\,58 & 1.0	\\
NGC\,4750 & -163\,$\pm$\,33 & 104\,$\pm$\,21 & 1.9	\\
\hline
\end{tabular}
\label{T_NaD}

\textit{Notes.}  \lq ID\rq: object designation as in Table\,\ref{T_sample}. Velocity (\lq V$_{\rm NaD}$\rq) and velocity dispersion (\lq $\sigma$$_{\rm NaD}$\rq ) of the neutral gas. As the fit was rather complicated leading to spurious solutions, we conservatively quoted the 20$\%$ uncertainty. The NaD doublet in NGC\,0266, has been modelled with two kinematic components whose values are reported in  a different line. \lq R$_{\rm NaD}$\rq \  indicates the ratio between the EWs of the two lines of NaD (Sect.\,\ref{Analysis_Abs_LF}).
\end{table}   
If we consider the individual narrow  components (red circles in Fig.\,\ref{Panel_kin_mix}) in  \textit{M}-models (both \textit{M1} and \textit{M2}), we found a general agreement within a tolerance of  $\pm$\,30\,km\,s$^{-1}$  (globally of the order of the instrumental dispersion of CAHA/TWIN data, i.e. $\sim$\,60\,km\,s$^{-1}$, Sect.\,\ref{Sample_data}) between the velocity and velocity dispersion of  narrow-[O\,I] and narrow-[S\,II].  The only exception in velocity dispersion is NGC\,4036 (Fig.\,\ref{Panel_kin_mix}; Table\,\ref{T_kin}). Two  exceptions  (NGC\,3642 and  NGC\,4450) are found for the velocities of the secondary component (diamonds in Fig.\,\ref{Panel_kin_mix}). However, of these two cases only NGC\,4450 should be considered a true outlier: as  the fit of the H$\alpha$-[N\,II] complex is less reliable (Table\,\ref{T_rms}) for  NGC\,3642. We found   less agreement when comparing velocity dispersion values for  the second component, with [O\,I] lines profiles being typically much broader than [S\,II]. The difference is up to a factor of $\sim$\,3 in the case of NGC\,4278 (Fig.\,\ref{Panel_kin_mix} right; Table\,\ref{T_kin}). \\
In these comparisons, taking into account the (larger) instrumental dispersion of NOT/ALFOSC data,  NGC\,1052 should not considered as outlier neither in velocity or in velocity dispersion.  \\
A broad  H$\alpha$ component (FWHM\,$>$\,1200 km\,s$^{-1}$) is required only in 7 out of 21 of the LINERs (i.e. 33$\%$, Table\,\ref{T_kin}). The BLR component is found mainly when we adopt \textit{S}- and \textit{M}-models (three cases each, Table\,\ref{T_sum_kin}), with the latter modelling giving less reliable measurements in two cases  (NGC\,0315 and NGC\,3642, Table\,\ref{T_rms}).\\
The measured FWHM of the broad component ranges from 1277\,km\,s$^{-1}$  (NGC\,2787) to 3158\,km\,s$^{-1}$  (NGC\,3642); the average value is 2401\,km\,s$^{-1}$ (Table\,\ref{T_sum_kin}). The velocity varies from  -228 (NGC\,3642)  to 374 km\,s$^{-1}$ (NGC\,0315)  though both from less reliable measurements; otherwise, excluding  those less reliable values, the velocity range would be $\sim$\,-50  to $\sim$\,330 km\,s$^{-1}$ (Table\,\ref{T_kin}). \\
For none of the 11 \textit{HST}/STIS spectra, the adopted best fit is obtained using the  \textit{O}-model. Excluding those four cases for which we were not able to test all the proposed models (Sect.\,\ref{Analysis_LF_space}, Table\,\ref{T_kin}), we found a slightly large prevalence of  best fits obtained with \textit{M}-models, i.e. 5/7 (Table\,\ref{T_summary}).  For all these five cases we adopted the  \textit{M1}-model. \\
\noindent In four cases (NGC\,2787, NGC\,3642, NGC\,4036 and NGC\,3998), one Gaussian per forbidden line and narrow H$\alpha$ (typically with $\sigma$ between  120 and 270 km\,s$^{-1}$, Table\,\ref{T_kin}) is adequate. In the remaining cases, two Gaussians   are required for a good fit of  forbidden lines, due to the presence of  broad wings in the  line profiles. \\
The broad component in \textit{HST}/STIS spectra  is ubiquitous (Table\,\ref{T_summary}). \\

\noindent Except for a few cases (see columns 6-9 in Table\,\ref{T_kin}), narrow components have velocities between -100 and 200  km\,s$^{-1}$. The velocity dispersion is 176\,km\,s$^{-1}$, on average (Table\,\ref{T_sum_kin}). Similarly,  the velocities of second component range from -200 to 150  km\,s$^{-1}$. These second components are however broader, with   velocity dispersion values  between 300 and 750 km\,s$^{-1}$ (Table\,\ref{T_kin}); the average value is 433 km\,s$^{-1}$ (Table\,\ref{T_sum_kin}). If we consider the individual narrow components in  \textit{M}-models, we found a trend similar to that obtained considering the same model  but with ground-based spectroscopy. Specifically, for the velocity and velocity dispersion of narrow components we found a general agreement with the same range of tolerance ($\pm$\,30\,km\,s$^{-1}$, globally of the order of the instrumental dispersion of \textit{HST}/STIS data, Sect.\,\ref{aux_data}) considered in  the spectral analysis of ground-based data (except NGC\,3398, Table\,\ref{T_kin}). Less agreement is found for the velocity and velocity dispersions of the second components. \\
The  measured FWHM of the broad H$\alpha$ components in \textit{HST} spectra  range from 2152\,km\,s$^{-1}$  (NGC\,5005, Fig.\,\ref{Panel_NGC5005}) to 7359\,km\,s$^{-1}$  (NGC\,4450, Fig.\,\ref{Panel_NGC4450}), with an average value of 3270\,km\,s$^{-1}$ (Table\,\ref{T_sum_kin}).\\
For the NaD absorption, in 7 out of 8 targets, a single kinematic component (a Gaussian pair) already gives a good fit, suggesting that if a second component exists in these galaxies, it is weak. The only exception is NGC\,0266, which  requires two Gaussian pairs (Fig.\,\ref{Panel_NaD}). Excluding the second component in NGC\,0266, velocities of the neutral gas components vary between  -165 and 165 km\,s$^{-1}$ (Table\,\ref{T_NaD}). This velocity range is similar but slightly larger than what was found for the narrow components in emission lines in both ground- and space-based spectroscopy (Table\,\ref{T_sum_kin}). Velocity dispersions values are in the range 104-335 km\,s$^{-1}$ (Table\,\ref{T_NaD}) and the average value is  220 km\,s$^{-1}$. These values are hence larger than those found for the narrow components of emission lines, but smaller than those found for second components in both sets of data. These  results will be discussed in Sections \ref{classification} and \ref{Disc_neutral}.\\
Considering all the  components listed in Table\,\ref{T_NaD},  the average values of R$_{\rm NaD}$ is $\sim$\,1.3 suggesting that the neutral gas is generally optically thick in these objects. There are only three LINERs (NGC\,0315, NGC\,0841 and NGC\,4750; Table\,\ref{T_NaD}) for which  R$_{\rm NaD}$\,$\geq$\,1.5 suggesting that the absorbing gas is optically thin. A more detailed analysis of the optical depth and column density of the neutral gas from   more complex line-profile modelling \citep{Rupke2005a} is beyond the scope of this paper.

\subsection{Ground versus space measurements}
\label{caha_hst_us}
For NGC\,2787  (Fig.\,\ref{Panel_NGC2787}) there is full agreement between the selected models\footnote{
In what follows, we consider  \textit{M1} and \textit{M2} models together within the mixed class  \textit{M}.} and components used to fit emission lines ground- and space-based data (Table\,\ref{T_kin}). However, this is not always the case.  Specifically, in seven cases (Table\,\ref{T_summary}) we select the same models in both  ground- and space-based spectra but the number of components used to model line profiles is different. In the unique case of NGC\,0315 (Fig.\,\ref{Panel_NGC0315}), we used the same combination of components in both ground- and space-based data sets but we employed different models. In three cases (NGC\,3998,  NGC\,3642 and NGC\,4143)  both  line models and components for narrow lines do not correspond (Table\,\ref{T_kin}).   \\
All the narrow components found for both ground- and space-based spectra have velocity dispersions generally lower than 300\,km\,s$^{-1}$. Nevertheless the distribution of the velocities is quite different, being that inferred from the \textit{HST}/STIS spectral modelling  skewed to positive velocities (Table\,\ref{T_kin}). A second component is needed to better model the \textit{HST}/STIS forbidden line profiles and narrow H$\alpha$ in 7 out of 11 cases. The velocity and velocity dispersion of this component is  quite different from case to case; we refer to Appendix\,\ref{App_comments_panels} for details. The properties of both narrow and second components for both data sets will be discussed in Sections \ref{classification}, and Sect.\,\ref{Disc_outflows}.\\
The broad H$\alpha$  component is ubiquitous when modelling the H$\alpha$ line profiles in  \textit{HST}/STIS spectra, at the contrary of what is found for  ground-based spectroscopic data (in 7 out of 11 a broad H$\alpha$ is seen only in \textit{HST}/STIS spectra, Table.\,\ref{T_kin}). When a broad component is seen in both \textit{HST} and CAHA data (four cases, Table.\,\ref{T_summary}) the FWHM measurements are consistent within the errors in the two cases of NGC\,0315 and NGC\,5077. 

\begin{table*}  
\caption[]{Average (median) and standard deviation measurements for velocity, velocity dispersion and FWHM in both ground- and space-based data.}
\begin{tabular}{l c c c c c c c c }
\hline 		
                 &           &\multicolumn{2}{|c|}{Ground}	& & \multicolumn{2}{|c|}{Space} \\			
 \hline 
         Component        &           &   Average\,(Median) & Standard Deviation &&  Average\,(Median) & Standard Deviation\\
          &              & km\,s$^{-1}$ & km\,s$^{-1}$ & & km\,s$^{-1}$ & km\,s$^{-1}$ \\
\hline 
N &  V        &     -10\,(-17) &   59  & &  84\,(105)  &  102 \\
                 &  $\sigma$ &     157\,(165) &   56  & &  176\,(172) & 53  \\
S&  V        &    -134\,(-108) &  137  & &  -28\,(-19)  &  99  \\
                 &  $\sigma$ &     429\,(446) &  179  & &  433\,(397)  & 130  \\
B  &  FWHM     &	  2401\,(2472) &   591 & & 3270\,(2689)  &  1509\\
\hline 		   		  				   
 \label{T_sum_kin}
\end{tabular}
\begin{flushleft}
\textit{Notes.} \lq N\rq \, ,\lq S\rq \, ,\lq B\rq \,  stand for narrow, second and broad components as in Table\,\ref{T_kin}. \lq V\rq \ and \lq $\sigma$\rq \ stand  for velocity and velocity dispersion.
\end{flushleft}
\end{table*}


\subsection{Comparison with previous   broad H$\alpha$ detections}
\label{Comparison_previous_BLR}

In Table\,\ref{T_FWHM} we summarise the measurement of the FWHM of the  broad H$\alpha$ components obtained in the present work (hereafter \textit{C18}) for ground- and space-based data and previous works. The  comparison between our measurements of the FWHM of the  broad H$\alpha$ component and those from previous works is discussed in what follows for both space (Sect.\,\ref{Disc_HST}) and ground- and space-based data (Sect.\,\ref{Disc_Palomar}).  The  possible factors generating agreement or discrepancy are discussed in Sect.\,\ref{Disc_BLR}.

\subsubsection{Space-based \textit{HST}/STIS measurements}
\label{Disc_HST}

\begin{figure}
\centering
\includegraphics[trim = 0.25cm .85cm 2.50cm 11.0cm, clip=true, width=.95\textwidth]{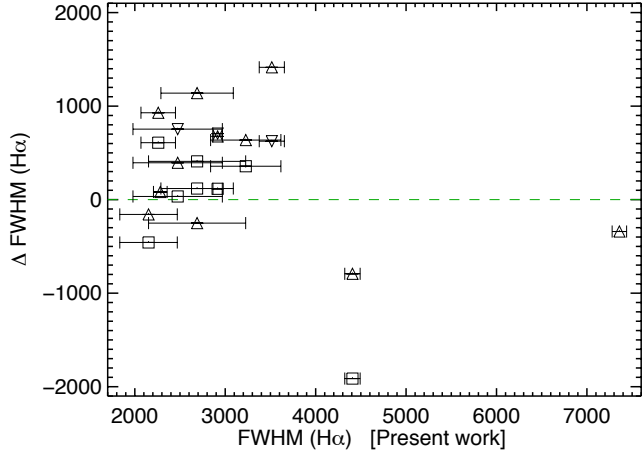}\\
\caption{The difference between the FWHM values of the H$\alpha$ broad component measured in space data in the present manuscript and those of previous works, i.e. $\Delta$ FWHM (H$\alpha$), against our measurements (Table\,\ref{T_FWHM}). Boxes and triangles mark the comparison with the FWHM listed in \textit{C15} and \textit{BC14}. More specifically, upward and downward triangles indicate the FWHM of the broad H$\alpha$  required by the fit using as template for the narrow component the [S\,II] and [O\,I] lines, respectively. An additional circle marks those galaxies with a less reliable H$\alpha$ fit (Table\,\ref{T_rms}). The green dashed line indicates the zero, as reference. Errors in y-axis are not available.}  \label{Panel_comp_fwhm} 		 
\end{figure}

The broad  component is required to model the H$\alpha$ line profile for all the 11 LINERs observed with \textit{HST}/STIS spectroscopy (Sect.\ref{Obs_res}).\\ 
In Fig.\,\ref{Panel_comp_fwhm}, our measurements of the  FWHM of the H$\alpha$ in space-based data are compared to those listed in previous works by   
\textit{BC14}  and \citet{Constantin2015} (hereafter \textit{C15}). The former comparison is particularly instructive as the results are obtained with different analysis of the same data set (Sect.\,\ref{aux_data}). The effects of the procedure for line modelling will be discussed in detail in Sect.\,\ref{Disc_modelling_eff}.\\ 
Of the 11 \textit{HST}/STIS spectra  analysed in the present work, only in two cases (NGC\,3398 and NGC\,4450) the BLR-component was confirmed by \textit{BC14}.   For NGC\,4450, the agreement of the FWHM measurements is rather good (within $\pm$\,200\,km\,s$^{-1}$). For NGC\,3998 our measurement is 15\,$\%$ smaller ($\sim$\,800\,km\,s$^{-1}$, Table\,\ref{T_FWHM}). However,  this is a particular case: despite the modelling reproduce well the line profiles (Table\,\ref{T_rms}), the [O\,I]$\lambda$6300 line is strong but partially truncated (Fig.\,\ref{Panel_NGC3998}). This compromises the detection of any putative second component in [O\,I], and hence in H$\alpha$ (Sect.\,\ref{Analysis_LF} ). \\
In the remaining nine sources, \textit{BC14} did not find  convincing evidence of a broad component. Specifically, although a broad component must generally be included to achieve a good fit, rather different FWHM values are inferred adopting either [O\,I] or [S\,II] lines as reference.\\
The \textit{HST}/STIS spectra of nine LINERs of the present sample were also analysed by \textit{C15}.  These authors presented the same dataset as that analysed in this paper but with a different treatment of data in terms of data reduction, stellar subtraction (Sec.\,\ref{Analysis_St_Sub}) and line modelling. In particular, \textit{C15} used only the [S\,II]  profiles as a template for  H$\alpha$-[N\,II]. However,  they allowed multiple components when modelling [S\,II] to take into account any possible asymmetries of the line profiles and broad wings.  In their work, they confirmed the broad H$\alpha$ detection (previously reported by \citealt{Ho1997b}) for all the nine LINERs in common with our sample (Table\,\ref{T_FWHM}). Of these, NGC\,4636 has not been analysed in the present work  (Tables \ref{T_kin} and \ref{T_FWHM}) as the data quality  is insufficient to proceed to any reliable analysis (see Fig.\,7 in \textit{BC14}).\\
We found a general agreement (within $\sim$\,400\,km\,s$^{-1}$) with the measurements by \textit{C15},  excluding those cases for which our H$\alpha$-[N\,II] modelling was less reliable (Table\,\ref{T_rms}). The match  is remarkably good in the cases of NGC\,4036 and NGC\,5077 (the difference in FWHM is less than 110\,km\,s$^{-1}$, Table\,\ref{T_FWHM} and Fig.\,\ref{T_FWHM}). A significant discrepancy of about 1900\,km\,s$^{-1}$ is found only for NGC\,3998. \\ 
In this comparison, it has to be noted that the broad component detected by \textit{C15} is located at large shift with respect to rest-frame  (up to $\sim$\,1200\,km\,s$^{-1}$ in the case of NGC\,0315) relative to the narrow H$\alpha$.  In contrast  the shift is generally lower, varying between -300 and 550 km\,s$^{-1}$ (Table\,\ref{T_kin}), in our measurements (considering both ground- and space-based data). \\

\subsubsection{Ground-based measurements}
\label{Disc_Palomar}
With our CAHA/NOT spectroscopy we confirm the presence of the broad H$\alpha$ component in   7 out of 21 (i.e. 33\,$\%$) LINERs-1 of the selected sample, by using mainly \textit{S}- and \textit{M}-models and allowing  multiple components to reproduce forbidden lines and narrow H$\alpha$ profiles (Sect.\,\ref{Obs_res}). Such a low detection rate of the broad H$\alpha$ component disagrees with that  from the analysis of spectra from the Palomar Survey (i.e. 100\,$\%$, \citealt{Ho1997b}, \textit{HFS97} hereafter). \\
In the common cases of broad-component detection, the measurement of the FWHM for such broad components are somewhat different (Table\,\ref{T_FWHM}). Specifically, in two cases (namely NGC\,2787 and NGC\,4636) the FWHM in Palomar observations is larger by a factor of $\sim$\,1.6 and $\sim$\,1.1, respectively (Table\,\ref{T_FWHM}). In four cases (NGC\,0315, NGC\,3718, NGC\,4750 and NGC\,5077) our measurements indicate a larger   FWHM of the broad component (by a factor $\sim$\,1.2).  In the remaining  case (NGC\,3642),  our FWHM-measurement is larger by a factor $\sim$\,2.5.

\begin{table}  
\caption[FWHM broad Halpha.]{FWHM of the H$\alpha$  broad component  from different works.}				
\begin{tabular}{l l c c c c }
\hline 					
ID &  Obs.    &  \textit{C18} & \textit{HFS97} & \textit{BC14} & \textit{C15} \\
    &              &    km\,s$^{-1}$ &    km\,s$^{-1}$ &   km\,s$^{-1}$ &    km\,s$^{-1}$\\
 \hline  
NGC\,0266  & CAHA $^{\dagger}$ & -	        & 1350  & -	      & - \\   
NGC\,0315 & NOT & [2474\,$\pm$\,389]         & 2000  & -	      & -  \\	     
          & HST $^{\dagger}$  & 3227\,$\pm$\,495 &  -    & 2590	      & 2870\\	    
NGC\,0841 & CAHA & -	        & 1350  & -	      & - \\   
NGC\,1052 & NOT  & [-]	        & 1950  & -	      & - \\  
          & HST  & [2916\,$\pm$\,52]         &  -	& 2240 (2210) & 2800 \\
NGC\,2681 & CAHA &	-        & 1550  & -	      & - \\   
NGC\,2787 & CAHA  & 1277\,$\pm$\,195         & 2050  & -	      & -\\
          & HST $^{\ddagger}$ & 2282\,$\pm$\,75         &  -	& 2200        & -\\ 
NGC\,3226 & CAHA & -	        & 2000  & -	      & - \\   
NGC\,3642 & CAHA  & [3158\,$\pm$\,565]         & 1250  & -	      & -\\ 
          & HST $^{\dagger}$  & [2259\,$\pm$\,191] &  -    & 1330        & 1650 \\       
NGC\,3718 & CAHA  & 2582\,$\pm$\,518         & 2350  & -	      & - \\
NGC\,3884 & CAHA $^{\dagger}$ & -	        & 2100  & -	      & - \\   
NGC\,3998 & CAHA  & [-]	        & 2150  & -	      & - \\
          & HST $^{\ddagger}$  & 4407\,$\pm$\,85         &  -	& 5200        & 6320 \\
NGC\,4036 & CAHA  & -	        & 1850  & -	      & - \\
 	  & HST  & 2474\,$\pm$\,595         &  -	& 2080 (1720) & 2440 \\
NGC\,4143  & CAHA $^{\ddagger}$ & -	        & 2100  & -	      & - \\
	  & HST  & 3515\,$\pm$\,139         &  -	& 2100 (2890) & - \\
NGC\,4278 & CAHA  & -	        & 1950  & -	      & -\\			
	  & HST $^{\ddagger}$  & 2689\,$\pm$\,537         &  -	& 2940        & 2280 \\
NGC\,4438 & CAHA & [-]	        & 2050  & -	      & - \\   
NGC\,4450 & CAHA  & -	        &  2300 & -	      & -\\ 
	  & HST  & 7359\,$\pm$\,80         &  -	& 7700        & - \\
NGC\,4636 & CAHA  & 2152\,$\pm$\,429         & 2450  & -	      & - \\
NGC\,4750 & CAHA  & 2367\,$\pm$\,106         & 2200  & -	      & -\\ 
NGC\,4772 & CAHA & -	        & 2400  & -	      & - \\   
NGC\,5005 & CAHA  & [-]	        & 1650  & -	      & -\\
          & HST $^{\dagger}$  & 2152\,$\pm$\,318 &  -    & 2310	      & 2610\\		
NGC\,5077 & CAHA  & 2797\,$\pm$\,560         & 2300  & -	      & -\\
          & HST $^{\dagger}$  & 2689\,$\pm$\,400 &  -    & 1550	      & 2570\\ 
\hline 
 \label{T_FWHM}
\end{tabular}

\textit{Notes.}   Measurements of the FWHM of the broad H$\alpha$ component. These are from: 
\textit{C18} (present work, see also Table\,\ref{T_kin}),  \textit{HFS97}  (ground Palomar data),  \textit{BC14} and  \textit{C15} (both from space \textit{HST}/STIS observations). For the measurements by \textit{BC14}, in parenthesis we indicate the FWHM of the broad H$\alpha$ required by a fit obtained using [O\,I] as template for narrow lines. The other measurements were done using [S\,II] lines as reference.  \lq ID\rq: object designation as in Table\,\ref{T_sample}. \lq Obs\rq \ stands for the origin of the optical data.  $^{\dagger}$ and $^{\ddagger}$ symbols and square brackets are as in Table\,\ref{T_kin}. 
\end{table}

\section{Discussion}
\label{discussion_results}

\subsection{Probing the BLR in type-1 LINERs}
\label{Disc_BLR}

The analysis of Palomar spectra by \citet{Ho1997b} indicated that all the LINERs in our selected sample show a broad H$\alpha$ component resulting in their classification as LINER-1 nuclei. Nevertheless, with our analysis of ground- and space-based data sets, we found discrepant results. On the one hand, with the analysis of ground-based observations, we found  the detection rate for the  broad H$\alpha$ component to be only 33\,$\%$ (Sect.\,\ref{Obs_res}). This result questions their  previous classification of type-1 LINERs. On the other hand, for all the LINERs with \textit{HST}/STIS spectra and in common with \textit{BC14}, the observed line profiles require the broad H$\alpha$ component for a satisfactory modelling.\\
The lack of  detection of the broad H$\alpha$ component in LINERs could be due to the absence of the BLR or its undetectability (e.g. the broad emission is too weak and/or highly contaminated by the starlight). If the BLR is present in LINERs, as it was supposed to be in our sample, its detectability (via the broad H$\alpha$ component) is sensitive to  two main factors: modelling and observational effects.

\subsubsection{Effect of different modellings}
\label{Disc_modelling_eff}

The two main factors related to modelling that might affect the detectability of the BLR component are: the stellar subtraction and the strategy adopted for the line modelling. The former determines the accuracy of achievement of a pure ISM spectra, which is particularly relevant for a proper kinematic decomposition.  This effect is expected to  be rather more relevant as the slit width increases (thus for our ground-based data). On the other hand, the strategy adopted for the emission line fitting is closely related to both the identification of the model that best fits the emission lines (i.e. in terms  of number of Gaussians) and the underlying physics of the assumed models (i.e. the choice to use [S\,II] or [O\,I] or both as template for H$\alpha$-[NII]). \\

\begin{figure}
\centering
\includegraphics[trim = 0.65cm 12.10cm 11.725cm 2.5cm, clip=true, width=0.465\textwidth]{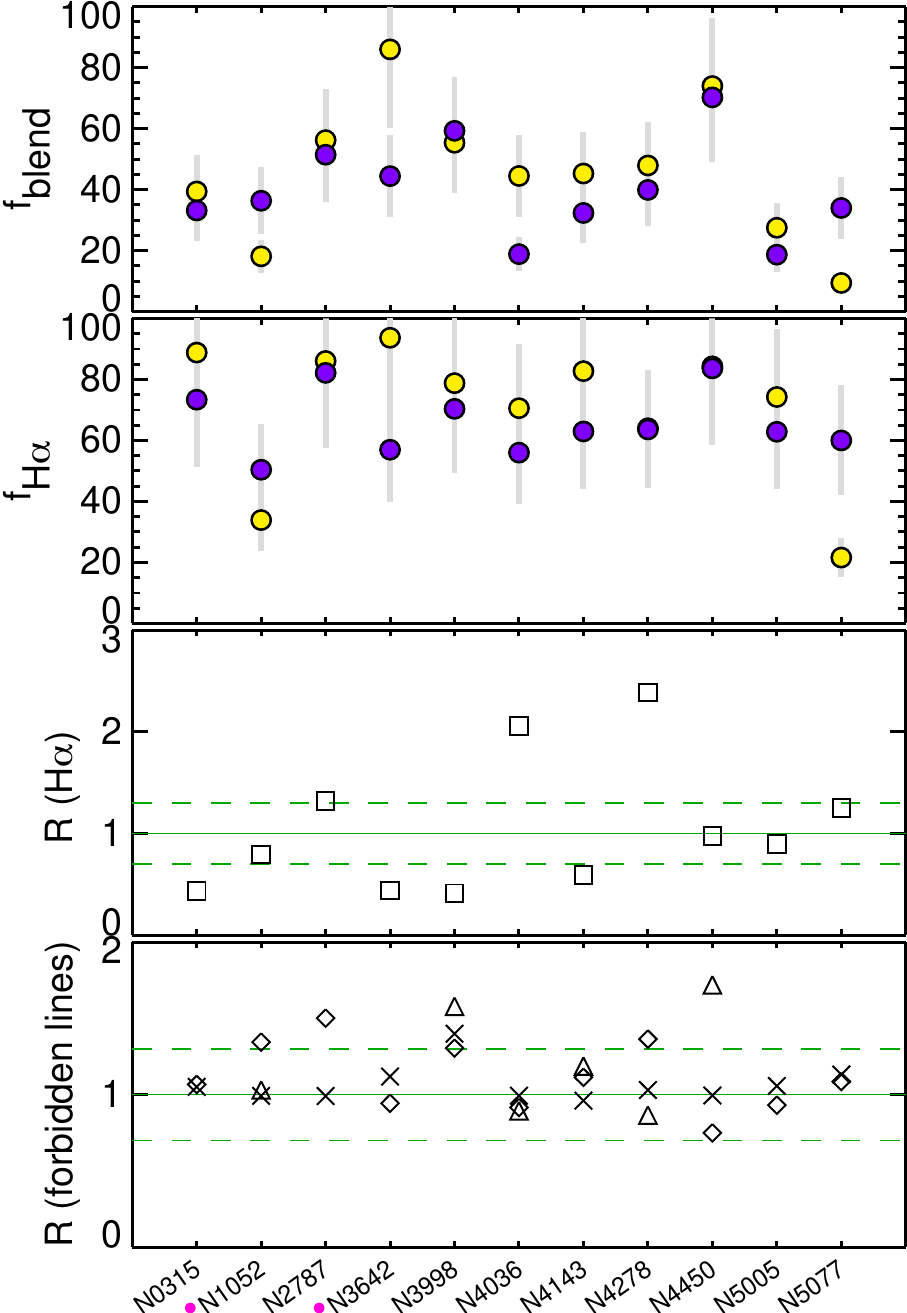}
\caption{Comparison between the results obtained from the proposed line fitting procedure applied to space-based data with the measurements from  \textit{BC14}  and  \citet{Balmaverde2016}.  From bottom to  top: the ratio (R) of the flux of the narrow component (plus the second component, if present)  used to model forbidden lines and narrow H$\alpha$ in the present manuscript and the above mentioned previous works, f$_{\rm H\alpha}$ and  f$_{\rm blend}$ (see text). These latter are expressed  in terms of a percentage. In the two bottom panels, the different species: [O\,I], [N\,II], [S\,II] and H$\alpha$ are  marked with triangles, diamonds, crosses and boxes, respectively. Green continuos and dotted lines indicate unity and  $\pm$\,30\,$\%$. In the  top panels, purple and yellow  mark the values of f$_{\rm H\alpha}$ and f$_{\rm blend}$ in the present work and those in the reference works, respectively.  Grey bars indicate  30\,$\%$  uncertainty.  Pink circles (near IDs on the x-axis) mark those cases with less reliable line-modelling (Table\,\ref{T_rms}).} 
 \label{space_ground} 		 
\end{figure}

\noindent The comparison between our measurements and those by \textit{BC14} provides the optimal framework to test our line fitting procedure (Sect.\,\ref{Analysis_LF}). Specifically, this comparison is not biased by any issue about the starlight subtraction or observational effects (e.g.  possible AGN variability, see also Sect.\,\ref{Disc_obs_eff}) as we stress that the results are obtained with two different analyses of the same data set. Hence, we checked if our procedure for the line modelling  is able to reproduce the fluxes listed in \textit{BC14} and \citet{Balmaverde2016} for H$\alpha$ and forbidden lines, respectively.  Fig.\,\ref{space_ground} presents the results of this test. \\
The comparison is rather positive as our measurements match those of \citet{Balmaverde2016} within  20\,$\%$ considering [S\,II] lines (crosses in Fig.\,\ref{space_ground}). Slightly less agreement ($\sim$\,30\,$\%$) is found for [O\,I] and [N\,II] and the largest scatter is seen for H$\alpha$ (Fig.\,\ref{space_ground}, bottom panels). Overall, the more evident exception is NGC\,3998 whose line modelling is somewhat complicated (Fig.\,\ref{Panel_NGC3998}). \\ 
We found a rather good agreement (within  30\,$\%$) also when comparing the different contribution of the broad component of H$\alpha$ to the H$\alpha$-[N\,II] blend and to the total H$\alpha$ emission i.e. f$_{\rm blend}$ and f$_{\rm H\alpha}$, respectively (Fig.\,\ref{space_ground}, top panels). These contributions are defined as:  f$_{\rm H\alpha}$\,=\, F(H$\alpha$$_{\rm  B}$)/F(H$\alpha$$_{\rm  N(+S)}$) and f$_{\rm blend}$\,=\, F(H$\alpha$$_{\rm  B}$)/F(H$\alpha$$_{\rm N+B(+S)}$+F([N\,II]$_{\rm N(+S)}$), where F is the line flux and \lq B\rq , \lq N\rq \ and \lq S\rq \ stand for broad, narrow and second (if present) components. The discrepancy is large ($\sim$\,40-50\,$\%$) for NGC\,3642   (due to an uncertain modelling, Fig.\,\ref{Panel_NGC3642}) and NGC\,5077 (Fig.\,\ref{Panel_NGC5077}).\\
Although, in the analysis by \textit{BC14}  they tested both the \textit{S}- and \textit{O}-models (Sect.\,\ref{Analysis_LF}) not finding a convincing piece of evidence supporting the presence of the BLR (Sect.\,\ref{Disc_HST}). However, it is important to note that they did not consider multiple Gaussians when fitting the forbidden lines used as template. In  the case of a broad and asymmetric profile of the selected template ([S\,II] and [O\,I])\footnote{The assumption of a single component in [S\,II] lines is critical as these lines are often blended (e.g. NGC\,1052, Fig.\,\ref{Panel_NGC1052}) contrary to [O\,I].}, one Gaussian is certainly an oversimplification. Therefore, the width of each single narrow Gaussian component  used for the selected template tend to be overestimated hampering the modelling of H$\alpha$ components (e.g. resulting in a spurious detection of the broad component). This might be partially at the origin of the mismatch seen in few cases (e.g. NGC5077) in Fig.\,\ref{space_ground}.\\
As mentioned above, the effects of the stellar subtraction are absent when comparing our results with those by \textit{BC14}. However, these effects (along with other factors Sect.\,\ref{Comparison_previous_BLR}) could still play a role in the comparison with the work by \textit{C15}, who removed  the stellar continuum from \textit{HST}/STIS spectra. Despite this, the nice agreement of the FWHM of the H$\alpha$ broad component (Fig.\,\ref{Panel_comp_fwhm} and Table\,\ref{T_FWHM}) seems to indicate that the starlight subtraction only moderately affects the measurements in \textit{HST}/STIS spectroscopic data.\\

\noindent We performed the same kind of comparison for our ground-based data  considering the results from  \textit{HFS97}. We mitigate possible slit-aperture effects by considering aperture-corrected fluxes (i.e. normalized to a 1\arcsec.0\,$\times$\,1\arcsec.0 aperture).  Nevertheless, we found a  discrepancy (of about 40-50\,$\%$, typically) between the measurements of forbidden lines and narrow H$\alpha$ in  different ground-based observations. The different detection rate of the broad H$\alpha$ component  between the present work and that by \textit{HFS97}  (i.e. 33\,$\%$ vs. 100\,$\%$, respectively, Sect.\ref{Obs_res}) prevents the one-to-one comparison of the measurement of  f$_{\rm H\alpha}$ and  f$_{\rm blend}$. \\
Moreover, the interpretation of such a comparison is not as straightforward as for space-based data, since besides the strategy adopted for the line modelling, a number of factors  may contribute to the possible discrepancies including: starlight subtraction, the difference in the slit-PAs and possible AGN variability. We already discussed the possible biases introduced with the modeling of the stellar continuum (Sect.\,\ref{Analysis_St_Sub}). Effects related to the differences in the slit-PAs and the possible AGN variability will be discussed in a dedicated section (Sect.\,\ref{Disc_obs_eff}). \\ %

\noindent Thus, our procedure for the line modelling is able to produce overall flux measurements  the BLR component consistent with previous works (e.g. no over/under-fitting). Nevertheless, the results obtained in the present comparison and those presented in Sect.\,\ref{Comparison_previous_BLR} indicate a significant impact of the line modelling on the detection and properties of the BLR component.

\subsubsection{Observational effects}
\label{Disc_obs_eff}

The broad H$\alpha$ component is expected to be relatively more dominant as the slit width decreases since for narrower slit observations the contamination of starlight and of narrow lines is reduced,  favoring the detection of the BLR-component. On the one hand, the BLR could be present but visible only at \textit{HST}/STIS scales, if dilution effects from the host galaxy have a dominant role.  On the other hand, the BLR could be intrinsically absent. In this case, finding a broad component could be the result of the combined effects of questionable starlight decontamination and of an inappropriate choice of the model for emission lines (Sect.\,\ref{Disc_modelling_eff}).\\
For the present work, aperture effects are likely more significant when comparing ground- and space-based data. Indeed, the \textit{HST}/STIS aperture is significantly smaller ($\leq$0.$\arcsec$2, \textit{BC14}) than the TWIN/ALFOSC aperture in our data (Table\,\ref{T_Obs_sample}).  Aperture effects are expected to be smaller when comparing ground-based data sets.  Specifically,   CAHA/TWIN and NOT/ALFOSC slit apertures are similar but slightly smaller (a factor 2-3 in size, typically, Table\,\ref{T_Obs_sample}) than the 2\arcsec.0$\times$4\arcsec.0 aperture in the Palomar Survey \citep{Filippenko1985}. \\ 
\noindent In Sect.\,\ref{caha_hst_us}, we briefly  compare ground- and space-based narrow and broad emission lines measurements finding differences in terms of the assumed model and the number of components (Table\,\ref{T_kin}). These differences could be due to  a sharp variation of velocities toward the SMBH that  could be related to the presence of non-rotational motions such as outflows. Indeed, the outer parts and cores of outflows (that are mapped in a different way in different slits) might behave otherwise leading to some velocity and velocity dispersions gradients. An alternative explanation is a complex stratification in density and ionization of the NLR. Observational effects could originate the different detection rate and FWHM measurements of the BLR component in our ground- and space-based data.\\

\noindent  Considering the ground-ground comparison, we mitigate aperture effects via flux normalization, as mentioned in Sect.\,\ref{Disc_modelling_eff}. Despite this, discrepancies remain when comparing narrow and  broad lines measurements. \\
In this comparison, the differences in the slit PA\footnote{Differences in the slit PA should be meaningless in the comparison with \textit{HST}/STIS results with the small nuclear aperture considering for obtaining the \textit{HST}/STIS spectra.} could be also relevant, as weak spectral features could be diluted or contaminated by potential nuclear H\,II regions.  Therefore, we checked if any trend exists  between the difference in the slit-PAs and those values of  f$_{\rm H\alpha}$ and  f$_{\rm blend}$, finding no correlations. The presence of either strong non-rotational motions or complex phenomena such outflows  might also complicate the emission lines fitting especially in case of severe H$\alpha$-[N\,II] or [S\,II] blending.\\

\noindent With the present data sets, it is not possible to quantify any effect related to possible AGN-variability which  does not occur at similar time-scale in different objects of the same AGN family (e.g. \citealt{Younes2011, HG2014, HG2016}).

\subsection{Classification of the velocity components}
\label{classification}

\begin{figure*}
\centering
\includegraphics[trim = .5cm 11.5cm 1.5cm 8.5cm, clip=true, width=1.\textwidth]{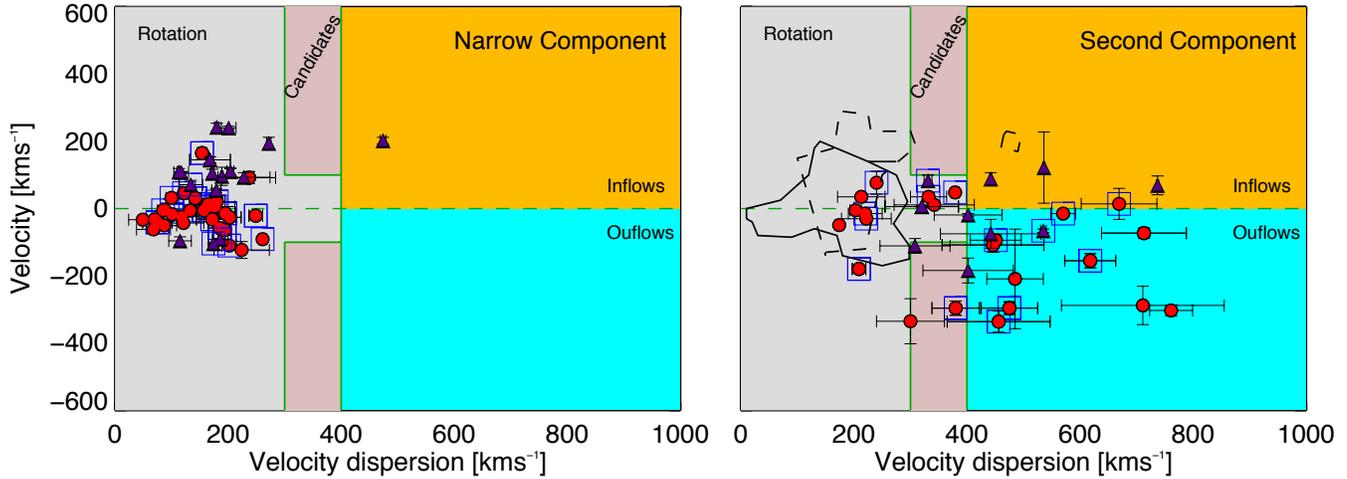}
\caption{Observed velocity dispersion - velocity ($\sigma$-V plane) plane for narrow (left) and second (right) components. Red circles an purple triangles mark the measurements from ground- and space-based data, respectively. An additional blue box marks those LINERs for which the fitting of the H$\alpha$ profile is less reliable (Table\,\ref{T_rms}).  In the right panel, we report  the measurements of the narrow component with contours (continuos and dashed lines are for ground- and space-based data, respectively). The different coloured areas marked with continuous and dashed green lines indicate the location of our measurements that could be explained by rotation (gray), outflows/inflow candidates (pink), outflows and inflows (orange and light blue, respectively).  See Sections \ref{classification} and \ref{Disc_outflows}  for details. Note that the number of objects with less reliable fit in this figure are different from those in Fig.7, where only \textit{HST} data are considered. For NGC\,1052, 4 points are used to represent the different model components (see Table\,3).}
 \label{Panel_sigmaV} 		 
\end{figure*}

To interpret the physics behind the line modelling proposed in Sect.\,\ref{Analysis_LF}, we consider the  measurements of the kinematics (velocity and velocity dispersion) from the nuclear   spectra in both ground- and space-based data sets (Table\,\ref{T_kin}). In Fig.\,\ref{Panel_sigmaV}, we present the distribution of the velocity for narrow and second components as a function of their velocity dispersion ($\sigma$-V plane). \\
We consider the kinematics of the narrow and second components from the lines used as template, i.e. [S\,II] and [O\,I] (Table\,\ref{T_kin}). We recall that for all the models proposed in Sect.\,\ref{Analysis_LF}, H$\alpha$ is tied to [O\,I], thus the components used to fit these emission lines share the same classification.  Nevertheless, the classification of the components used to model [N\,II] lines is not always linked to that of either [S\,II] and [O\,I] as \textit{M}-models encompass two different alternatives for [N\,II] (Sect.\ref{Analysis_LF}). Therefore, in our classification scheme we therefore neglect the kinematical information inferred from [N\,II],  for simplicity.\\ 
For a more accurate analysis, we combine these results with the measurements of  velocities from  the 2D spectra (before the extraction of the nuclear one, i.e. measuring individual sections). Specifically, we inferred the amplitude of the velocity-field of the ionised gas using the position-velocity diagrams presented in  Fig.\,\ref{Panel_PVD} (Appendix\,\ref{App_comments_panels}). \\
Since the observations were performed  without any orientational prescription (Sect.\,\ref{Sample_data}), it is possible that the kinematic components associated to the regular rotation might be present in the line of sight. Thus, we estimate the maximum broadening of emission lines due to rotation, as the peak-to-peak amplitude of the ionised-gas velocity rotation curve (Fig.\,\ref{Panel_PVD}), which is  generally lower than $\sim$\,400\,km\,s$^{-1}$. This is a conservative estimate as, generally, larger values (i.e., amplitudes of $\sim$\,300-400\,km\,s$^{-1}$) are observed outside the region used to obtain the nuclear spectra (Table\,\ref{T_Obs_sample} and Fig.\,\ref{Panel_PVD}). Indeed,  considering only the individual spectra used to obtain the final (nuclear) one, the  velocity amplitude of the ionised gas rotation is generally smaller (up to $\sim$\,300\,km\,s$^{-1}$). These estimates are in fair agreement with the spectroscopic measurements of the  velocity amplitude of the ionised gas rotation for samples of nearby spirals (GHASP survey, \citealt{Epinat2010}) and early type galaxies (ATLAS$^{\rm 3D}$, \citealt{Ganda2006, Cappellari2007}).\\
The typical values of  velocity dispersion for the narrow components are well below the two limits, 300 and 400 km\,s$^{-1}$,  estimated in the case of line-broadening due to rotation (Fig.\,\ref{Panel_sigmaV} right); velocities are typically rest-frame (i.e.  0\,$\pm$\,50\,km\,s$^{-1}$) except for a few cases (see Table\,\ref{T_kin}). Therefore, a simple interpretation is that the line widths of the narrow component can be explained by rotation. \\
The distribution of the kinematic measurements for the second component is more spread in the $\sigma$-V plane than that for the narrow component (Fig.\,\ref{Panel_sigmaV} left). This behaviour is also evident in Fig.\,\ref{Panel_kin_models}.  Those kinematic components with $\sigma$\,$\sim$\,400-800 \,km\,s$^{-1}$ (that cannot be produced by rotation) are likely associated to turbulent non-rotational motions. Similarly,  those components with 300\,$<$\,$\sigma$\,$<$\,400 km\,s$^{-1}$, with  velocities larger  than $\pm$\,100\,km\,s$^{-1}$ are  difficult to interpret as due to rotation. Therefore, these components are considered as candidates for non-rotational motions. All the emission line components associated to non-rotational motions (and candidates) will be discussed in Sect.\ref{Disc_outflows}.\\
Finally, the broad H$\alpha$ component is interpreted as a signature of the presence of the BLR (as mentioned already in Sect.\,\ref{Analysis_LF}). \\

\noindent For the sake of homogeneity, we adopted the same criterion to classify the velocity components for both  ionised and neutral gas phases.   The motivations are twofold. First,  the NaD detection is robust only in a minor fraction of our LINERs (i.e. 8/21, 38\,$\%$, Sect.\,\ref{Analysis_Abs_LF}). Second, the individual spectra in the region used to extract the nuclear one lack  the required S/N for stellar modelling and line fitting. Hence,  with the present data set it is impossible to  determine the neutral gas rotation curves, as done for emission lines (Fig.\,\ref{Panel_PVD}). Therefore, according to the classification scheme for the velocity components in emission lines, all the neutral gas  kinematics can be interpreted as due to rotation, except  for the case of NGC\,0315 (candidate for non-rotational motion).\\
The neutral gas velocity dispersion is  typically larger  than that of the ionised-gas (narrow component). This might indicate the presence of either  mild non-rotational motions of the neutral gas in these LINER-nuclei (as discussed in Sect.\,\ref{Disc_neutral}) or two different rotating discs (ionised and neutral). Specifically, the neutral gas disc rotation does not necessarily  follow that of the ionised disc; with the former possibly either lagging or being dynamically hotter and thicker compared to the latter  (as seen in nearby U/LIRGs, \citealt{Cazzoli2014, Cazzoli2016}). \\
We discuss individual cases in Appendix\,\,\ref{App_comments_panels}, taking into account the possible differences and analogies of ionised and neutral gas kinematics. 

\subsection{Non-rotational motions}
\label{Disc_outflows}
 
We identify four areas in the $\sigma$-V plane (Fig.\,\ref{Panel_sigmaV}) occupied by the kinematic measurements for the ionised gas that we associate with:
 \begin{itemize}
\item[--] rotation: in the region limited by  $\sigma$\,$<$\,300 km\,s$^{-1}$, and by 300\,$<$\,$\sigma$\,$<$\,400 km\,s$^{-1}$ and 100\,$<$\,$\mid$\,V\,$\mid$ km\,s$^{-1}$  (grey); 
\item[--] candidates for non-rotational motions: in the regions limited by 300\,$<$\,$\sigma$\,$<$\,400 km\,s$^{-1}$ and $\mid$\,V\,$\mid$\,$<$ 100 km\,s$^{-1}$ (pink);
\item[--] non-rotational motions  within the region limited by $\sigma$\,$>$\,400\,km\,s$^{-1}$, at both negative (blue) and positive velocities (yellow).
\end{itemize}
These are conservative criteria considering both the 2D rotation curves (Fig.\,\ref{Panel_PVD}) and  the kinematics of all the components in our nuclear spectroscopy (see Sect.\,\ref{classification} for details). \\
In Table\,\ref{class_sum}, we summarise  the present classification. In the following, we explore the possibility of rotation (\textit{R}), candidates for non-rotational motions (\textit{C}) and of those non-rotational motions being produced by outflows (\textit{O}) or inflows (\textit{I}) having negative and positive velocities, respectively. It is worth to note that Monte Carlo simulations based on the 3D outflow models by \citet{Bae2016} indicate that intermediate-width lines with positive velocities can also be interpreted as outflow (hence not inflow). This is a consequence of dust obscuration and special prescriptions of the outflow geometry (e.g. biconical outflows). In this paper we assume the outflow/inflow interpretation and classified the line-components accordingly.  However, with the present data set it is not possible to infer outflow opening angles and geometries and hence producing a detailed model for the outflows is beyond the aim of this work. \\

\noindent The kinematics of the narrow component can be interpreted as rotation in all cases (Fig.\,\ref{Panel_sigmaV} left). 
Although, the data points for the second component can be found in all the four areas (Fig.\,\ref{Panel_sigmaV} right) encompassing all  the proposed possibilities with [O\,I] and [S\,II] classifications being different in 5 cases (NGC\,1052, NGC\,3642, NGC\,4143, NGC\,4278 and NGC\,4450; Table\,\ref{class_sum}). Taking into account the various possibilities for second components, the final adopted criterion for claiming a possible interpretation as outflow/inflow (rotation) is that both [S\,II] and [O\,I] second component are classified as outflow/inflow (rotation) in Fig.\,\ref{Panel_sigmaV} (see also Table\,\ref{class_sum}). In those 5 cases (Table\,\ref{class_sum}) where the wavelength coverage of our data  does not allow to probe either [S\,II] or [O\,I], the final classification is set equal to that of the only second component of the forbidden line adopted for the line fitting. If non-rotational motions (either outflow or inflow) or a candidate are seen in only  one component we consider these putative interpretation as less reliable.  The final classification is listed in column 5 of Table\,\ref{class_sum}.

\noindent For ground-based data, this criterion clearly identifies  6 and 3 cases out of 15  with outflows (40\,$\%$) and rotation (20\,$\%$), respectively (Table\,\ref{class_sum}).  Only one robust case (NGC\,4750, Table\,\ref{class_sum}) of candidate for non-rotational motions has been found. In the remaining five cases (33\,$\%$, Table\,\ref{class_sum}), the classification for [O\,I] and [S\,II] is different.
Specifically, in all of these five cases, the classification as non-rotational motions (outflows) or candidate held only for [O\,I] with [S\,II] being classified as rotation.   Conversely, the alternative for which [S\,II] is classified as non-rotational motions and [O\,I] rotation never occurs. Such a  large prevalence of non-rotational motions or candidate found in [O\,I] with respect to [S\,II] (i.e. 10 vs. 5 cases, considering only those cases for which both forbidden lines are present) might indicate that LINERs are only able to power low-ionisation emission-line outflows.  A detailed and possibly 2D study with IFS data is needed to investigate further this issue.\\
There are not systematic studies of outflows in type-1 LINERs via ground-based spectroscopy, but only works of individual cases  (e.g. \citealt{Dopita2015,Brum2017}). The detection rate of outflows in ground-based data is lower than what found in studies of sample of nearby (z\,$<$\,0.2) type-2 luminous AGNs and quasars (e.g. \citealt{Harrison2014} and \citealt{Villar2014}) and local (z\,$<$\,0.1) U/LIRGs (e.g. \citealt{Arribas2014}) which is typically larger than  70\,$\%$.\\
We found that, generally, the outflow-components have velocities varying from -15\,km\,s$^{-1}$ to -340\,km\,s$^{-1}$, and velocity dispersions in the range of 450-770\,km\,s$^{-1}$ (on average, 575\,km\,s$^{-1}$). \\
In our data set, the components interpreted as outflow have intermediate kinematic properties between outflows observed in AGNs  and U/LIRGs. More specifically, the range of velocity is typically smaller than what found in outflows in AGNs but the velocity dispersion values are consistent with e.g. those of turbulent outflowing ionised gas in type-2 AGNs found by \citet{Harrison2012} using IFS observations of [O\,III]. The kinematics of these outflowing components is more extreme than that of outflows observed in U/LIRGs (e.g. \citealt{Rupke2005c, Arribas2014}) being the velocities more negative and velocities dispersion much larger.\\

\noindent  The same classification criterion applied to the results from space-based data implies a null detection rate for outflows, whereas we found 2 cases out of 7  of rotation and 2 cases of candidates for non-rotational motions (Table\,\ref{class_sum}). The outflow interpretation is less reliable for NGC\,1052 (Fig.\,\ref{Panel_NGC1052}) while NGC\,4450 (Fig.\,\ref{Panel_NGC4450}) is the only LINER for which the components are classified  as outflow ([S\,II]) and inflow ([O\,I]). However, in this case the outflow classification is less reliable as the H$\alpha$ broad component contaminates the [S\,II]$\lambda$6716 line. Data at higher spectral resolution are needed to confirm such intriguing outflow-inflow scenario. We found only one case  (NGC\,4278, Fig.\,\ref{Panel_NGC4278}) of inflow (Table\,\ref{class_sum}).\\

\noindent Excluding the particular case of NGC\,4450, the number of detections of non-rotational motions ($\sim$\,30$\%$) is lower than what found for the second components in ground-based data analysis. This is in  partial agreement with the  study of the H$\alpha$ morphology at \textit{HST} scales in LINERs by \citet{Masegosa2011} which indicates that outflows are frequent (but not ubiquitous) in LINERs. For the present sample of LINERs, the direct comparison between our results and the H$\alpha$ morphology studied in \citet{Masegosa2011} can be done only in four cases (NGC\,0315, NGC\,1052, NGC\,4278 and NGC\,5005).  In two cases (NGC\,1052 and NGC\,5005) the H$\alpha$ emission shows morphological evidence of nuclear outflows (see also \citealt{Pogge2000}). Strong outflows are also seen in the \textit{HST}/STIS data via multiple velocity components as reported by \citet{Walsh2008} (e.g. NGC\,1052). \\
A possible explanation of the low detection rate of outflows is that their extended nature (e.g. a nuclear expanding bubble) is not fully captured by \textit{HST}/STIS spectroscopy with narrow slit  ($\leq$\,0.2$^{\prime}$$^{\prime}$).
The detection of line asymmetries interpreted as outflows could be relatively more difficult as the slit width decreases. Slit-orientation effects might also play a role. Furthermore, issues associated to the wavelength-coverage (Sect.\,\ref{Analysis_LF_space}) may affect the detection of line asymmetries in space-based data, especially if these are more frequent in [O\,I] rather in [S\,II] as seen from the results obtained in ground-based data (Table\,\ref{class_sum}).\\

\begin{table}  
\caption[class_sum]{Summary of the classification for the second component.}				
\begin{tabular}{l l c  c c }
\hline 					
ID &  Obs.    & [S\,II] & [O\,I] &   Final \\
 \hline  
NGC\,0266 & CAHA $^{\dagger}$   & - &  O & O \\
NGC\,0315 & NOT 		& O & O& O \\
          & HST $^{\dagger}$  	& R &  - & R \\
NGC\,1052 & NOT  		& R & O& [O]  \\
          & HST  		& R & O&  [O] \\
NGC\,2681 & CAHA 		& R &  R & R \\
NGC\,3226 & CAHA 		& O &  O & O \\
NGC\,3642 & CAHA  		& R & C&  [C] \\
          & HST $^{\dagger}$  	& - & -& -  \\
NGC\,3884 & CAHA $^{\dagger}$	& - & O  & O  \\
NGC\,3998 & CAHA  		& O & O&  O \\
          & HST $^{\ddagger}$  	& - &  - & -  \\
NGC\,4036 & CAHA 		& R &  R & R  \\
 	  & HST  		& - & -  & -  \\
NGC\,4143 & CAHA $^{\ddagger}$ 	& R &  O & [O]  \\
	  & HST  		& R &  R &  R \\
NGC\,4278 & CAHA  		& R &  I & [I]  \\	
	  & HST $^{\ddagger}$  	& I &  I &   I \\
NGC\,4438 & CAHA 		& R & R & R  \\
NGC\,4450 & CAHA  		& R &  O & [O]  \\
	  & HST  		& O &  I &   [O] \\
NGC\,4750 & CAHA  		& C &  C &  C \\
NGC\,5005 & CAHA  		& O & O& O  \\
          & HST $^{\dagger}$  	& C &  - &  C \\
NGC\,5077 & CAHA  		& - & -  &  - \\
          & HST $^{\dagger}$  	& C &  - & C \\
\hline  			      
 \label{class_sum}		      
\end{tabular}

\textit{Notes.} \lq ID\rq: object designation as in Table\,\ref{T_sample}. We list only those LINERs for which a second component is needed to model the observed line profiles either in ground- and/or space-based spectroscopy. \lq Obs\rq \ stands for the origin of the optical data.  $^{\dagger}$ and $^{\ddagger}$ symbols are as in Table\,\ref{T_kin}. \lq R\rq , \lq O\rq , \lq I\rq \ and \lq C\rq \ indicate: rotation, outflow, inflow and candidate for non-rotational motions, respectively (see text). The last column report the final classification, where square-brackets indicate a less reliable outflow, inflow or candidate  for non-rotational classification.
\end{table}
\noindent  There are two cases (NGC\,1052 and NGC\,4278) in which the same classification holds for both the ground- and space-based data.  Both  second components required to model NGC\,1052 share the same classification in the two data sets. The second component in [O\,I] is classified as outflow, in agreement with the outflow interpretation by \citet{Walsh2008} (with \textit{HST}/STIS data) and by \citet{Dopita2015} (ground-based IFS data). For NGC\,4278 (Fig.\,\ref{Panel_NGC4278}) the inflow classification holds for all the emission lines  in both ground- and space-based data sets (except [S\,II] in ground-based data, Table\,\ref{class_sum}). This is particularly interesting as the signature of inflows as a second intermediate component at positive (redshifted) velocities are elusive (e.g. \citealt{Coil2011, Rubin2012}) whereas outflows have been observed in a variety of galaxies and at any redshift (\citealt{Veilleux2005} and \citealt{King2015} for reviews). \\
We refer to Appendix\,\ref{App_comments_panels} for detailed comments on individual cases.

\begin{figure*}
\centering
\includegraphics[trim = 2.2cm 15.5cm 1.95cm 6cm, clip=true, width=1.\textwidth]{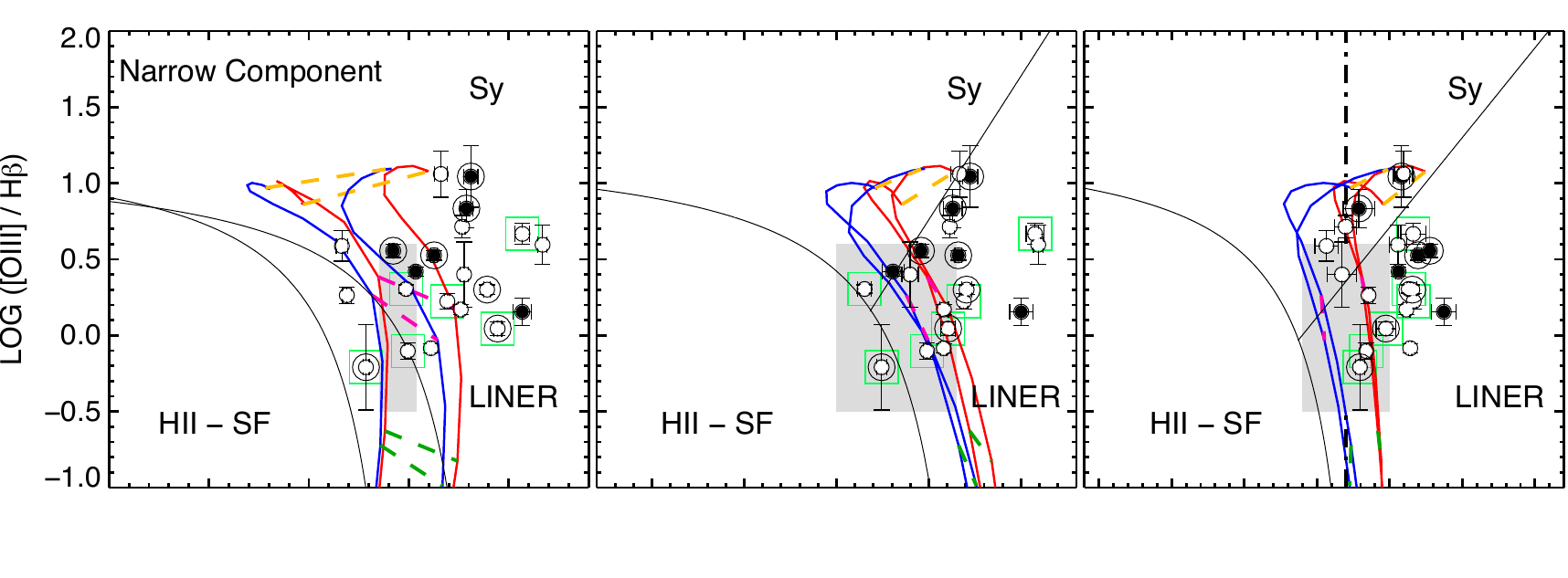}\\
\vspace{-3.35cm}
\includegraphics[trim = 2.2cm 15.5cm 1.95cm 3.75cm, clip=true, width=1.\textwidth]{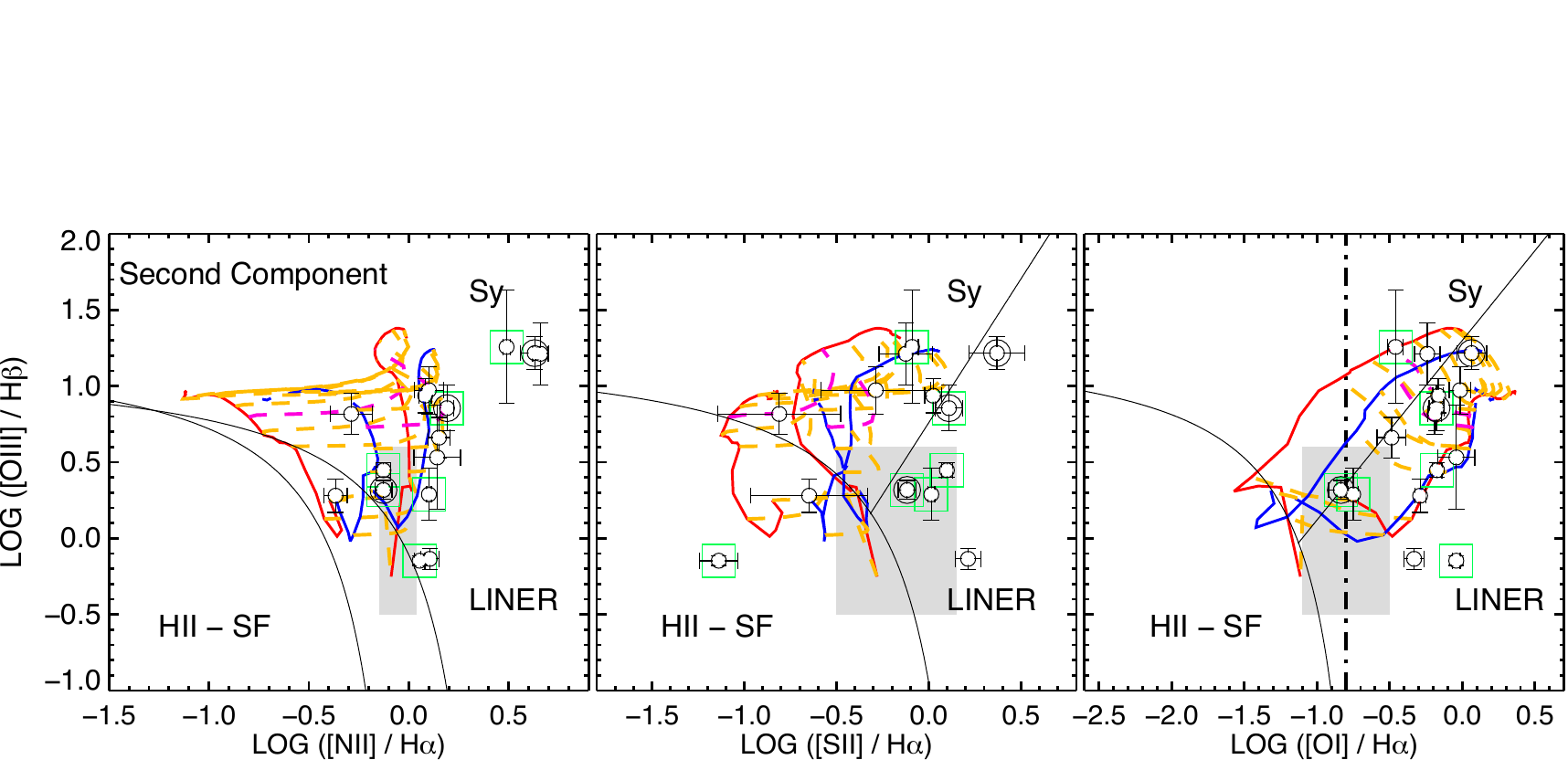}
\caption{Optical standard BPT diagrams for both narrow and second components (top and bottom panels, respectively) obtained for our ground-based spectroscopy.  Circles mark the data points presented in this paper. In the top panels, filled symbols mark those LINERs for which forbidden lines are  modelled well with only one Gaussian (i.e. the second component is absent, see Table\,\ref{T_kin}). In all the panels an additional circle marks those LINERs for which  a broad component is required to optimally reproduce the H$\alpha$ line profile (see Table\,\ref{T_kin}). An additional light-green square marks those galaxies with less reliable fit of H$\alpha$ (Table\,\ref{T_rms}).  Black lines in all diagrams represent the dividing curves between H\,II star-forming regions, Seyferts, and LINERs from \citet{Kewley2006} and \citet{Kauffmann2003}.  Gray boxes show the predictions of photoionization models by pAGB stars for Z\,=\,Z$_{\sun}$, a burst age of 13 Gyr  \citep{Binette1994} and ionization parameter values (log\,U) between -3 and -4. Log\,U is typically -3.5 in LINERs \citep{Netzer2015}. The predictions of AGN and shock  -ionization models are overlaid in each diagram. \textit{Top:}  Dusty AGN photoionization grid from \citet{Allen2008} with n$_{\rm el}$\,=\,100\,cm$^{-3}$ (Sect.\,\ref{Disc_models}), for Z$_{\odot}$ and 2\,Z$_{\odot}$, for different power-law spectral indices ($\alpha$\,=\,-1.7, -1.4, blue and red lines) and ionization parameter values (log\,U\,=\,-3.6, -3.0, 0.0, green, pink and yellow dashed-lines).  \textit{Bottom:}  Shock+precursor grids from \citet{Groves2004}  with Z\,=\,Z$_{\odot}$ and for different n$_{\rm el}$. Blue and red curves correspond to models with n$_{\rm el}$\,=\,100\,cm$^{-3}$ and n$_{\rm el}$\,=\,1000\,cm$^{-3}$, respectively (Sect.\,\ref{Disc_models}). We marked the values corresponding to the minimum and maximum preshock magnetic field allowed in each model. Shock-velocities are from 100 to 800 km\,s$^{-1}$ (yellow dashed lines, Sect.\,\ref{Disc_models}). In pink is indicated the average shock-velocity for the second component (i.e. $\sim$\,450 km\,s$^{-1}$, Table\,\ref{T_sum_kin}). The dividing line between weak-[O\,I] and strong-[O\,I] LINERs \citep{Filippenko1992} is marked in black with a dot-dashed line.}
 \label{Panel_bpt} 		 
\end{figure*}

\begin{figure*}
\centering
\includegraphics[trim = 2.2cm 15.65cm 1.95cm 6.25cm, clip=true, width=1.\textwidth]{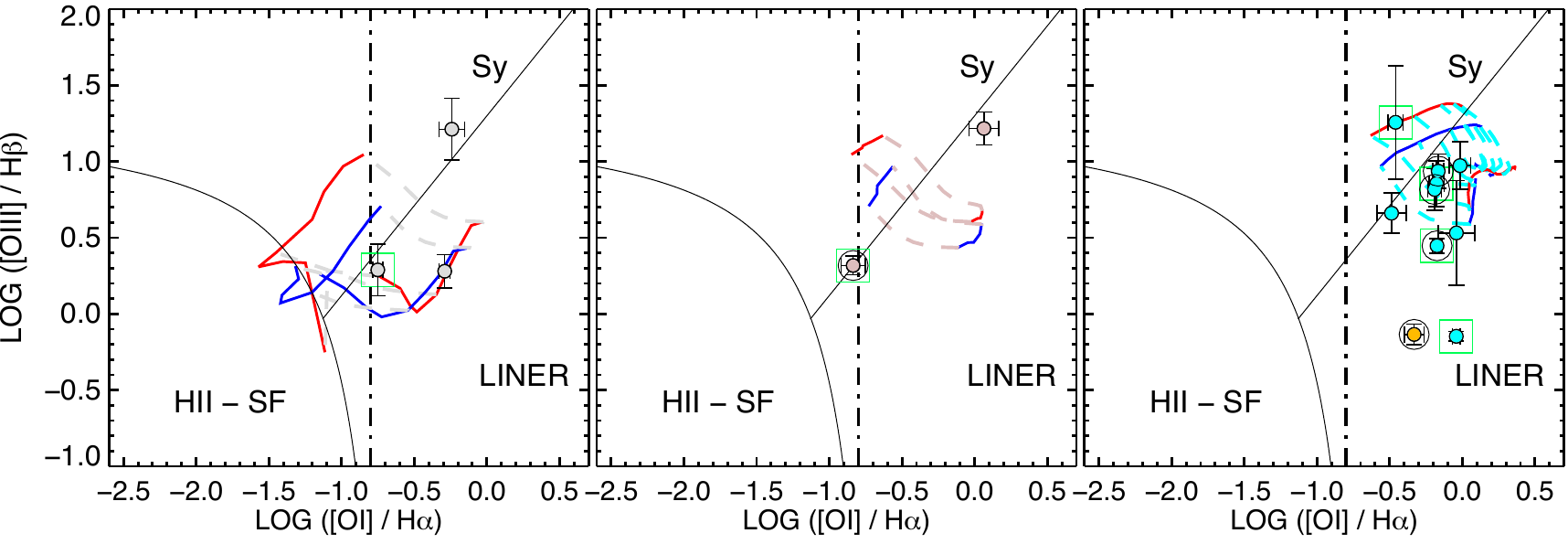}\\
\caption{Optical standard BPT diagrams. Coloured circles mark the data point for the second component presented in this paper (as in Fig.\,\ref{Panel_bpt}, bottom-right). Additional symbols as light-green square and black lines are as in Fig.\ref{Panel_bpt}. We consider predictions of shock-ionization models with Z\,=\,Z$_{\odot}$ and for different n$_{\rm el}$: (100\,cm$^{-3}$ and 1000\,cm$^{-3}$) as in Fig.\ref{Panel_bpt}.   The  shock+precursor grids are overlaid in each diagrams considering different shock-velocities. Specifically, from left to the right,  we consider only shock-velocities relative to the rotation, candidate for non-rotational motions and non-rotational motions, respectively (Fig.\,\ref{Panel_sigmaV}). Iso-velocities are marked with dashed-lines and colour coded according to the four areas in Fig.\,\ref{Panel_sigmaV} (see also Sect.\ref{Disc_outflows}). Similarly, for the colour coding for the observed line ratios (being outflows and inflows marked in light blue and orange, respectively). An additional circle marks those cases for which the final classification is less reliable (Table\,\ref{class_sum}).}
 \label{Panel_bpt_cc} 		 
\end{figure*}

\subsection{The underlying physics of the different  components}
\label{Disc_models}

The ionizing mechanism of optical emission lines in LINERs is widely debated and three  alternative scenario have been proposed: AGN, pAGB stars and shocks (see \citealt{Ho2008} and \citealt{Netzer2015} for reviews). \\
Our data set allows us to classify narrow and second components used to model emission lines combining line-fluxes and kinematics. In Fig.\,\ref{Panel_bpt}, we investigate the location of the line ratios for the narrow and second emission line components onto standard \lq BPT-diagrams\rq \ \citep{Baldwin1981}. These are empirically-derived diagrams based on optical emission line ratios (selected to be essentially unaffected by reddening) that allow to discriminate different ionizing mechanisms. We consider the \citet{Kewley2006} and \citet{Kauffmann2003} curves that allow to distinguish between the locus occupied by star forming systems  and AGNs, these separates also the locus between LINERs and and Seyferts. We use the BPT classification to infer the dominant mechanism in LINERs and compare our results with models.  Additionally, we consider the criterion  for dividing weak-[O\,I] and strong-[O\,I] LINERs proposed by \citet{Filippenko1992}. In this pioneering work, Filippenko and collaborators separate those two classes of LINERs on the basis of the [O\,I]/H$\alpha$ ratio which is smaller (larger) than 0.16 for weak-[O\,I] (strong-[O\,I], genuine AGN) LINERs. This criterion is particularly helpful as the emission-line spectrum of gas photoionised by the stellar continuum of O-stars can mimic that of LINERs only in the weak-[O\,I] case. Then, we  take into account the kinematical information for each component used to model emission lines and its kinematical classification (Table\,\ref{class_sum}). \\
We inferred the electron density (n$_{\rm el}$) from the relative flux of the [S\,II] doublet i.e. by interpolating the curve in Fig.\,5.3 of  \citealt{Osterbrock1989} for T\,=\,10$^{4}$\,K (see also \citealt{Draine2011}). We found  for the narrow component: n$_{\rm el}$\,$\leq$\,600\,cm$^{-3}$ (typically $\sim$\,100\,cm$^{-3}$), and for the second component: 100\,$\leq$\,n$_{\rm el}$\,$\leq$\,1000\,cm$^{-3}$. Moreover, we consider shock-velocities from 100 to 800 km\,s$^{-1}$, as larger values of velocity dispersions  are not observed in these LINERs (Table\,\ref{T_kin}, Figures \ref{Panel_kin_models} and \ref{Panel_sigmaV}).\\
The proposed approach is twofold and is particularly helpful as the different models corresponding to the three possible scenarios (AGN, pAGB stars and shocks) might suffer degeneracies when displayed onto BPTs.\\

\noindent For ground-based data, the line ratios for the narrow component are generally consistent with those observed in AGNs (either  Seyfert  or LINERs) considering the dividing curves proposed by \citet{Kewley2006} and \citet{Kauffmann2003}, excluding as dominant ionization mechanism the star-formation. This is particularly evident in the [O\,I]/H$\alpha$ diagram (Fig.\,\ref{Panel_bpt} top-right). The only two exceptions are NGG\,3642 and NGC\,3884; with the additional case of NGC\,3998 whose line ratios lie onto the boundaries between AGNs and star-formation. However, these outliers are  mainly seen in the [N\,II]/H$\alpha$ diagram (Fig.\,\ref{Panel_bpt}, top-left) which suffers of potential problems associated with Nitrogen abundances (\citealt{PerezMontero2009}, \citealt{Maiolino2017} and references therein).\\
We consider the grids of AGN photoionisation models having n$_{\rm el}$ inferred from the data and adopting  a range of 
metallicities, ionisation parameter and power-law spectral indices, similarly to  \citet{Kehrig2012}. We found that these models reproduce well  the observed line ratios for the narrow component in $\sim$\,10 cases considering all the diagrams (Fig.\,\ref{Panel_bpt}, top panels). \\
We then explore, the possibility of (dominant) ionisation by pAGB stars by comparing the observed data points with  the models proposed by \citet{Binette1994}. We note that these models cannot reproduce the large [O\,III]/H$\beta$ ratios (e.g. between 0.6-1.2, in log units) for the narrow component in six cases; and that a large overlap between  pAGBs-stars and  AGN models exists. This overlapping might result in an ambiguous classification in at least six cases (Fig.\,\ref{Panel_bpt}, top panels). Hence, we attempt to break such as degeneracy introducing the  additional criterion for  weak-[O\,I] and strong-[O\,I] LINERs proposed by \citet{Filippenko1992}.  By  combining the selected pAGBs-models and the [O\,I]/H$\alpha$-criterion,  the area corresponding to the stars as genuine dominant mechanism of ionisation is thus reduced. The observed line ratios of the narrow component are generally above the 0.16 threshold (-0.8 in log units) for identifying strong-[O\,I] LINERs. The only exception is NGC\,0266; with two additional cases (NGC\,2681 and NGC\,4143) whose [O\,I]/H$\alpha$ ratios are $\sim$\,0.16 (lying onto strong/weak -[O\,I] dividing line, Fig.\,\ref{Panel_bpt} top-right).\\
As a further test, we also checked the possibility of the shock-ionization as the dominant mechanism for producing optical lines in LINERs using the shock+precursor models by \citet{Groves2004} with n$_{\rm el}$\,=\,1\,cm$^{-3}$ and n$_{\rm el}$\,=\,100\,cm$^{-3}$. However, these shock-models are able to reproduce the observed line ratios only in a smaller number of cases (typically less than five).  This indicates that shocks are not the dominant mechanism but they could be however important in LINERs. \\
The kinematics of the narrow component can be explained by rotation in all cases (Sect.\ref{Disc_outflows}). \\
All together these results strongly exclude the pAGBs stars scenario in favor of the AGN-ionisation as dominant mechanism of ionisation for the narrow component in these LINERs with shocks being however relevant.\\

\noindent The location onto BPTs of  the line ratios  for the second component  excludes the star-formation as the dominant mechanism of  ionisation (Fig.\,\ref{Panel_bpt} bottom). This is especially evident in the [O\,I]/H$\alpha$ diagram (as for the narrow component) with the majority of the data points lying in the LINER-area (Fig.\,\ref{Panel_bpt}, bottom-right). The only exceptions are NGGC\,4036 in both [N\,II]/H$\alpha$ and [S\,II]/H$\alpha$, and NGC\,3998 only in the [N\,II]/H$\alpha$ one (Fig.\,\ref{Panel_bpt}, bottom panels). \\
The [O\,III]/H$\beta$ line ratios for the second component are larger with respect to those of the narrow component being generally above  0.6  (in log units) disfavoring the pAGB scenario. In addition, the [O\,I]/H$\alpha$ ratios of the second component are well above the 0.16-threshold supporting the strong-[O\,I] nature of these LINERs.\\
We attempted to reproduce the observed line ratios for the second component with the shock-models by \citet{Groves2004} with n$_{\rm el}$\,=\,100\,cm$^{-3}$ and n$_{\rm el}$\,=\,1000\,cm$^{-3}$ (Fig.\,\ref{Panel_bpt}, bottom) as tested for the narrow component. The match between the  shock models and data points occurs differently for low and high ionization lines (e.g. \citealt{Chisholm2017}). Therefore, we mainly focus in the [O\,I]/H$\alpha$ diagram  as it is the most reliable among standard optical BPTs for studying shocks  \citep{Allen2008}. We found that generally the  selected models are able to reproduce the observed data points (except two cases, Fig.\,\ref{Panel_bpt}, bottom-right).\\
\noindent The kinematic classification of the second component shows different possibilities from rotation  to strong non-rotational motions, with the latter being the most frequent (Fig.\,\ref{classification}, Sect.\ref{Disc_outflows} and Table\,\ref{class_sum}). In the following, we explore the connection between the kinematics of the second component and its dominant ionization mechanism by plotting colour-coded BPT-diagrams. Specifically in Fig.\,\ref{Panel_bpt_cc}, we combine the data points and models for the [O\,I]/H$\alpha$ diagram  with the kinematical classification shown in Fig.\,\ref{Panel_sigmaV} (as colour coding) and summarised in Table\,\ref{class_sum}.\\
Shocks models (grid at low velocities, Fig.\,\ref{Panel_bpt_cc} left) are able to reproduce the observed line ratios in two (NGC\,4036 and NGC\,4438) out of three cases for which the  component is classified as rotation. This possibly indicates the presence of mild-shocks associated to the ionised gas rotation (which could be somewhat perturbed). In the case of NGC\,2681, the grid of shock models could still reproduce the observed line ratios (Fig.\ref{Panel_bpt}, bottom-right) but at velocities of $\sim$\,600-700\,km\,s$^{-1}$ hence much larger than observed (Table\,\ref{T_kin}, Fig.\,\ref{Panel_bpt_cc} left).\\
Similarly to the case of NGC\,2681, shocks models (Fig.\ref{Panel_bpt} bottom-right) are able to reproduce the observed line ratios for those two cases classified as candidate for non-rotational motions (NGC\,3642 and NGC\,4750). However, their line ratios are found at either  lower or larger velocities than observed  (Fig.\,\ref{Panel_bpt_cc} middle).\\
\noindent The selected shocks models are generally able to reproduce the observed line ratios within errors of the second component with the only exception of NGC\,3998  (Fig.\,\ref{Panel_bpt_cc}, right).\\
These results are in agreement with studies based on IFS observations of nearby LINERs (e.g. NGC\,1052, \citealt{Dopita2015}) and local U/LIRGs (e.g.\,\citealt{HoTing2014}) showing that outflows are usually associated with shocks that excite optical emission lines with enhanced line ratios.\\
For the unique case of inflow (NGC\,4278, Table\,\ref{class_sum}), the selected models cannot reproduce the observed line ratios (Fig.\,\ref{Panel_bpt_cc}, right). However, the corresponding data points, at low [O\,III]/H$\beta$ and large [O\,I]/H$\alpha$ ratios, are well reproduced  by shock-models with the same parameters as those considered in this analysis but without precursor. These shock-models are not shown in Fig.\ref{Panel_bpt_cc} for clarity, therefore we refer to \citet{Groves2004}.\\

\noindent Unfortunately, we cannot use the BPT diagrams to study the possible ionization mechanism of the component found in the analysis of \textit{HST}/STIS data, as only the red bandpass is available to us. Despite this, the [N\,II]/H$\alpha$, [S\,II]/H$\alpha$ and [O\,I]/H$\alpha$ ratios give helpful limits to constraint the dominant mechanism of ionization considering the dividing lines between ionization from AGN and star formation. \\
On the one hand, the ranges of values for the  [N\,II]/H$\alpha$ and [S\,II]/H$\alpha$ ratios are particularly large for attempting to constraint the dominant ionisation mechanism. Specifically, these ranges are: -0.4\,-\,0.8 and -0.6\,-\,0.1 for [N\,II]/H$\alpha$ and [S\,II]/H$\alpha$ ratios, respectively, for the narrow component; and -0.5\,-\,0.6 for the second one for both ratios. On the other hand, the ranges for the [O\,I]/H$\alpha$ ratio are -0.7\,-\,0.1 and -0.4\,-\,0.1, for the narrow and second component, respectively.  Thus,  these  [O\,I]/H$\alpha$ ratios derived from the analysis of \textit{HST}/STIS data are outside the region which identifies  star formation as the dominant ionization mechanism (favoring the AGN scenario). Moreover, they  also indicate that those nuclei are already strong-[O\,I] LINERs (according to the criterion proposed by \citealt{Filippenko1992}) at \textit{HST}/STIS scales.

\subsection{A lack of neutral gas outflows in type-1 LINERs?}
\label{Disc_neutral}

In Sect.\,\ref{Analysis_Abs_LF} we described for the first time the detection of the neutral gas in the nuclear region of the selected type-1 LINERs via the modelling and the analysis of the NaD absorption. This is novel as our sample is composed by low-luminosity AGNs not classified neither as  starbursts  or U/LIRGs, having SFR \,$\leq$\,0.5 (calculated from H$\alpha$ luminosity, Table\,2 in \textit{HFS97}) and log(L$_{\rm IR}$/L$_{\sun}$)\,$\leq$\,8-9 in many cases (e.g. \citealt{Sturm2006}). The large majority of  previous works focused on starbursts (e.g. \citealt{Heckman2000}) or U/LIRGs (e.g. \citealt{Rupke2005c, Cazzoli2016}) on the basis of long-slit and IFS observations. The analysis of the neutral gas component of ISM (via NaD) in AGNs has been done only for small samples of U/LIRGs with a Seyfert nucleus or quasars (e.g. \citealt{Rupke2005b}, \citealt{Krug2010} and \citealt{Villar2014}). \\
 According to the classification scheme proposed in Sect.\,\ref{classification} all the neutral gas kinematic components (except one) could be interpreted as rotation. The neutral gas velocities (Table\,\ref{T_NaD}) are roughly consistent with those of neutral gas discs (detected via NaD absorption) in e.g. U/LIRGs \citep{Cazzoli2016} but with a larger velocity dispersion considering average values (87 vs. 220 km\,s$^{-1}$).\\
The possible explanation of the lack of neutral gas outflows, or more in general of non-rotational motions,  is twofold. First, the neutral component in outflows is possibly less significant  in AGNs or quasars  than in starbursts galaxies and U/LIRGs, as argued by \citet{Villar2014}. Secondly, such null detection rate might be a consequence of the conservative limits we assumed for kinematic components in emission lines (Sect.\,\ref{Analysis_Abs_LF}). As ionised and neutral gas correspond to different phases of the outflows, these may have a different kinematics (e.g. \citealt{Cazzoli2016}). Neutral gas kinematical components with  large velocity dispersions ($>$\,300-400\,km\,s$^{-1}$) are found only in Seyferts or the most luminous ULIRGs (\citealt{Krug2010, Martin2005}). Furthermore, the velocity thresholds assumed to identify outflows are generally lower (of the order of 50-100 km\,s$^{-1}$). If the adopted limits are relaxed, and we assume  those for identifying outflows assumed in \citet{Cazzoli2016}, i.e. V\,$<$\,-60\,km\,s$^{-1}$ and $\sigma$\,$>$\,90\,km\,s$^{-1}$, we would have found three cases for possible outflows (NGC\,0266, NGC\,1052 and NGC\,4750). Likewise,  NGC\,0315 would have been classified as inflow.

\section{Summary and conclusions}
\label{summary_conclusions}

We analysed a sample composed by the 22 nearby (z\,$<$\,0.025) type-1 AGN-LINERs from the Palomar Survey, covering the best-defined sample of this kind (\textit{HFS97}). We  studied the properties of the ionised  and neutral gas in these LINERs on the basis of CAHA/TWIN, NOT/ALFOSC, and \textit{HST}/STIS optical long-slit spectroscopy. \\
First, for each LINER observed via ground-based spectroscopy we modeled and subtracted the stellar continuum to obtain a pure ISM spectrum. Second, we modeled a set of emission lines in both ground- and space-based data testing various models in order to explore the presence of an H$\alpha$ broad component (indicative of an AGN) and to study the kinematics and the ionisation mechanism for the ionised gas. Finally,  we modelled the NaD ISM-absorption doublet with Gaussian components to investigate the neutral gas in these LINER nuclei.\\
We excluded NGC\,4203 from the analysis as its very broad and double peaked H$\alpha$ profile requires  a more detailed modelling.\\
The main conclusions can be summarised as follows:
\begin{enumerate}

\item[1. ---] \textit{The AGN nature of type-1 LINERs.} The H$\alpha$ broad component indicative of the presence of the AGN is elusive in our ground-based spectroscopy with a detection of only 7 out of 21 cases (33\,$\%$ of detection rate). The measured FWHM of the broad component ranges from 1277\,km\,s$^{-1}$  to 3158\,km\,s$^{-1}$ (2401\,km\,s$^{-1}$, on average).  Conversely, the broad component in \textit{HST}/STIS spectra  is ubiquitous, and its FWHM values are between  2152\,km\,s$^{-1}$  and 7359\,km\,s$^{-1}$ (3270\,km\,s$^{-1}$, on average). The broad component is seen in both ground- and space-based data  in four cases. The  detectability of the BLR-component is sensitive to  modelling and observational effects. 
On the one hand,  a broad component could be the result of the combined effects of questionable starlight decontamination and of an inappropriate choice of the template for the H$\alpha$-[N\,II] blend. Additionally, a single Gaussian fit of template forbidden lines is an oversimplification in many cases. This assumption has a significant impact on the detection and the determination of the properties of the BLR component.  However, from our analysis, it seems that aperture and slit-PA differences are not very relevant when comparing ground-based data.\\ 

\item[2. ---] \textit{Multiple component fitting of forbidden lines and  stratification of the NLR.}  For ground-based data, in 15 out of 21 cases the modelling of forbidden lines requires two components. Similarly, for space-based data, we found that two components in forbidden lines are needed to reproduce the observed line profiles in 7 out of 11 cases. However, we do not always find full agreement between the selected models and components used to fit emission lines for the two data sets. When we are able to test all the proposed models, in 9/19 (5/7) cases for ground (space) -based data, we found a prevalence  of best fits obtained with the Mixed models (those with different properties for [S\,II] and [O\,I], see Sect.\,\ref{Analysis_LF}) that take into account differences in critical densities and shock-enhancements. This  suggests that the NLR stratification is often present in type-1 LINERs. [O\,I] lines profiles are typically broader than [S\,II] (up to a factor of $\sim$\,3, considering second components). \\

\item[3. ---] \textit{2D Ionized gas rotation and rotational broadening.} The  peak-to-peak velocity  amplitude of the ionised gas rotation is generally lower than $\sim$\,300-400\,km\,s$^{-1}$ in fair agreement with the spectroscopic measurements of the  velocity amplitude of the ionised gas rotation for samples of nearby spirals and early type galaxies.  The broadening due to the possible superposition of several rotation components is (conservatively) estimated to be up to 400 km\,s$^{-1}$. \\

\item[4. ---] \textit{Kinematics of narrow and second components of emission lines. }  For the narrow component, velocities are typically rest-frame (i.e.  0\,$\pm$\,50\,km\,s$^{-1}$, except a few cases) and velocity dispersion values are smaller than 300\,km\,s$^{-1}$.  For the second components the velocity range is large (from -350 km\,s$^{-1}$ to 100 km\,s$^{-1}$). The velocity dispersions vary  between 150 and 800 km\,s$^{-1}$  (429\,km\,s$^{-1}$, on average) being generally broader than  narrow components. Generally, for space \textit{HST}/STIS data, except for a few cases, narrow components have velocities between -100 and 200  km\,s$^{-1}$, being the average velocity dispersion 176\,km\,s$^{-1}$. The velocity of the second component ranges from -200 to 150  km\,s$^{-1}$ with   velocity dispersion values  between 300 and 750 km\,s$^{-1}$ (433 km\,s$^{-1}$, on average).\\

\item[5. ---] \textit{Kinematical classification of the emission line components.} Four areas in the $\sigma$-V plane are defined for different kinematical interpretations with the following limits (expressed in km\,s$^{-1}$): a) rotation: either $\sigma$\,=\,[0, 300] or $\sigma$\,=\,[300, 400] and V\,=\,[-100, 100], b) candidate for non-rotational motions: $\sigma$\,=\,[300, 400] and $\mid$\,V\,$\mid$\,$>$\,100 and c) non-rotational-motions (as outflows or inflows): $\sigma$\,$>$\,400. We adopt the interpretation that blue/redshifted lines might be produced by inflows/outflows. However integral field spectroscopy would be required to confirm/discard such a scenario. For both ground- and space-based data, the kinematics of the narrow component can be explained with rotation in all cases whereas that of the second components encompass all possibilities. Furthermore, the classification might be different for [S\,II] and [O\,I], being non-rotational motions more frequently traced by [O\,I]. Taking these results into account, the final adopted criterion for claiming the possible interpretation as an  outflow/inflow (rotation) is that both [S\,II] and [O\,I] second component are classified as outflow/inflow (rotation). \\

\item[6. ---] \textit{Non-rotational motions.}  From our ground-based data, we identified 6 out of 15 (40\,$\%$) cases that may be interpreted as reliable outflows. Outflow-components have velocities varying from -15\,km\,s$^{-1}$ to -340\,km\,s$^{-1}$, and velocity dispersions in the range of 450-770\,km\,s$^{-1}$ (on average, 575\,km\,s$^{-1}$). These values are intermediate between those for outflows observed in AGNs and starbursts (as U/LIRGs).  We did not interpreted as outflows any case in \textit{HST}/STIS data. These results  partially disagree with studies of the H$\alpha$ morphology in LINERs which indicate that outflows are frequent in LINERs. The possible explanation is twofold. First, the extended nature of outflows is not fully captured by the \textit{HST}/STIS spectra. Second, if outflows are more frequent in [O\,I]  than [S\,II] (as seen for ground-based data), the absence or truncation of [O\,I] in 8 of the space-based spectra hinders the interpretation. We found one case (NGC\,4450) for which the inflow interpretation held for all the emission lines (except [S\,II]) in both ground- and space-based data sets.\\

\item[7. ---] \textit{Ionization mechanisms.}  By combining the location of line ratios onto BPTs and the empirical dividing curves between H\,II star-forming regions, Seyfert and LINERs, theoretical models of AGNs, pAGBs and shocks -ionisation and the weak/strong -[O\,I] classification, we exclude the pAGB scenario in favor of the  AGN as the dominant mechanism of ionisation of the narrow component in these LINERs. The [O\,I]/H$\alpha$ values for space-based spectra  also support this result. Generally, the line ratios for the second component are well reproduced  by shock-models. In particular, we combined line ratios with our kinematical classification for the   most reliable optical BPT diagram for shocks (i.e. [O\,I]/H$\alpha$) to confirm the presence of shocks associated to outflows.\\

\item[8. ---] \textit{Neutral gas in type-1 LINERs.} For the NaD absorption lines, in 7 out of 8 cases, a single kinematic component already gives a good fit, suggesting that if a second component exists in these galaxies, it is weak. The  velocities of the narrow components vary between  -165 and 165 km\,s$^{-1}$; the velocity dispersions are in  the range 104-335 km\,s$^{-1}$ (220 km\,s$^{-1}$, on average). The range of velocities and velocity dispersions of the neutral gas are  slightly larger than what we found for narrow components in emission lines in both ground- and space-based spectroscopy, but smaller than those found for second components in both sets of data.  The result of our line modelling suggests that the neutral gas is generally optically thick in these objects. Generally, the neutral gas kinematical components could be interpreted as rotation. The neutral gas velocities are roughly consistent with those of neutral gas discs (detected via NaD absorption) in U/LIRGs but with a larger velocity dispersion considering average values. The possible explanation of the lack of neutral gas outflows, or more in general of non-rotational motions,  is twofold. First, the neutral component in outflows is possibly less significant  in AGNs or quasars  than in starbursts galaxies and U/LIRGs. Second, such null detection rate might be a consequence of the conservative limits we assumed for kinematic components in emission lines, as the different phases of outflows might have different kinematics (see Sect.\,\ref{Disc_neutral}). \\
\end{enumerate}

\noindent Therefore, the kinematics and fluxes that we derive from our spectroscopic data  provided valuable insight into our general understanding of LINERs, in terms of AGN-classification, dominant ionisation mechanism, kinematics  and the possible presence of multiphase-outflows.

\section*{Acknowledgments} 
The manuscript is mainly based on observations collected at the Centro Astron{\'o}mico Hispano Alem{\'a}n (CAHA) at Calar Alto, operated jointly by the Max-Planck Institut f{\"u}r Astronomie and the Instituto de Astrof{\'i}sica de Andaluc{\'i}a (CSIC). \\
Part of the data presented here were obtained  with ALFOSC, which is provided by the Instituto de Astrof{\'i}sica de Andalucia under a joint agreement with the University of Copenhagen and NOTSA.\\
We acknowledge financial support by the Spanish MEC under grants AYA2013-42227-P and AYA2016-76682-C3. MP acknowledges financial support from the Ethiopian Space Science and Technology Institute (ESSTI) under the Ethiopian Ministry of Science and Technology (MoST). OGM acknowledges support from the PAPIIT project IA103118. LHG acknowledges partial support from FONDECYT through grant 3170527. \\
The predictions of AGN and shock  -ionization models were downloaded from the web page \url{http://www.strw.leidenuniv.nl/~brent/itera.html}. This paper made use of the plotting package \textsc{jmaplot}, developed by Jes{\'u}s Ma{\'i}z-Apell{\'a}niz \url{http://jmaiz.iaa.es/software/jmaplot/ current/html/jmaplot_overview.html}. \\
This research has made use of the NASA/IPAC Extragalactic Database (NED), which is operated by the Jet Propulsion Laboratory, California Institute of Technology, under contract with the National Aeronautics and Space Administration.\\
We acknowledge the usage of the HyperLeda database (\url{http://leda.univ-lyon1.fr}).\\
The authors acknowledge very much the anonymous referee for her/his instructive comments that helped to improve the presentation of this paper.

\bibliographystyle{mn2e}
\bibliography{Bibliography.bib}

\appendix
\section[A]{Stellar modelling}
\label{App_templates}
\begin{figure*}
\vspace{2cm}
\includegraphics[trim = 3.1cm 12.875cm 3.4cm 7.225cm, clip=true, width=.8440\textwidth]{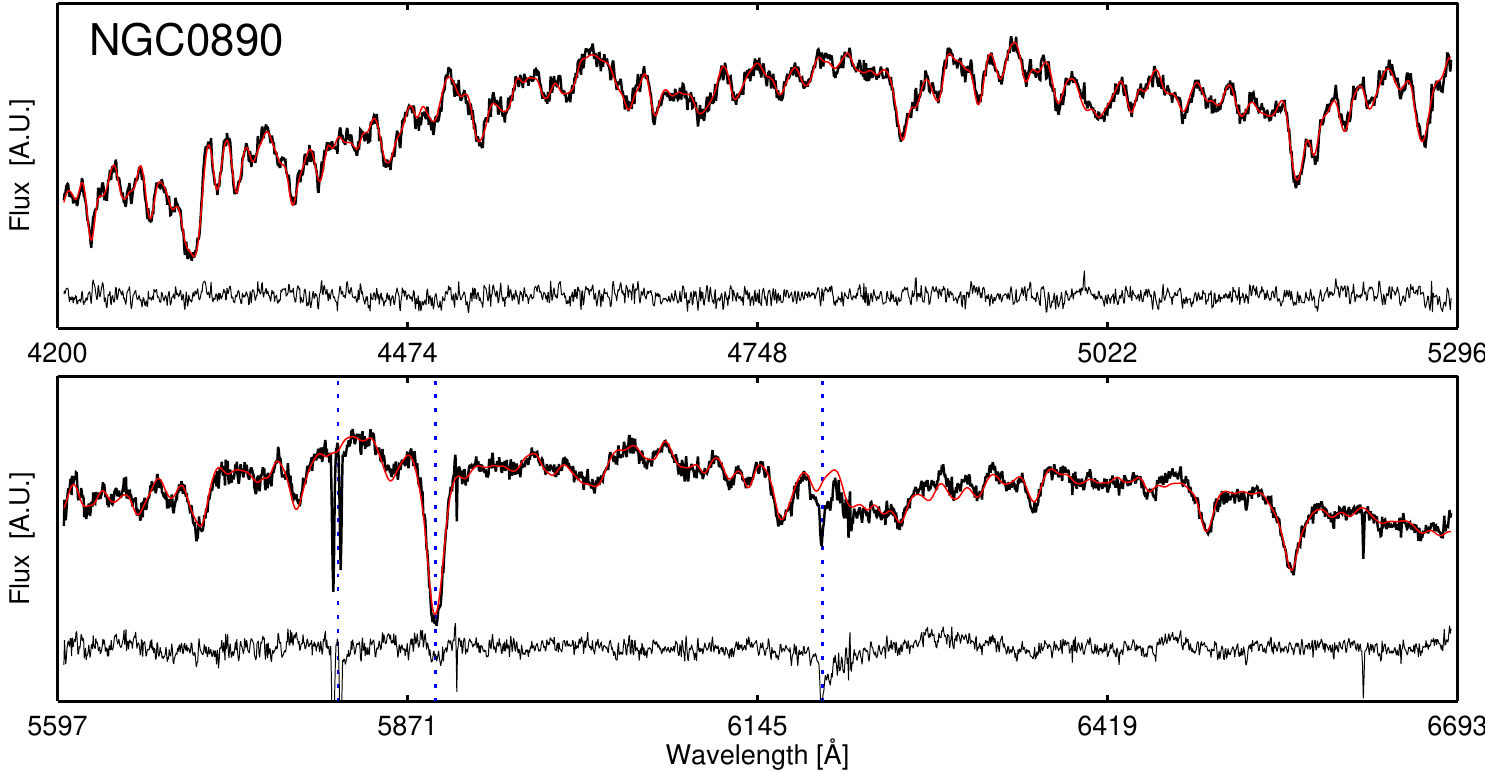}\\
\caption{The optical spectra for the template-galaxies (labeled on top-left) with overlaid its best-fitting stellar spectrum (red line) derived with the \textsc{pPXF} approach (Sect.\,\ref{Analysis_St_Sub}) along with the residuals. The location of the spectral features masked out during the fitting are marked with blue dotted lines.}
\label{Panel_templates} 		 		 
\end{figure*}

\begin{figure*}

\includegraphics[trim = 3.1cm 13.9cm 3.4cm 3.160cm, clip=true, width=.8440\textwidth]{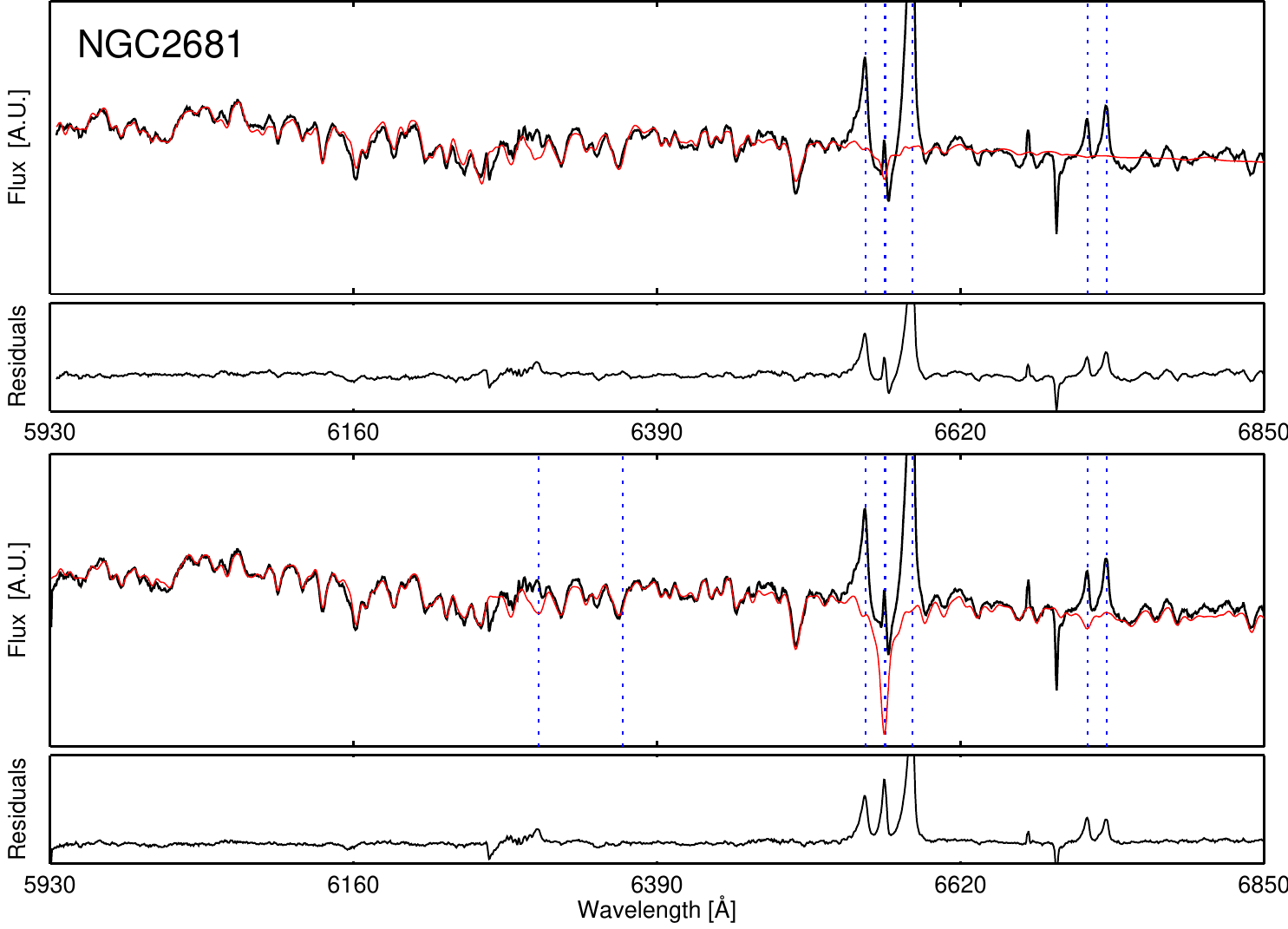}\\
\vspace{.35cm}
\includegraphics[trim = 3.1cm 13.9cm 3.4cm 3.160cm, clip=true, width=.8440\textwidth]{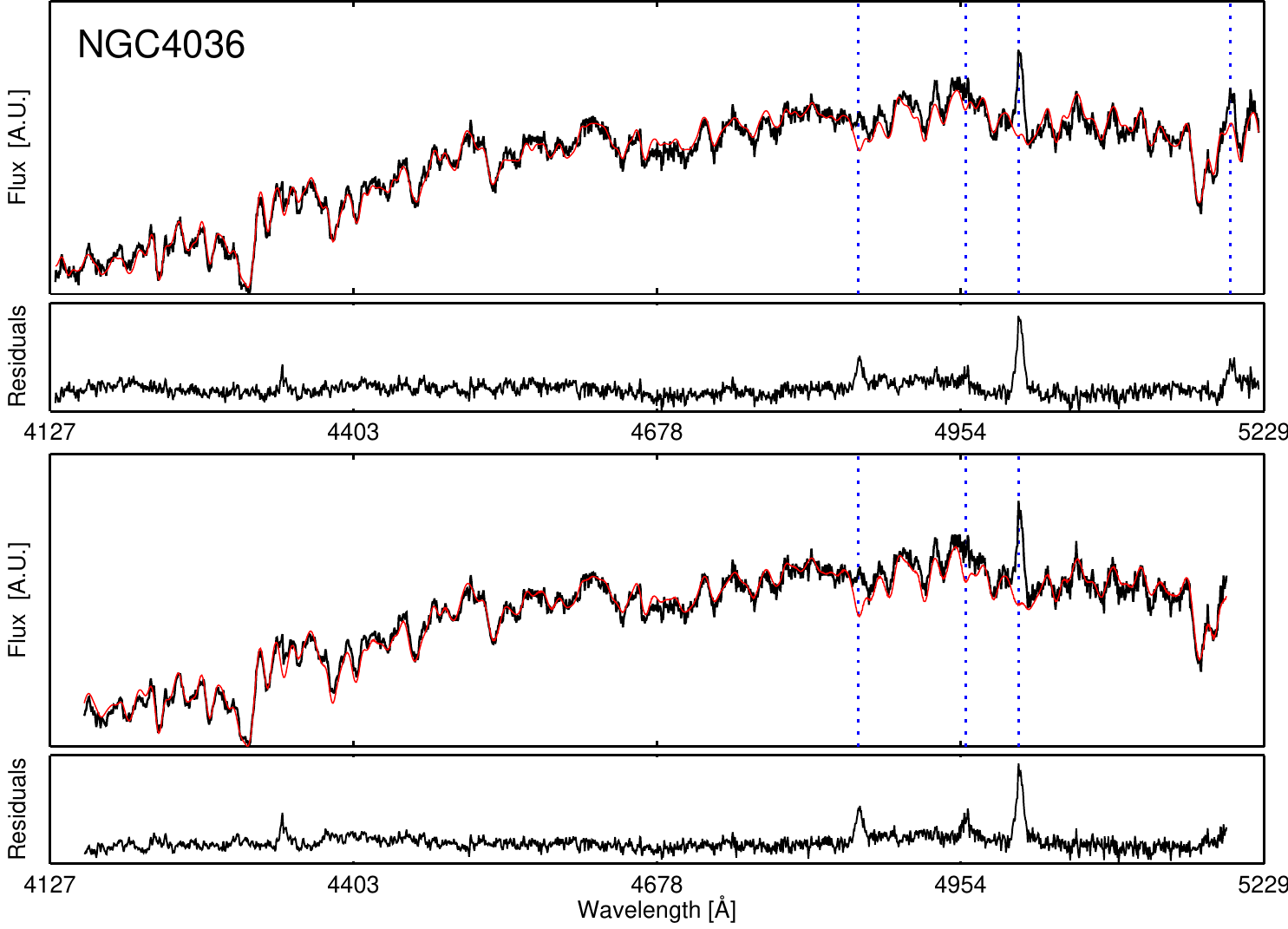}\\
\caption{Stellar continuum modelling performed with different methods for those LINERs with questionable \textsc{pPXF} modelling (Sect.\,\ref{Analysis_St_Sub}). The red line indicates the modelled stellar spectrum that matches the observed continuum, obtained applying the \textsc{pPXF} (top) and \textsc{STARLIGHT} (bottom) methods (Sect.\,\ref{Analysis_St_Sub}).  The location of the spectral features masked out during the fitting are marked with blue dashed lines. The spectra are shown with a zoomed view to highlight weak continuum features. Moreover the region redward to $\sim$\,6850\,\AA \ is not considered in these plots as dominated by atmospheric absorption. The lower panels show  the residuals from the stellar subtraction after the stellar continuum modeling. For all these cases the selected model  is \textsc{STARLIGHT}  (see also Table\,\ref{T_kin}). }
\label{Panel_comparison_stars} 		 		 
\end{figure*}

Table\,\ref{T_Obs_templ} summarises the details of the optical observations for the ten non-active galaxies  selected from the work of \textit{HFS97} (See Sections \ref{Sample_data} and \ref{Analysis_St_Sub}).\\
Fig.\,\ref{Panel_templates} presents  the stellar continuum modelling  of the spectra of template galaxies obtained with the \textsc{pPXF} procedure. The spectrum and the continuum model for NGC\,4026 are shown in Fig.\,\ref{Panel_template}, thus the correspondent panel is omitted here.\\
In Fig.\,\ref{Panel_comparison_stars} we compare the results of the stellar continuum fitting from  \textsc{pPXF} and \textsc{STARLIGHT} (Sect.\,\ref{Analysis_St_Sub}). We show this comparison only for those LINERs for which the  \textsc{pPXF} modelling is questionable hence we finally selected the \textsc{STARLIGHT} method (Sect.\,\ref{Analysis_St_Sub}). Fig.\,\ref{Panel_stsub} presents the stellar continuum modelling for the  LINER NGC\,4203.\\
Finally, Table\,\ref{T_rms} summarises all the values considered to prevent overfitting emission lines (Sect.\,\ref{Analysis_LF}).
\begin{table*}
  \caption{Optical observations details for normal galaxies used as template to test starlight subtraction.}
  \begin{tabular}{@{}lc c c c c c c  c c @{}}
  \hline
ID                 & RA               & DEC              & Night             & EXP & Air mass  &  Seeing  & Slit PA &   Nuclear aperture& Emission lines\\
                &                &             &              & (s) &   & ($^{\prime}$$^{\prime}$) & ($^{\circ}$) &   ($^{\prime}$$^{\prime}$) & \\
   \hline
NGC\,0890 & 02 22 01.0  & +33 15 58     & 22 Dec 2014 & 3\,$\times$\,1800 & 1.10 &  0.6 & 67 & 1.2\,$\times$\,2.2 & no \\
NGC\,1023 & 02 40 24.0  & +39 03 48    &  03 Dec 2012& 3\,$\times$\,1200   & 1.43  & 1.5 &  250 &  1.2\,$\times$\,1.7 & no\\
NGC\,2950 & 09 42 35.1  & +58 51 05    & 04 Dec 2012 & 3\,$\times$\,1200     & 1.08  & 1.0 &  270 & 1.2\,$\times$\,1.7 & yes \\
NGC\,3838 & 11 44 13.7  & +57 56 54    & 11 Apr 2013 & 3\,$\times$\,2000   & 1.10  & 1.6 &  247 & 1.5\,$\times$\,3.4 & yes \\
NGC\,4026 & 11 59 25.2  & +50 57 42    &  10 Apr 2013 & 2\,$\times$\,1500   &  1.20 & 1.2 &  102 & 1.2$\times$\,2.8  & yes \\	      
NGC\,4339 & 12 23 34.9  & +06 04 54    & 13 Apr 2013 & 3\,$\times$\,1800 & 1.17   & 1.2 &  170  &  1.2\,$\times$\,2.2 &no \\
NGC\,4371 & 12 24 55.4  & +11 42 15    &  12 Apr 2013& 3\,$\times$\,1500 &  1.11 & 1.0 &     159 & 1.2\,$\times$\,2.2 & yes \\
NGC\,4382 & 12 25 24.1  & +18 11 29    & 15 Apr 2013 & 4\,$\times$\,1200 &    1.72 & 1.5   & 55& 1.2\,$\times$\,1.7 & yes \\
NGC\,5557 & 14 18 25.7  & +36 29 37    & 14 Apr 2013 & 4\,$\times$\,1500 &    1.21 & 1.5  &75 & 1.2\,$\times$\,2.2 & no\\
NGC\,7332$^{*}$  & 22 37 24.5  & +23 47 54   &  29 Sep 2013 & 3\,$\times$\,1200& 1.15 &  0.7 & 107 & 1.2\,$\times$\,3.8 & yes \\ 
\hline
\end{tabular}
\begin{flushleft}
\textit{Notes.}  \lq ID\rq: object designation. \lq RA\rq \ and  \lq DEC\rq \  are the coordinates. \lq Night\rq: date the object was observed.  \lq EXP\rq: exposure time for the observations.  \lq Air mass\rq: full air mass range of the observations.  \lq Slit PA\rq : slit position angle of the observations (as measured from North and eastwards on the sky). \lq Nuclear aperture\rq :  indicated as the angular size of the nuclear region corresponding to the spectra presented in this work. The nuclear aperture is indicated as:  slit width $\times$ selected region during the extraction of the final spectrum. The last column indicate if any emission lines originated in the ISM are present in the stellar subtracted spectra. $^{*}$  marks the galaxy-template observed with ALFOSC/NOT instead of TWIN/CAHA.\\
\end{flushleft}
\label{T_Obs_templ}
\end{table*}

\begin{figure*}
\vspace{-0.1cm}
\includegraphics[trim = 3.1cm 12.875cm 3.4cm 7.225cm, clip=true, width=.8440\textwidth]{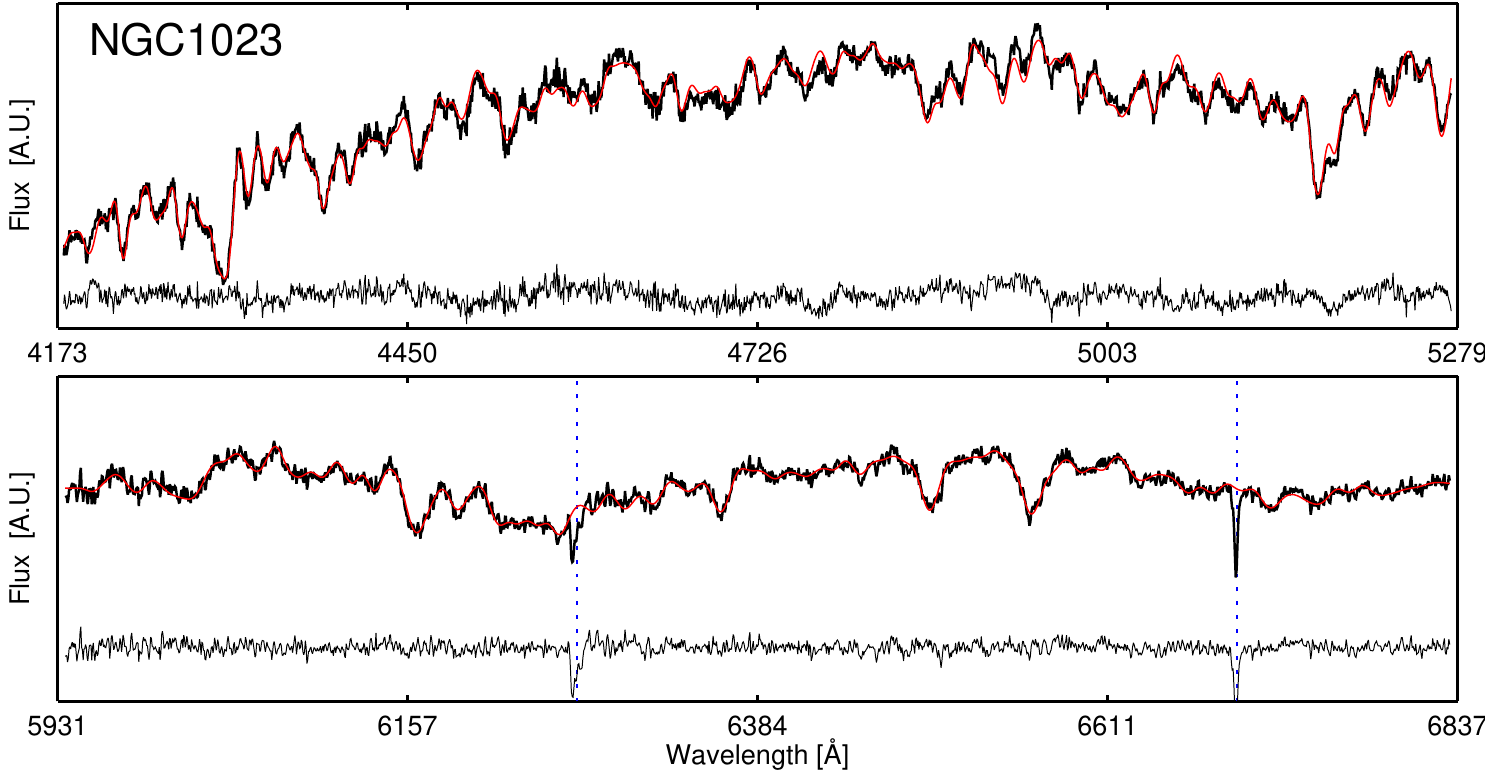}\\
\vspace{.1cm}
\includegraphics[trim = 3.1cm 12.875cm 3.4cm 7.225cm, clip=true, width=.8440\textwidth]{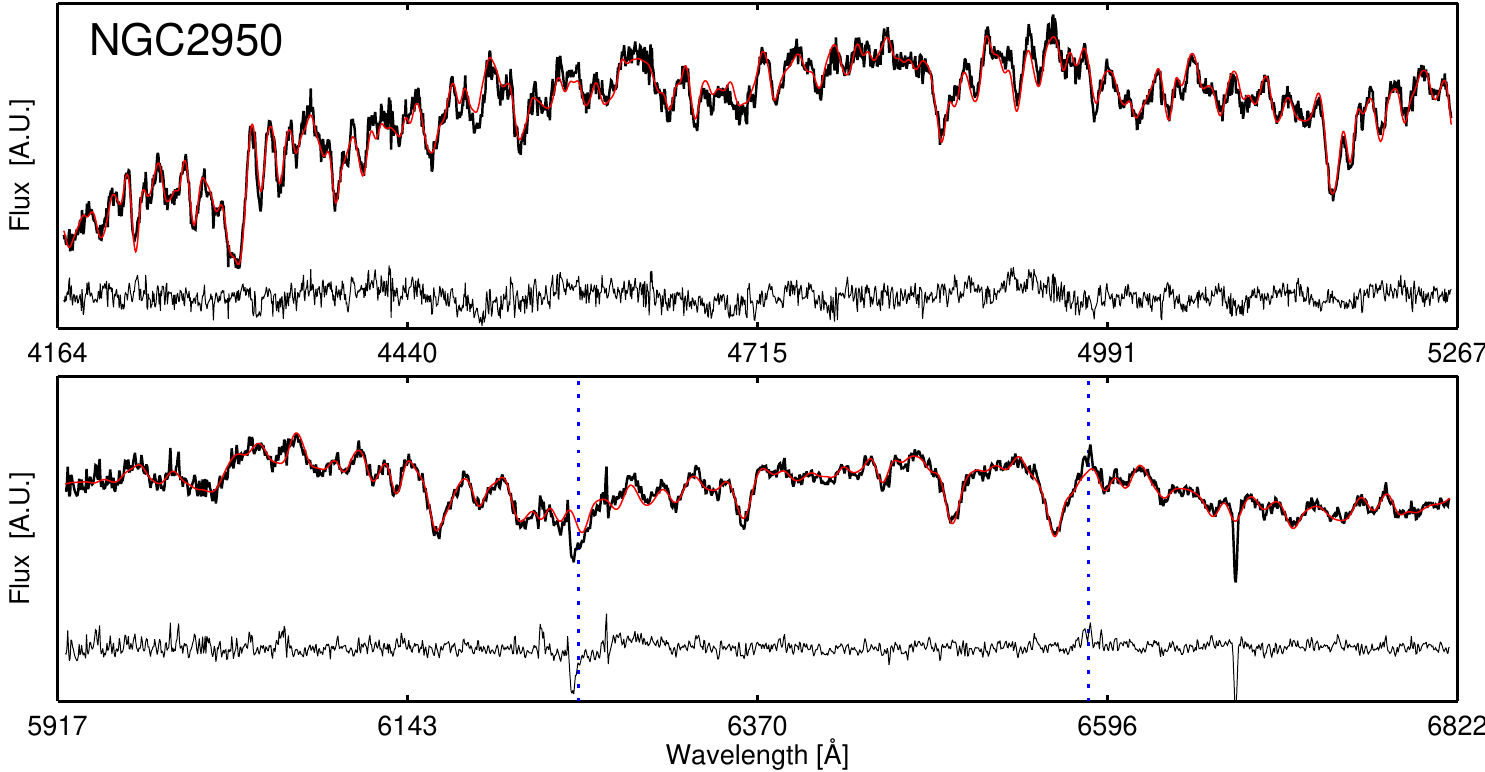}\\
\vspace{.1cm}
\includegraphics[trim = 3.1cm 12.875cm 3.4cm 7.225cm, clip=true, width=.8440\textwidth]{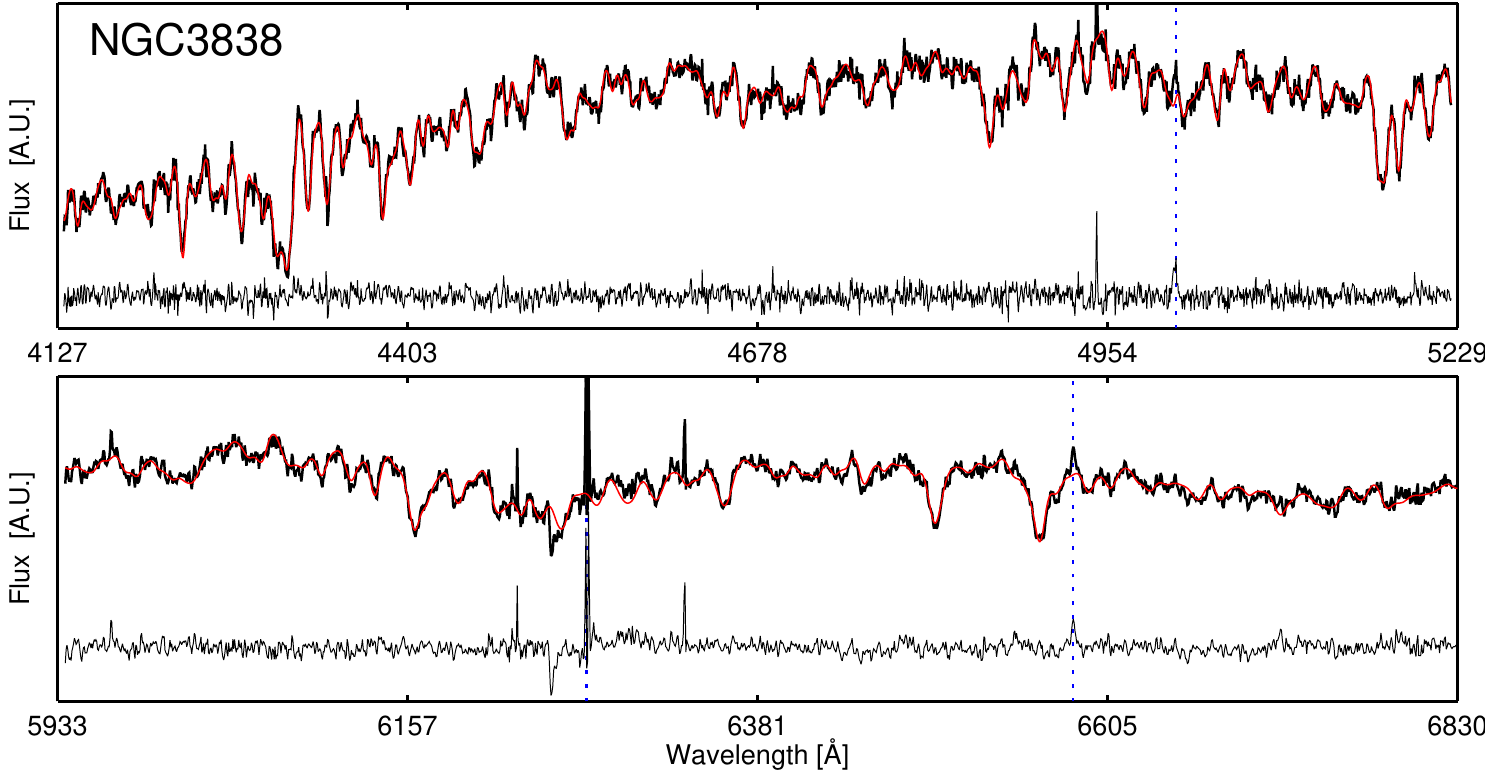}\\
 \captionsetup{labelformat=empty}{Fig.\,\ref{Panel_templates}\,--\,Continued.} 	
\end{figure*}

\begin{figure*}
\includegraphics[trim = 3.1cm 12.875cm 3.4cm 7.225cm, clip=true, width=.8440\textwidth]{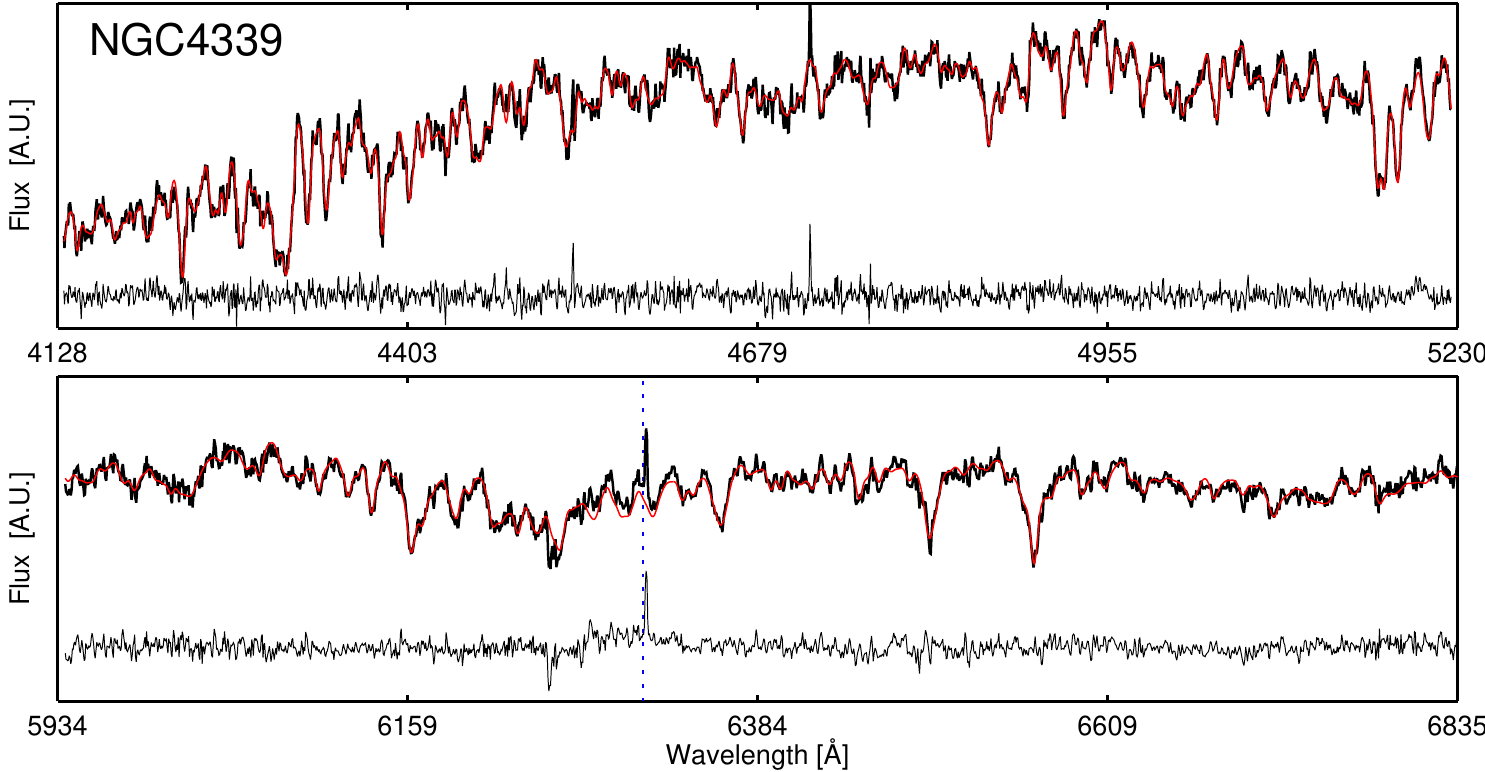}\\
\vspace{.2cm}
\includegraphics[trim = 3.1cm 12.875cm 3.4cm 7.225cm, clip=true, width=.8440\textwidth]{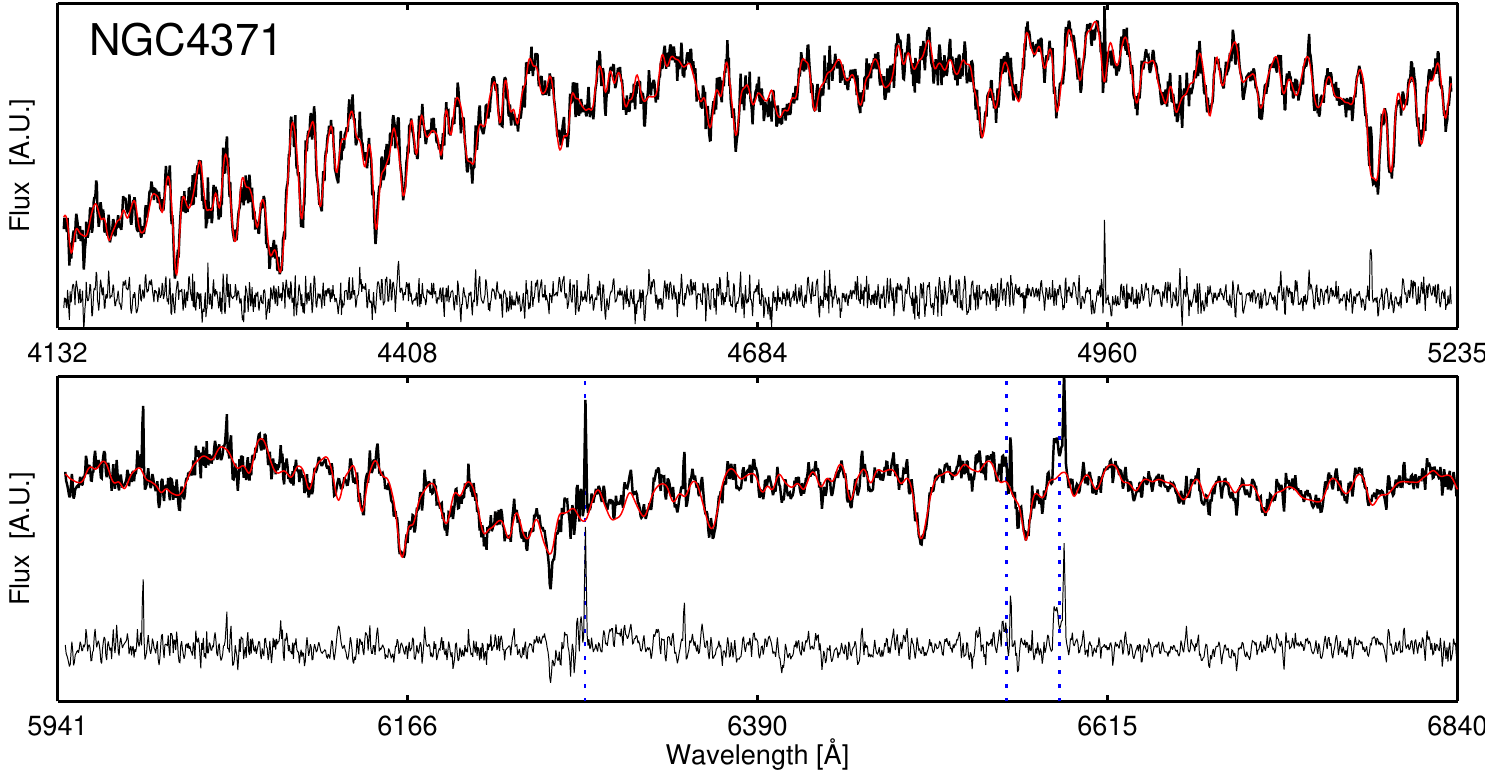}\\
\vspace{.2cm}
\includegraphics[trim = 3.1cm 12.875cm 3.4cm 7.225cm, clip=true, width=.8440\textwidth]{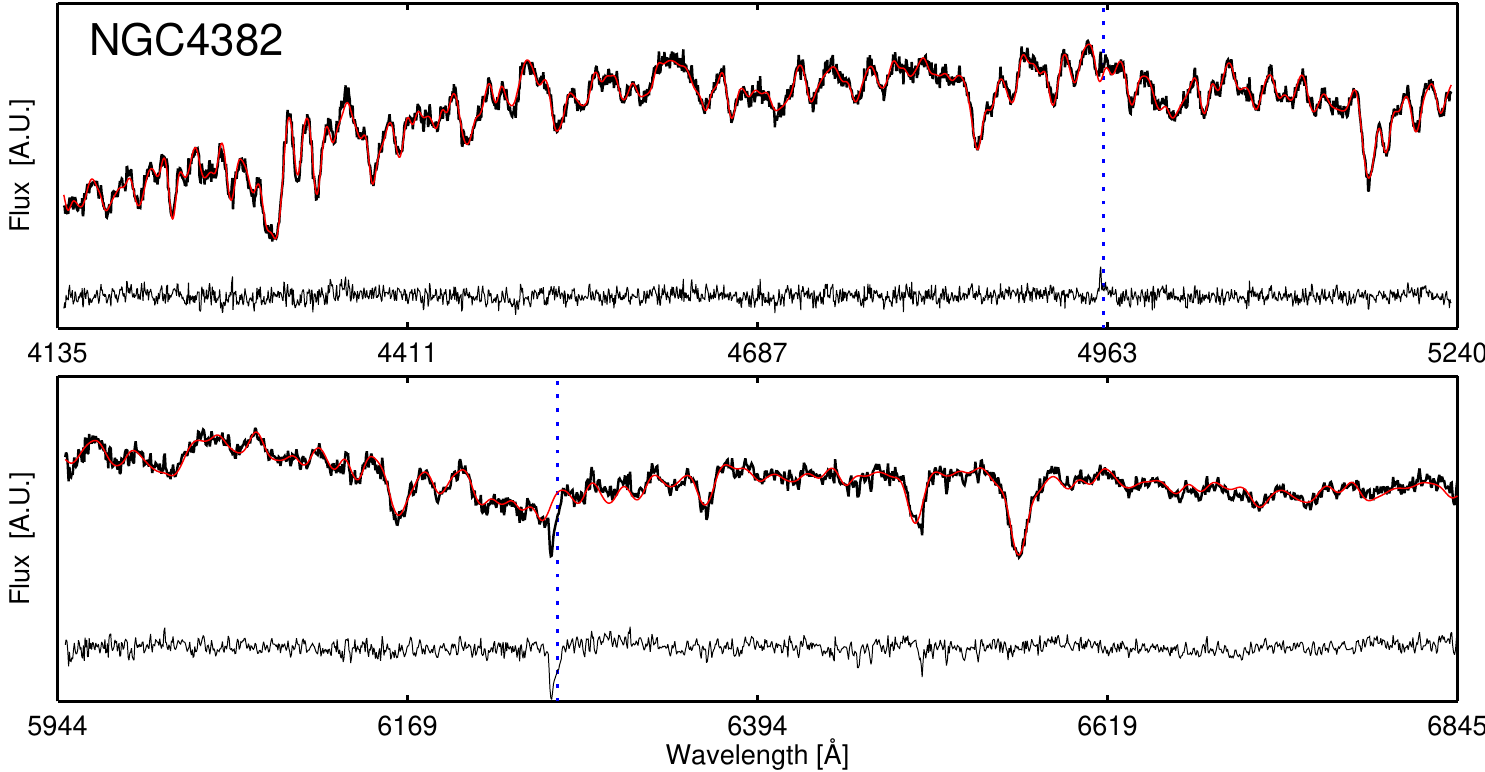}\\
 \captionsetup{labelformat=empty}{Fig.\,\ref{Panel_templates}\,--\,Continued.} 			 
\end{figure*}

\begin{figure*}
\includegraphics[trim = 3.1cm 12.875cm 3.4cm 7.225cm, clip=true, width=.8440\textwidth]{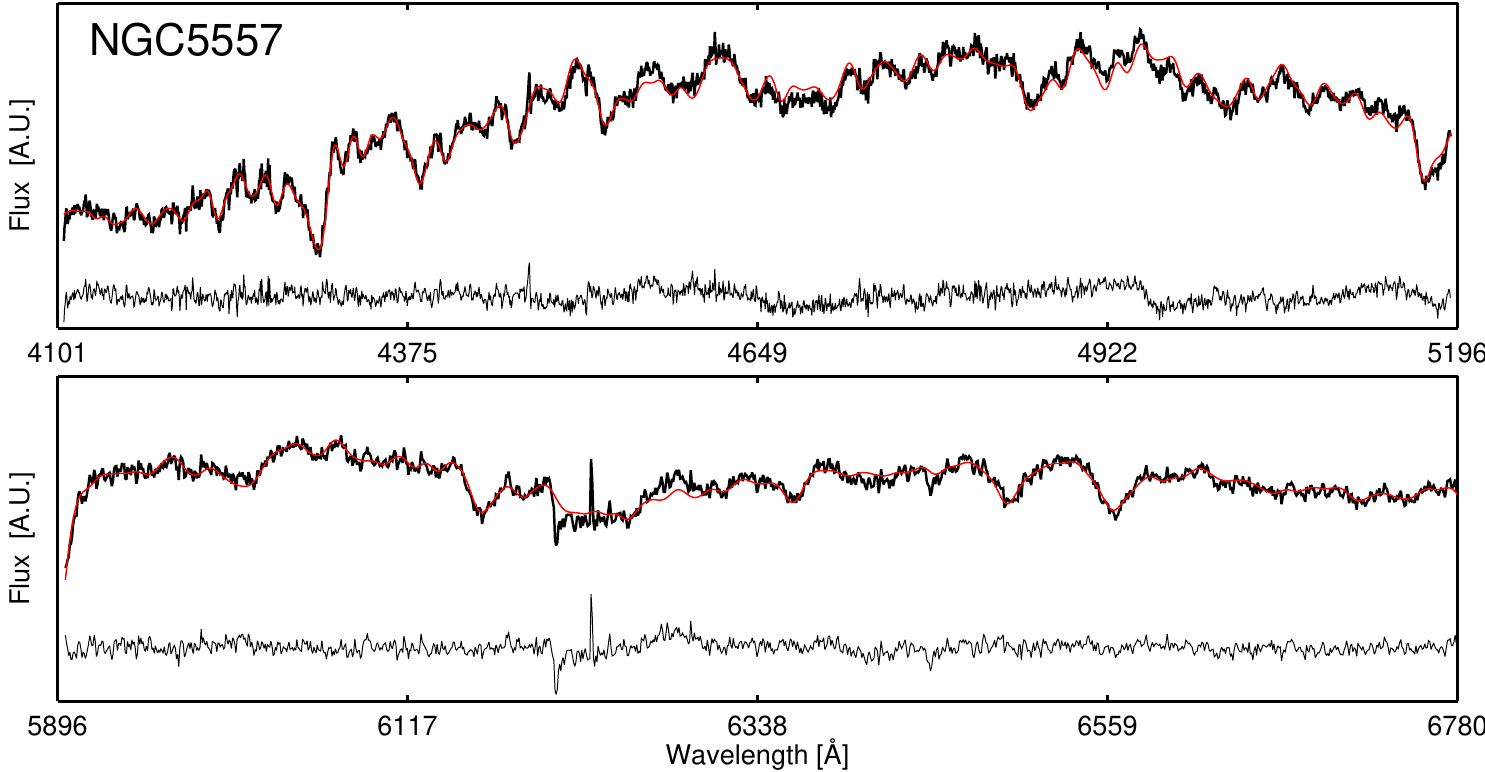}\\
\vspace{.2cm}
\includegraphics[trim = 3.1cm 12.875cm 3.4cm 7.225cm, clip=true, width=.8440\textwidth]{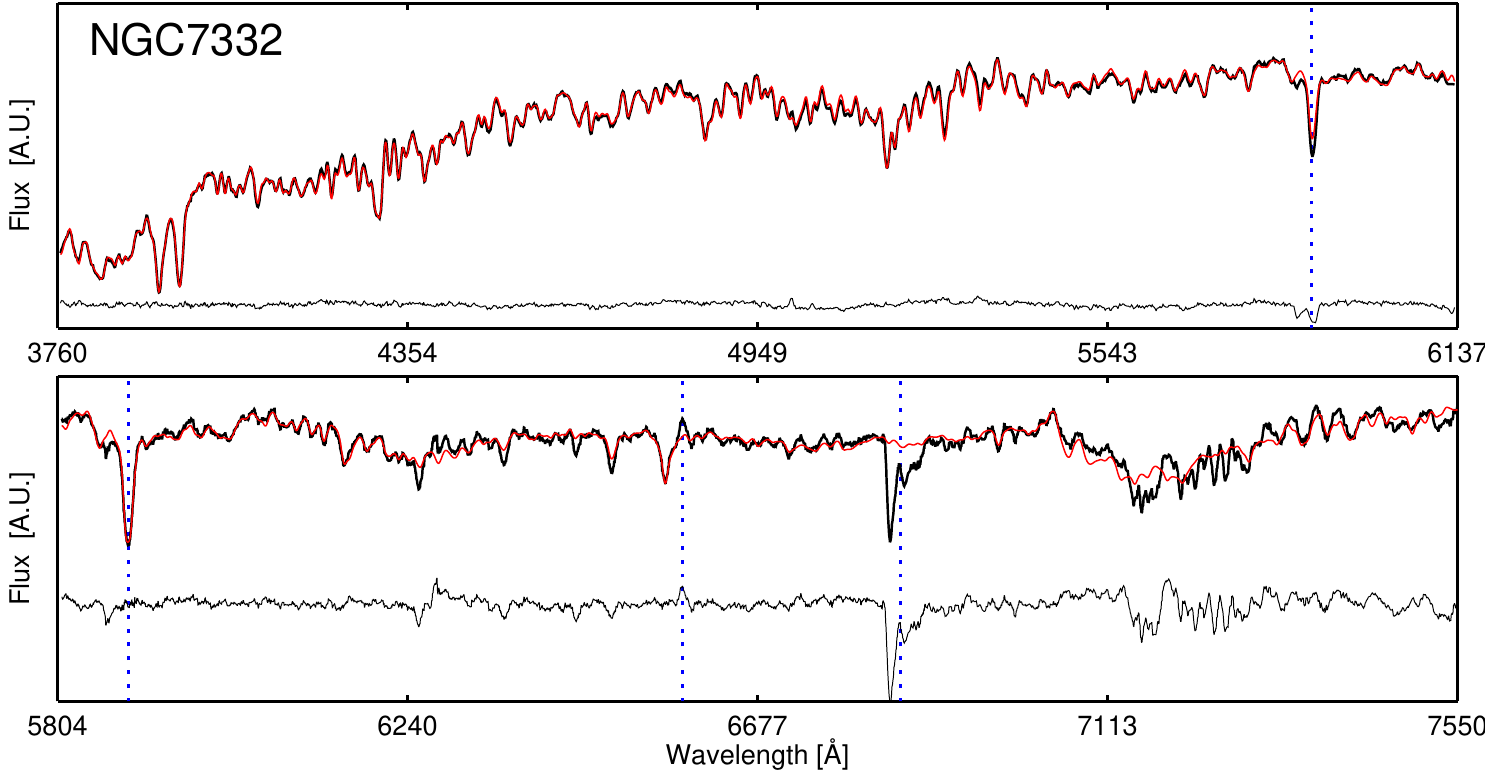}\\
 \captionsetup{labelformat=empty}{Fig.\,\ref{Panel_templates}\,--\,Continued.} 	 		 
\end{figure*}
\clearpage

\begin{figure*}
\includegraphics[trim = 3.1cm 13.9cm 3.4cm 3.160cm, clip=true, width=.8440\textwidth]{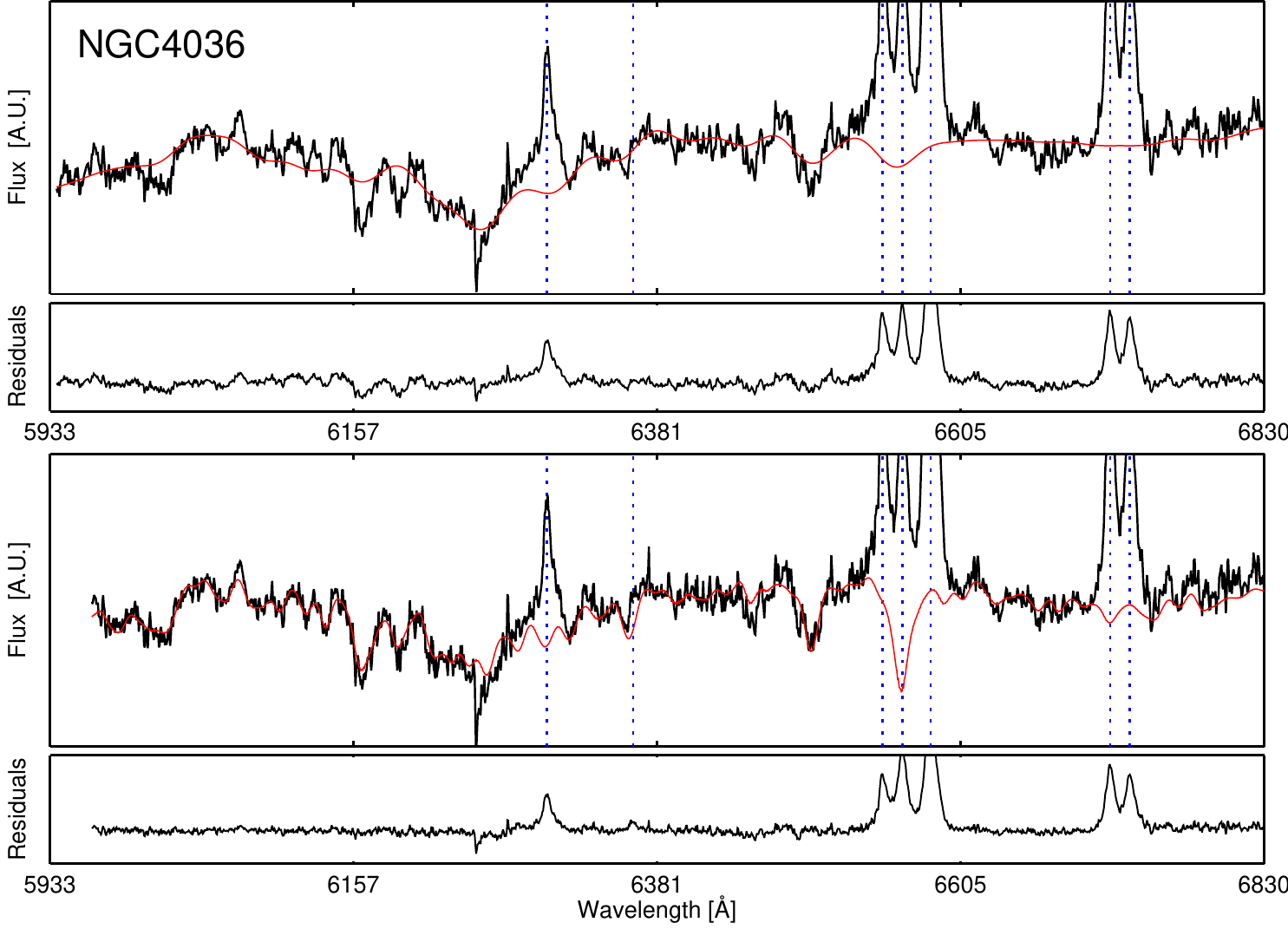}\\
\vspace{.35cm}
\includegraphics[trim = 3.1cm 13.9cm 3.4cm 3.160cm, clip=true, width=.8440\textwidth]{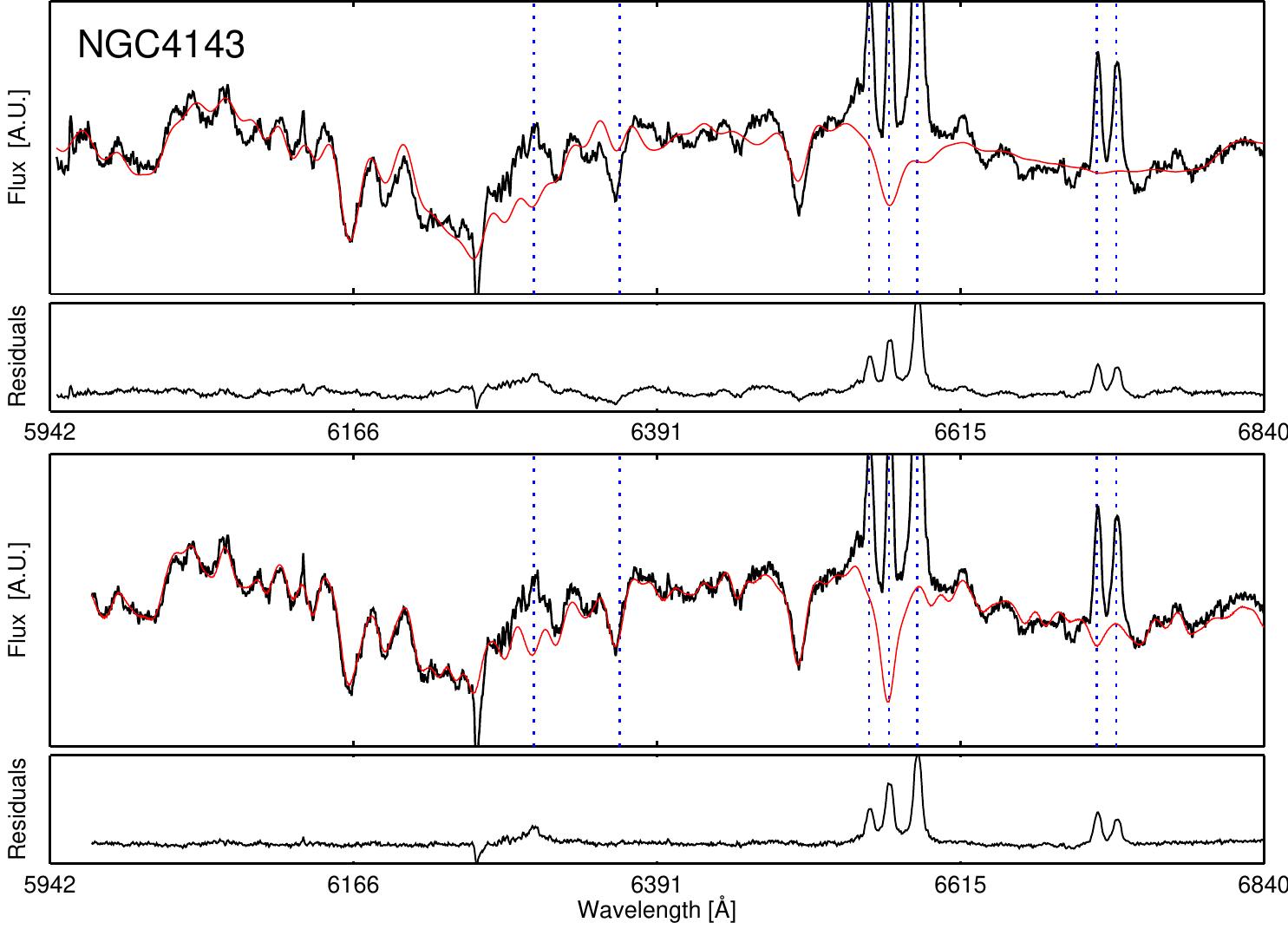}\\
 \captionsetup{labelformat=empty}{Fig.\,\ref{Panel_comparison_stars}\,--\,Continued.} 			 
\end{figure*}

\begin{figure*}
\includegraphics[trim = 3.1cm 13.9cm 3.4cm 3.160cm, clip=true, width=.8440\textwidth]{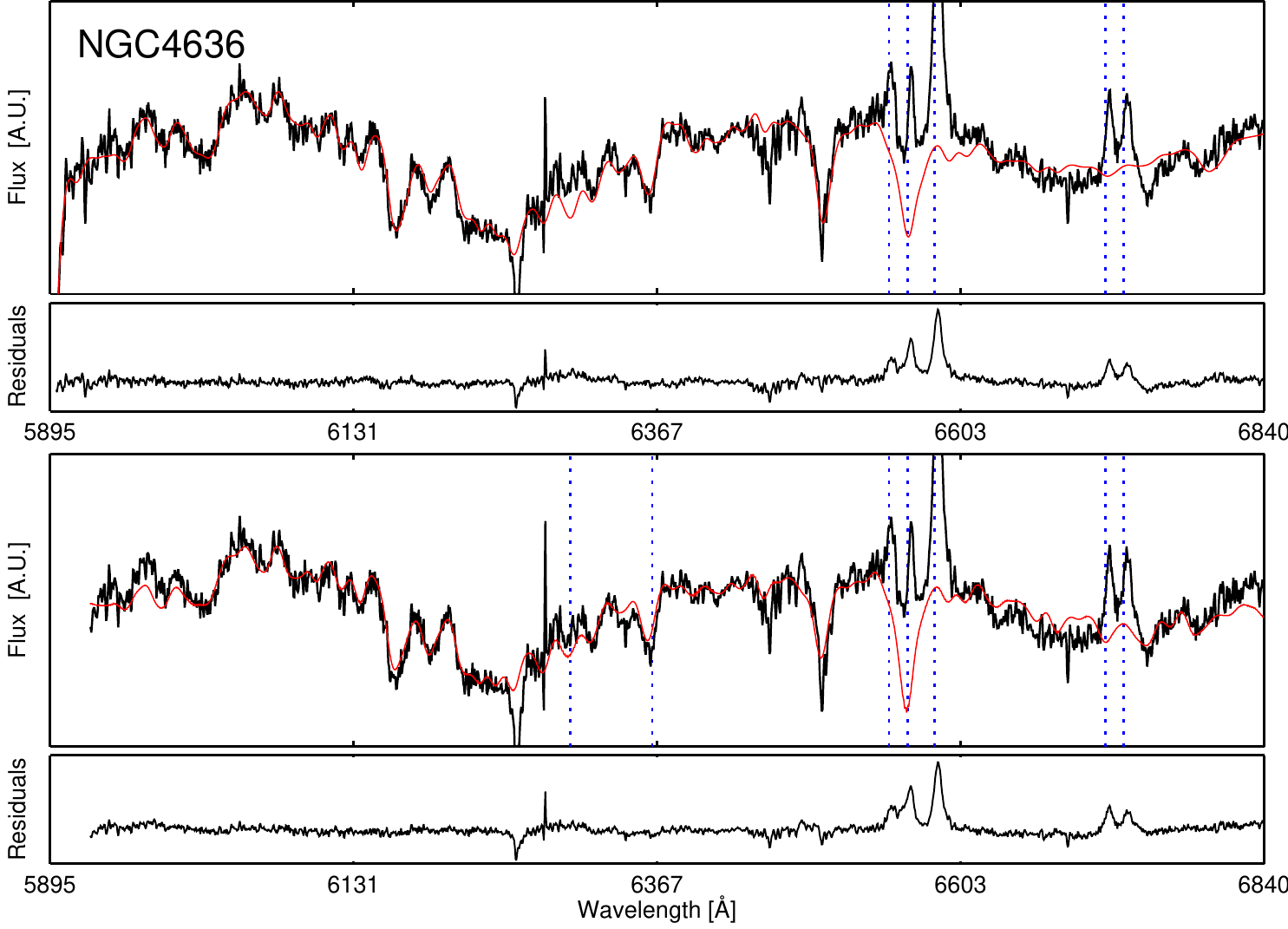}\\
\vspace{.35cm}
\includegraphics[trim = 3.1cm 13.9cm 3.4cm 3.160cm, clip=true, width=.8440\textwidth]{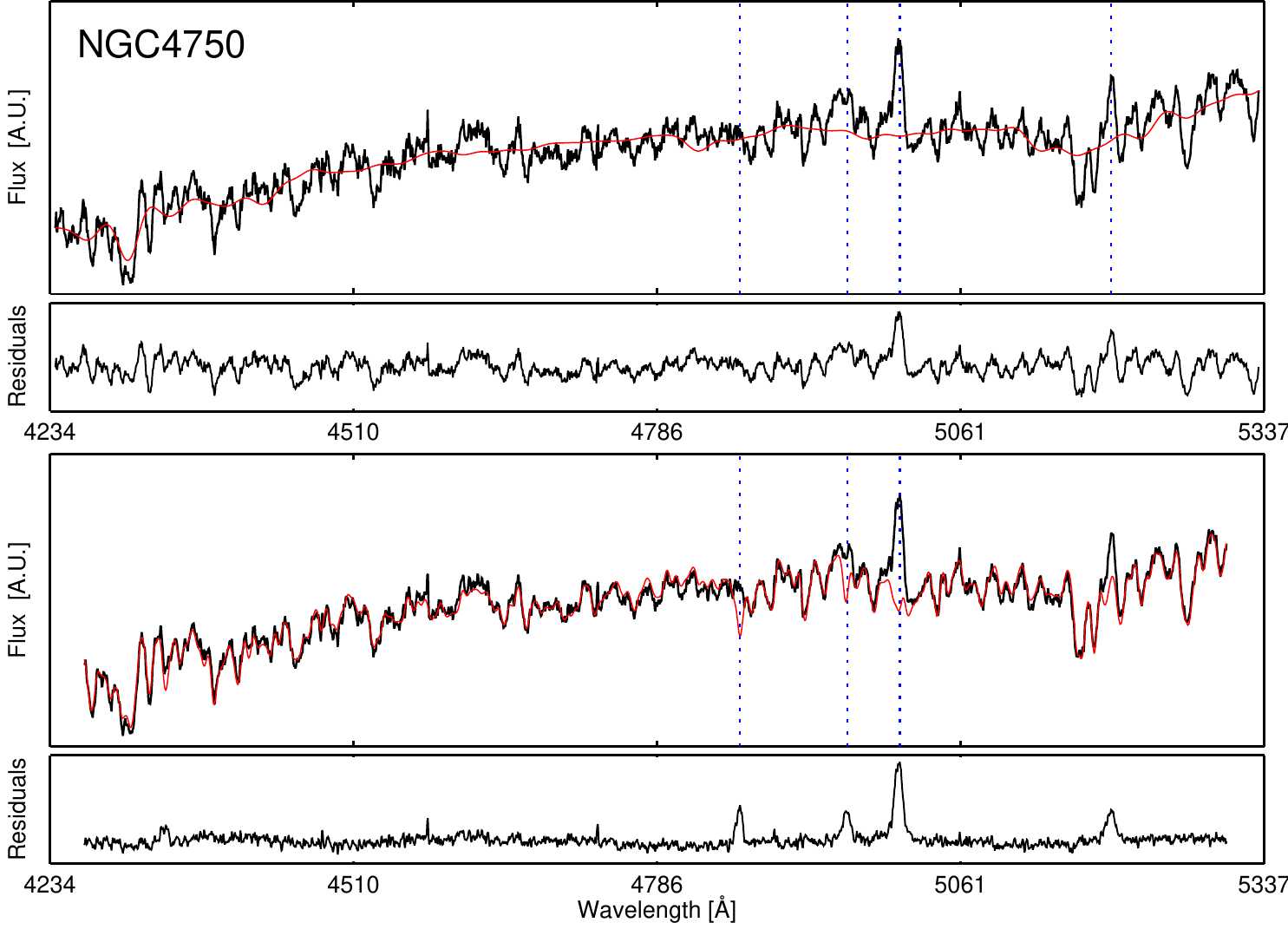}\\
 \captionsetup{labelformat=empty}{Fig.\,\ref{Panel_comparison_stars}\,--\,Continued.} 			 
\end{figure*}

\begin{figure*}
\includegraphics[trim = 3.1cm 13.9cm 3.4cm 3.160cm, clip=true, width=.8440\textwidth]{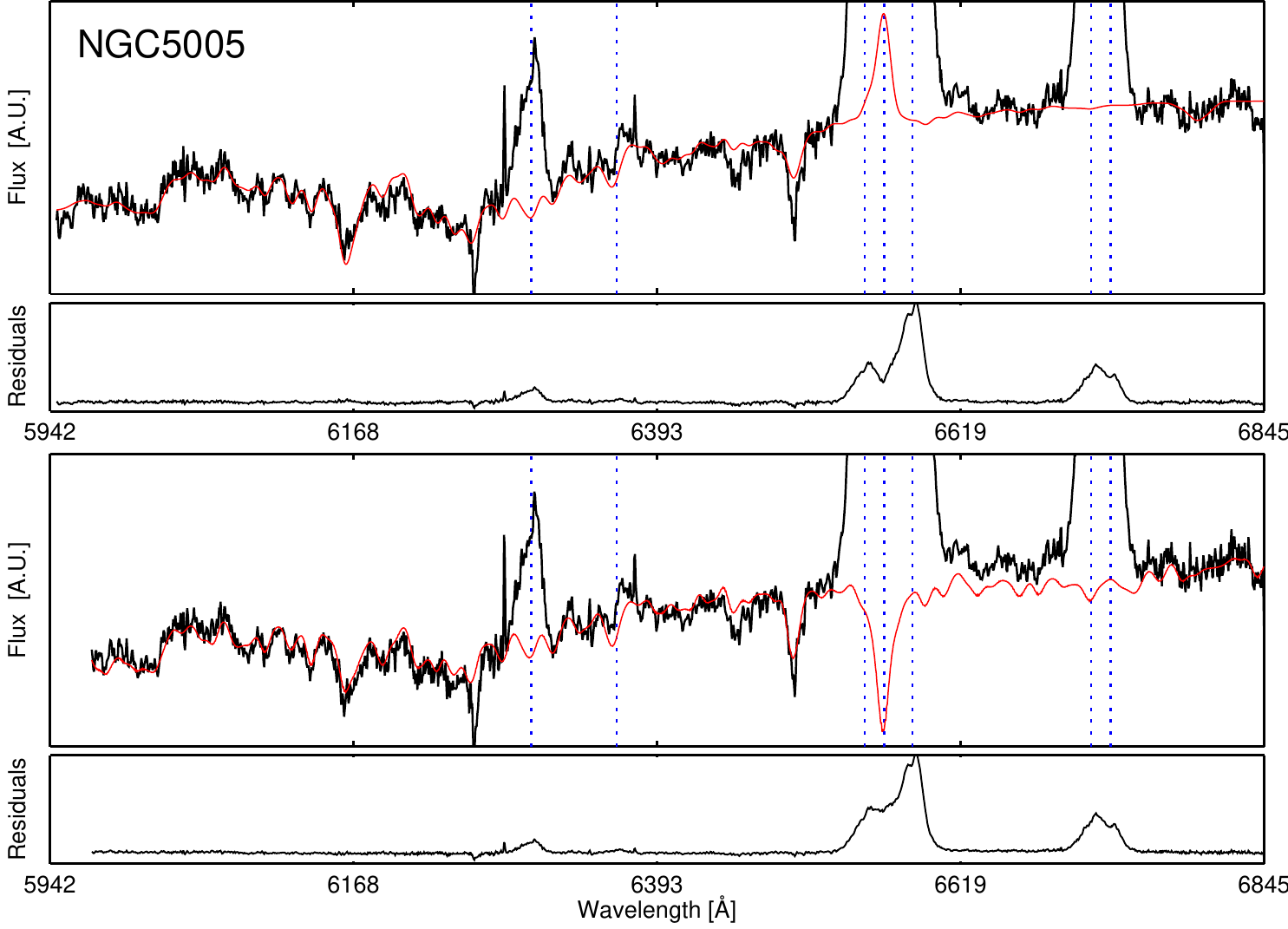}\\
\vspace{.35cm}
\includegraphics[trim = 3.1cm 13.9cm 3.4cm 3.160cm, clip=true, width=.8440\textwidth]{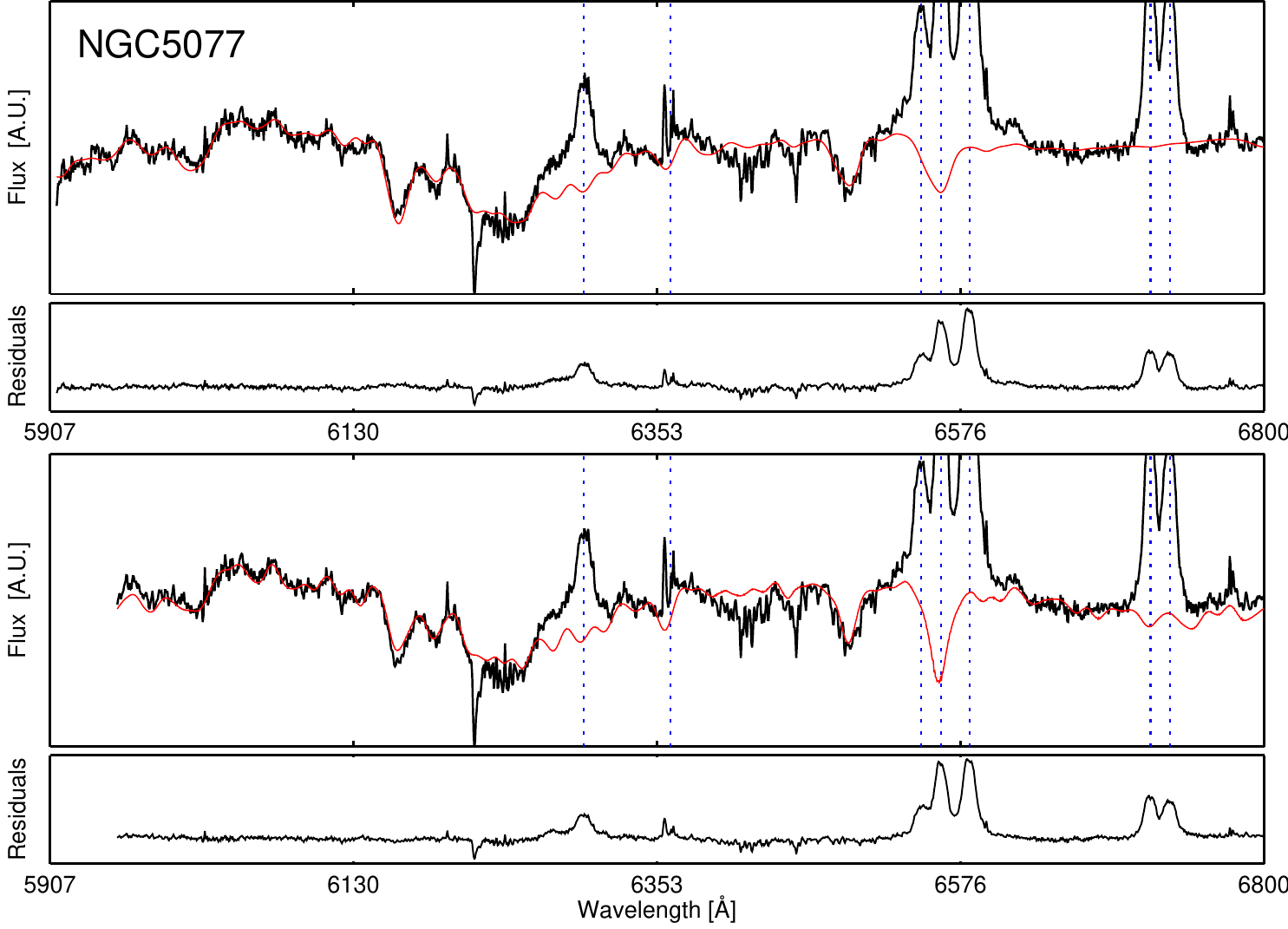}\\
 \captionsetup{labelformat=empty}{Fig.\,\ref{Panel_comparison_stars}\,--\,Continued.} 	 		 
\end{figure*}
\clearpage

\begin{table*}
\caption[Overfitting]{Summary of the values evaluated to prevent overfitting emission lines.}
\begin{tabular}{l l c c c c c c c c c }
\hline 						
ID        &  Obs.             & Mod. & Comp. & $\varepsilon_{\rm c}$ & $\varepsilon$$^{\rm [O\,I]}$$_{\rm N}$& $\varepsilon$$^{\rm [O\,I]}$$_{\rm N+S}$ &  $\varepsilon$$^{\rm [S\,II]}$$_{\rm N}$& $\varepsilon$$^{\rm [S\,II]}$$_{\rm N+S}$ & $\varepsilon$$^{\rm H\alpha}$$_{\rm N(+S)}$&$\varepsilon$$^{\rm H\alpha}$$_{\rm N+B(+S)}$\\
\hline 
NGC\,0226 & CAHA$^{\dagger}$  & O & N\,+\,S	  & 0.0143 & 1.3893 & 1.5088 &  ---   & ---    & 2.3668 & ---    \\
NGC\,0315 & [NOT]              & M & N\,+\,S\,+\,B & 0.0061 & 2.9184 & 1.9224 & 4.2010 & 1.9160 & 5.0966 & 5.4522 \\
          & HST $^{\dagger}$  & S & N\,+\,S\,+\,B & 12.764 &  ---   &  ---   & 3.6023 & 2.1042 & 7.6608 & 2.2244 \\	 	  	
NGC\,0841 & CAHA              & S & N		  & 0.0179 & 1.2525 & ---    & 0.7633 & ---    & 1.0398 & ---    \\
NGC\,1052 & [NOT]    	      & M & N\,+\,S 	  & 0.0436 & 5.2044 & 2.6391 & 4.8007 & 1.9587 & 7.6481 & ---    \\
  	  & [HST]               & M & N\,+\,S\,+\,B & 10.535 & 7.3194 & 2.5107 & 5.6332 & 2.5163 & 11.893 & 5.4754 \\
NGC\,2681 & CAHA   	      & O & N\,+\,S 	  & 0.0042 & 2.6604 & 2.5589 & ---    & 1.4102 & 2.3355 & 1.0018 \\
NGC\,2787 & CAHA              & S & N\,+\,B 	  & 0.0221 & 1.3256 & ---    & 1.0097 & ---    & 1.3392 & ---    \\
 	  & HST $^{\ddagger}$ & S & N\,+\,B 	  & 61.142 & 0.8707 &  ---   & 0.8906 &   ---  & 1.7578 & 1.0298 \\
NGC\,3226 & CAHA 	      & O & N\,+\,S 	  & 0.0134 & 1.3636 & 1.1753 & ---    & 1.7574 & 2.5608 &  \\
NGC\,3642 & [CAHA]              & M & N\,+\,S\,+\,B & 0.0163 & 1.7874 & 1.7838 & 3.5604 & 1.2728 & 4.4198 & 4.3211  \\
          & [HST] $^{\dagger}$  & S & N\,+\,B  	  & 24.277 &  ---   &  ---   & 1.5359 &   ---  & 8.0335 & 6.2624 \\          
NGC\,3718 & CAHA              & M & N\,+\,B 	  & 0.0152 & 1.7210 & ---    & 1.0180 & ---    & 1.8414 & 1.4384	 \\
NGC\,3884 & CAHA $^{\dagger}$ & O & N\,+\,S 	  & 0.0066 & 1.3960 & 1.2535 & ---    & ---    & 1.8567 & ---	 \\
NGC\,3998 & [CAHA]              & O & N\,+\,S 	  & 0.0520 & 3.9890 & 2.5499 & ---    & 2.4168 & 3.7271 & ---	 \\
          & HST $^{\ddagger}$ & M & N\,+\,B 	  & 313.70 & 1.9269 &  ---   & 1.1459 &  ---   & 7.983  & 3.0695 \\
NGC\,4036 & CAHA 	      & M & N\,+\,S 	  & 0.0311 & 1.1460 &1.0271 & 1.0590 & 1.0142 & 1.3522 & ---\\
          & HST               & M & N\,+\,B 	  & 9.0111 & 0.9509 &  ---   & 0.8204 &  ---   & 2.6777 & 2.4079    \\	  
NGC\,4143 & CAHA 	      & M & N\,+\,S 	  & 0.0047 & 2.3230 & 2.0726 & 1.7070 & 1.4728 & 3.2108 & ---	 \\
          & HST $^{\ddagger}$ & S & N\,+\,S\,+\,B & 45.540 &   ---  & 1.4536 & 1.2072 & 1.1482 & 2.3801 & 1.5893   \\
NGC\,4203 & CAHA              & M & N\,+\,S\,+\,B & 0.0099 & 2.0212 & 1.3327 & 1.7502 & 1.1678 & 6.0657 & 2.8283 \\
          & [HST]               & M & N\,+\,S\,+\,B & 31.058 & 4.7849 & 2.1168 & 4.0582 & 1.6267 & 18.517 & 5.7687   \\
NGC\,4278 & CAHA              & M & N\,+\,S 	  & 0.0116 & 2.5285 & 2.1568 & 2.2769 & 1.8593 & 3.0817 & ---	 \\
          & HST $^{\ddagger}$ & M & N\,+\,S\,+\,B & 6.6675 & 1.6094 & 1.4464 & 1.3310  & 0.9954 & 2.3561 & 1.9355    \\	   
NGC\,4438 & [CAHA]              & M & N\,+\,S 	  & 0.0095 & 2.9692 & 2.4282 & 3.7552 & 1.1571& 4.3757 & ---	 \\
NGC\,4450 & CAHA	      & M & N\,+\,S 	  & 0.0284 & 1.4971 & 1.1393 & 1.0932 & 0.5911 & 1.3655 & ---	 \\
          & HST               & M & N\,+\,S\,+\,B & 33.585 & 2.0183 & 1.5548 & 2.4561 & 1.8462 & 7.9210 & 2.7456 \\
NGC\,4636 & CAHA	      & S & N\,+\,B 	  & 0.0088 & 1.9763 & ---    & 1.1856 & ---    & 2.2206 & 1.8684\\
NGC\,4750 & CAHA 	      & O & N\,+\,S\,+\,B & 0.0180 & 1.6476 & 1.4759 & ---    & 1.0711 & 2.1151 & 2.3040 \\
NGC\,4772 & CAHA              & S & N		  & 0.0121 & 1.6562 & ---    & 1.8778 & ---    & 2.0312 & ---	 \\
NGC\,5005  & [CAHA] & S & N\,+\,S	  & 0.0299 & ---    & 2.0176 & 2.1457 & 2.4765 & 3.5698 & ---	 \\
          & HST $^{\dagger}$              & S & N\,+\,S\,+\,B & 31.407 &  ---   &  ---   & 1.9060 & 1.3951 & 2.0732 & 1.1658   \\	  
NGC\,5077  &CAHA  & S & N\,+\,B 	  & 0.0202 & 1.2003 & ---    & 0.5672 & ---    & 1.5123 & 0.8314 \\
          & HST $^{\dagger}$              & S & N\,+\,S\,+\,B & 20.042 &  ---   &  ---   & 1.4208 & 1.2749 & 1.9169 & 1.5632     \\	       						     
\hline  					   
 \label{T_rms}
\end{tabular}

\begin{flushleft}
 \textit{Notes.}  \lq ID\rq: object designation as in Table\,\ref{T_sample}.  \lq OBS\rq:  origin of the optical data. \lq Mod.\rq: best-fitting model for emission lines.  \lq S\rq,   \lq O\rq \ and   \lq M\rq \ stand for models based on [S\,II] or [O\,I] or mixed -type.  \lq Comp.\rq: components used to achieve the best-fitting model. \lq N\rq \ and \lq B\rq \, stand for narrow and broad, component, while \lq S\rq \ indicates the second component when present (see Sect.\,\ref{Analysis_LF}, for details). The values of the standard deviation calculated in the residual spectra considering a line-free continuum ($\varepsilon_{\rm c}$) and emission lines ($\varepsilon$$^{\rm [O\,I]}$, $\varepsilon$$^{\rm [S\,II]}$ and $\varepsilon$$^{\rm H\alpha}$ in until of $\varepsilon_{\rm c}$) in the residual spectrum after the fitting, for different components.  $^{\dagger}$ and $^{\ddagger}$ symbols are as in Table\,\ref{T_kin}. Square-brackets indicate the data for which  the fit of  the H$\alpha$-[N\,II] emission is not well constrained ($\varepsilon$$^{\rm H\alpha}$$>$\,3\,$\varepsilon_{\rm c}$).	
  \end{flushleft}
\end{table*}
\clearpage

\section[B]{Comments, spectra and images of individual sources}
\label{App_comments_panels}
The first part of this appendix is devoted to shortly comment the characteristics of each LINER. \\

\noindent In the second part, we show the \textit{HST}-image (in the F814W-band  if available) from the Hubble Legacy Archive and the spectra analysed in this manuscript for each LINER. \\
More specifically, we present \lq sharp-divided\rq \ \textit{HST} images. A sharp-divided image  \citep{Marquez1999, Marquez2003}  is obtained by dividing the original image, \textit{I},  by a filtered version of it, \textit{BI}, i.e. \textit{I\,/\,BI}. Its result is very similar to that of the unsharp masking technique (which subtracts instead of dividing, i.e. \textit{I\,-\,BI}), but for the former the levels are around unity, which facilitates the comparison among different galaxies. Features departing from axisymmetry, together with those with sizes close to the size of the filter are better seen in the sharp-divided images. In our case, the images are median filtered with the \textsc{iraf} command  \lq median\rq \ using a box of 30 pixels.  As shown in the figures of this Appendix, the results clearly show asymmetric structures in the centres, that cannot be appreciated in the original images.  \\
We present rest-frame blue and red spectra  and the corresponding stellar continuum-model and the Gaussian fits to emission lines profiles after stellar-subtraction. When available, the line-modelling to \textit{HST}/STIS line profiles is also shown. \\
The images and spectra are arranged as follows:\\
 \textit{Top:} \textit{HST} image, the orientation is North up, East to the left. Overlaid to these \textit{HST}-images, we mark the slit used to obtain the data analysed in this work and the major axis of rotation (from the HyperLeda database) with continuos and dashed lines, respectively. The white bar shows a linear scale of 100 pc (see also column 3 in Table\,\ref{T_sample}). For the LINERs at z\,$\geq$\,0.005 (Table\,\ref{T_sample}), we indicate a larger linear scale (either 200, 300 or 400 pc). If available, the line-modelling to \textit{HST}/STIS line profiles is also shown.\\
\textit{Middle:} the 2 lines show the ground-based, rest-frame spectra and their continuum-model for the blue and red ranges. The residuals (i.e. data - model) are also presented. The red line indicates the modelled stellar spectrum that matches the observed stellar continuum (Sect.\,\ref{Analysis} and Table\,\ref{T_kin}). The central wavelength of the most relevant spectral features masked-out in the procedure for the stellar continuum modelling (as in Fig.\,\ref{Panel_stsub}) are  marked with vertical lines. Specifically, in green and in blue those emission lines considered and excluded for the analysis, respectively.\\
 \textit{Bottom:} Gaussian fits to emission lines profiles (after stellar continuum subtraction) in both blue (\textit{left}) and red (\textit{right}) spectra. We marked with different colours the Gaussian components required to model the emission lines (same colours mark the same kinematic components). These components are named on the top-right  and summarised in Table\,\ref{T_kin} (see also Sect.\,\ref{Analysis}, for details). The red curve  shows the total contribution coming from the Gaussian fit.  Residuals from the fit are in the small lower-panels in which grey lines indicate the rest frame wavelengths of the emission lines.\\
For each LINER, the caption include comments about the line profiles in ground-based data and their modelling. We refer to \textit{BC14} (and references therein) for similar comments but for emission lines and their fit for \textit{HST}/STIS spectra. \\

\noindent  Finally, the third part of this appendix is devoted to the present position-velocity diagrams (PV curves) for  each LINERs.\\

\begin{figure*}
\vspace{-0.8cm} 
\includegraphics[trim = 1.10cm 1.05cm 11.0cm 17.75cm, clip=true, width=.36\textwidth]{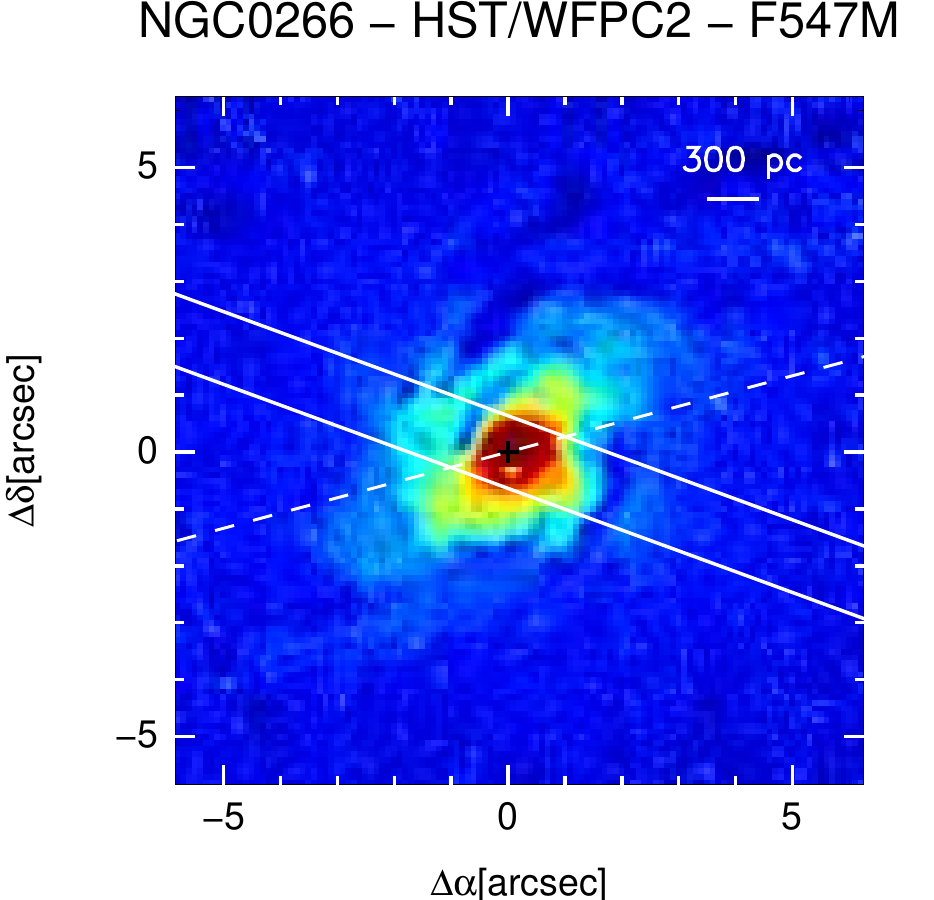} 
\vspace{-0.35cm} 
\hspace{-0.3cm} 
\includegraphics[trim = 2.4cm 19.75cm 2.7cm 3.2cm, clip=true, width=.90\textwidth]{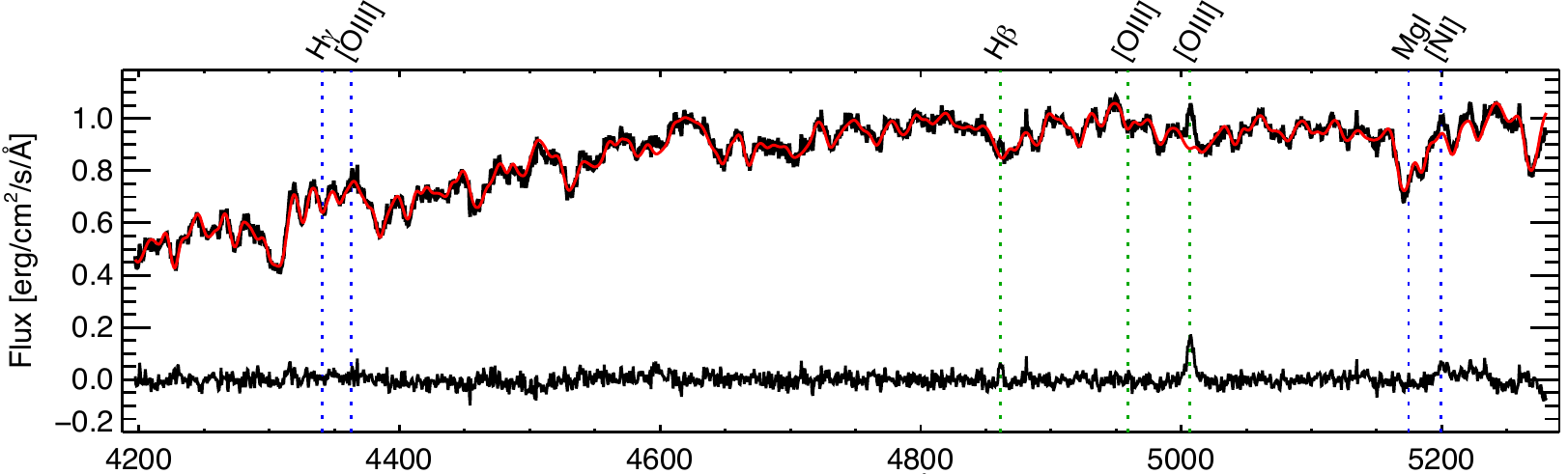} \\
\vspace{-0.10cm}
\includegraphics[trim = 2.4cm 18.75cm 2.7cm 3.15cm, clip=true, width=.895\textwidth]{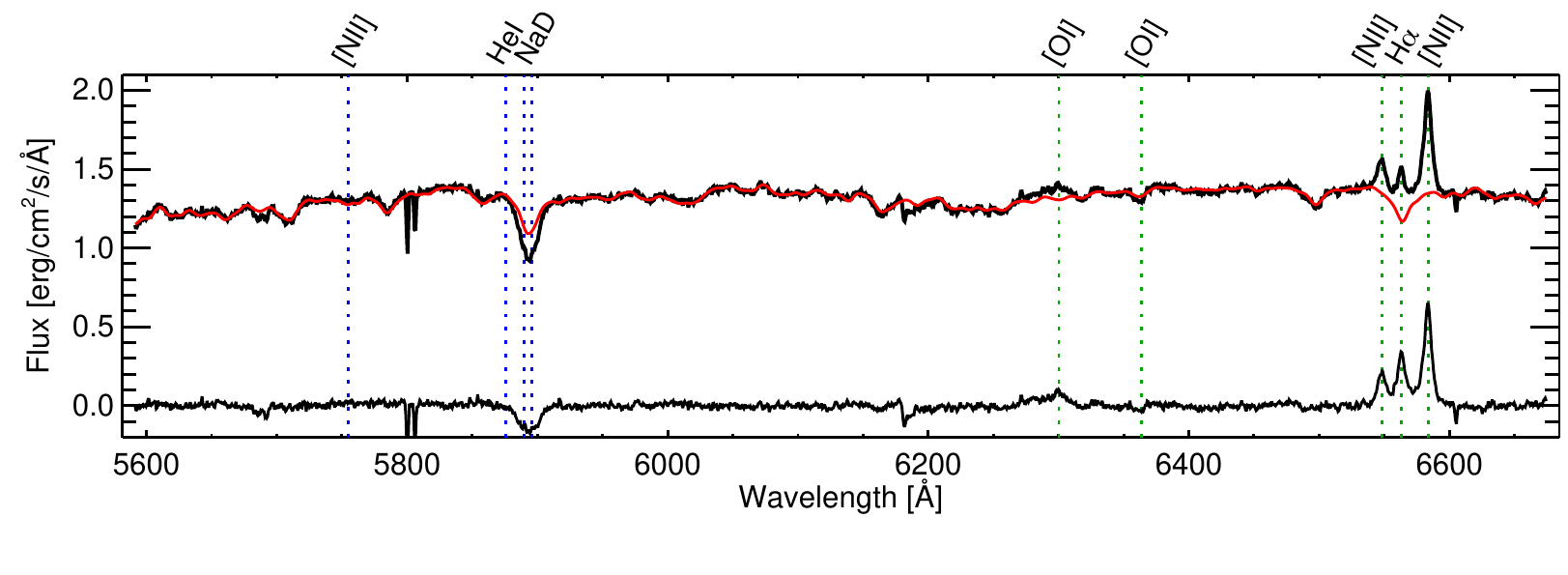} \\
\vspace{-0.45cm}
\includegraphics[trim = 4.9cm 13.415cm 5.25cm 6.3cm, clip=true, width=.4600\textwidth]{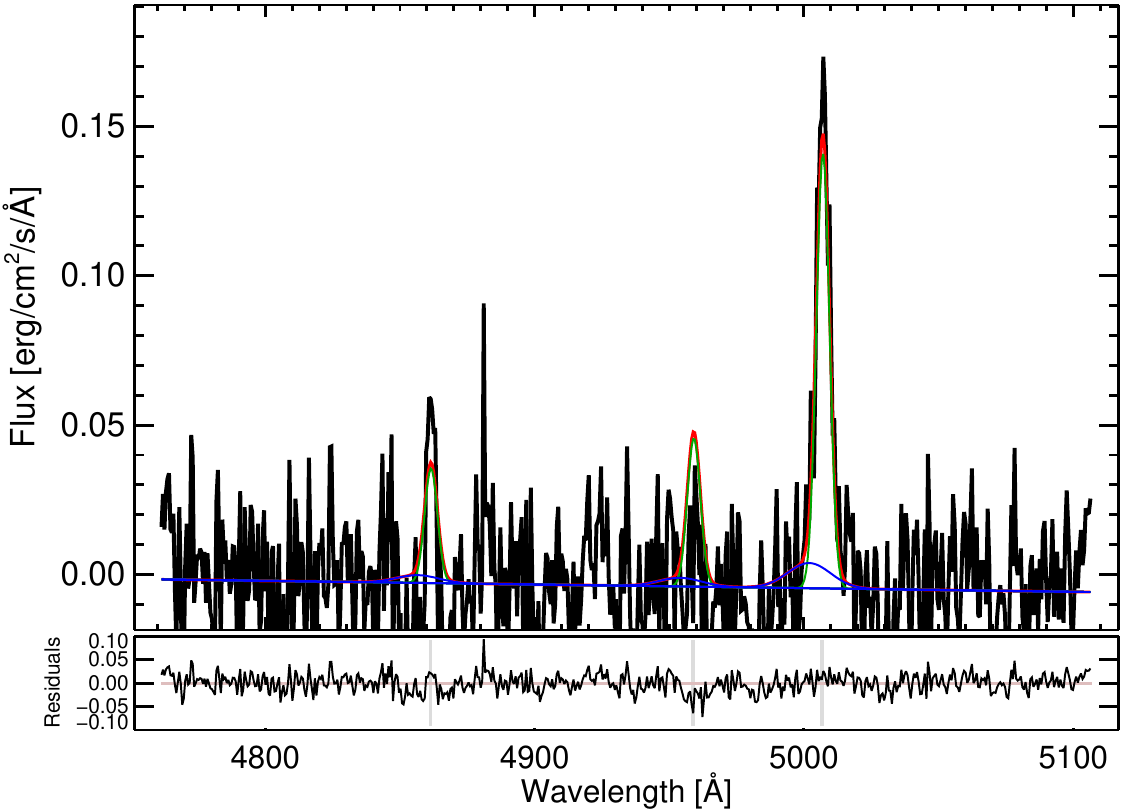}
\hspace{0.1cm} 
\includegraphics[trim = 5.55cm 13.415cm 5.25cm 6.3cm, clip=true, width=.4335\textwidth]{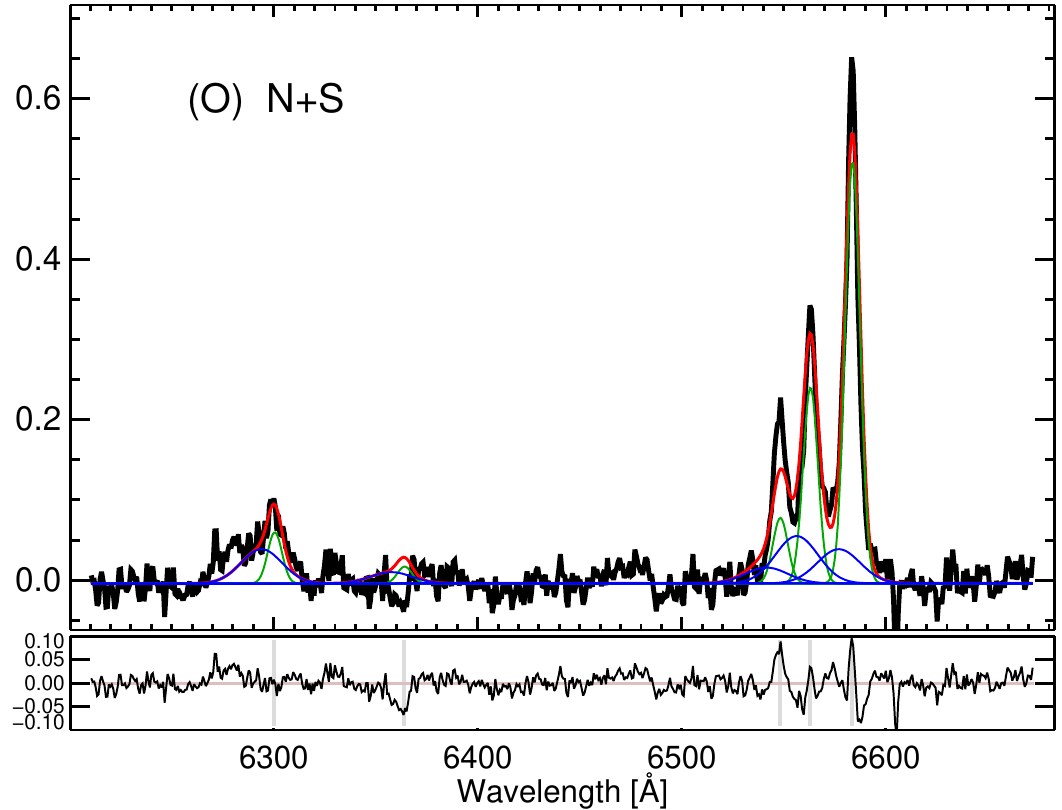}  
\caption{\textit{First row}: sharp divided \textit{HST} image.  The two continuous white lines represent the slit used in our ground-based spectroscopy. The PA of the major axis is marked with a dashed line. When an \textit{HST}/STIS spectrum is available, the line modelling is shown. \textit{Second and third rows}: rest-frame ground-based blue and red spectra, respectively; the red line corresponds to the continuum reproducing the stellar population, the vertical lines mark the most relevant spectral features masked out before the continuum fitting. \textit{Forth row}: Gaussian fit to the stellar population subtracted spectra in the H$\beta$ (left) and H$\alpha$ (right) regions. We marked with different colours the components, named on the top-right,  required to model the emission lines.  The red curve shows the total contribution from the fit. Residuals from the fit are in the lower-panels in which grey lines mark the rest frame wavelengths of the spectral features. \\
NGC\,0266: The H$\alpha$-[NII] lines are unblended in the observed spectrum. An additional second component is evident in the narrow lines only after starlight subtraction. The single-component model of [O\,I] has lower residuals than the double Gaussians one (Table\,\ref{T_rms}). However, such single component would have a width of $\sim$\,750\,km\,s$^{-1}$ which is unrealistically large.  We did not find strong evidence of a BLR component in contrast of what found by \textit{HFS97}.}
 \label{Panel_NGC0266} 		 		 
\end{figure*}

\begin{figure*}
\includegraphics[trim = 2.cm 17.0cm 2.cm 1.5cm, clip=true, width=1.\textwidth]{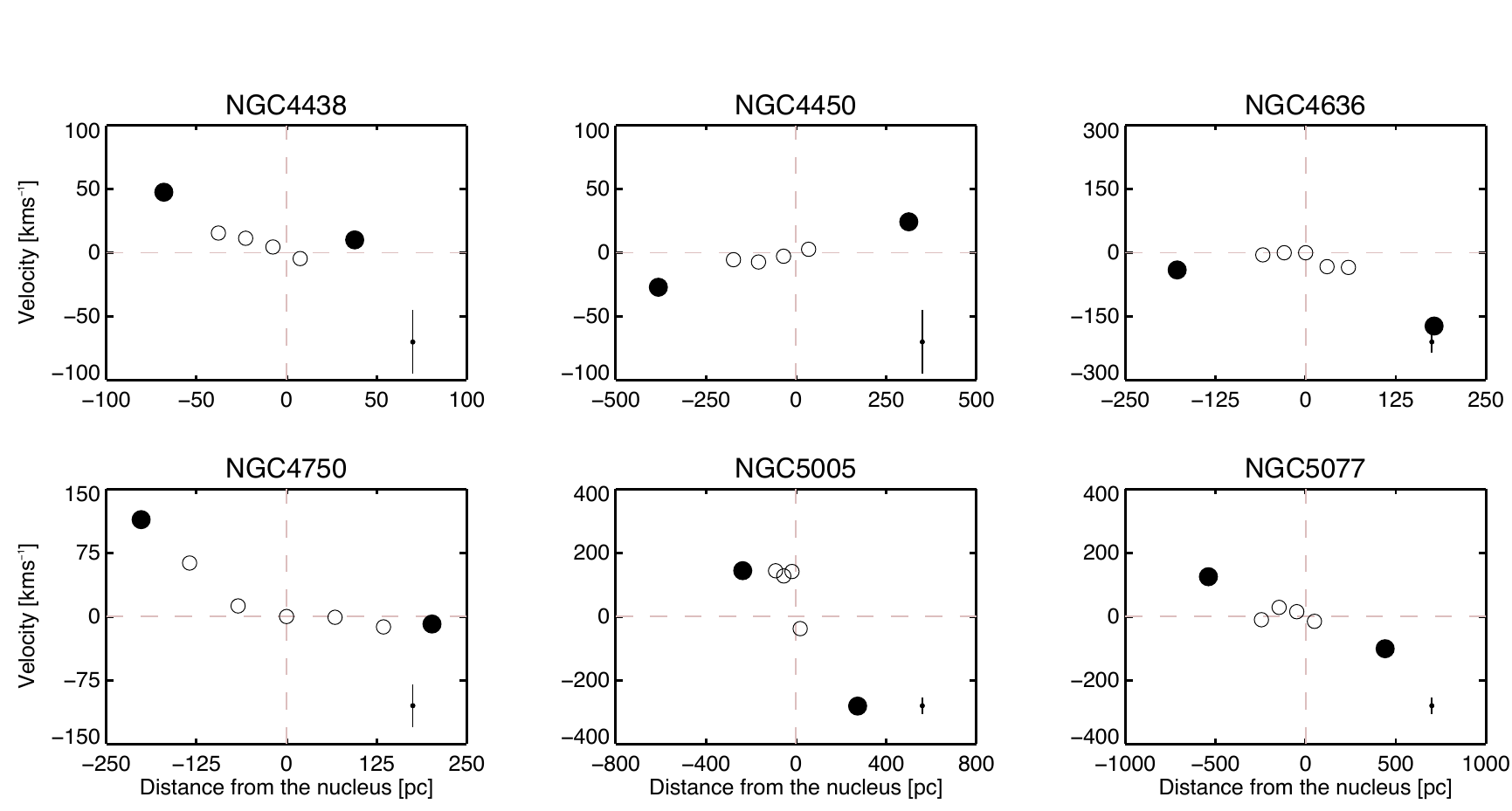} 
\caption{Position-velocity diagram for all the LINERs in the sample except NGC\,4772. For this LINER the velocity distribution is very chaotic with multiple kinematic components  already  in the innermost regions, preventing to constrain any putative rotation pattern. The empty circles mark those data-points obtained considering the peak velocity of the most intense emission line (generally [N\,II]$\lambda$6584) in the individual spectra corresponding to the nuclear extraction. As a reference, we show the velocities corresponding to the individual spectra at larger distances from the center. Positions and velocities for a given galaxy are referred to that of the central spectrum in its nuclear extraction (the lines tracing the origin of coordinates are plotted as pink, dashed lines. At the bottom-right of each panel we show an error-bar representative for all the measurements.}	 	
\label{Panel_PVD} 	 
\end{figure*}

\noindent \textbf{NGC\,0266.} This LINER-nucleus is hosted by a spiral galaxy with a well studied bar \citep{Font2017}. \citet{Garrido2005}, GHASP survey,  found that the rotation curve exhibits a strong velocity gradient in the inner $\sim$\,5\arcsec \,(i.e. $\sim$1.6\,kpc), reaching a maximuum rotational velocity of 363\,kms$^{\rm -1}$ at a larger distance (26\,kpc). The CAHA spectroscopic data also show evidence for rotation (Fig.\,\ref{Panel_PVD}) but with lower rotational velocity than what found by  \citet{Garrido2005} since our PA is not placed along the major axis (Fig.\,\ref{Panel_NGC0266}). We also report for the first time the detection of a kinematic component likely  associate to a possible outflow.  \textit{HFS97} reported  a BLR component with a  FWHM of 1350 km\,s$^{-1}$, but we did not. The red spectrum of this LINER does not cover the [S\,II] doublet (Sect.\,\ref{Analysis_LF}) and the blue spectra has a low S/N.  The  NaD absorption has been modelled with two kinematic components (Fig.\,\ref{Panel_NaD}). The narrow component shares the same velocity dispersion with the same component of the ionised gas, but the velocities of the two  ISM-phases are somewhat different (Table\,\ref{T_kin}). The second, and broader one, is different than the same component of the ionised gas, though both have blueshifted velocities (Table\,\ref{T_kin}).\\

\noindent \textbf{NGC\,0315.} The nucleus of this  elliptical galaxy shows an unresolved source of ionised gas on top of a dusty disc of $\sim$\,200\,pc visible in the \textit{HST} image \citep{Masegosa2011}. Two-sided, well-resolved radio jets are associated to the X-ray jet  at PA\,= -49$^{\circ}$ \citep{Nagar2005, Kharb2012}. Therefore this jet-structure is not aligned with the possible outflow found in our spectroscopic study (Table\,\ref{class_sum}).  Such an outflow would have a complex ionization structure as the properties of low and high ionization lines are different  (Table\,\ref{T_kin}). The broad H$\alpha$ component is detected in both our ground- and space-based spectroscopy and in previous works (Table\,\ref{T_FWHM}). Unfortunately, such component is not well constrained  in our fit with  ALFOSC/NOT data (Table\,\ref{T_rms}). Part of the discrepancy of measurements at different epochs may be due to AGN variability on scales of years (as reported by \citealt{Younes2011, HG2014}). The 2D rotation is very perturbed (Fig.\,\ref{Panel_PVD}).  The neutral gas kinematics is very different from that of the ionised gas with redshifted velocity and with a velocity dispersion  larger (smaller) than that of the narrow (second) component in emission lines. The latter consideration reflects the general behaviour of the neutral gas in our sample of LINERs (Sect.\,\ref{Analysis_Abs_LF}).\\

\begin{figure*}
\vspace{-0.25cm} 
\includegraphics[trim = 1.10cm .85cm 11.0cm 17.75cm, clip=true, width=.40\textwidth]{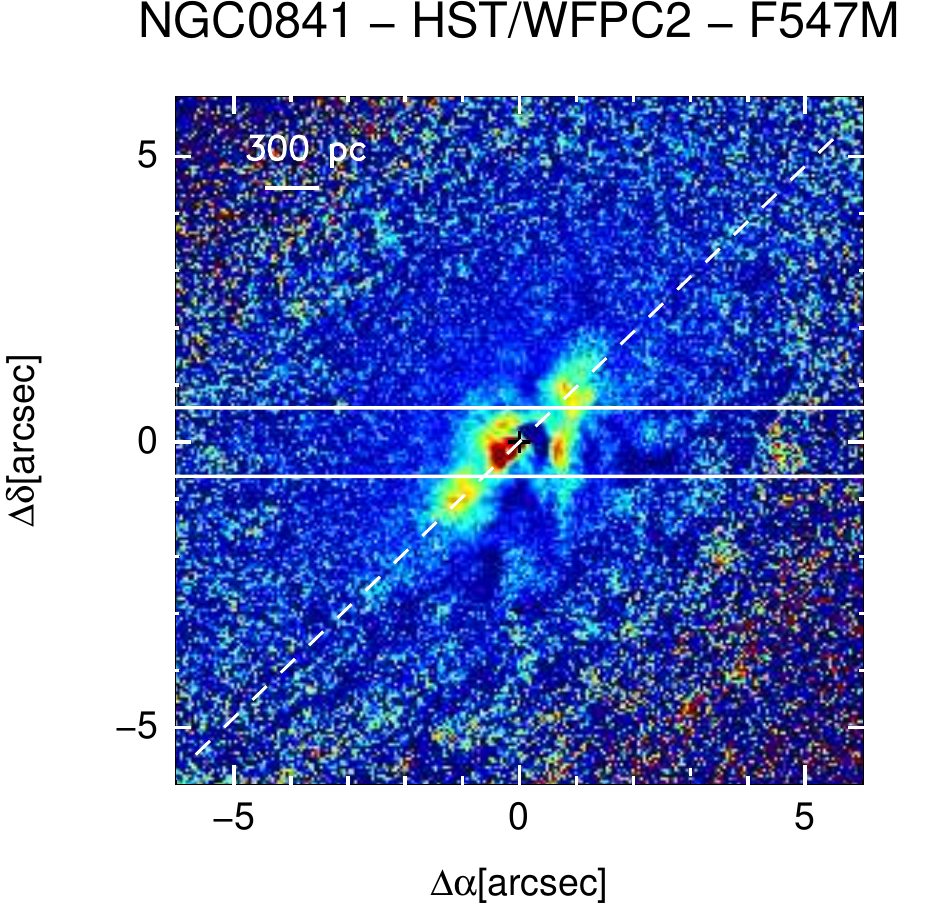} 
\hspace{-0.3cm} 
\includegraphics[trim = 2.4cm 19.75cm 2.7cm 3.75cm, clip=true, width=.915\textwidth]{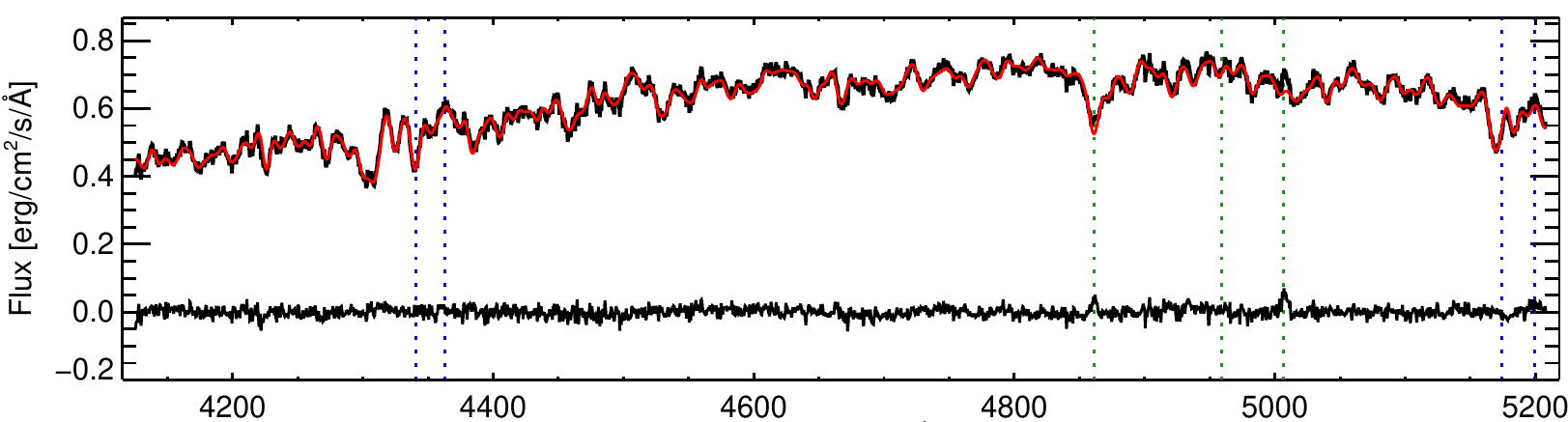} \\
\vspace{-0.10cm}
\includegraphics[trim = 2.4cm 18.75cm 2.7cm 3.75cm, clip=true, width=.91\textwidth]{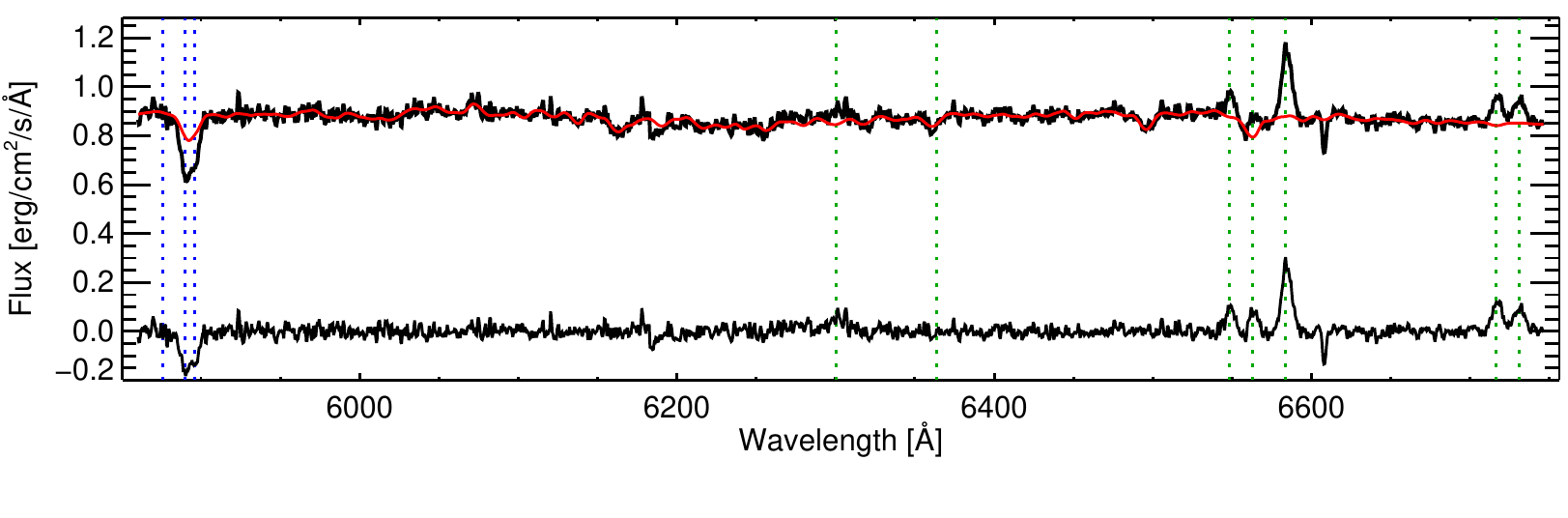} \\
\vspace{-0.45cm}
\includegraphics[trim = 4.9cm 13.25cm 5.25cm 6.3cm, clip=true, width=.4715\textwidth]{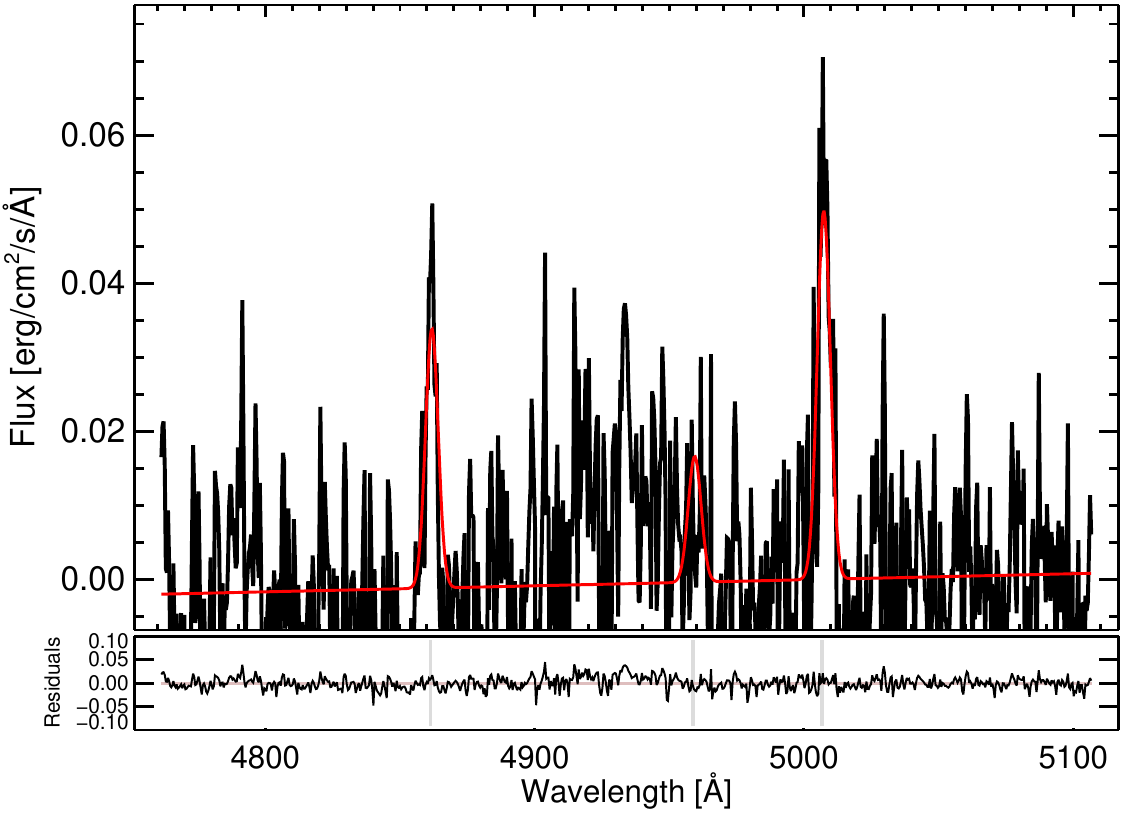}
\hspace{0.1cm} 
\includegraphics[trim = 5.55cm 13.25cm 5.25cm 6.3cm, clip=true, width=.445\textwidth]{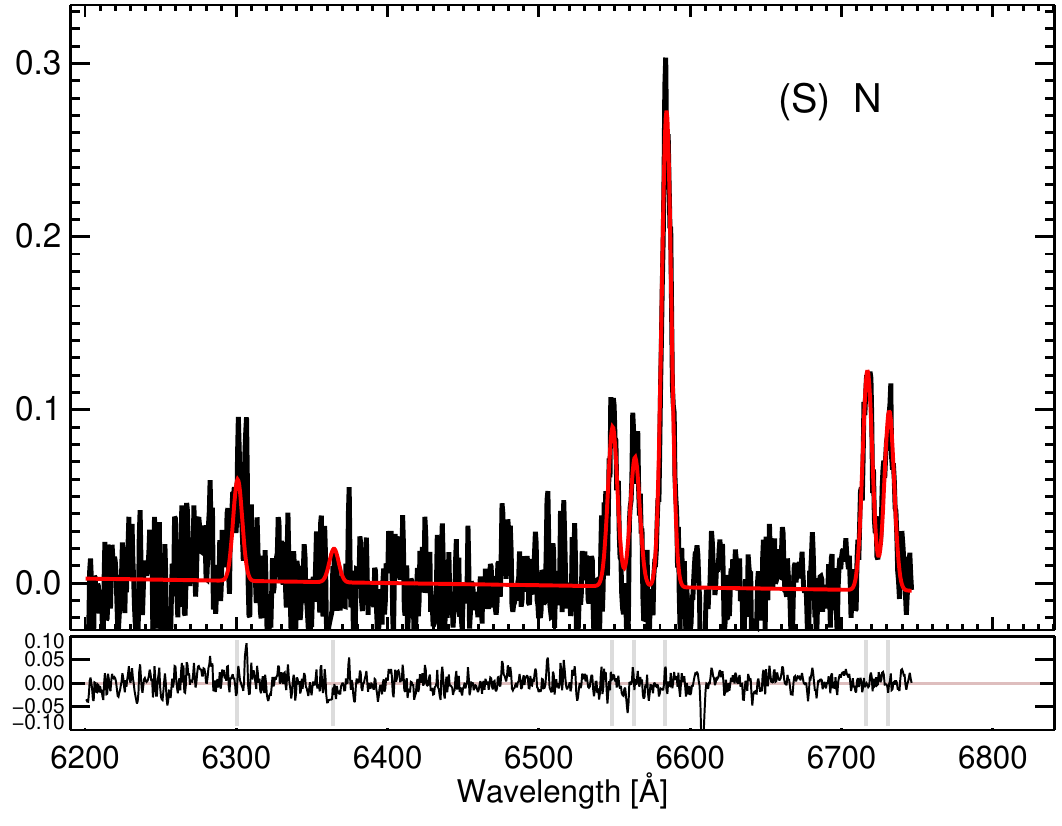}  
\caption{(General description as in Fig.\,\ref{Panel_NGC0266}.) NGC\,0841: H$\alpha$ is strongly affected by stellar absorption in the observed spectrum (a good starlight subtraction is essential). All emission lines are well modelled  with a single Gaussian component, no evidence for neither broad nor second components (H$\alpha$ and [N\,II] are well resolved).  A visual inspection indicates that lines profiles are very different from those in the Palomar spectra (\textit{HFS97}).}
 \label{Panel_NGC0841} 		 		 
\end{figure*}

\noindent \textbf{NGC\,0841.}  It is classified as a grand design nuclear spiral with a spiral dust lane down to the centre by \citet{GonzalezDelgado2008}. Rotation is clearly detected (though with some distortion, Fig.\,\ref{Panel_PVD}) even if our slit is not aligned with the major axis (Fig.\,\ref{Panel_NGC0841}). A BLR component is not needed to model, in our ground-based spectrum, the H$\alpha$ line profile (which is not blended with [N\,II]), though previously reported by \textit{HFS97}.  The NaD doublet seems to originate in a neutral gas rotating disc whose kinematics is rather similar to that of the ionised gas (velocity dispersions are the same within uncertainties, Tables \ref{T_kin} and \ref{T_NaD}).\\

\noindent \textbf{NGC\,1052.} An elliptical galaxy with large scale dusty structures harbors the prototypical LINER nucleus. It is an X-ray absorbed source with little optical obscuration and broad lines in polarized light \citep{Barth1999} indicating a peculiar absorbing geometry \citep{Burtscher2016}. Space-and ground-based red optical spectra are remarkably similar (with a severe H$\alpha$-[N\,II]  blend), except for [S\,II]. \citet{Onori2017} modelled the [N\,II] emission an \textit{HST}/FOS spectrum (R\,=\,2800) with one Gaussian though with rather large FWHM, i.e. 682\,km\,s$^{-1}$ that may indicate the presence of unresolved outflow components.  The BLR component in their analysis is of $\sim$\,2200\,km\,s$^{-1}$, a slightly larger value is found considering the near-IR He\,I$\lambda$1.083$\mu$m line i.e. $\sim$\,2400\,km\,s$^{-1}$. The broad H$\alpha$ component was also detected by \textit{BC14}  and \textit{C15}, in fair agreement (Table\,\ref{T_FWHM}).  We found evidence for the BLR only in \textit{HST}/STIS spectra. This could be partially due to a less reliable fit to the H$\alpha$ emission line in ground- and space-based datasets (Table\,\ref{T_rms}). Second components in Oxygen and Hydrogen lines are interpreted as a possible outflow (Table\,\ref{class_sum}) as already suggested in previous works \citep{Pogge2000, Walsh2008, Dopita2015, Sugai2005}. [S\,II] behaves otherwise than [O\,I] in both ground- and space-based spectroscopy (Tables \ref{T_kin} and \ref{class_sum}). The neutral gas probed via NaD  shares the same kinematics with the narrow component of the [O\,I] (Tables \ref{T_kin} and \ref{T_NaD}).\\

\noindent \textbf{NGC\,2681.}  It is a S0 galaxy with large-scale spiral dust lanes \citep{Masegosa2011}. The second component fitted to the emission lines is relatively narrow (209\,$\pm$\,12\,km\,s$^{-1}$) and blueshifted (-180\,$\pm$\,8\,km\,s$^{-1}$). This might indicate a complex nuclear rotation pattern, in fair agreement with the complicated and multi-component structure already reported via optical and near-IR imaging by \citet{Dullo2016} and \citet{Laurikainen2005}. 
Interestingly, both star-formation and AGN optical BPT line ratios are observed by the nucleus with the first increasingly significant when the aperture size is increased \citep{Dullo2016}.\\

\noindent \textbf{NGC\,2787.} The most noticeable feature of this lenticular galaxy is a set of concentric elliptical dust rings  \citep{Masegosa2011}. There is a good match between the modelling of ground- and space-based data, in terms of the adopted model and number of components. The FWHM of the BLR components are nevertheless quite different (Table\,\ref{T_FWHM}) being $\sim$\,1000\,km\,s$^{-1}$ lower in ground-based data than that from \textit{HST}/STIS spectroscopy in present and previous works. This could be due to short-term variability \citep{HG2014}. Based on integral field spectroscopic data, \citet{Brum2017} probed a pure rotation velocity field with high projected amplitude ($\sim$\,250\,km\,s$^{-1}$) which is also partially captured by CAHA observations (Fig.\,\ref{Panel_PVD}). Additionally they found a mild AGN outflow and nuclear bar (see also \citealt{Erwin2003}). We did not find any sign of complex kinematics associated to these features  probably due to a different slit orientation and the contribution from the bar and a putative outflow. The component used to model the  neutral gas NaD absorption shares similar kinematics to that of the narrow component of the ionised gas (Tables \ref{T_kin} and \ref{T_NaD}). However, the doublet is affected by  absorption telluric lines making the line modelling uncertain (Fig.\,\ref{Panel_NaD}), and avoiding the detection of a putative blueshifted second component. \\

\noindent \textbf{NGC\,3226.} This elliptical galaxy is characterised by a  bright and compact nucleus in a  dusty environment. The kinematics of the second component indicate the presence of a possible outflow (at a velocity of 155\,$\pm$\,21 km\,s$^{-1}$) consistent with the  outflow-like structure emerging from the nucleus seen in \textit{HST} images  \citep{Masegosa2011}. Though evidence for the presence of an AGN has been found  via X-ray by  \citet{Capetti2006},  the optical H$\alpha$ line-profile modelling does not require any BLR-originated component.   The NaD doublet was modelled with a single kinematic component  (Fig.\,\ref{Panel_NaD}) found at the systemic velocity (2\,$\pm$\,2 km\,s$^{-1}$). Its velocity dispersion (212\,$\pm$\,42 km\,s$^{-1}$, Table\,\ref{T_NaD}) agrees within the uncertainties with that of the narrow component of the ionised gas (185\,$\pm$\,20 km\,s$^{-1}$Table\,\ref{T_kin}). However, as for NGC\,2787, the blue wing of the absorption doublet is affected by the residual of skyline subtraction and telluric absorption (Fig.\,\ref{Panel_NaD}).\\

\noindent \textbf{NGC\,3642.} A strong point source is surrounded by some diffuse circumnuclear H$\alpha$ emission in this LINER \citep{Pogge2000}. It is hosted by a spiral galaxy with large-scale dusty structures  and several H\,II regions \citep{Chiaberge2005}. The second kinematic component  is classified as a candidate for a possible outflow, this is probably a consequence of our conservative limits. Indeed,  the kinematical properties of this component (V\,=\,-335\,$\pm$\,67\,km\,s$^{-1}$ and $\sigma$\,=\,300\,$\pm$\,60\,km\,s$^{-1}$)  are most likely not due to the rotation which shows a small amplitude ($<$\,100\,km\,s$^{-1}$, Fig.\,\ref{Panel_PVD}). The BLR detection is confirmed with both ground- and space-based optical spectroscopy (Table\,\ref{T_FWHM}), although the H$\alpha$-fit is less reliable (Table\,\ref{T_rms}). Only some X-ray properties of this LINER in ROSAT observations are present (\citealt{Komossa1999} and references therein).
The stellar subtraction leaves some residuals that can be interpreted as resonant NaD emission.  Unfortunately, the low S/N in this wavelength region makes highly questionable the detection of such peculiar emission.\\

\noindent \textbf{NGC\,3718.}  This spiral galaxy has a prominent dust lane, which runs across the entire stellar bulge, and a warped molecular and atomic gas disc \citep{Sparke2009, Krips2005}. It also shows signs of a past interaction (probably with NGC\,3729, \citealt{Markakis2015}). The  elongated  structures seen in the \textit{HST} images and e-MERLIN data  are indicative of a small-scale bipolar jets or outflows. As for the case of NGC\,0315, different measurements of the BLR components (Table\,\ref{T_FWHM}) could be due to AGN variability on scales of years  \citep{Younes2011}.  The region of NaD is strongly affected by telluric absorption compromising the possibility to infer the neutral gas kinematics.\\


\noindent \textbf{NGC\,3884.}  The optical emission line profiles observed  in this spiral galaxy have been modelled with two kinematical components. The narrow component is interpreted as rotation which has a small amplitude as seen in the corresponding PV-curve (Fig.\,\ref{Panel_PVD}). The second component is rather broad and blueshifted, and it is interpreted as a possible outflow (Table\,\ref{class_sum}). As for the case of NGC\,3642,  the detection of a resonant NaD emission is uncertain.\\

\begin{figure*}
\vspace{-0.25cm} 
\includegraphics[trim = 1.10cm .85cm 11.0cm 17.75cm, clip=true, width=.40\textwidth]{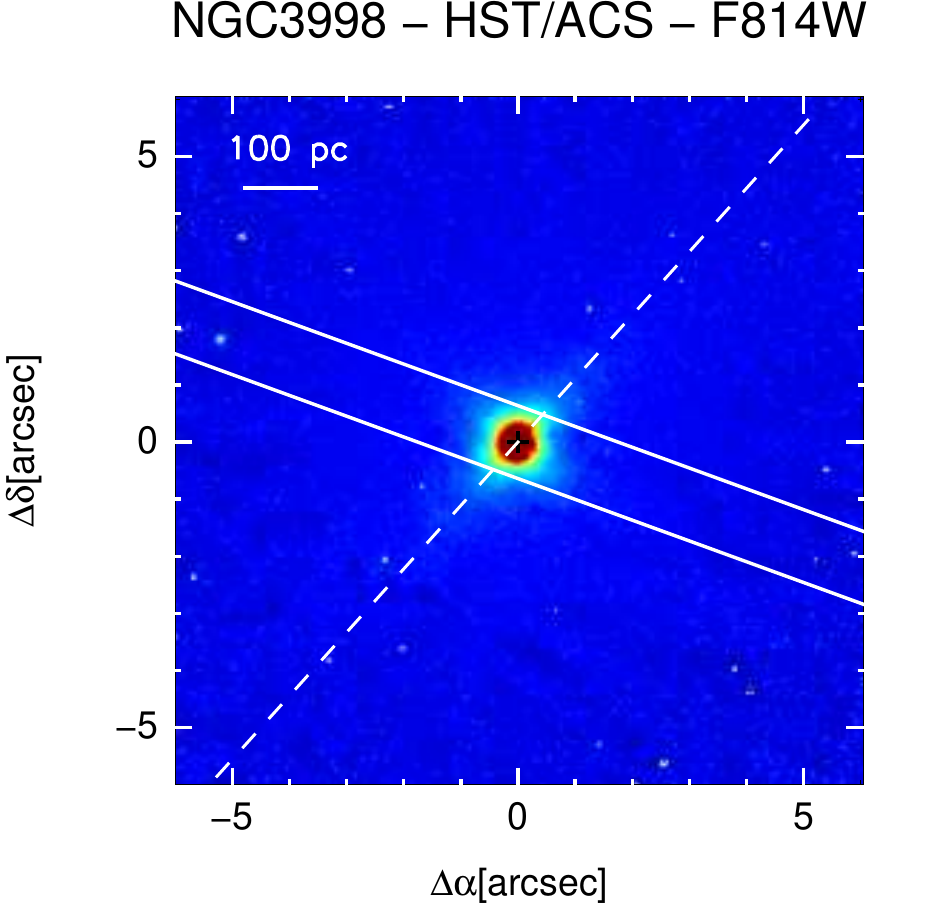} 
\hspace{-0.3cm} 
\includegraphics[trim = 4.5cm 13.cm 5.25cm 6.25cm, clip=true, width=.475\textwidth]{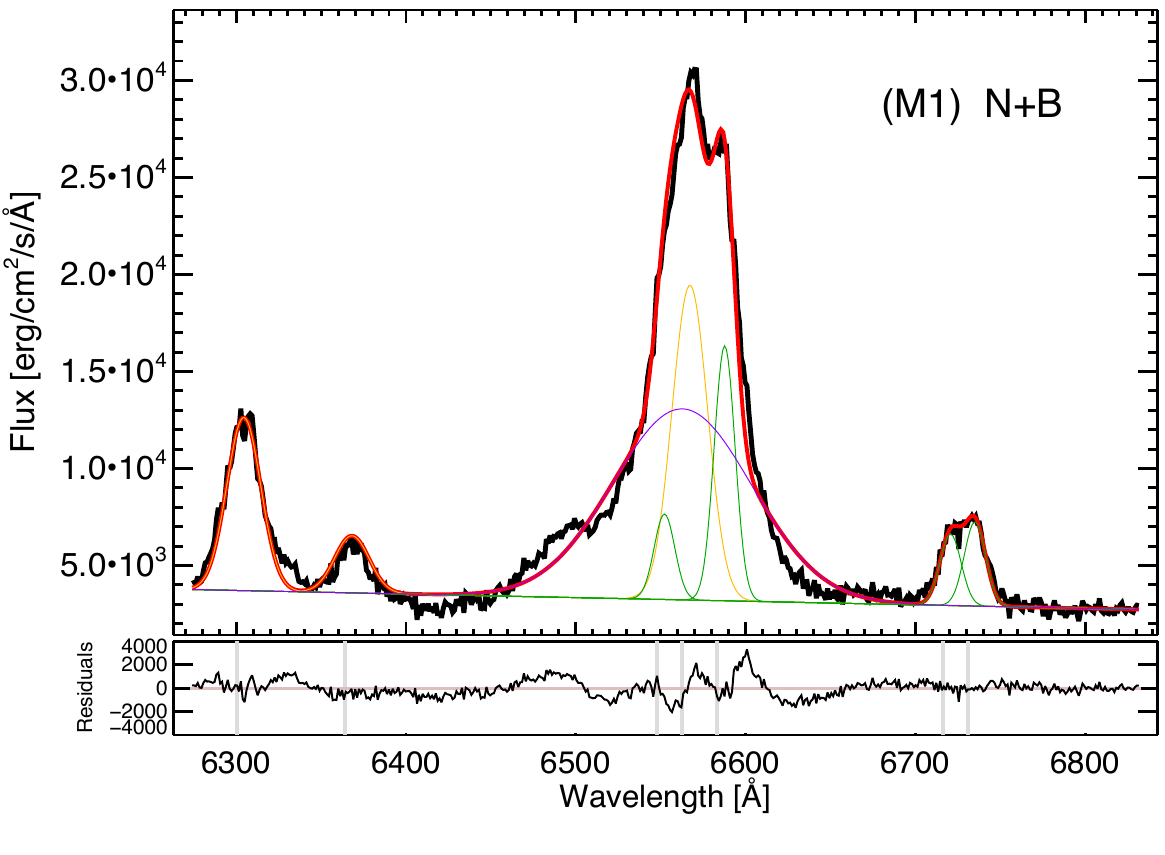}  \\
\vspace{-0.10cm}
\includegraphics[trim = 2.4cm 19.75cm 2.7cm 3.75cm, clip=true, width=.915\textwidth]{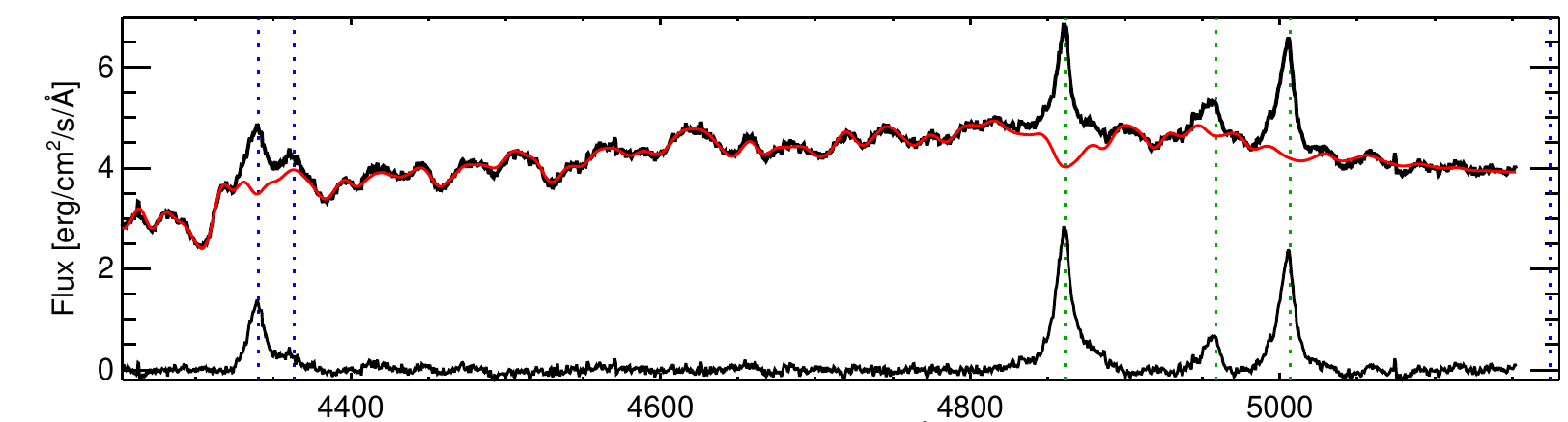} \\
\vspace{-0.10cm}
\includegraphics[trim = 2.4cm 18.75cm 2.7cm 3.75cm, clip=true, width=.91\textwidth]{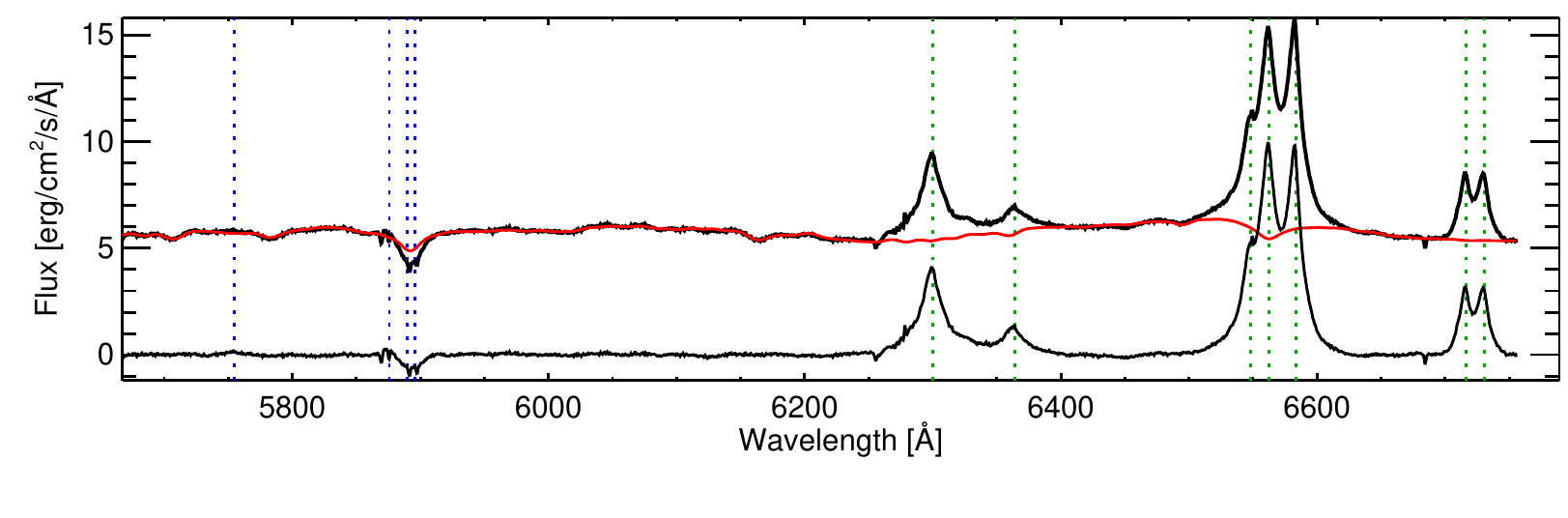} \\
\vspace{-0.45cm}
\includegraphics[trim = 5cm 13.25cm 5.25cm 6.3cm, clip=true, width=.465\textwidth]{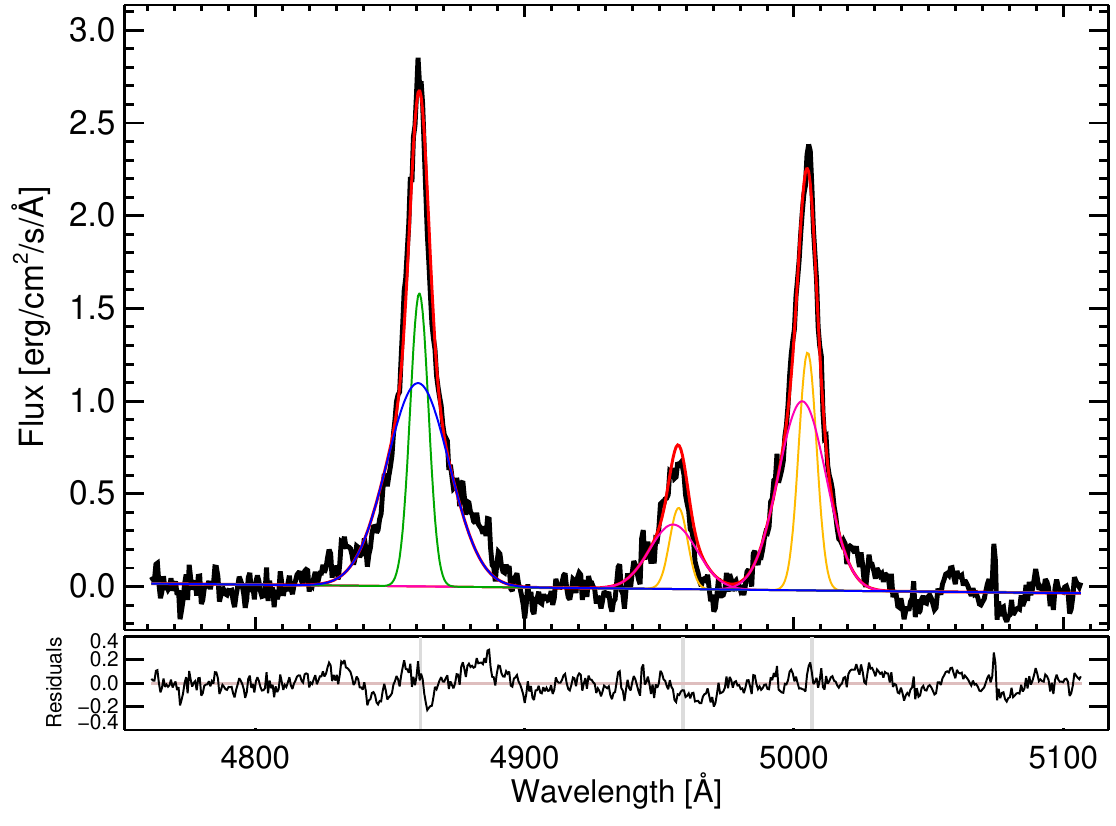}
\hspace{0.1cm} 
\includegraphics[trim = 5.55cm 13.25cm 5.25cm 6.3cm, clip=true, width=.445\textwidth]{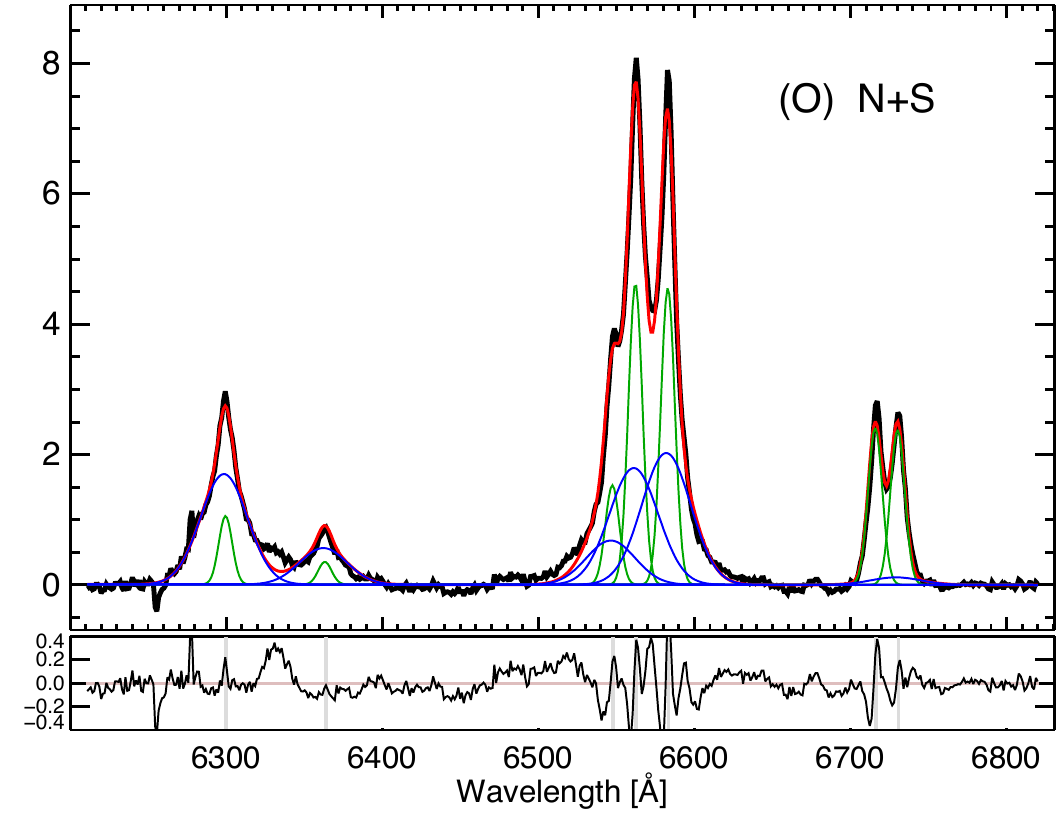} 
\caption{(General description as in Fig.\,\ref{Panel_NGC0266}.) NGC\,3998: a second component is evident in the [O\,I] line being rather weak in [S\,II]. Issues related to the stellar subtraction might be at the origin of the small bump between Oxygen lines. The [O\,III] lines have different kinematics (light and pink curves) with respect to the other lines.} 
\label{Panel_NGC3998} 		 		 
\end{figure*}

\noindent \textbf{NGC\,3998.} This S0 galaxy has a  disc-like weak emission with  a diameter of $\sim$\,100 pc surrounding a compact nucleus and little indication of dust in the nuclear region \citep{Pogge2000, Masegosa2011}. The 2D gas kinematics is disc dominated by disc rotation, classified as a near-oblate, face-on fast rotator by \citet{Boardman2016}. Some anisotropies and  signs of warps are found at large galactocentric distance  likely related to a past interaction \citep{Boardman2017}. The disc kinematics is not fully captured by our ground-based spectroscopy likely due to the slit orientation with respect to the major axis (Fig.\,\ref{Panel_NGC3998}). X-ray spectroscopy at 0.1-100 keV data provides a strong indication for an AGN origin of the LINER activity in this galaxy \citep{Pellegrini2000}. The nucleus is a low-power radio AGN with a kpc-size one-sided jet and  S-shaped lobes with a total extent of about 20\,kpc \citep{Frank2016}. We do find evidence of a BLR-originated component only in \textit{HST}/STIS spectra; this could be partially due to a not fully reliable fit to the H$\alpha$ emission line in ground-based data (Table\,\ref{T_rms}). Components of intermediate width in ground-based spectroscopy are interpreted as a possible outflow (Table\,\ref{class_sum}).  Broad forbidden lines are also found in the space \textit{HST}/STIS spectrum but their fit is rather uncertain.   The  NaD doublet has been modelled with a single kinematic component (Fig.\,\ref{Panel_NaD}) found at the systemic velocity but with rather high velocity dispersion ($\sigma$\,=\,292\,$\pm$\,58\,km\,s$^{-1}$), intermediate between those of the narrow and second components of the emission lines (Table\,\ref{T_kin}).  Similarly to  the cases of NGC\,2787 and NGC\,3226, the doublet is affected by skylines and telluric absorption making the line modelling less reliable (Fig.\,\ref{Panel_NaD}).\\

\noindent \textbf{NGC\,4036.} This E-S0 galaxy is characterised by a well-identified nucleus and a complex filamentary and clumpy structure with several  filaments \citep{Pogge2000, Masegosa2011, Walsh2008}.  Both narrow and second components of emission lines are identified with rotation (Sect.\,\ref{classification} and Table\,\ref{class_sum}). However, the second broader component could be associated to the complex H$\alpha$ emission structure present in the inner 3\arcsec. 2D spectroscopic data at higher spectral and spatial resolution are needed to study this feature in detail. 
A broad component is not required to fit the H$\alpha$-profile in our ground-based spectra in contrast to what was found by \textit{HFS97}. Such a broad component is essential to model  H$\alpha$  in the \textit{HST}/STIS spectrum.\\

\noindent \textbf{NGC\,4143.} For this E-S0 galaxy, we modelled  the forbidden line profiles with two kinematic components  in both ground- and space-based spectra. However, only the second component of Oxygen lines in ground-based data, indicates the possible presence of an outflow. A broad component is needed for modelling the H$\alpha$ line only in HST/STIS spectra.  The FWHM of this component is larger  than 1000\,km\,s$^{-1}$ than that measured by \textit{HFS97} based on [S\,II] as reference; a much smaller difference is found when using [O\,I] as reference in   \textit{BC14}  (Table\,\ref{T_kin}).\\

\noindent \textbf{NGC\,4203.} It is a nearly face-on SAB with a large-scale dusty structures \citep{Chiaberge2005}. This galaxy is surrounded by a very large, low-column-density and distorted H\,I disc and a  disrupted companion dwarf galaxy is present at the east of the galaxy \citep{Mustafa2015}. The nucleus is variable in the optical \citep{Devereux2011}, UV \citep{Maoz2005}, and radio \citep{Nagar2002}. A possible explanation for the double-peaked broad H$\alpha$ emission line   \citep{Ho1997b, Shields2007, Balmaverde2014} is the presence of an inflow associated with the accretion disc, since the former, presumably, fuels the latter \citep{Devereux2011}. This would be consistent with the presence of the possible inflow-components found in both ground- and space-based data. A component  coming from a weak outflow was also detected in [S\,II] in ground-based data. However, the broad H$\alpha$  is contaminating these line profiles,  therefore their interpretation is rather uncertain.\\
The fit to the H$\alpha$ emission line in  both ground- and space-based data is not reliable (Table\,\ref{T_rms}) since a more elaborate physical modelling would be needed to fit the very broad and double-peaked emission line profile (likely originated in the outer parts of the accretion disc surrounding the SMBH, e.g. \citealt{StorchiBergmann2017}).  \\

\noindent \textbf{NGC\,4278.}  It is a relatively isolated elliptical galaxy with complex and irregular dust structure in its core with knots and filaments and a two-sided pc-scale jet \citep{Pellegrini2012}. The narrow components are interpreted as rotation in agreement with \citet{Sarzi2006} and \citet{Morganti2006} who already found a kinematical structure typical of an extended and regular disc. The second component is interpreted as a possible inflow and could be consistent with a small accretion region present at the centre. A more detailed, possibly 2D-IFS, study of the interplay between the ISM and nuclear activity is needed to confirm this scenario. A broad H$\alpha$ component is extracted from the complicated blend observed  in \textit{HST}/STIS data, only.\\

\noindent \textbf{NGC\,4438.} This spiral galaxy has a disturbed morphology with strong dust lanes and  a ring-like structure \citep{Masegosa2011}. The latter may be a consequence of environmental effects or the close interaction with M86 and/or NGC\,4435  \citep{Cortese2010}. Both narrow and second components are interpreted as rotation (as for the case of NGC\,2681) but the 2D rotation curve has a very low amplitude and is perturbed (Fig.\,\ref{Panel_PVD}). The second (broader) components ($\sigma$\,$\sim$\,200\,km\,s$^{-1}$, Table\,\ref{T_kin}) could be either related to its  complex nuclear morphology or related to the H$\alpha$ outflow-like structure in  \citet{Masegosa2011}. We did not find evidence for a BLR component but our modelling of H$\alpha$ is not fully reliable (Table\,\ref{T_rms}). Overall, the peculiar properties  of NGC\,4438 make it an ideal target for IFS observations.\\

\begin{figure*}
\vspace{-0.25cm} 
\includegraphics[trim = 1.10cm .85cm 11.0cm 17.75cm, clip=true, width=.40\textwidth]{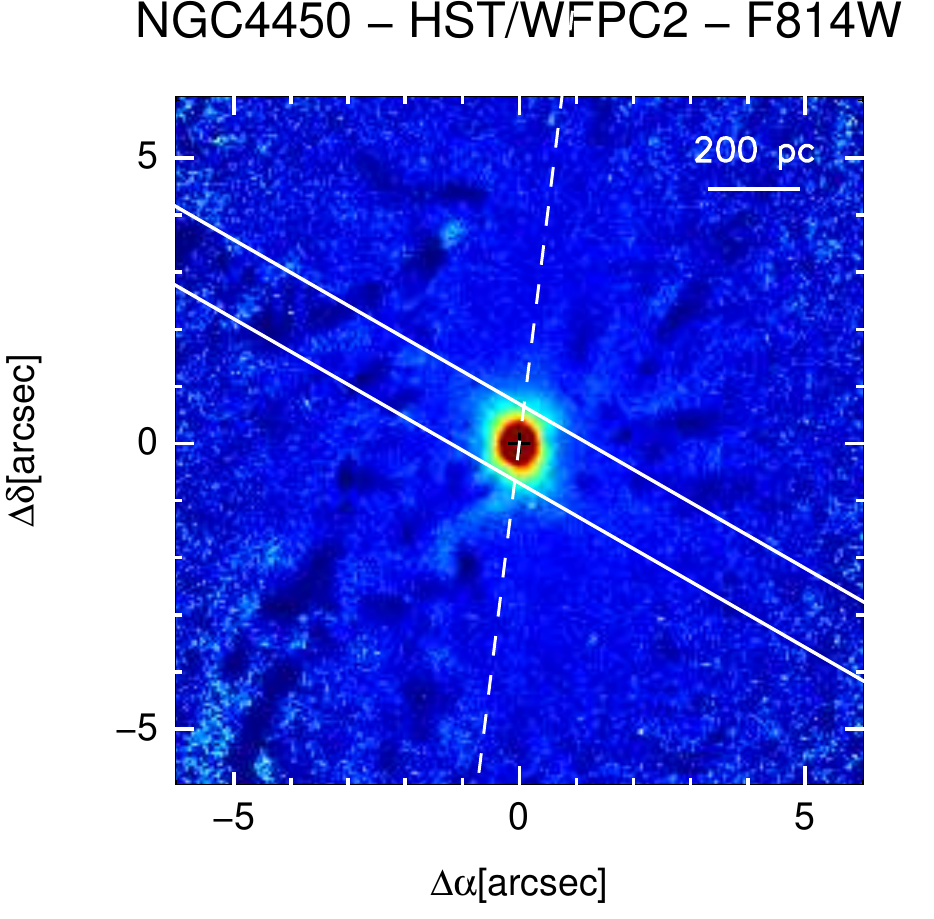} 
\hspace{-0.3cm} 
\includegraphics[trim = 4.5cm 13.cm 5.25cm 6.25cm, clip=true, width=.475\textwidth]{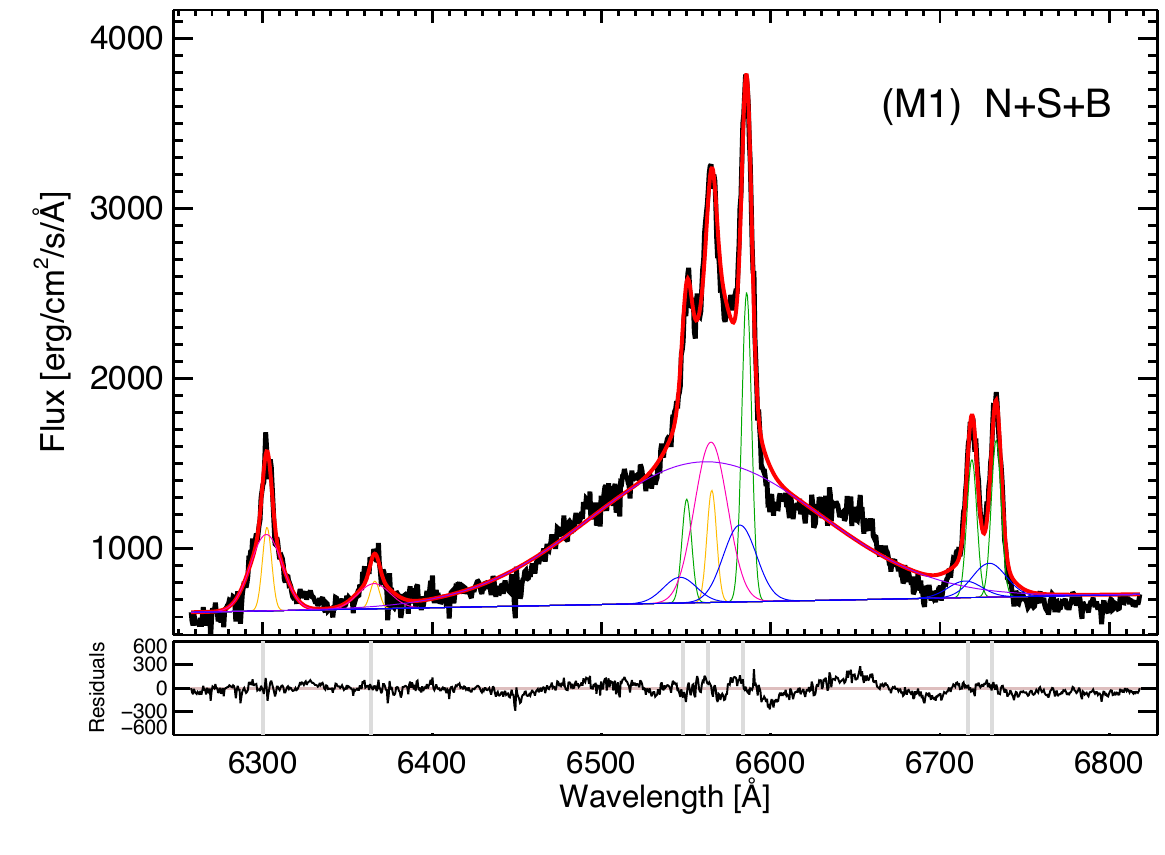}  \\
\vspace{-0.10cm}
\includegraphics[trim = 2.4cm 19.75cm 2.7cm 3.75cm, clip=true, width=.915\textwidth]{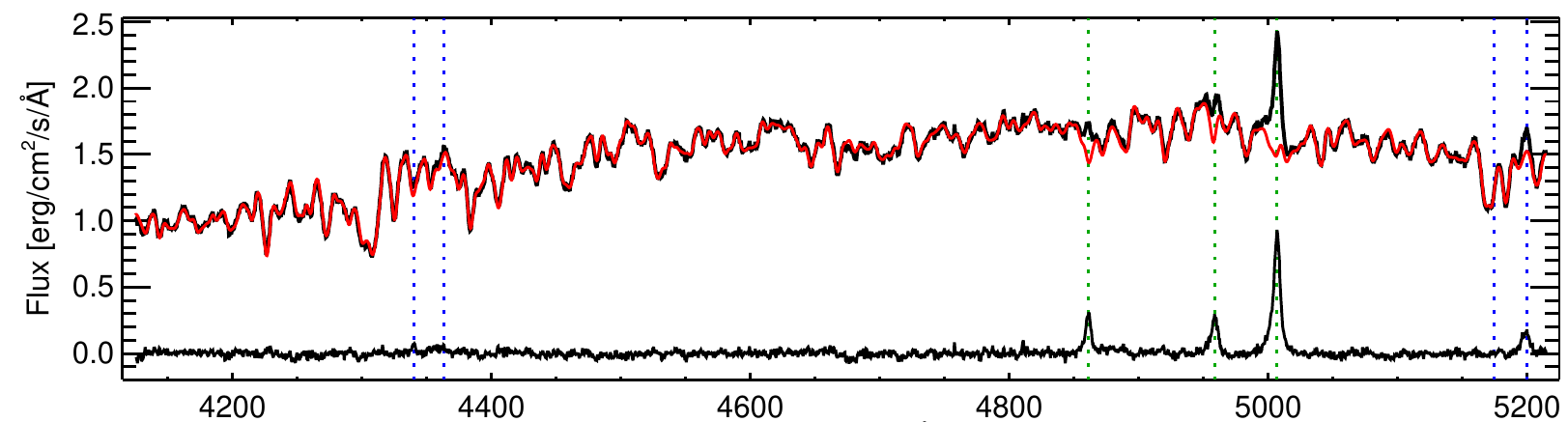} \\
\vspace{-0.10cm}
\includegraphics[trim = 2.4cm 18.75cm 2.7cm 3.75cm, clip=true, width=.91\textwidth]{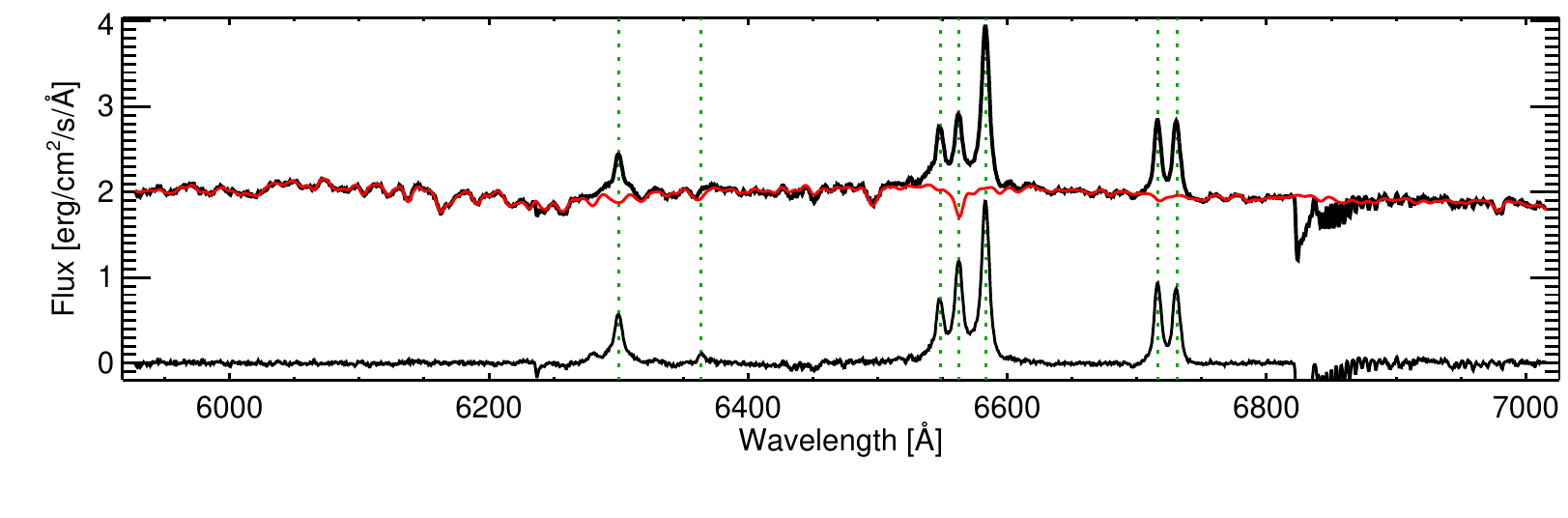} \\
\vspace{-0.45cm}
\includegraphics[trim = 5cm 13.25cm 5.25cm 6.3cm, clip=true, width=.465\textwidth]{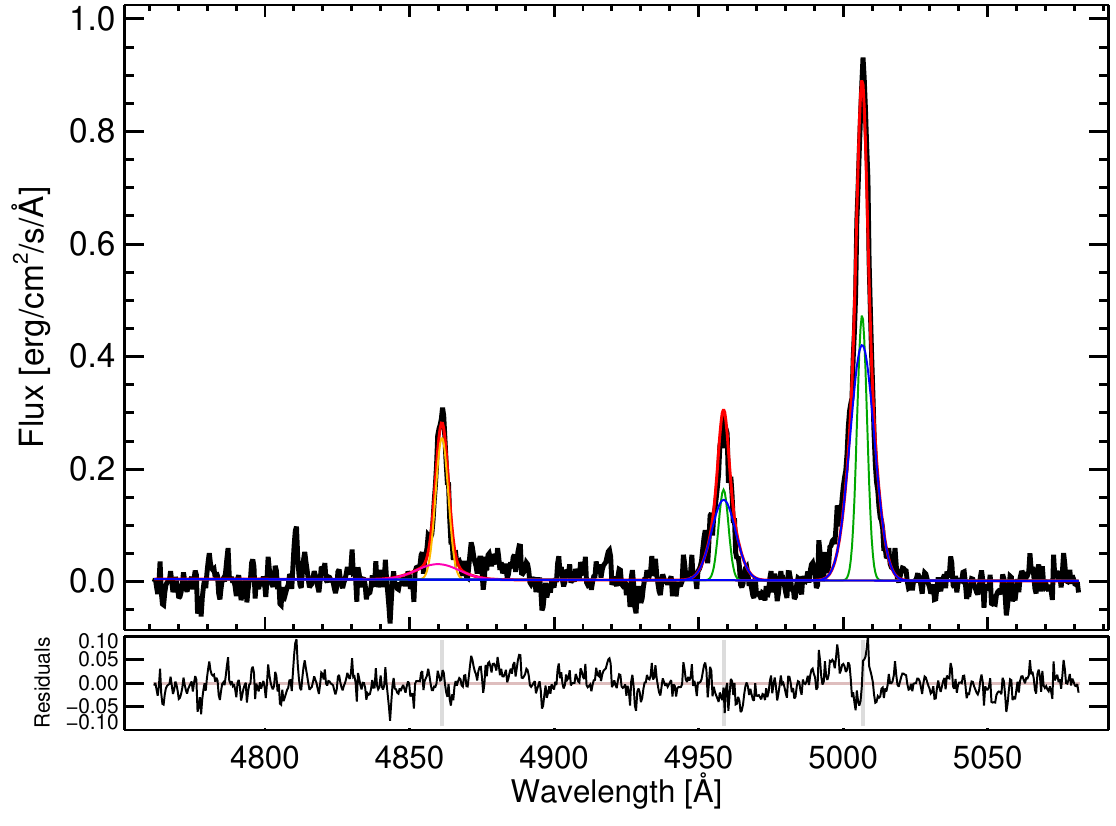}
\hspace{0.1cm} 
\includegraphics[trim = 5.55cm 13.25cm 5.25cm 6.3cm, clip=true, width=.445\textwidth]{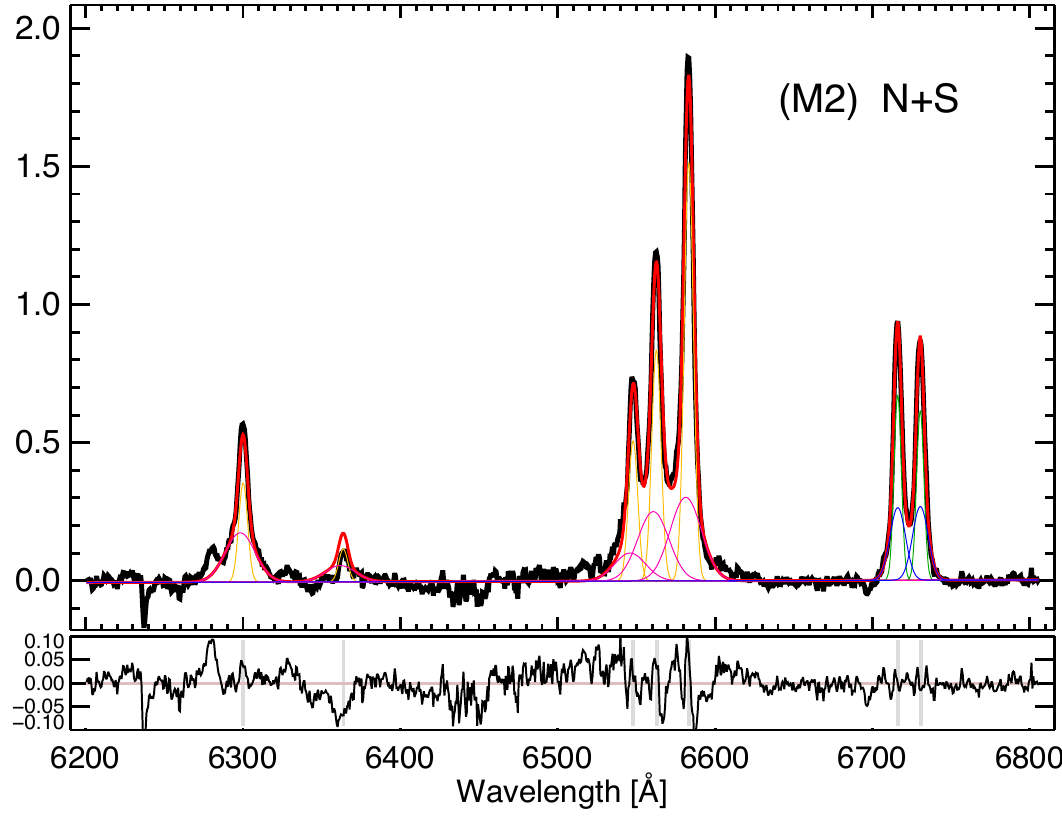} 
\caption{(General description as in Fig.\,\ref{Panel_NGC0266}.) NGC\,4450. Forbidden lines are rather broad even before the stellar subtraction. [O\,I] ([S\,II]) line-profiles  have blue (red) wings.} \label{Panel_NGC4450} 		 		 
\end{figure*}

\noindent \textbf{NGC\,4450.}  \citet{Brum2017} found a clear rotating disc pattern with the line of nodes oriented along the North-South direction with some deviations. The flat and low amplitude rotation curve presented in Fig.\,\ref{Panel_PVD} partially disagrees with that obtained by \citet{Cortes2015} probably due to difference in slit orientations (Fig.\,\ref{Panel_NGC4450}). The same authors interpreted the perturbation of the rotation curve either as due to non-circular motions or related to an external gas acquisition or a minor merger event (see also \citealt{Coccato2013}). All this could be in agreement with the complex outflow/inflow scenario suggested by the second components of the forbidden lines (Table\,\ref{class_sum}); but spatially resolved data at higher spectral resolution are needed to confirm this hypothesis. A double-peaked broad H$\alpha$ line profile (interpreted as the signature of the outer parts of a relativistic accretion disc) has been reported by \citet{Ho2000}. Interestingly, this feature is completely absent in our ground-based spectroscopic data being only evident in \textit{HST}/STIS data.\\

\noindent \textbf{NGC\,4636.}  It is one of the most well-known nearby elliptical galaxies.  It has a central compact source and a very faint ring-like structure \citep{Masegosa2011}. The prominent emission towards the North with a clear outflow-like morphology  \citep{Masegosa2011} is consistent with X-ray observations \citep{OSullivan2005}. These X-ray data confirm the presence of a cavity and a plume  that appear to be the product of past AGN activity (AGN outbursts), with the AGN being quiescent at present.  This could explain the relatively low  FWHM of the broad H$\alpha$ component (considering our sample). Unfortunately, the outflow-like structure is not fully captured in our CAHA spectroscopy (probably due to to slit orientation).\\

\noindent \textbf{NGC\,4750.}  This spiral galaxy has a very bright nucleus with some arc or shell-like structures \citep{Mason2015}. The position-velocity curve shows some clear asymmetries at large radii (Fig.\,\ref{Panel_PVD}). 
The presence of a faint blue wing in Oxygen lines is interpreted as possibly being originated by an outflow  (Tables \ref{T_kin} and \ref{class_sum}). The measurement of the FWHM of the BLR-component is in fair agreement with  that reported by \textit{HFS97}. Blueward of the NaD, the spectrum is rather noisy  due to some residual telluric absorption. Despite this, we were able to model the weak doublet feature with one kinematical component (Fig.\,\ref{Panel_NaD}); it has a completely different  kinematics with respect to the ionised gas (Tables \ref{T_kin} and \ref{T_NaD}), with a low velocity dispersion and a rather large blueshifted velocity.\\

\noindent \textbf{NGC\,4772.} The optical emission line profiles of this spiral galaxy (known as  \lq eye galaxy\rq) are well modelled with only one kinematic component. The velocity is rest frame, the velocity dispersion is the largest  among our measurements for the narrow component (i.e. 249\,$\pm$\,12\,km\,s$^{-1}$). Such broadening is likely due to the chaotic kinematics (i.e. a superposition of  multiple components) present in the individual spectra in our nuclear aperture. This could be consistent with the results obtained by \citet{Hayne2000}, who found that the stars and the ionised gas in the centre of this galaxy are counter-rotating (often taken as an indication of a merger). Many knots are observed inside the slit we used to obtain the present  ground-based data; preventing a reliable position-velocity distribution.\\

\noindent \textbf{NGC\,5005.}  The distinguish features of this weakly barred spiral galaxy  are the strong dust lane crossing the galaxy from East to West (offset from the nucleus), the fan-shaped filaments, the compact nuclear clumps and its obscured nucleus \citep{Pogge2000}. The presence of a weak AGN is confirmed by the detection of a compact hard X-ray source \citep{GonzalezMartin2006} which is variable on scales of months \citep{Younes2011}. The second component in the CAHA spectrum is interpret as a possible outflow. In the \textit{HST}-spectrum a second component is also present and is interpreted as a candidate outflow. This component is at the same velocity of the outflow seen in ground-based data ($\sim$\,110\,km\,s$^{-1}$), the velocity dispersion is slightly lower (308 vs. 446 km\,s$^{-1}$) but consistent within uncertainties. These results are  consistent with the interpretation of the H$_{2}$$\lambda$2.122$\mu$m,  [Fe\,II]$\lambda$1.644$\mu$m and  Br$\gamma$$\lambda$2.166$\mu$m near-IR flux measurements  by  \citet{Bendo2004}. The BLR component is only detected  in the \textit{HST}/STIS spectroscopic data, possibly due to a less reliable fit to the H$\alpha$ emission line in ground-based data (Table\,\ref{T_rms}). \\

\noindent \textbf{NGC\,5077.} This elliptical galaxy has a compact unresolved source in the nucleus and dust lanes on top of a smooth gas distribution.  The observed position-velocity curve (Fig.\,\ref{Panel_PVD}) and the narrow component in the the CAHA/TWIN spectra are consistent with gas in regular rotation around the galaxy centre (as seen also by \citealt{DeFrancesco2008}). A second component that may be produced by an outflow is present only in the \textit{HST}/STIS spectrum (Table\,\ref{class_sum}). A broad component is required for modelling the H$\alpha$ line profile in both ground- and space-based spectroscopic data (Table\,\ref{T_FWHM}).

\clearpage
\begin{figure*}
\vspace{-0.25cm} 
\includegraphics[trim = 1.10cm .85cm 11.0cm 17.75cm, clip=true, width=.40\textwidth]{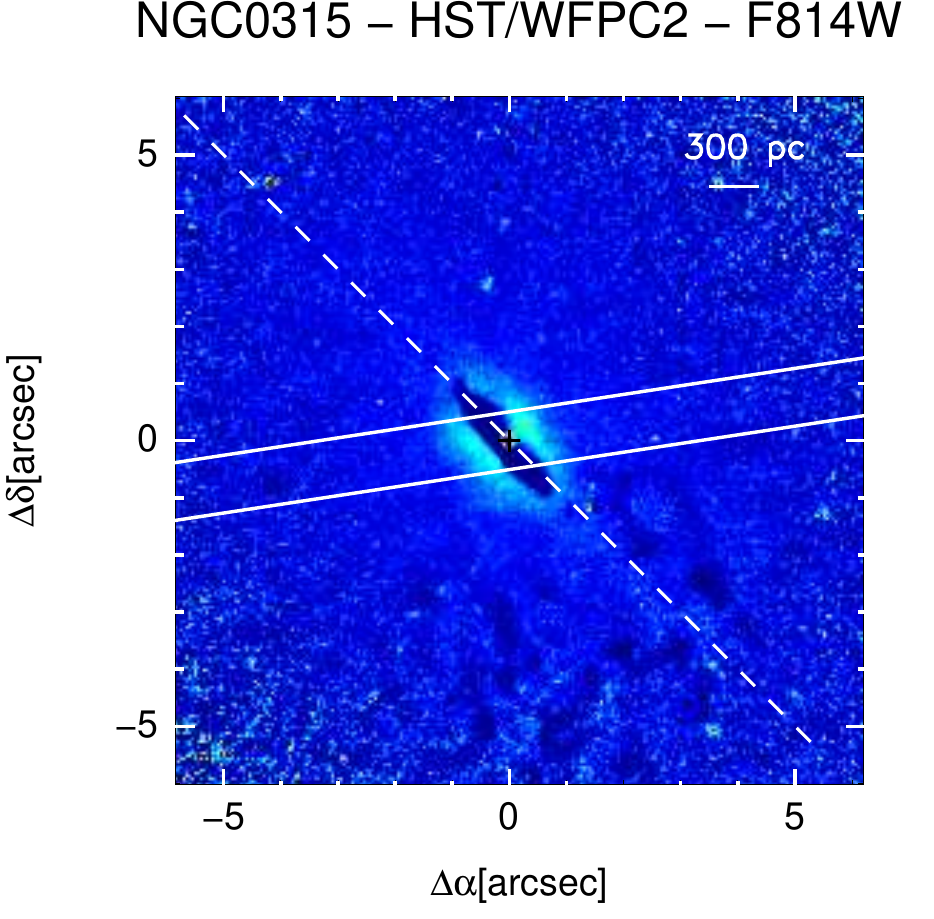} 
\hspace{-0.3cm} 
\includegraphics[trim = 4.5cm 13.cm 5.25cm 6.25cm, clip=true, width=.475\textwidth]{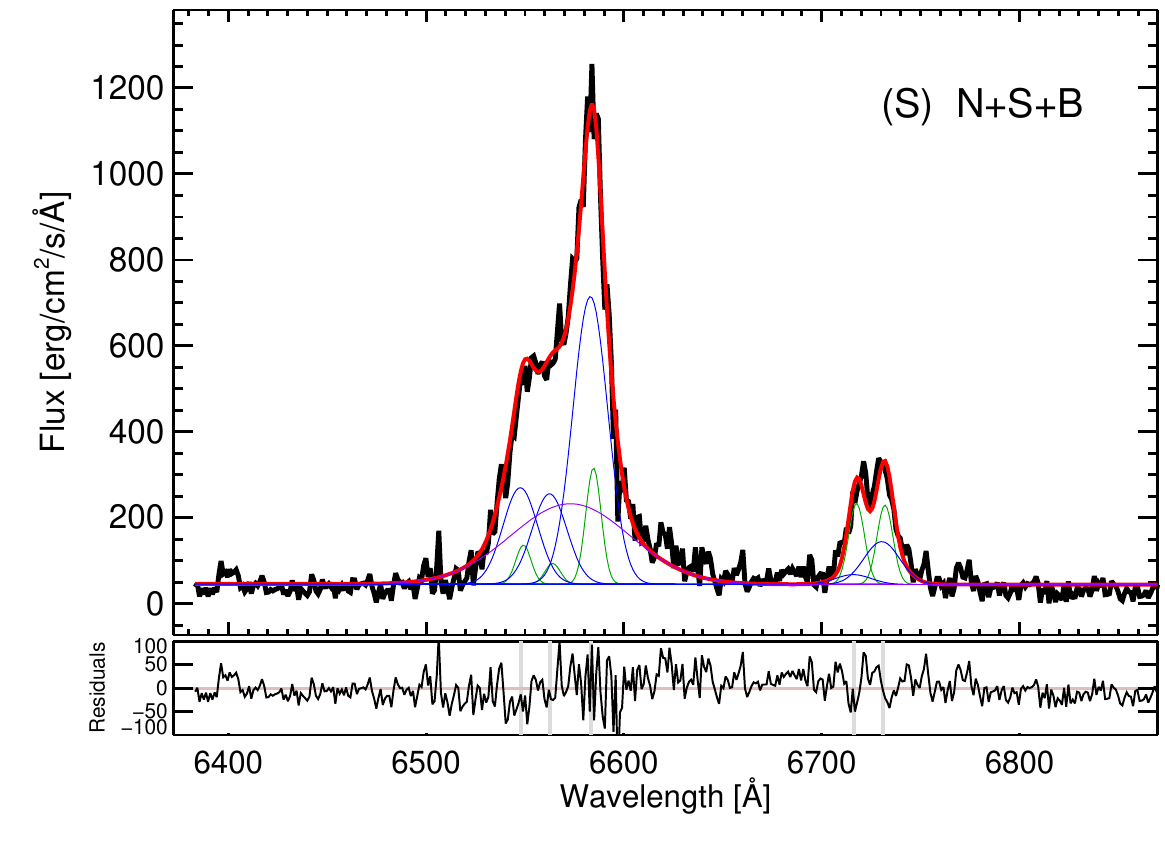}  \\
\vspace{-0.10cm}
\includegraphics[trim = 2.4cm 19.75cm 2.7cm 3.15cm, clip=true, width=.87\textwidth]{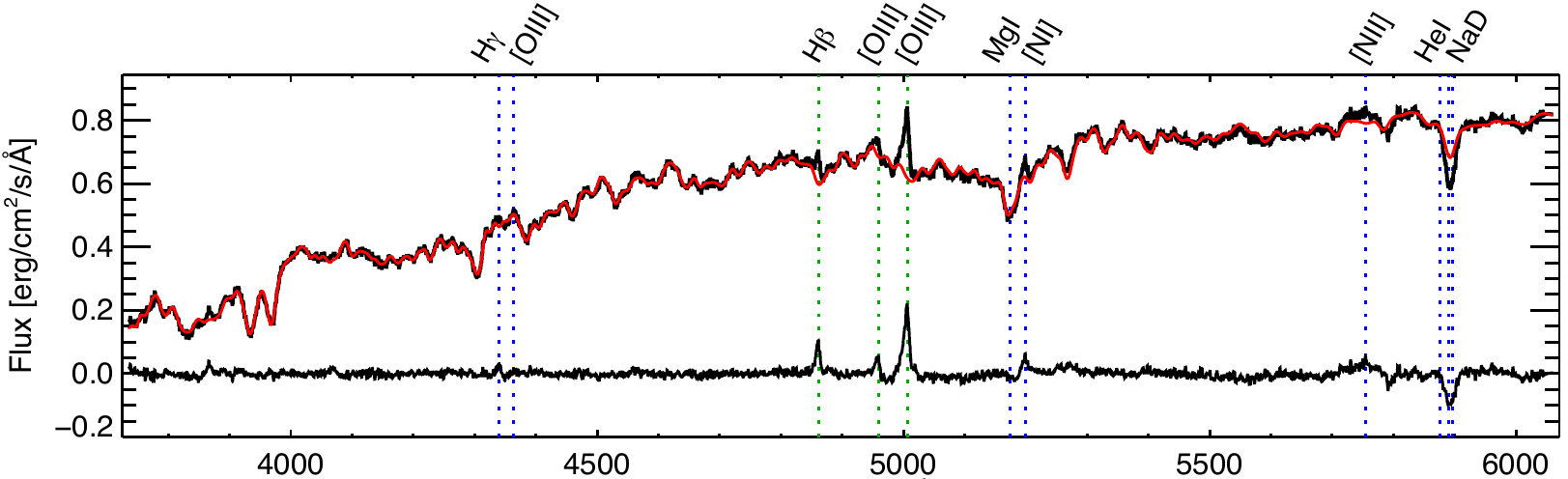} \\
\vspace{-0.10cm}
\includegraphics[trim = 2.4cm 18.75cm 2.7cm 3.15cm, clip=true, width=.87\textwidth]{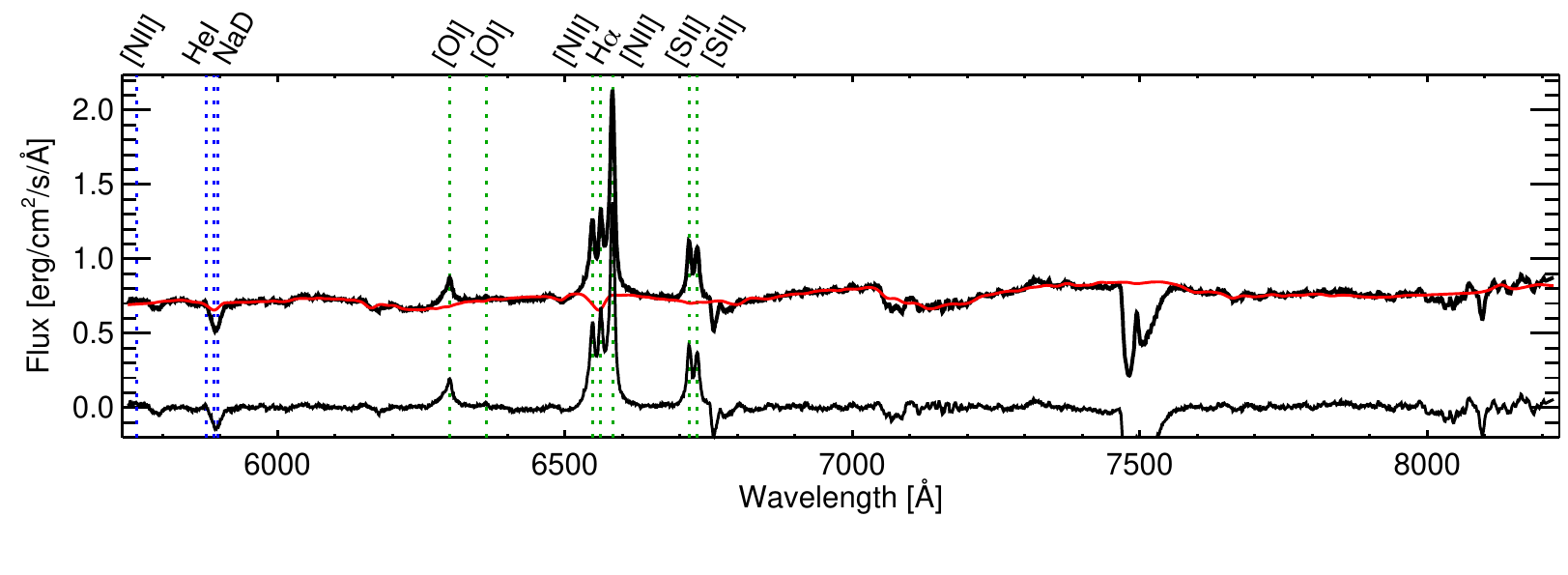} \\
\vspace{-0.45cm}
\includegraphics[trim = 4.9cm 13.25cm 5.25cm 6.3cm, clip=true, width=.44\textwidth]{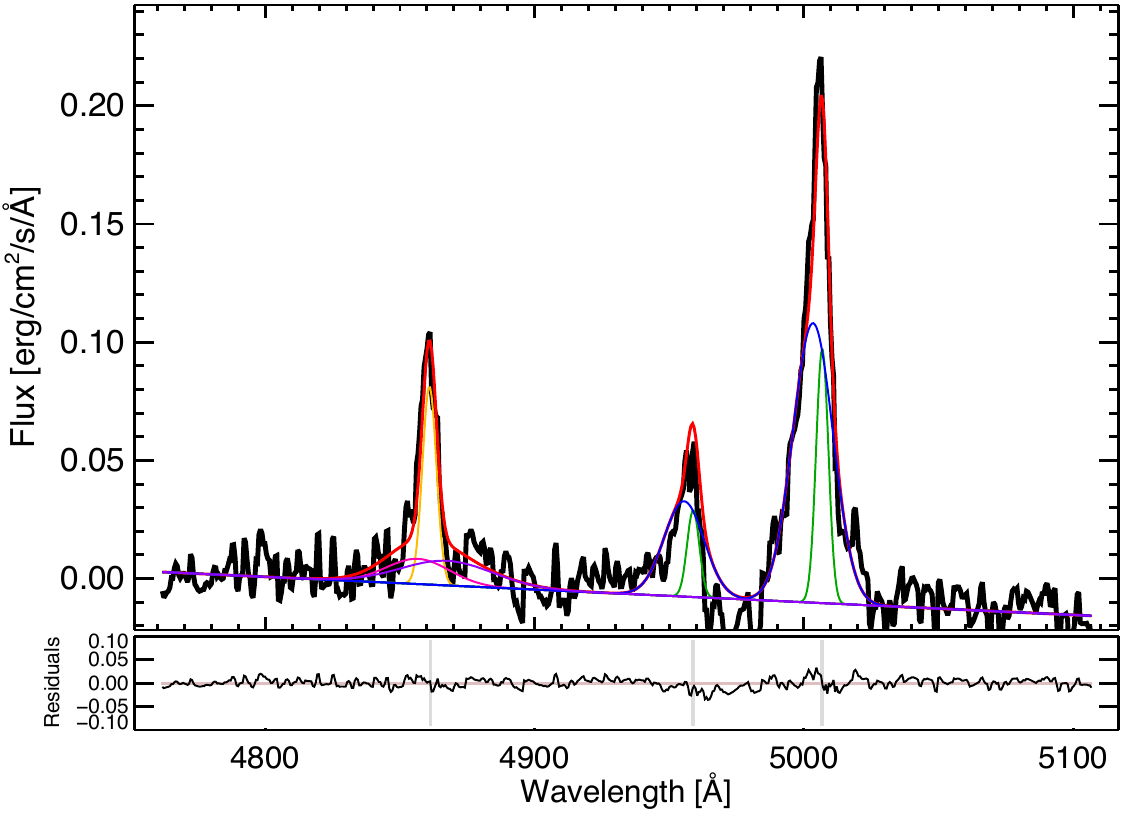}
\hspace{0.1cm} 
\includegraphics[trim = 5.55cm 13.25cm 5.25cm 6.3cm, clip=true, width=.415\textwidth]{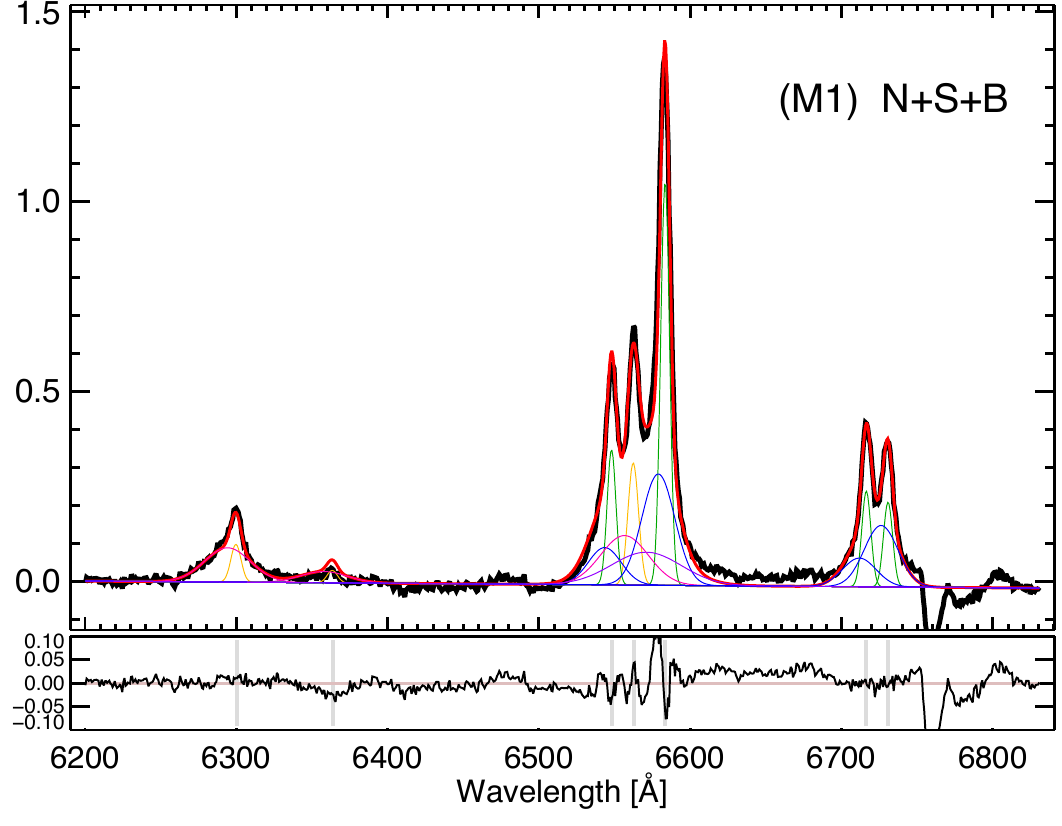}  
\caption{(General description as in Fig.\,\ref{Panel_NGC0266}.) NGC\,0315: [O\,I] lines are  asymmetric and  the [S\,II] doublet is rather broad. These considerations suggest the presence of two different second components both blueshifted but with different widths. We did not find a clear improvement of the standard deviation in H$\alpha$ for the ground-based data fitting when adding the broad  component (Table\,\ref{T_rms}). However, the presence of a red wing near the base of [N\,II]$\lambda$6584 (absent in the other template lines) makes the broad component necessary to adequately model the H$\alpha$-[N\,II] complex. Thus a broad component is needed for a good fit, but its presence is not obvious and it is rather weak  (in  good agreement with  \textit{HFS97}).}
 \label{Panel_NGC0315} 		 		 
\end{figure*}
\clearpage

\clearpage

\begin{figure*}
\vspace{-0.25cm} 
\includegraphics[trim = 1.10cm .85cm 11.0cm 17.75cm, clip=true, width=.40\textwidth]{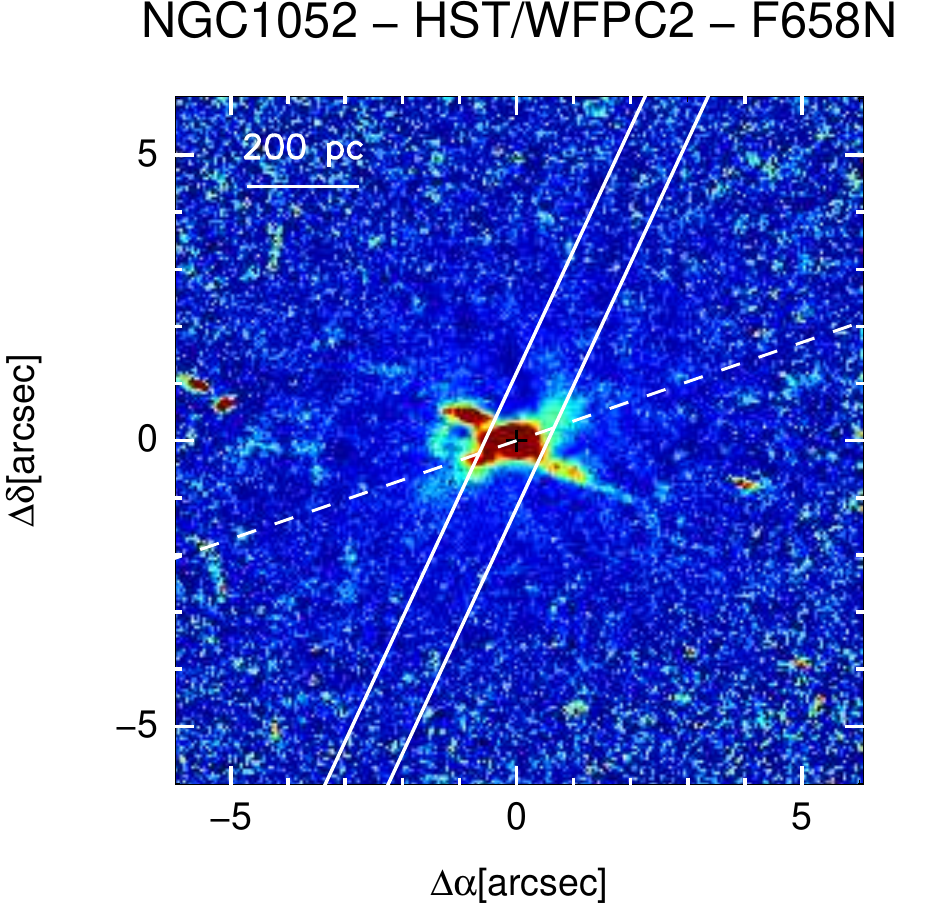} 
\hspace{-0.3cm} 
\includegraphics[trim = 4.5cm 13.cm 5.25cm 6.25cm, clip=true, width=.475\textwidth]{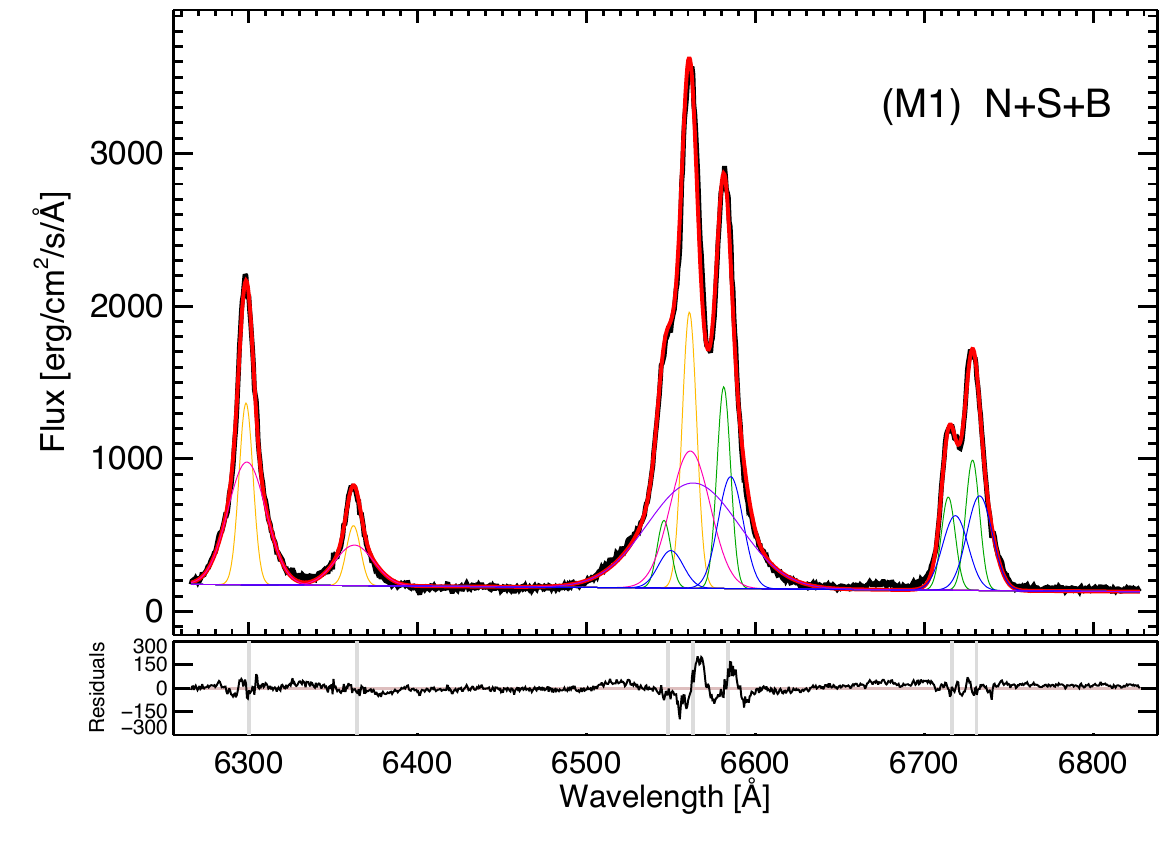}  \\
\vspace{-0.10cm}
\includegraphics[trim = 2.4cm 19.75cm 2.7cm 3.75cm, clip=true, width=.915\textwidth]{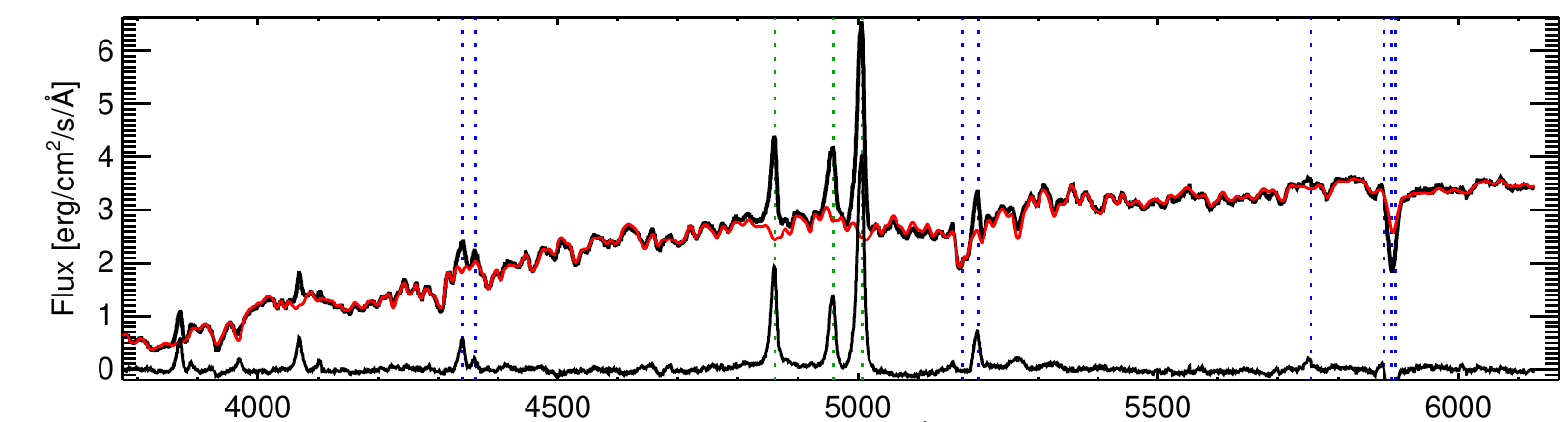} \\
\vspace{-0.10cm}
\includegraphics[trim = 2.4cm 18.75cm 2.7cm 3.75cm, clip=true, width=.91\textwidth]{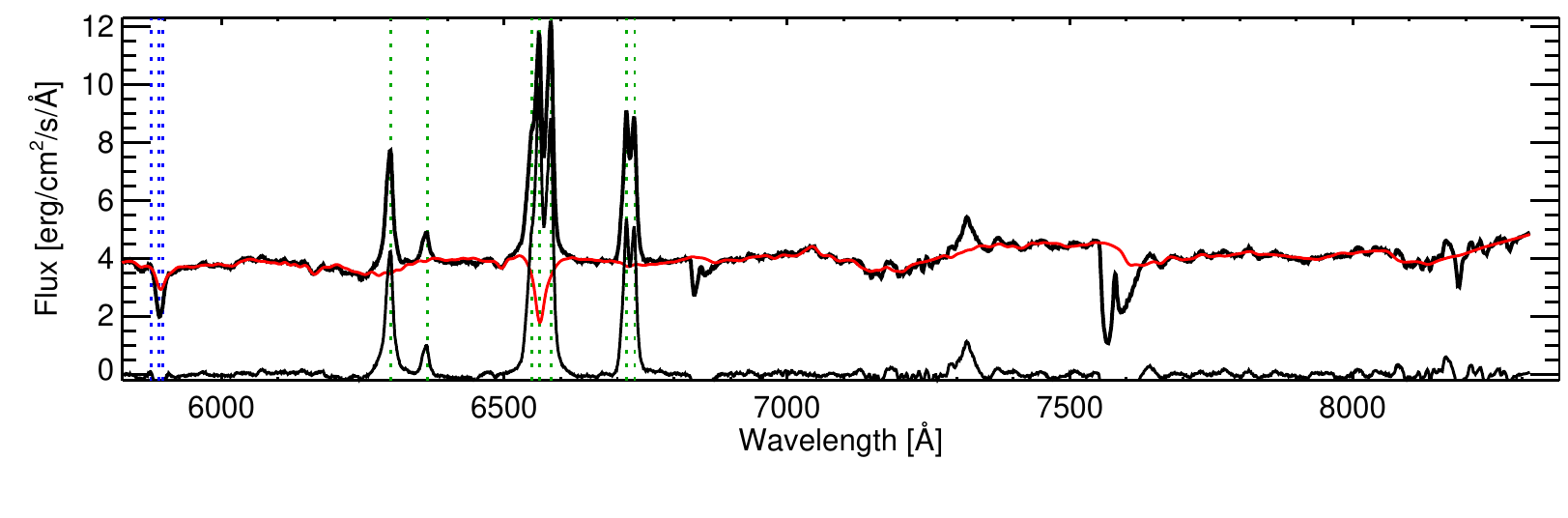} \\
\vspace{-0.45cm}
\includegraphics[trim = 4.9cm 13.25cm 5.25cm 6.3cm, clip=true, width=.4715\textwidth]{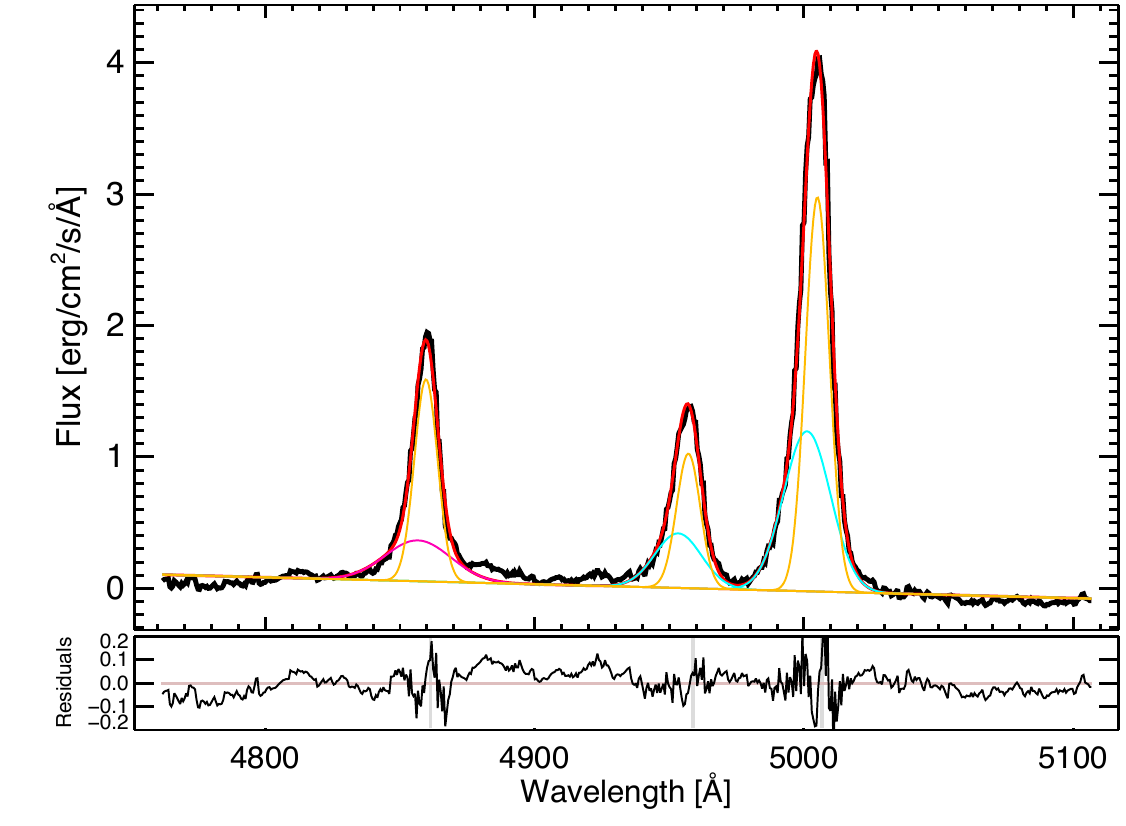}
\hspace{0.1cm} 
\includegraphics[trim = 5.55cm 13.25cm 5.25cm 6.3cm, clip=true, width=.445\textwidth]{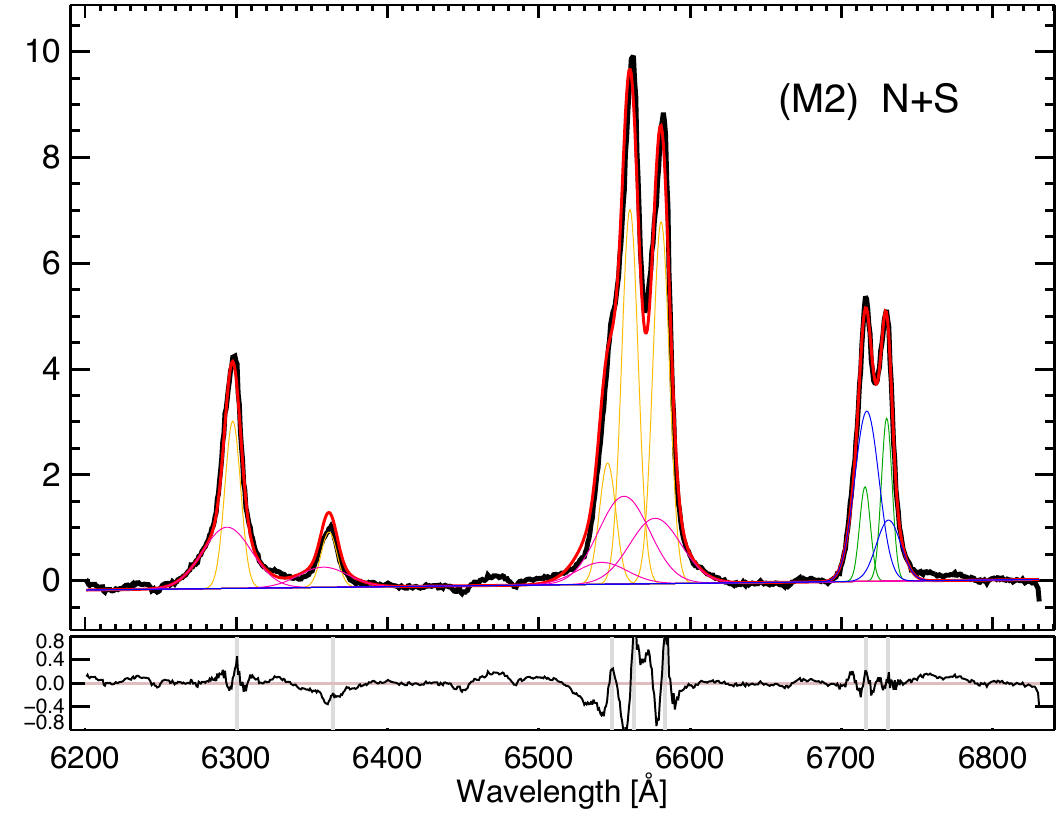}  
\caption{(General description as in Fig.\,\ref{Panel_NGC0266}.) NGC\,1052: broad blue wings are evident in [O\,I] but not in [S\,II].  The second component in [O\,III] (light blue curves) seems to behave otherwise compared to all the other emission lines. We also report the detection of [Ca\,II]$\lambda$$\lambda$7291,7324 and [O\,II]$\lambda$$\lambda$7318,7319,7330,7331 emission lines which are strongly blended.} 
\label{Panel_NGC1052} 		 		 
\end{figure*}
\clearpage

\begin{figure*}
\vspace{-0.25cm} 
\includegraphics[trim = 1.10cm .85cm 11.0cm 17.75cm, clip=true, width=.40\textwidth]{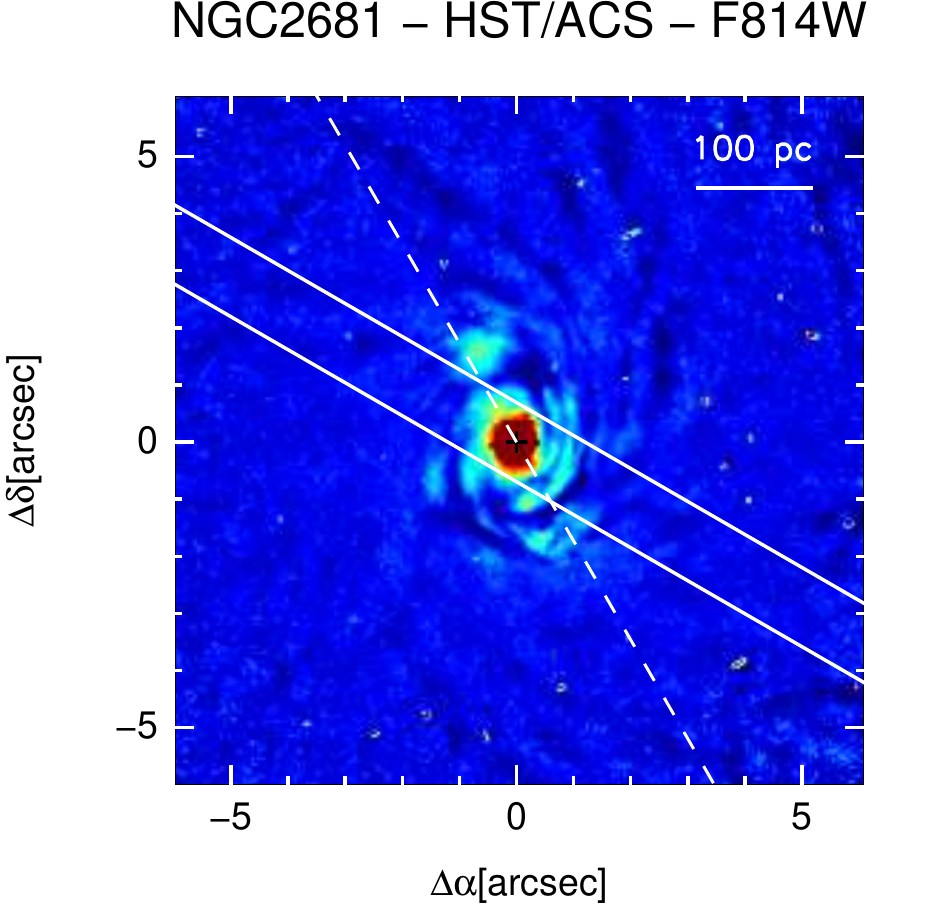}
\hspace{-0.3cm} 
\includegraphics[trim = 2.4cm 19.75cm 2.7cm 3.75cm, clip=true, width=.915\textwidth]{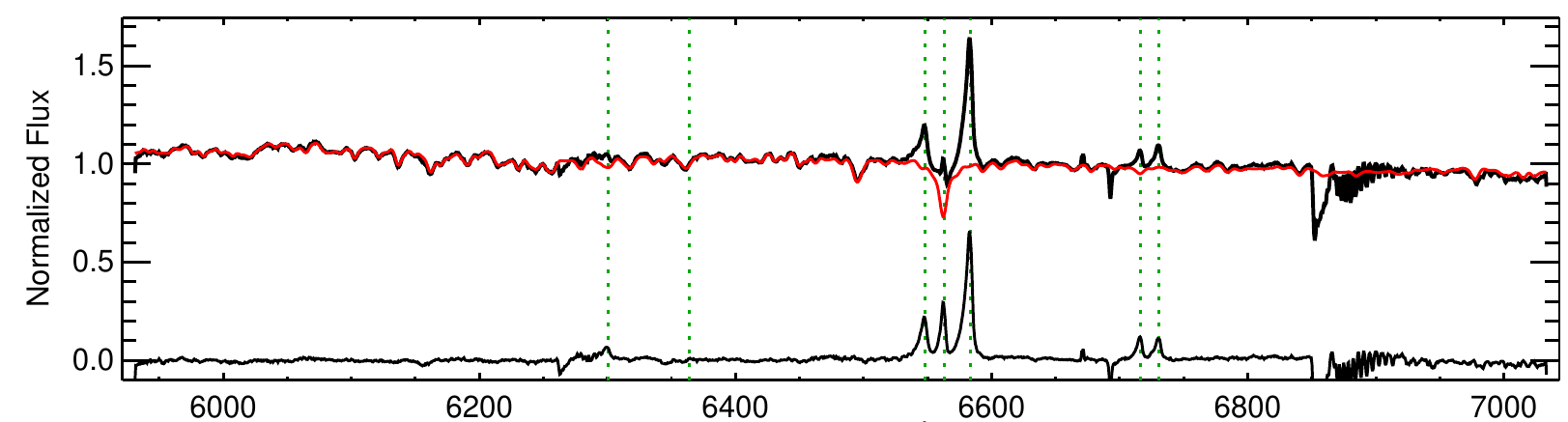}\\
\vspace{-0.10cm}
\includegraphics[trim = 2.4cm 18.75cm 2.7cm 3.75cm, clip=true, width=.91\textwidth]{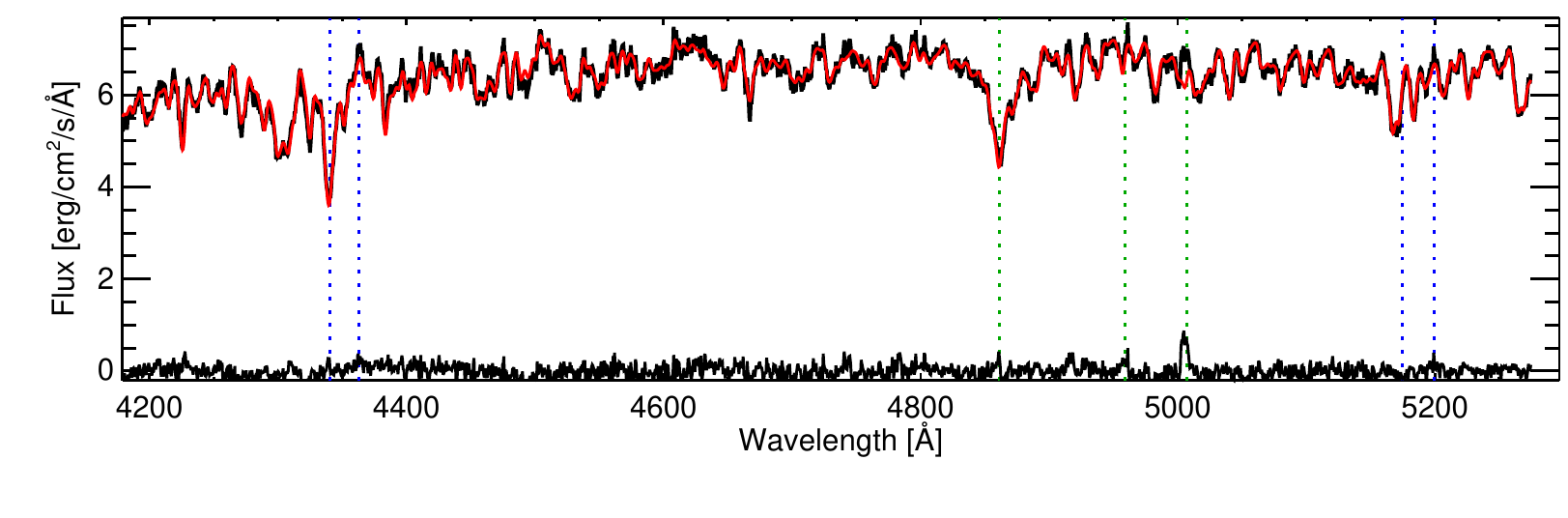} \\
\vspace{-0.45cm}
\includegraphics[trim = 5cm 13.25cm 5.25cm 6.3cm, clip=true, width=.465\textwidth]{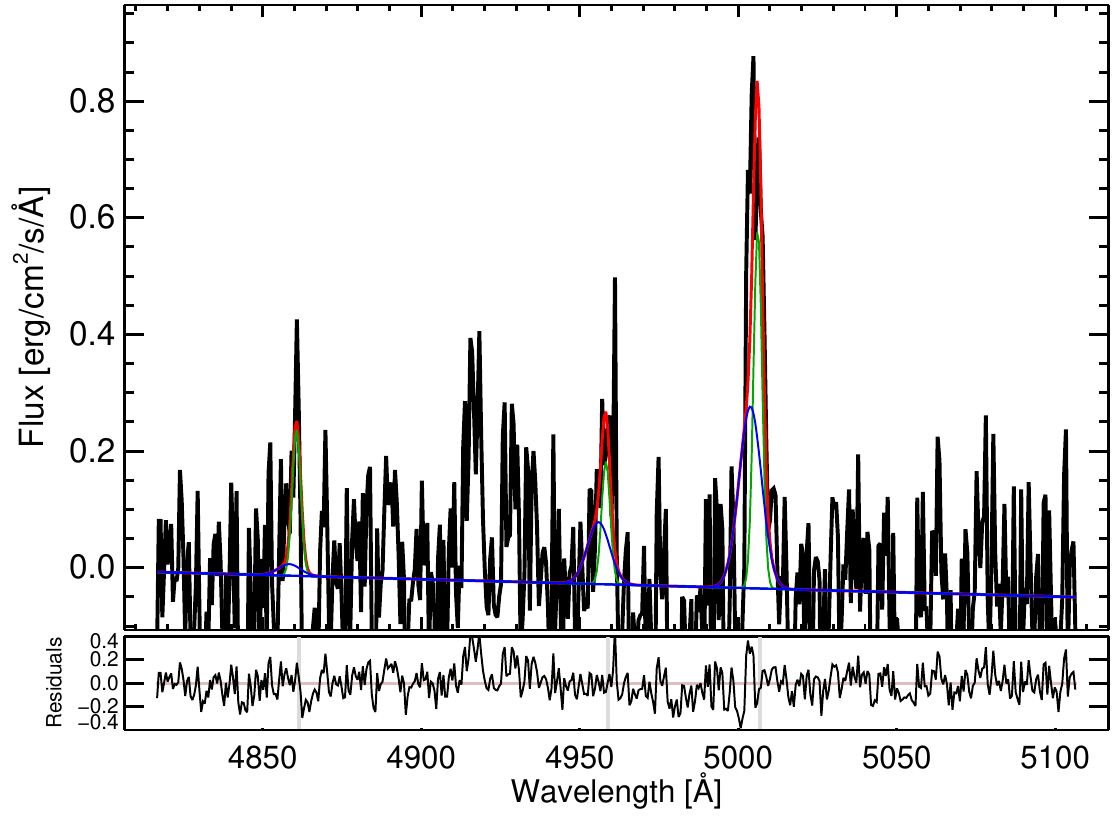}
\hspace{0.1cm} 
\includegraphics[trim = 5.55cm 13.25cm 5.25cm 6.3cm, clip=true, width=.445\textwidth]{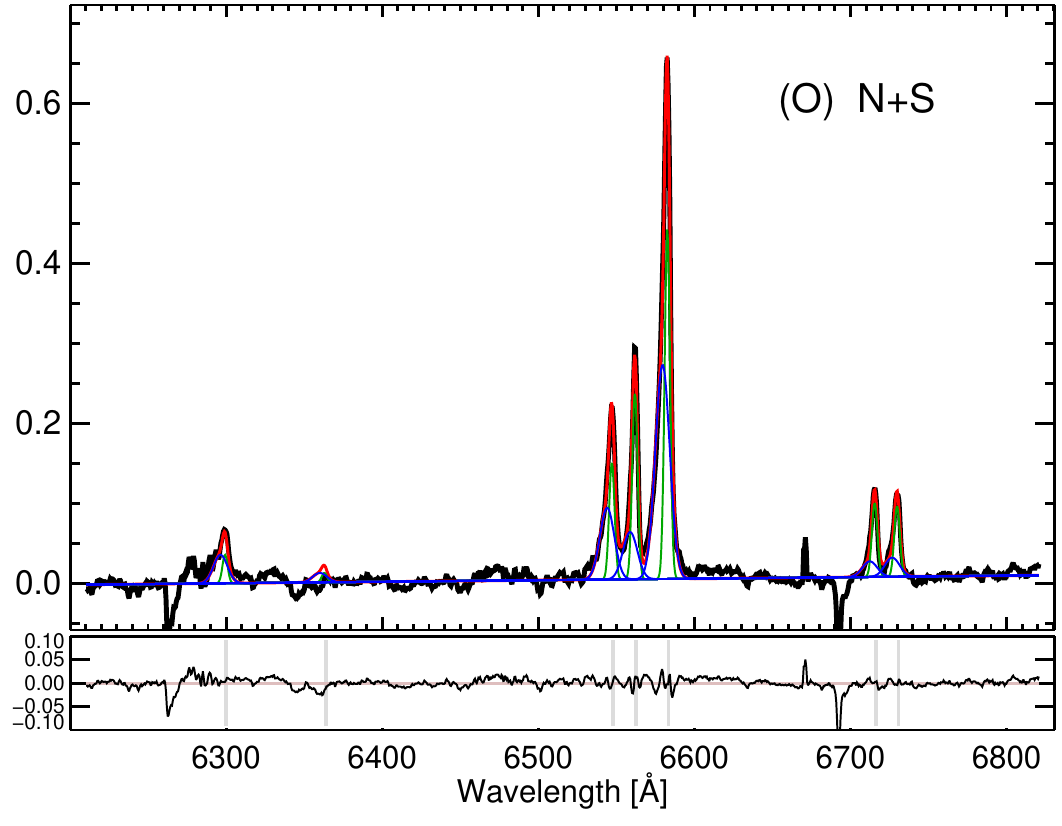}  
\caption{(General description as in Fig.\,\ref{Panel_NGC0266}.) NGC\,2681: as in the case of NGC\,0841, H$\alpha$ is strongly affected by stellar absorption in the observed spectrum. All emission lines are rather narrow, a  dip is seen blueward of both [O\,I] and [S\,II].  After applying the procedure of the line modelling and by visual inspecting  all the results (Sect.\,\ref{Analysis_LF}),  the broad component appears to be absent. Contrary to \textit{HFS97}, we did not find any tail (indicative of a broad component) in [N\,II] and  the Fe\,I$\lambda$6523 feature is absent in our spectrum.}
 \label{Panel_NGC2681} 		 		 
\end{figure*}
\clearpage

\begin{figure*}
\vspace{-0.25cm} 
\includegraphics[trim = 1.10cm .85cm 11.0cm 17.75cm, clip=true, width=.40\textwidth]{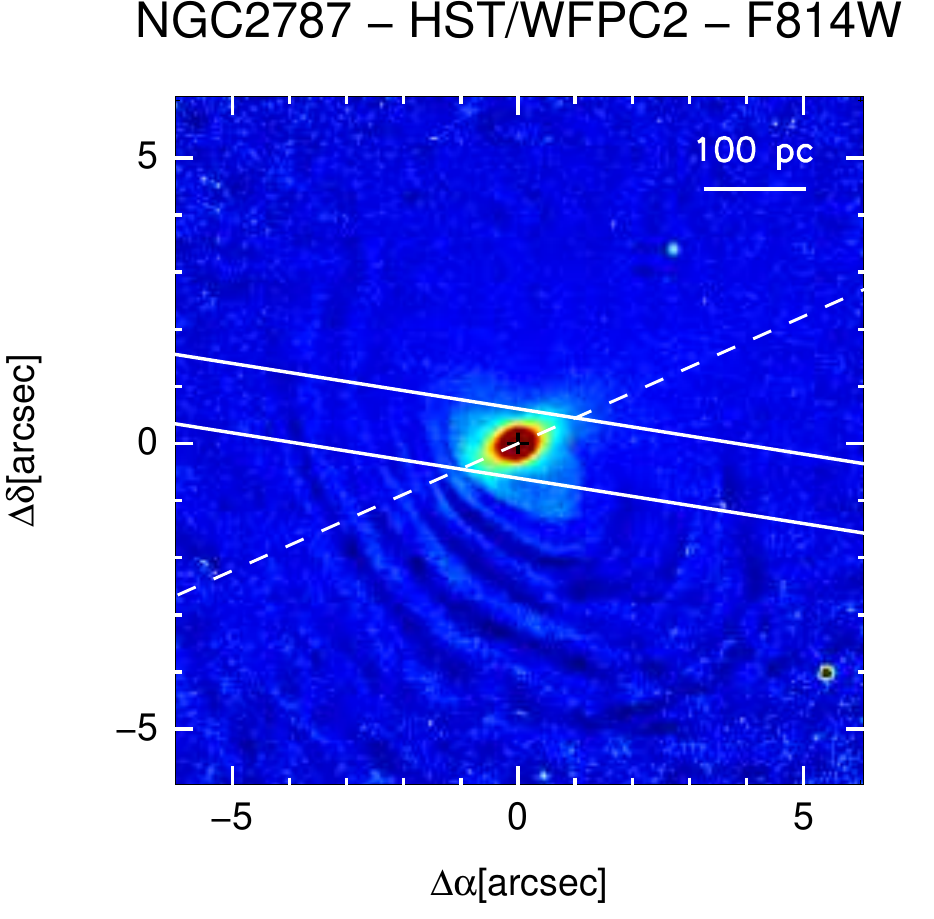} 
\hspace{-0.3cm} 
\includegraphics[trim = 4.5cm 13.cm 5.25cm 6.25cm, clip=true, width=.475\textwidth]{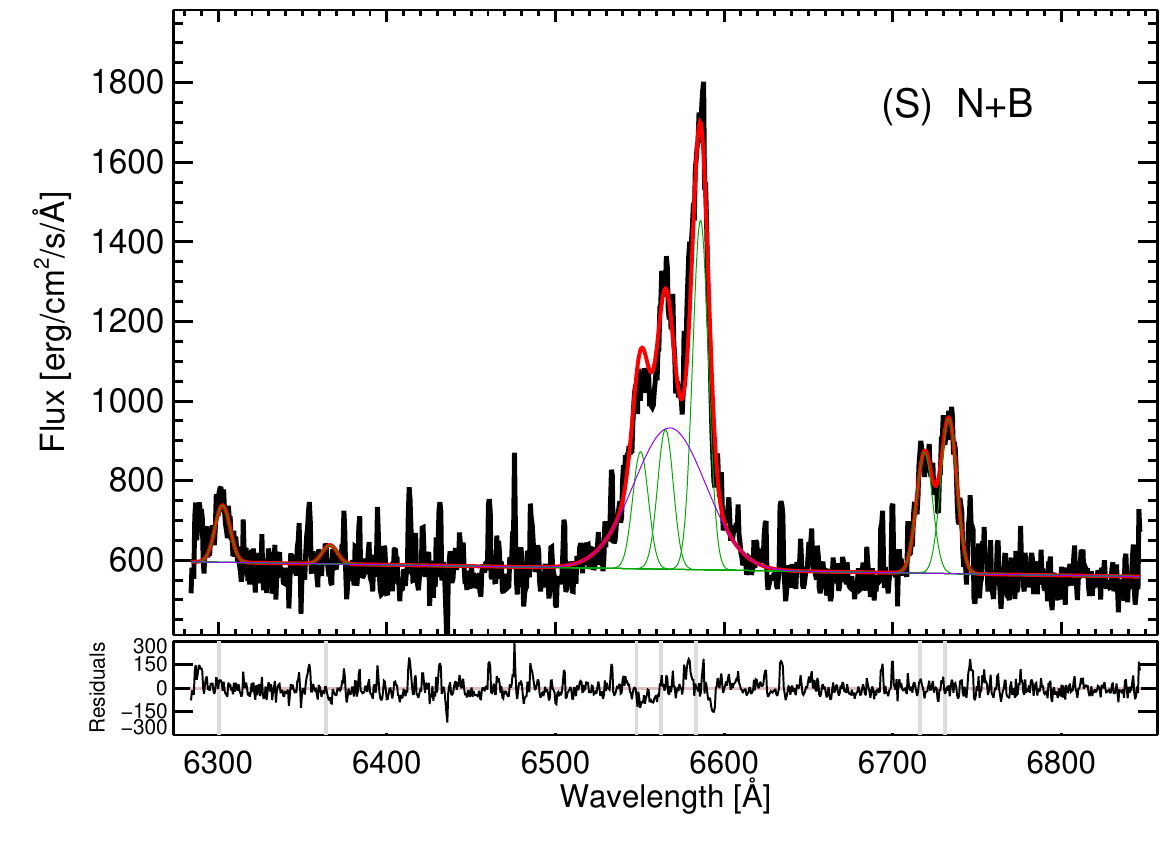}  \\
\vspace{-0.10cm}
\includegraphics[trim = 2.4cm 19.75cm 2.7cm 3.75cm, clip=true, width=.915\textwidth]{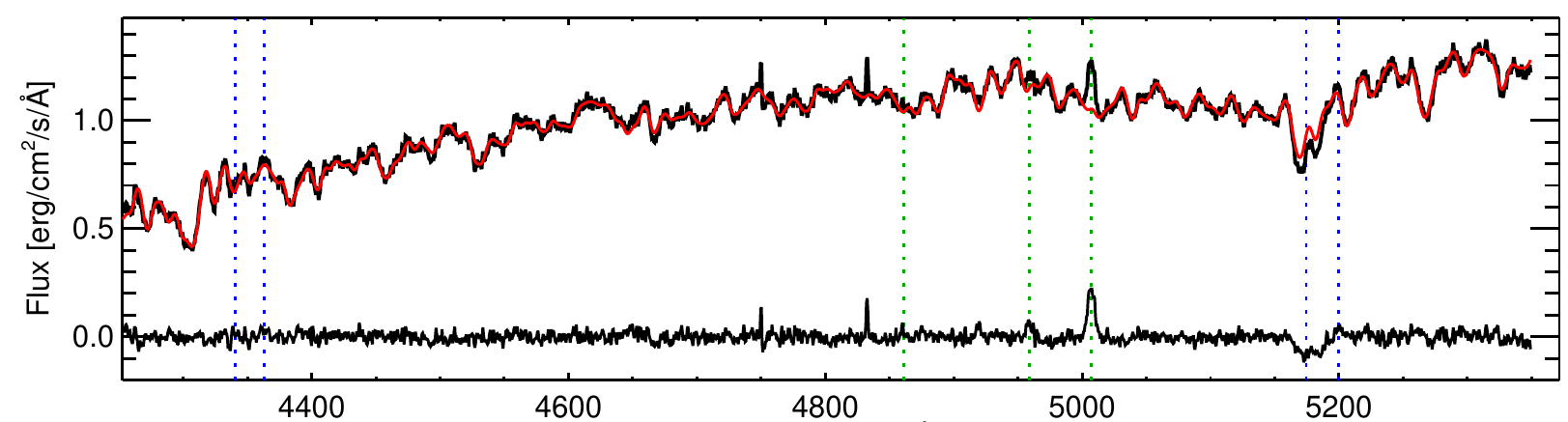} \\
\vspace{-0.10cm}
\includegraphics[trim = 2.4cm 18.75cm 2.7cm 3.75cm, clip=true, width=.91\textwidth]{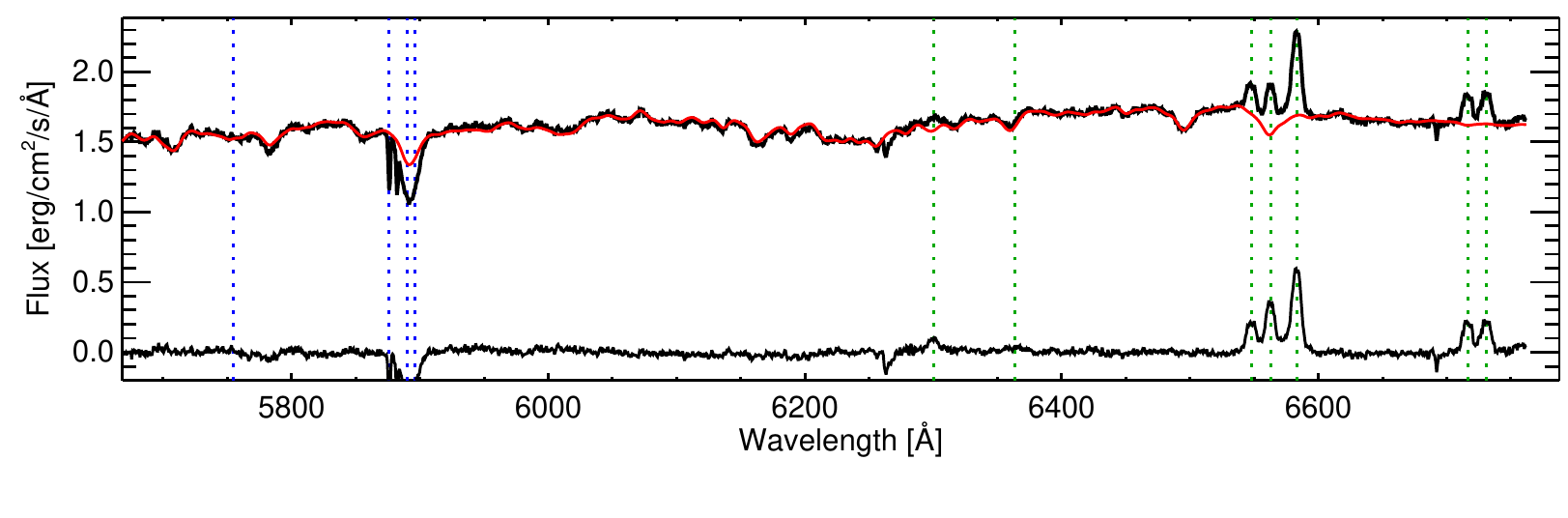} \\
\vspace{-0.45cm}
\includegraphics[trim = 4.9cm 13.25cm 5.25cm 6.3cm, clip=true, width=.4715\textwidth]{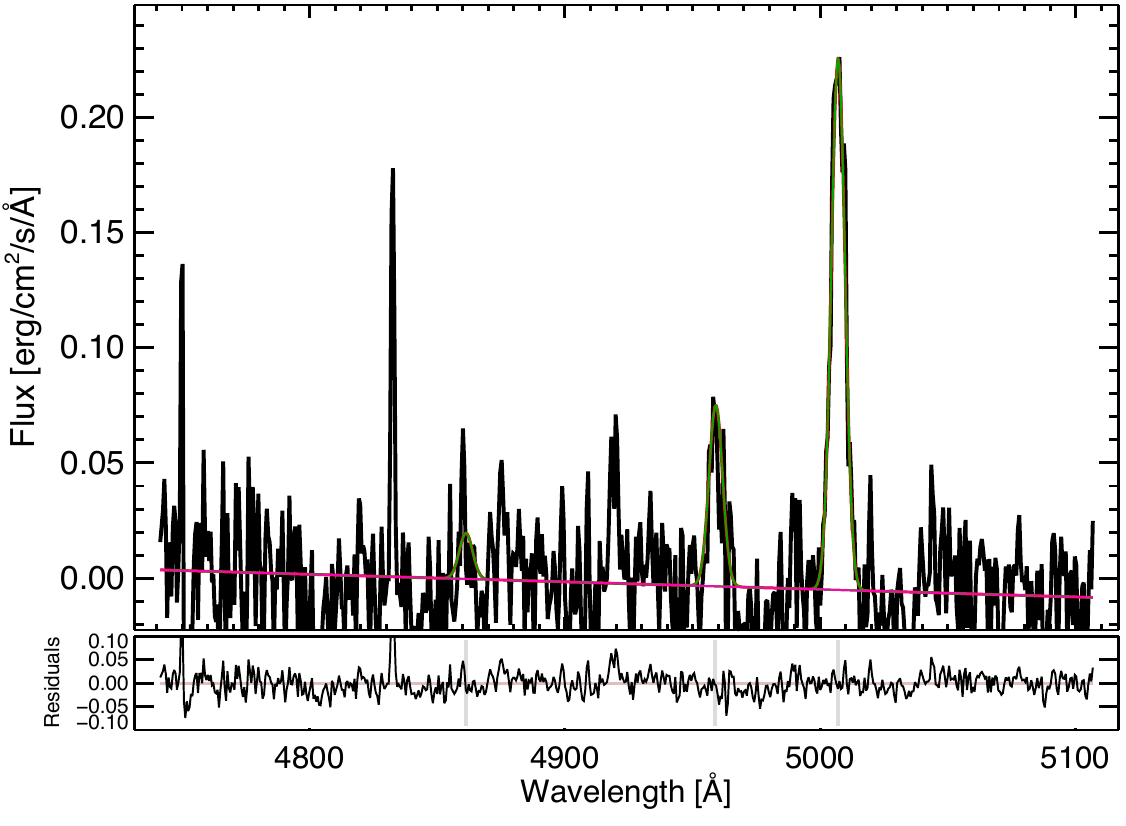}
\hspace{0.1cm} 
\includegraphics[trim = 5.55cm 13.25cm 5.25cm 6.3cm, clip=true, width=.445\textwidth]{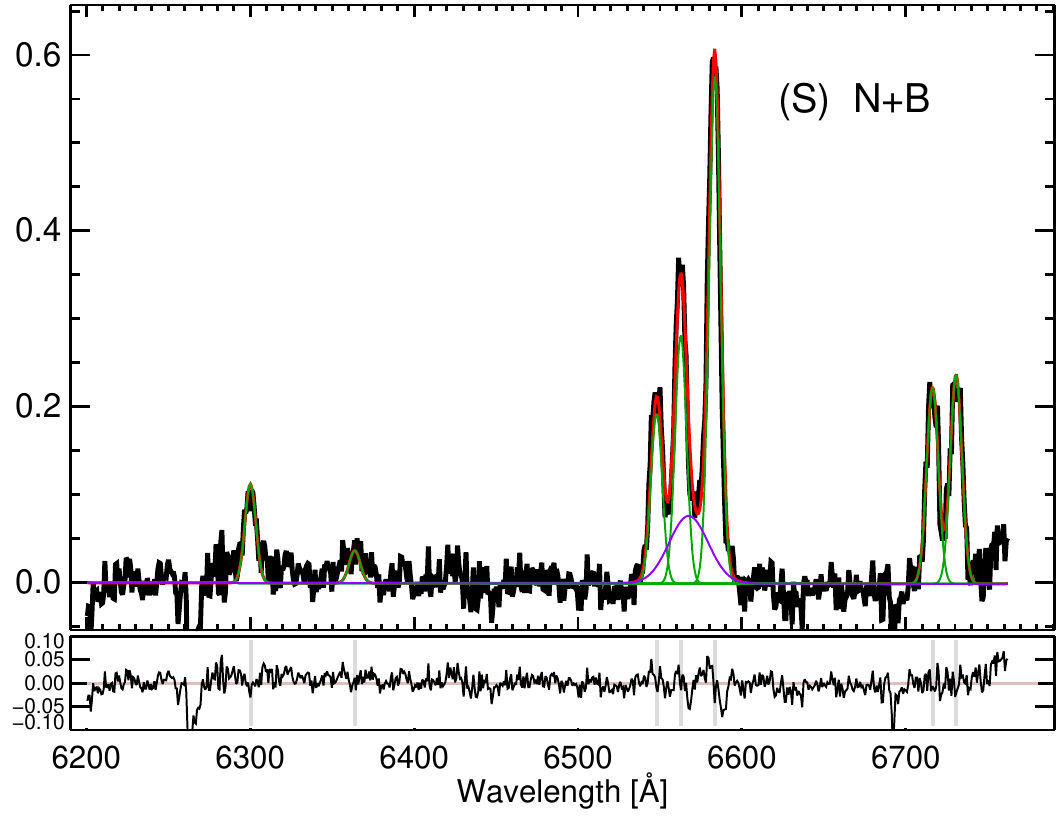}  
\caption{(General description as in Fig.\,\ref{Panel_NGC0266}.) NGC\,2787: the H$\alpha$-[N\,II] line profiles are quite different from those in  the Palomar spectra (\textit{HFS97}). Specifically, there are not any indications of an extended red wing. Narrow lines are  well modelled  with one component, an additional broad  H$\alpha$ component (which is fairly prominent, as for the Palomar spectrum) is needed to well reproduce the line profiles in our red spectrum.  The S/N of the blue spectrum is low and H$\beta$ is very weak, what hamper the detection of an eventual broad H$\beta$ component.} \label{Panel_NGC2787} 		 		 
\end{figure*}
\clearpage
\begin{figure*}
\vspace{-0.25cm} 
\includegraphics[trim = 1.10cm .85cm 11.0cm 17.75cm, clip=true, width=.40\textwidth]{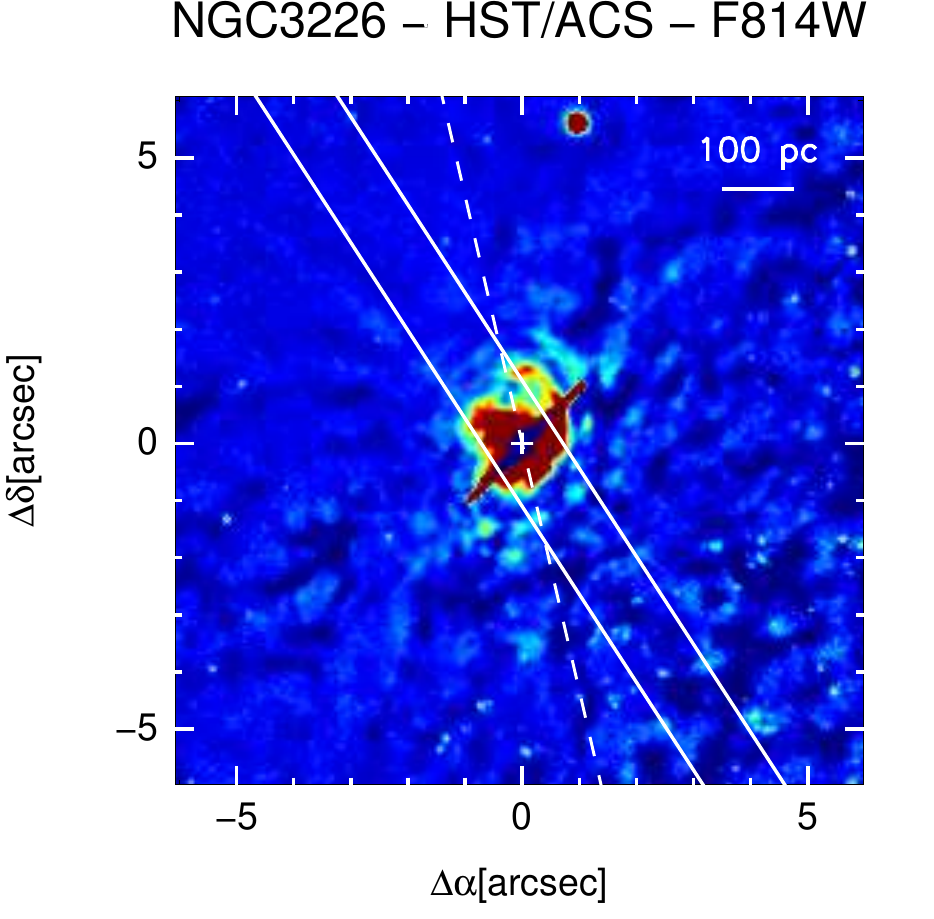} 
\hspace{-0.3cm} 
\includegraphics[trim = 2.4cm 19.75cm 2.7cm 3.75cm, clip=true, width=.915\textwidth]{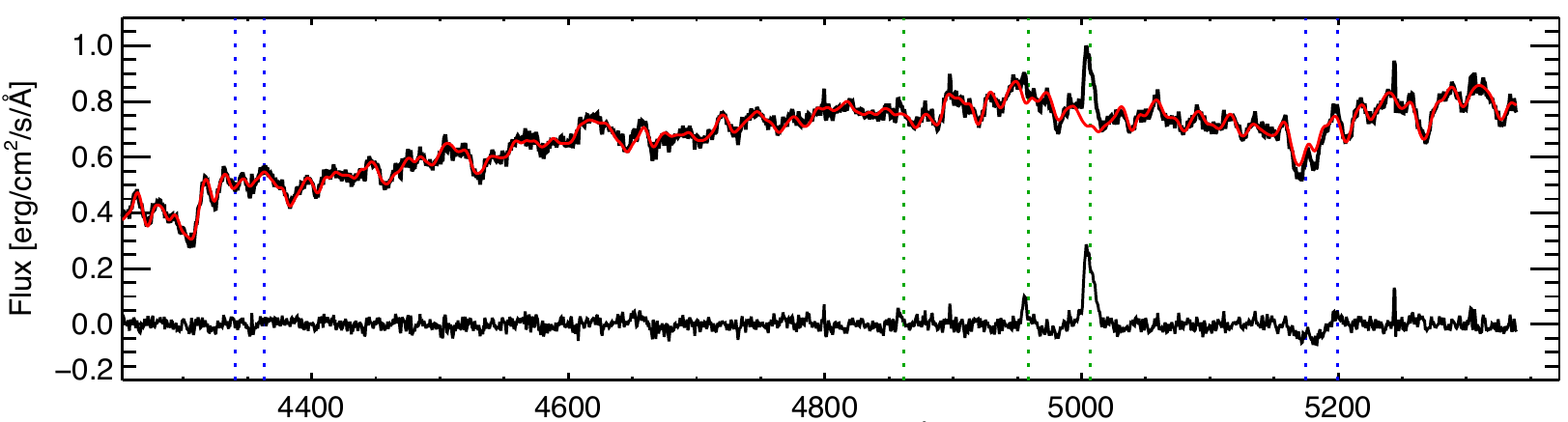} \\
\vspace{-0.10cm}
\includegraphics[trim = 2.4cm 18.75cm 2.7cm 3.75cm, clip=true, width=.91\textwidth]{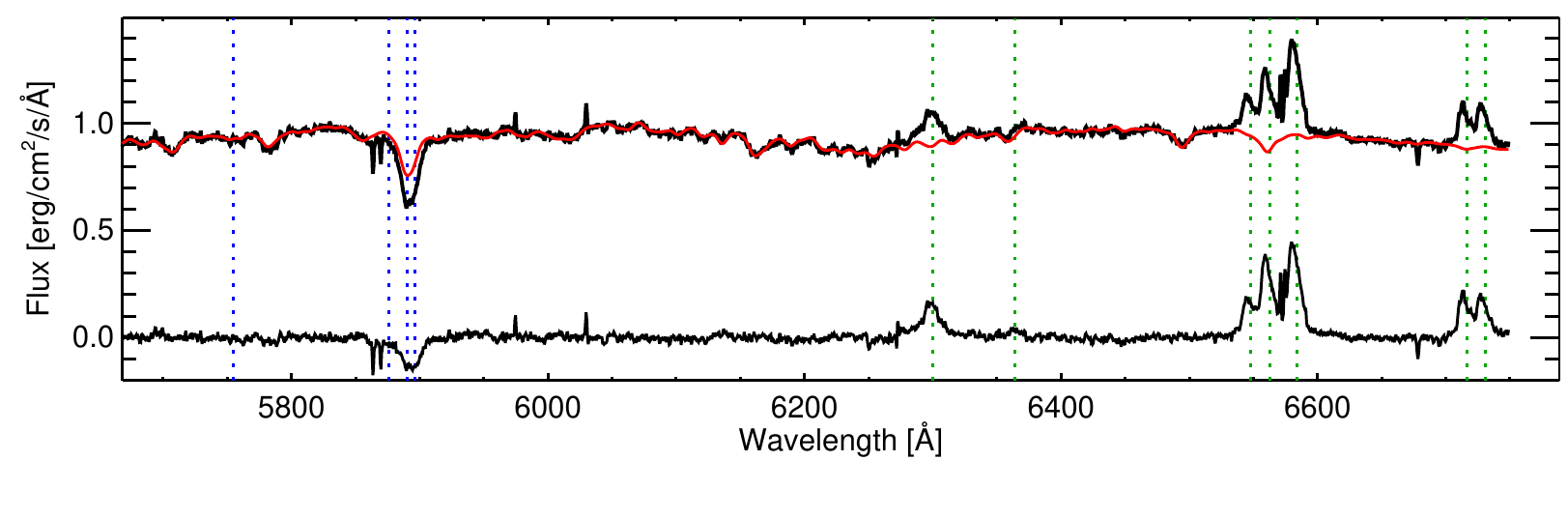} \\
\vspace{-0.45cm}
\includegraphics[trim = 4.9cm 13.25cm 5.25cm 6.3cm, clip=true, width=.4715\textwidth]{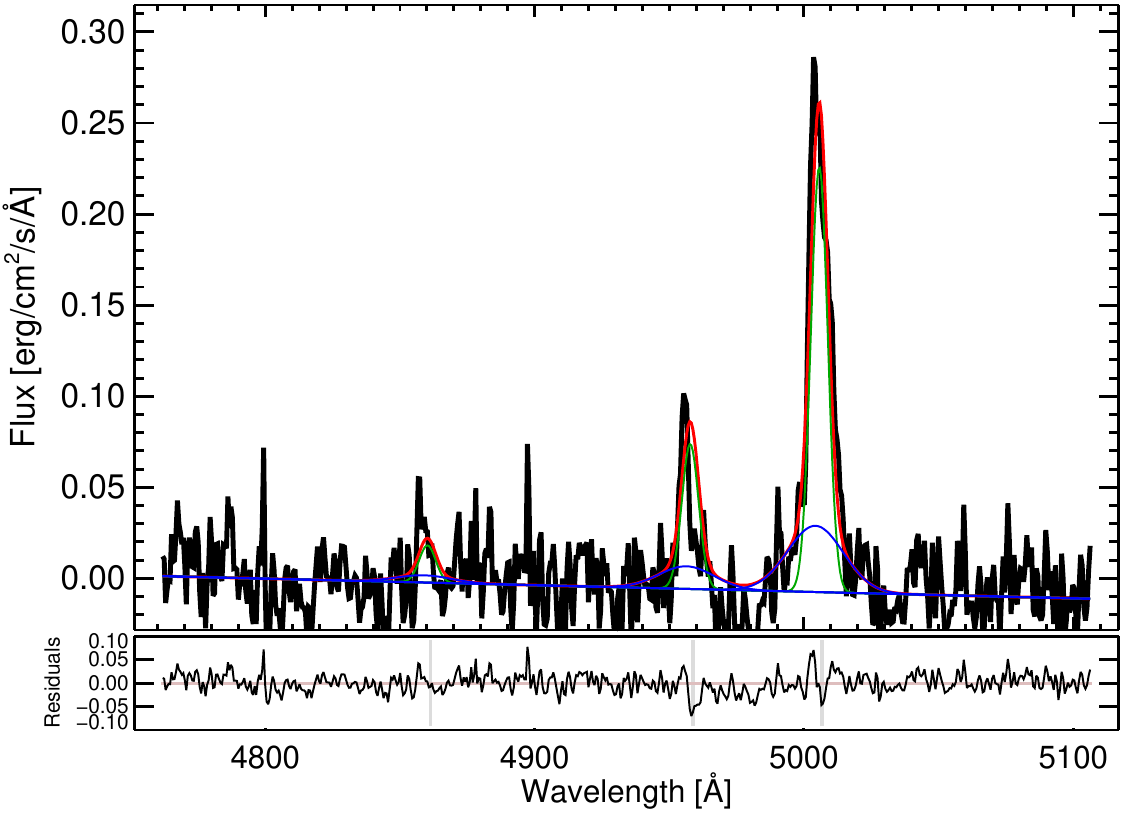}
\hspace{0.1cm} 
\includegraphics[trim = 5.55cm 13.25cm 5.25cm 6.3cm, clip=true, width=.445\textwidth]{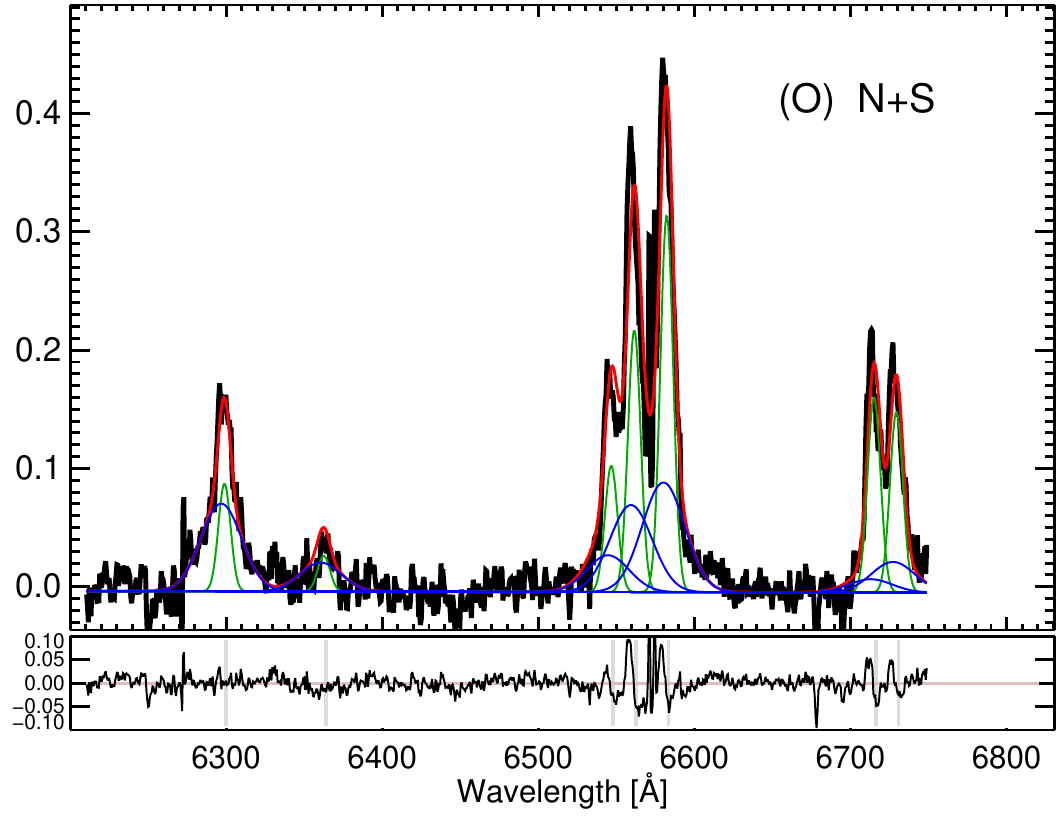}  
\caption{(General description as in Fig.\,\ref{Panel_NGC0266}.) NGC\,3226: in spite of the [S\,II] lines being at the red end of the spectrum, a satisfactory line fitting is reached. Even if a  single-component model of [O\,I] produces  lower residuals than the double Gaussians one, its resulting width ($\sim$\,500\,km\,s$^{-1}$) is unrealistically large, what made us conclude that a two-components models is better in this case.}
 \label{Panel_NGC3226} 		 		 
\end{figure*}
\clearpage

\begin{figure*}
\vspace{-0.25cm} 
\includegraphics[trim = 1.10cm .85cm 11.0cm 17.75cm, clip=true, width=.40\textwidth]{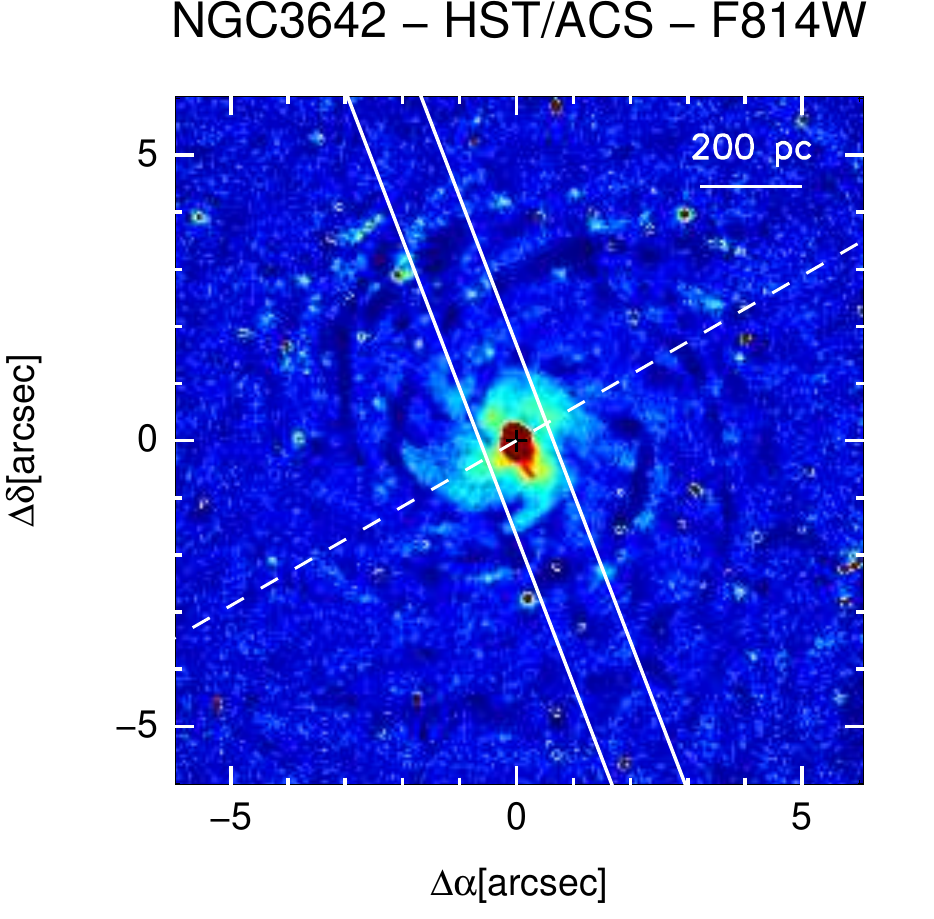} 
\hspace{-0.3cm} 
\includegraphics[trim = 4.5cm 13.cm 5.25cm 6.25cm, clip=true, width=.475\textwidth]{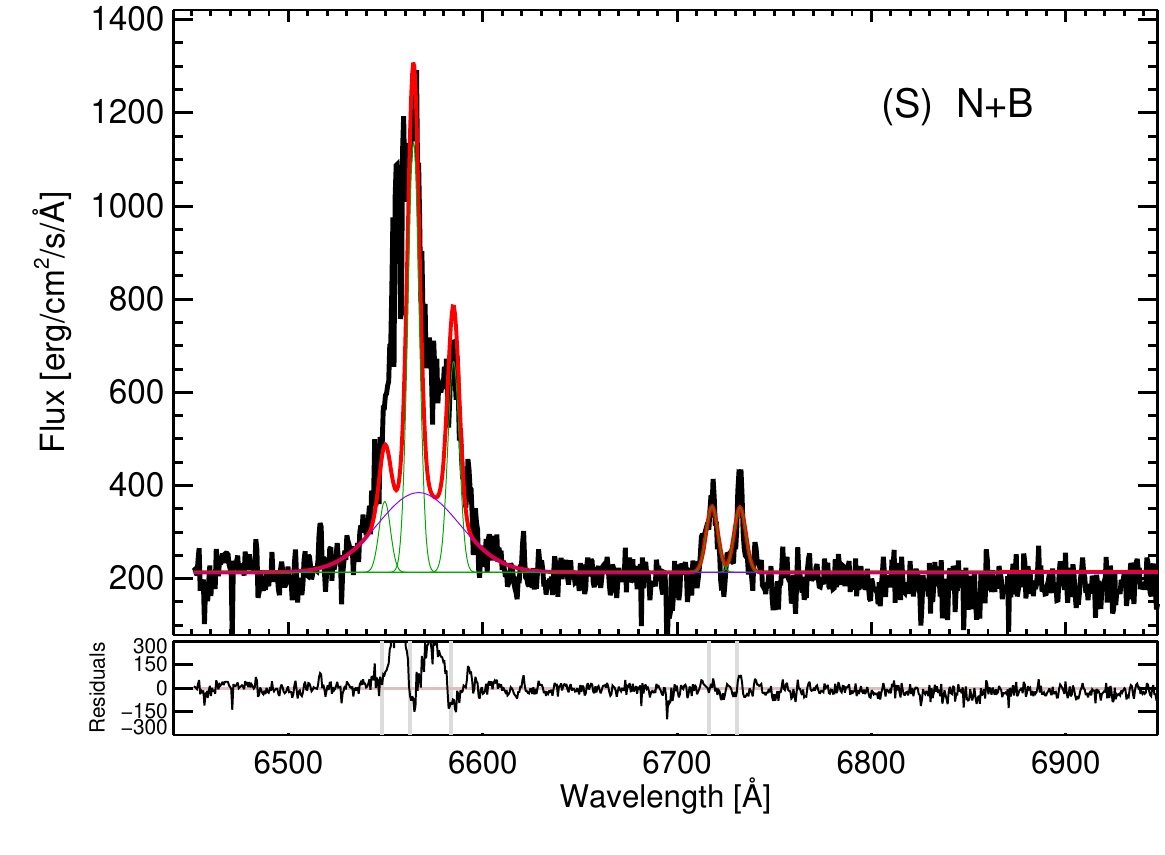}  \\
\vspace{-0.10cm}
\includegraphics[trim = 2.4cm 19.75cm 2.7cm 3.75cm, clip=true, width=.915\textwidth]{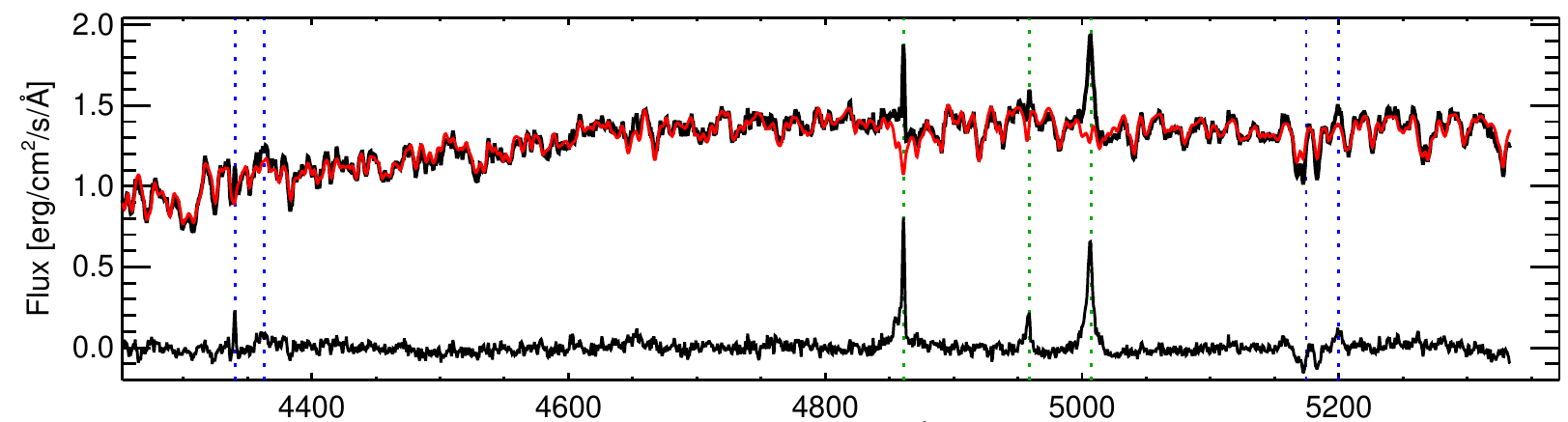} \\
\vspace{-0.10cm}
\includegraphics[trim = 2.4cm 18.75cm 2.7cm 3.75cm, clip=true, width=.91\textwidth]{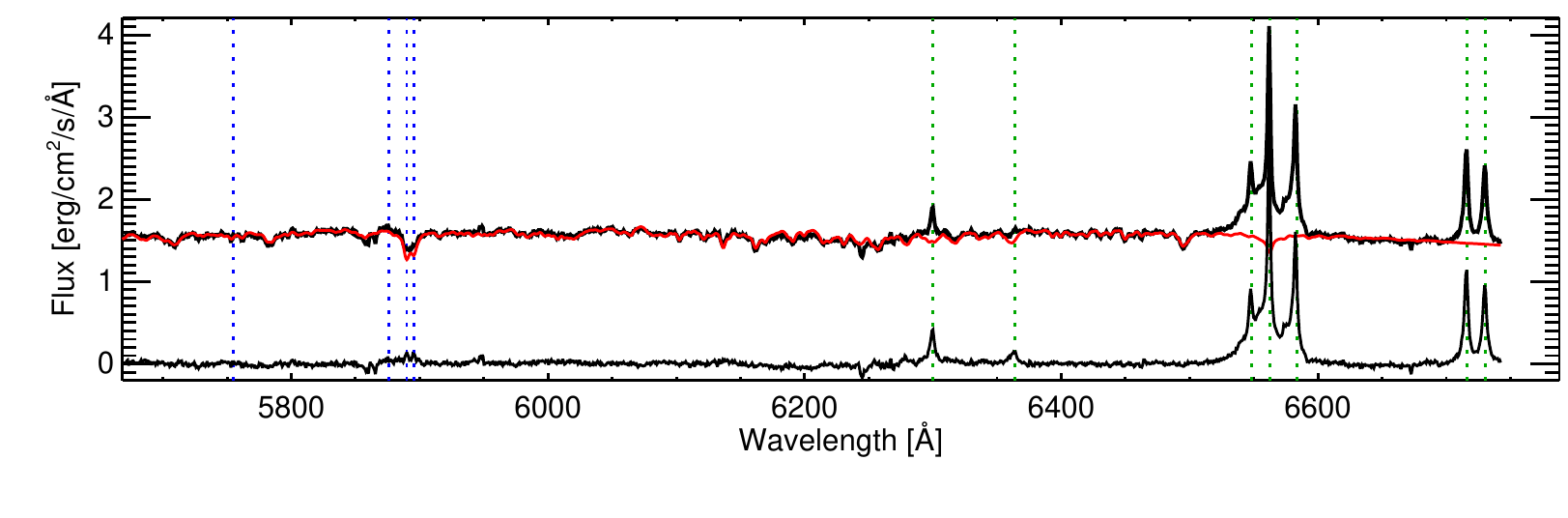} \\
\vspace{-0.45cm}
\includegraphics[trim = 5cm 13.25cm 5.25cm 6.3cm, clip=true, width=.465\textwidth]{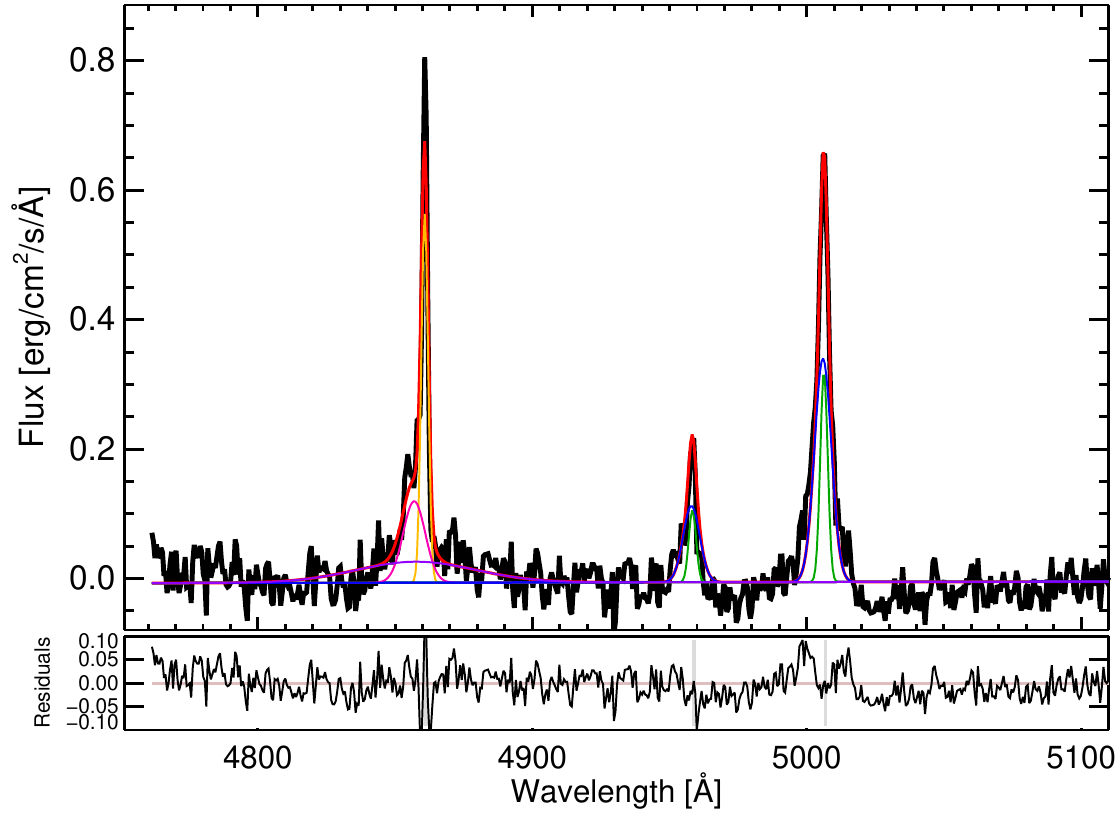}
\hspace{0.1cm} 
\includegraphics[trim = 5.55cm 13.25cm 5.25cm 6.3cm, clip=true, width=.445\textwidth]{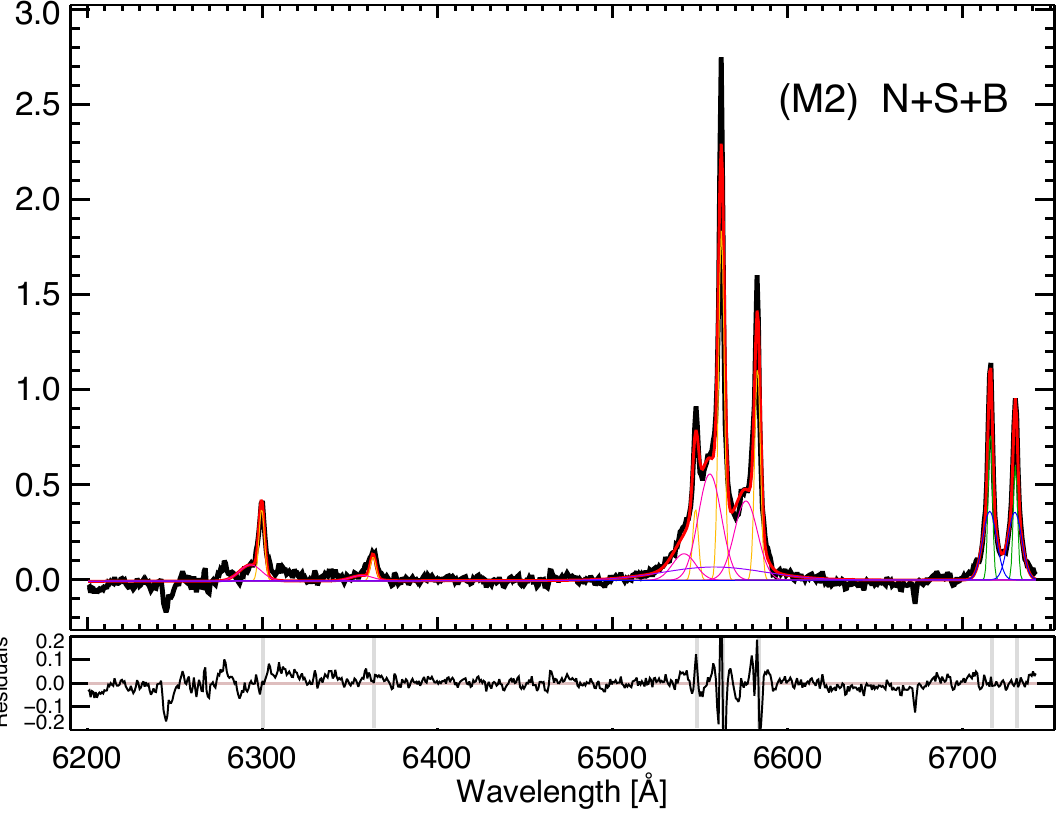}  
\caption{(General description as in Fig.\,\ref{Panel_NGC0266}.) NGC\,3642: narrow emission lines have small widths and  broad wings  as found for the Palomar spectrum (\textit{HFS97}). A small improvement of $\varepsilon$$^{\rm [O\,I]}$ (0.0036, Table\,\ref{T_rms}) is found when using two components instead of one. The presence of wings in all emission lines suggests that the second component is true and not an artifact. The broad component is present but is relatively weak. We did not find difference in H$\alpha$ and [N\,II] line profiles contrary to the modelling proposed by \textit{HFS97}. [O\,III]$\lambda$5007 seems to require a broader second component, but it is not clear for  [O\,III]$\lambda$4959.} 
\label{Panel_NGC3642} 		 		 
\end{figure*}
\clearpage
\begin{figure*}
\vspace{-0.25cm} 
\includegraphics[trim = 1.10cm .85cm 11.0cm 17.75cm, clip=true, width=.40\textwidth]{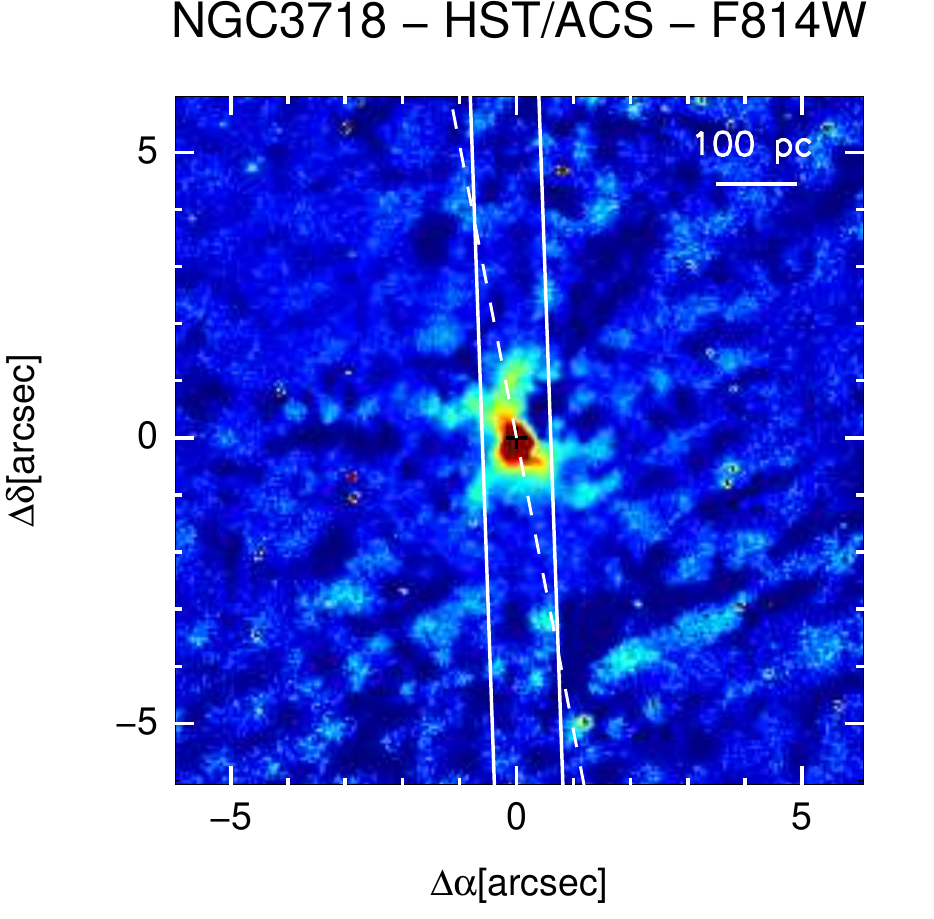} 
\hspace{-0.3cm} 
\includegraphics[trim = 2.4cm 19.75cm 2.7cm 3.75cm, clip=true, width=.915\textwidth]{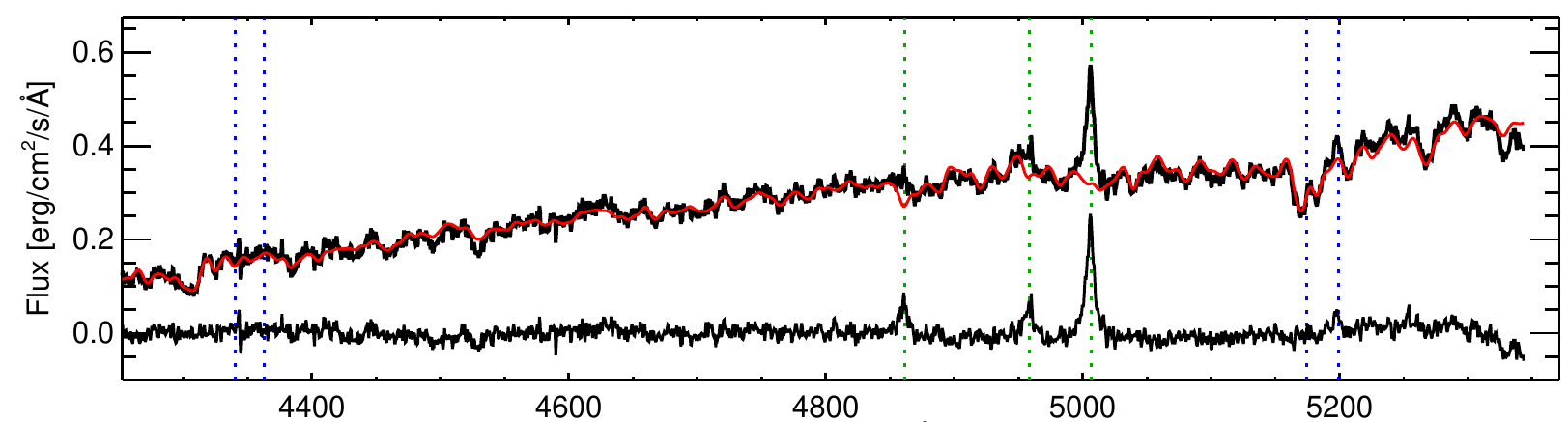} \\
\vspace{-0.10cm}
\includegraphics[trim = 2.4cm 18.75cm 2.7cm 3.75cm, clip=true, width=.91\textwidth]{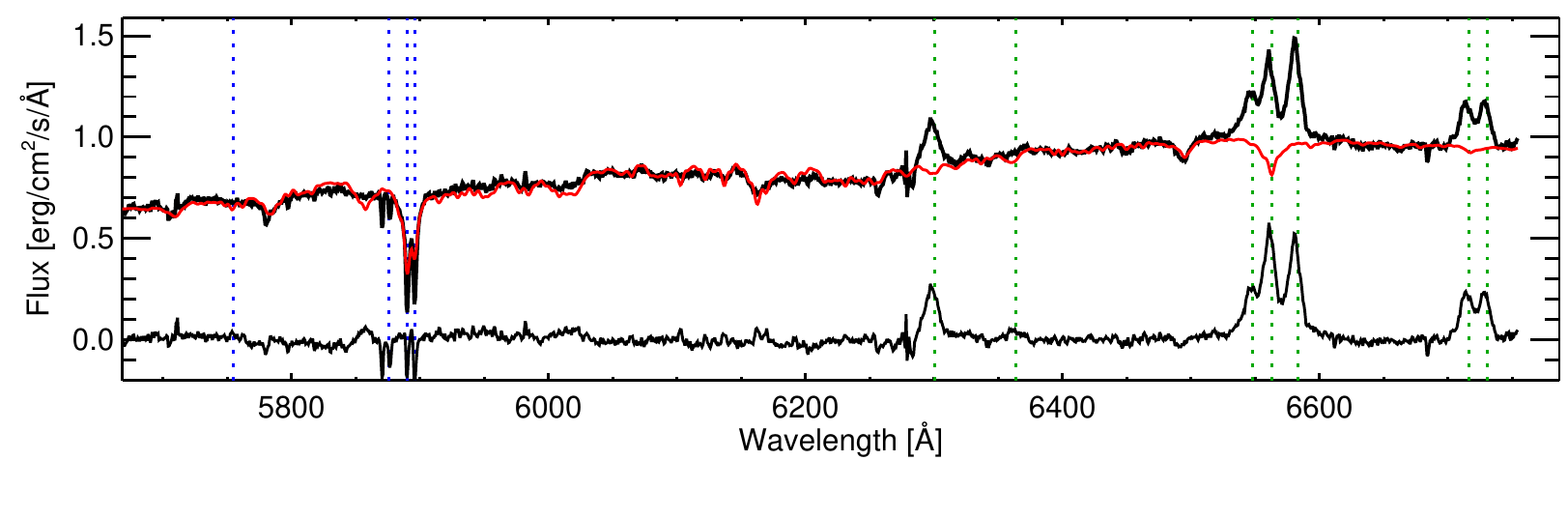} \\
\vspace{-0.45cm}
\includegraphics[trim = 4.9cm 13.25cm 5.25cm 6.3cm, clip=true, width=.4715\textwidth]{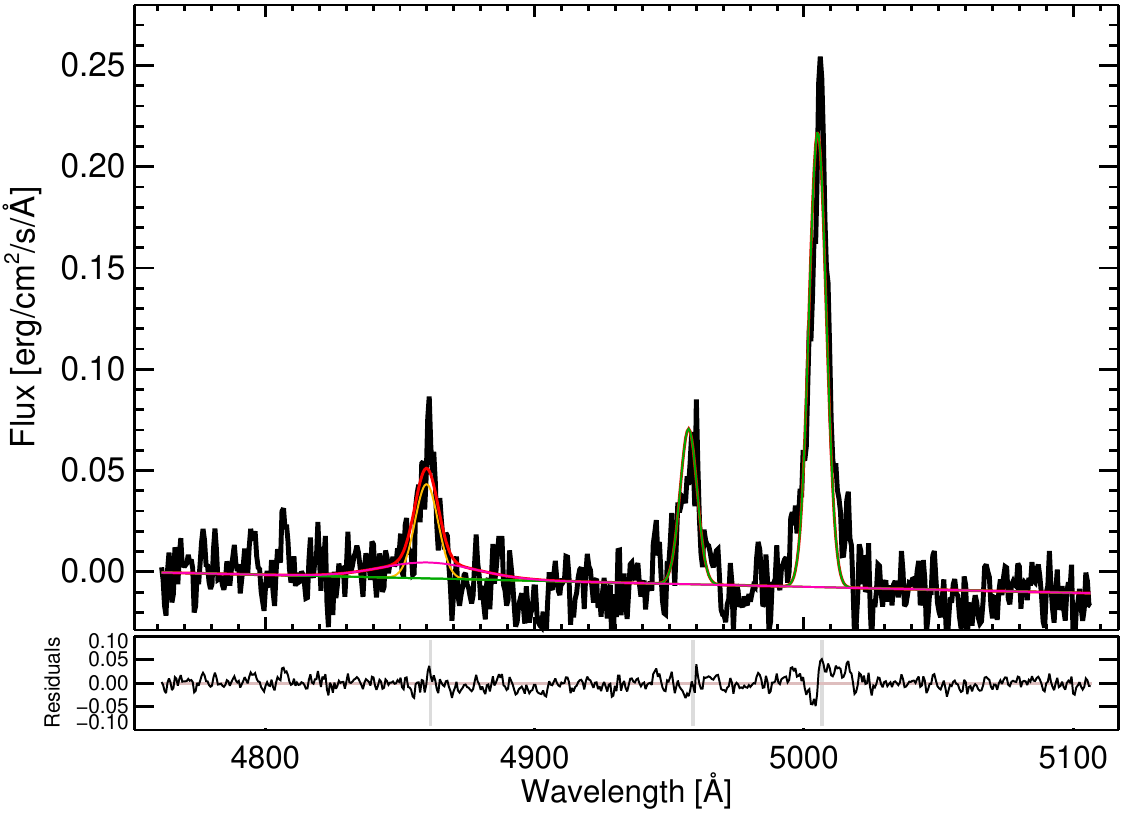}
\hspace{0.1cm} 
\includegraphics[trim = 5.55cm 13.25cm 5.25cm 6.3cm, clip=true, width=.445\textwidth]{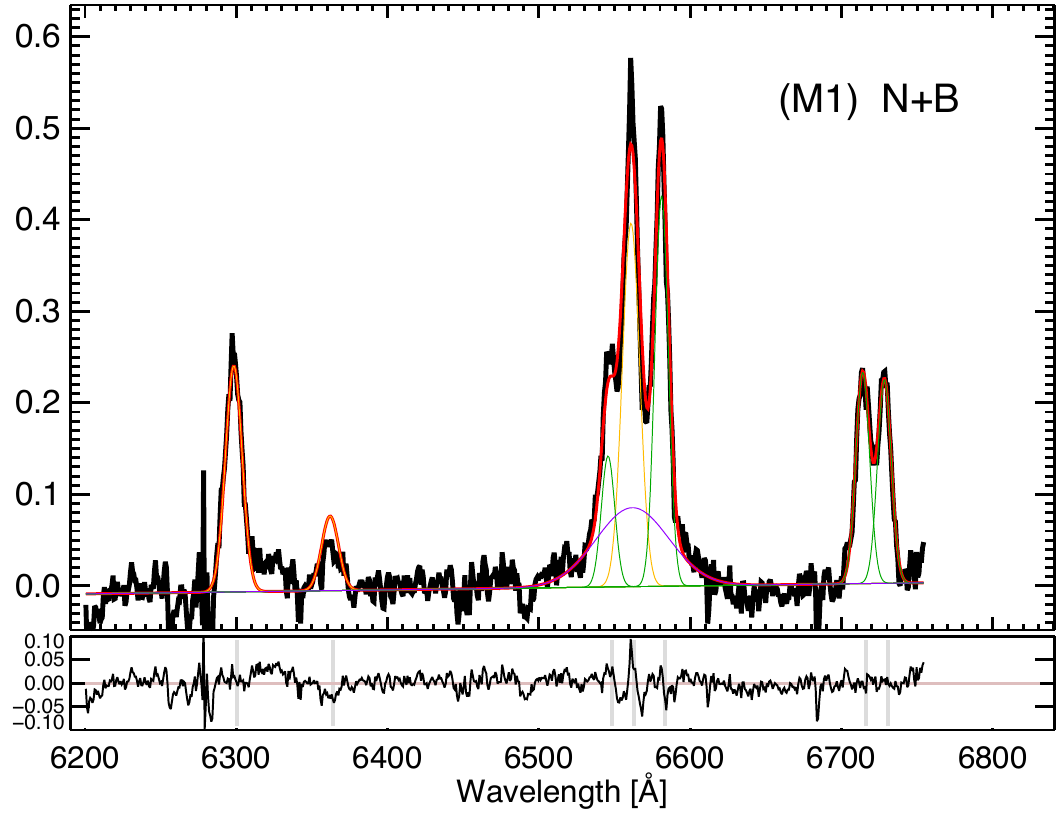}  
\caption{(General description as in Fig.\,\ref{Panel_NGC0266}.) NGC\,3718: even if both [O\,I] and [S\,II] regions have some problems (i.e. dips, edge of the wavelength range) and H$\alpha$ and [N\,II] are strongly blended, the resulting fit is still satisfactory (Table\,\ref{T_rms}).}
 \label{Panel_NGC3718} 		 		 
\end{figure*}
\clearpage
\begin{figure*}
\vspace{-0.25cm} 
\includegraphics[trim = 1.10cm .85cm 11.0cm 17.75cm, clip=true, width=.40\textwidth]{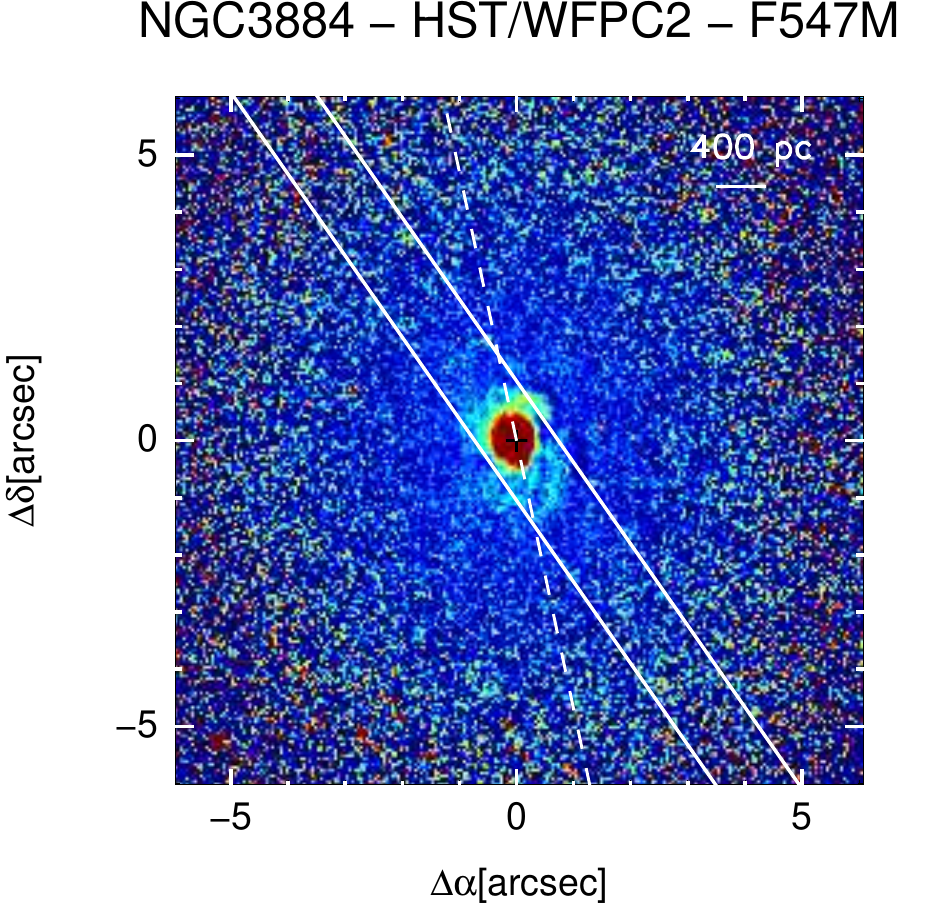} 
\hspace{-0.3cm} 
\includegraphics[trim = 2.4cm 19.75cm 2.7cm 3.75cm, clip=true, width=.915\textwidth]{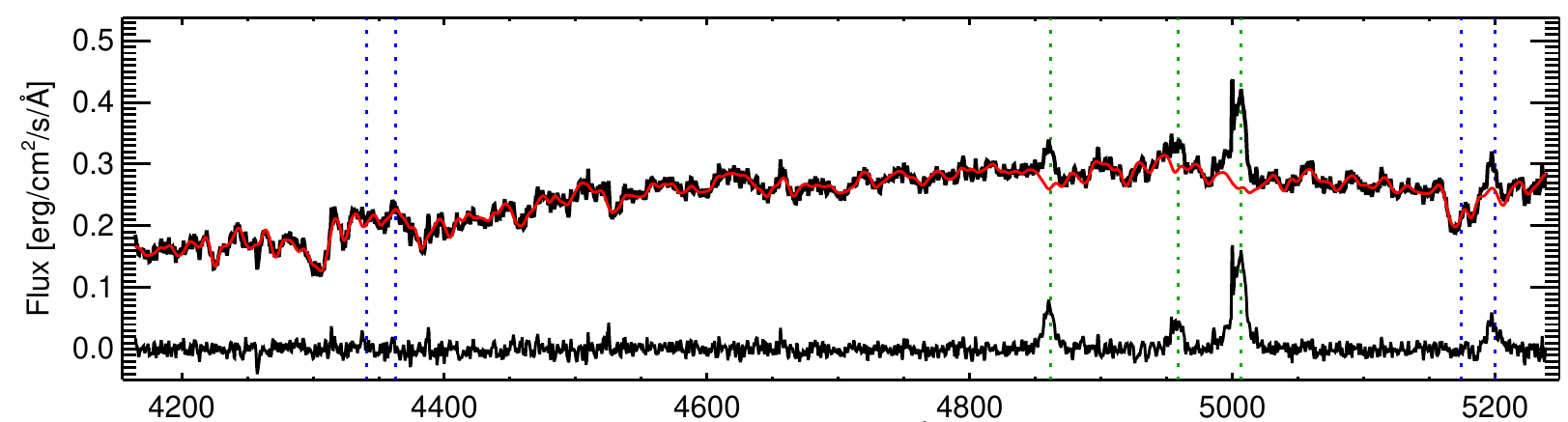} \\
\vspace{-0.10cm}
\includegraphics[trim = 2.4cm 18.75cm 2.7cm 3.75cm, clip=true, width=.91\textwidth]{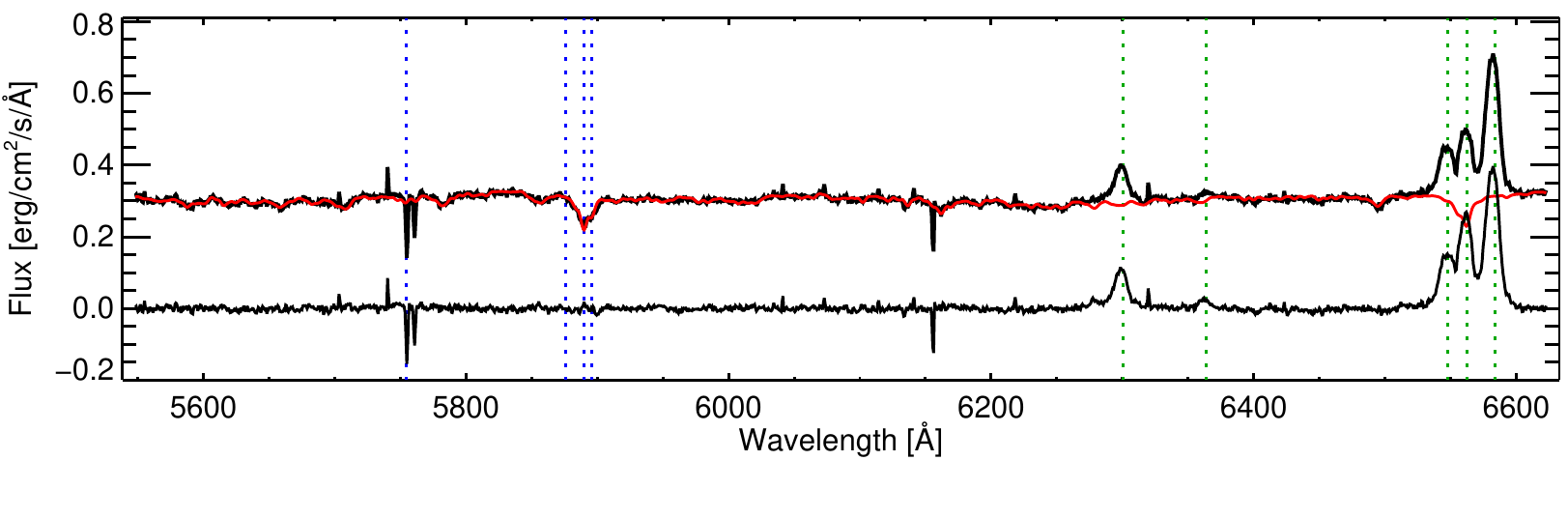} \\
\vspace{-0.45cm}
\includegraphics[trim = 4.9cm 13.25cm 5.25cm 6.3cm, clip=true, width=.4715\textwidth]{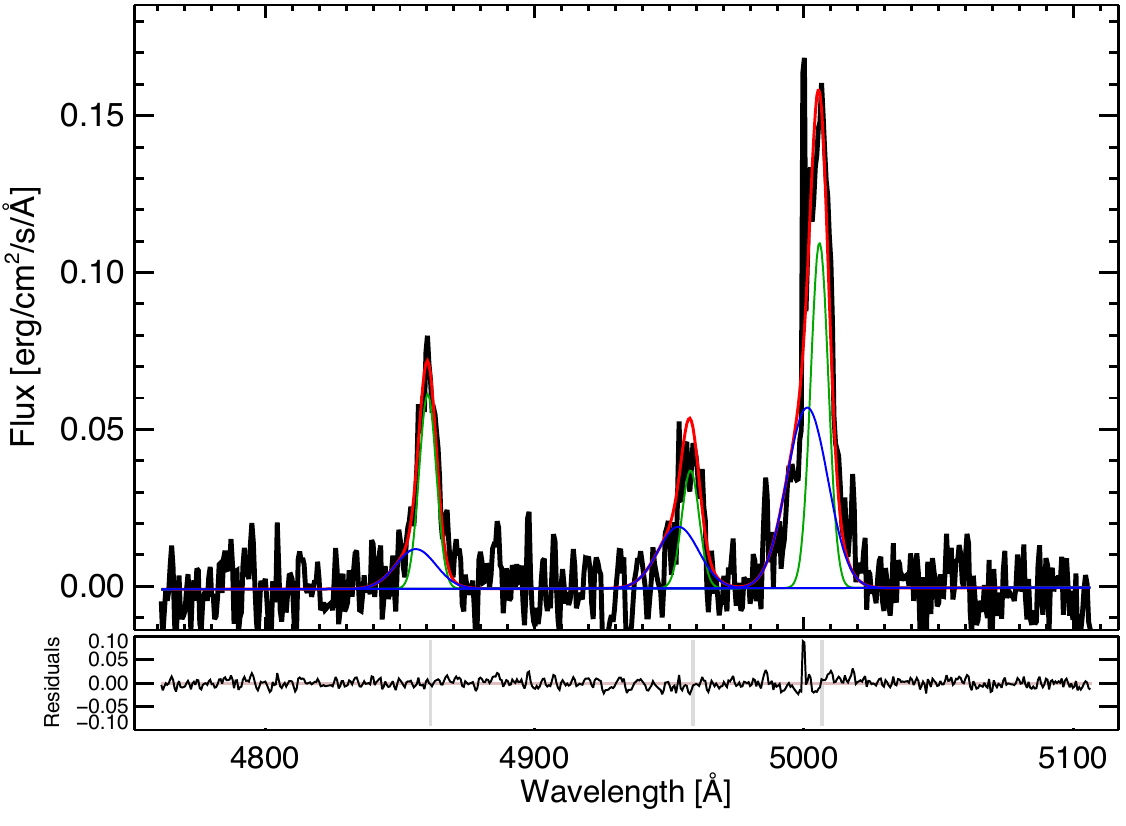}
\hspace{0.1cm} 
\includegraphics[trim = 5.55cm 13.25cm 5.25cm 6.3cm, clip=true, width=.445\textwidth]{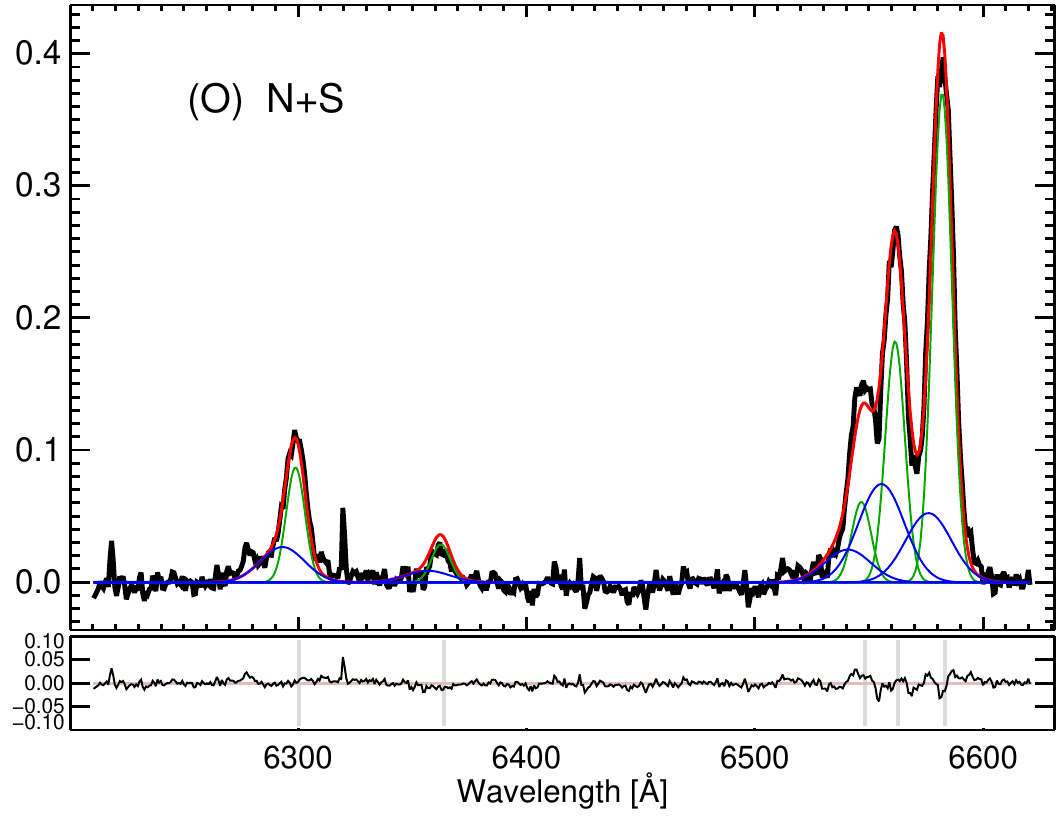}   
\caption{(General description as in Fig.\,\ref{Panel_NGC0266}.) NGC\,3884: a blue wing is evident in the [O\,I] line. A broad component is not needed for the modelling of H$\alpha$. Tails and asymmetries are found by \textit{HFS97} when analyzing H$\alpha$-[N\,II] in the Palomar spectrum for this LINER but these are absent in our spectrum.}
 \label{Panel_NGC3884} 		 		 
\end{figure*}
\clearpage

\clearpage


\begin{figure*}
\vspace{-0.25cm} 
\includegraphics[trim = 1.10cm .85cm 11.0cm 17.75cm, clip=true, width=.40\textwidth]{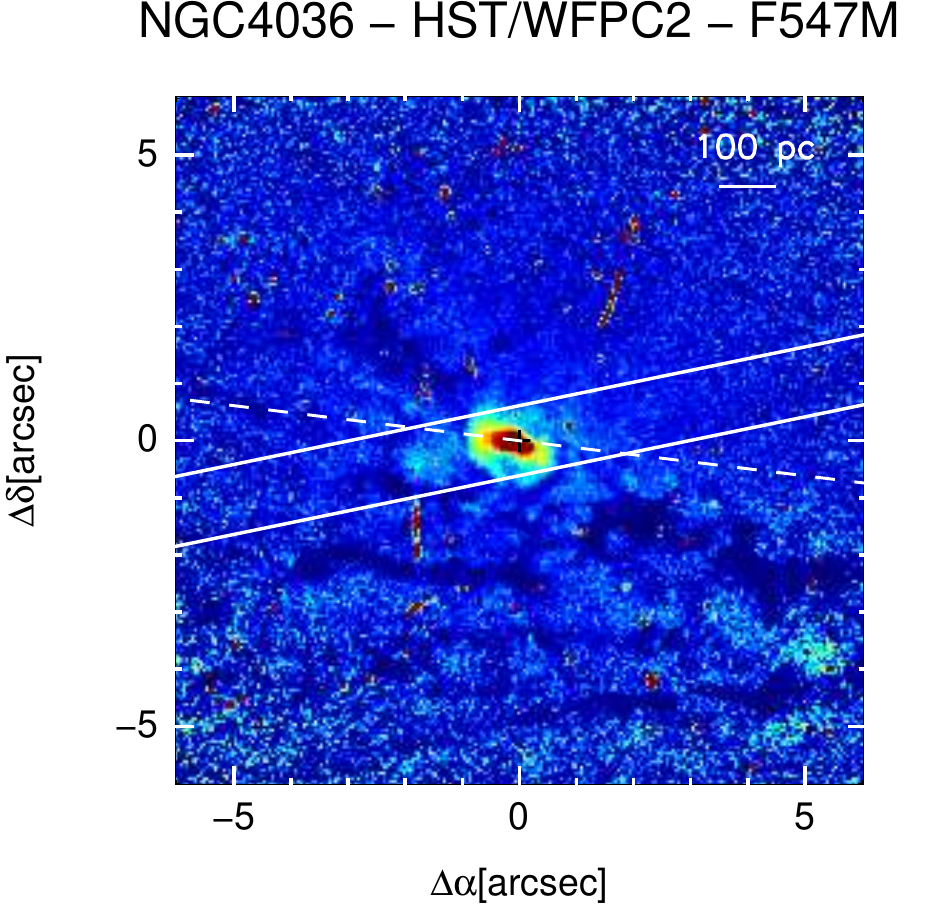} 
\hspace{-0.3cm} 
\includegraphics[trim = 4.5cm 13.cm 5.25cm 6.25cm, clip=true, width=.475\textwidth]{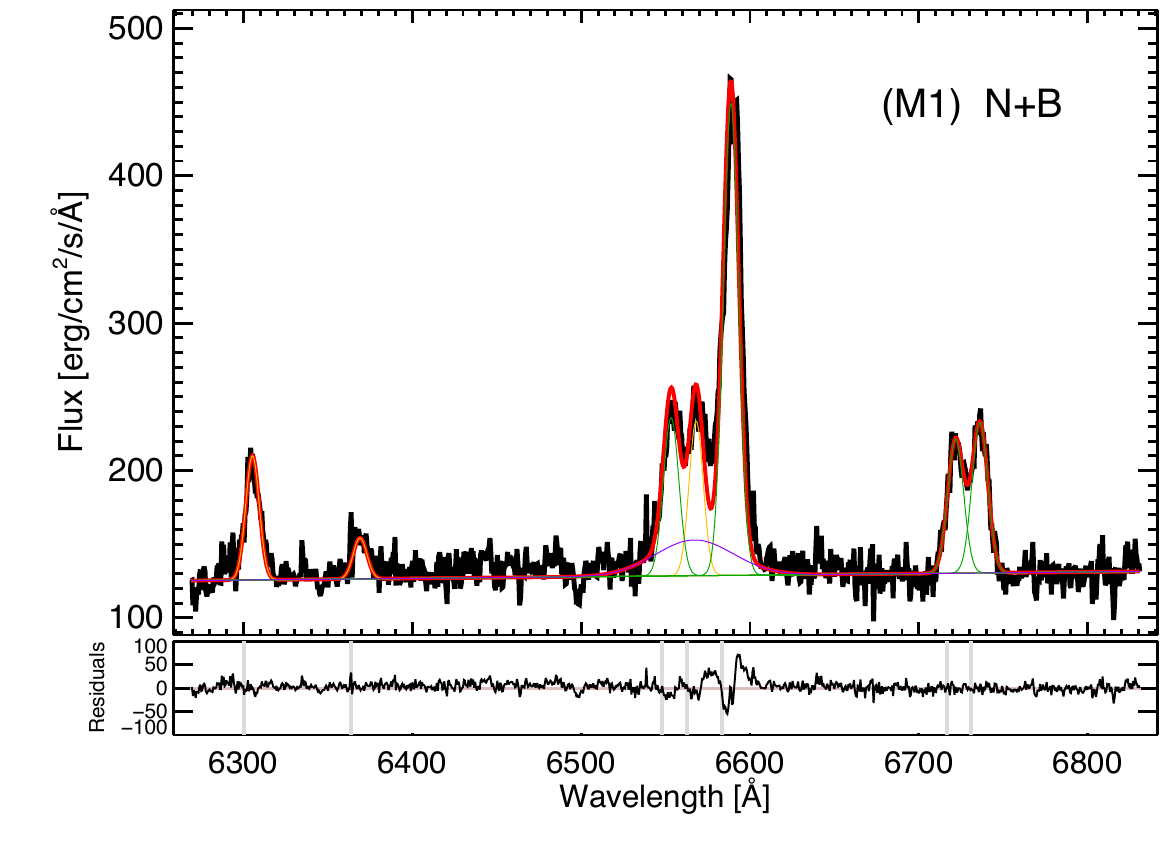}  \\
\vspace{-0.10cm}
\includegraphics[trim = 2.4cm 19.75cm 2.7cm 3.75cm, clip=true, width=.915\textwidth]{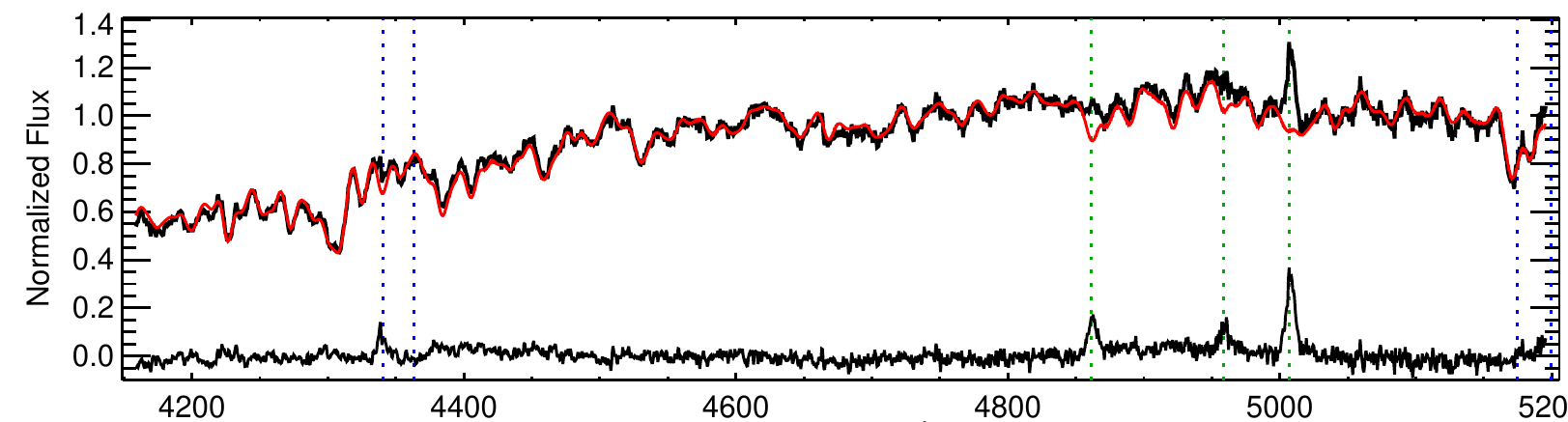} \\
\vspace{-0.10cm}
\includegraphics[trim = 2.4cm 18.75cm 2.7cm 3.75cm, clip=true, width=.91\textwidth]{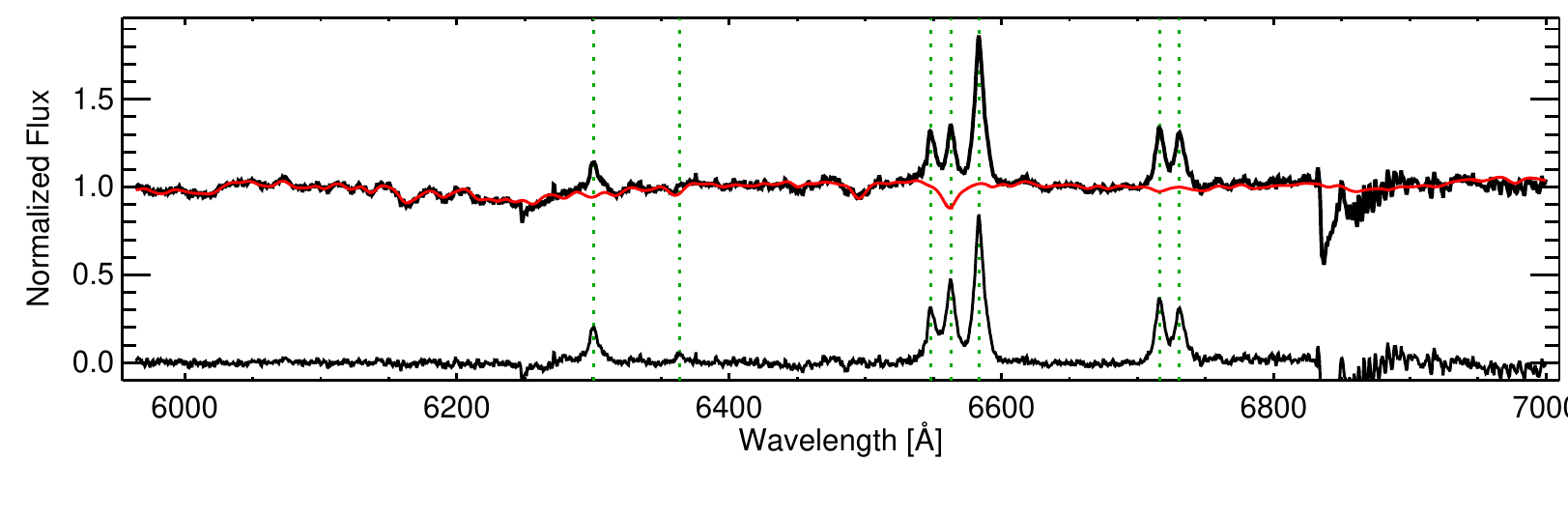} \\
\vspace{-0.45cm}
\includegraphics[trim = 5cm 13.25cm 5.25cm 6.3cm, clip=true, width=.465\textwidth]{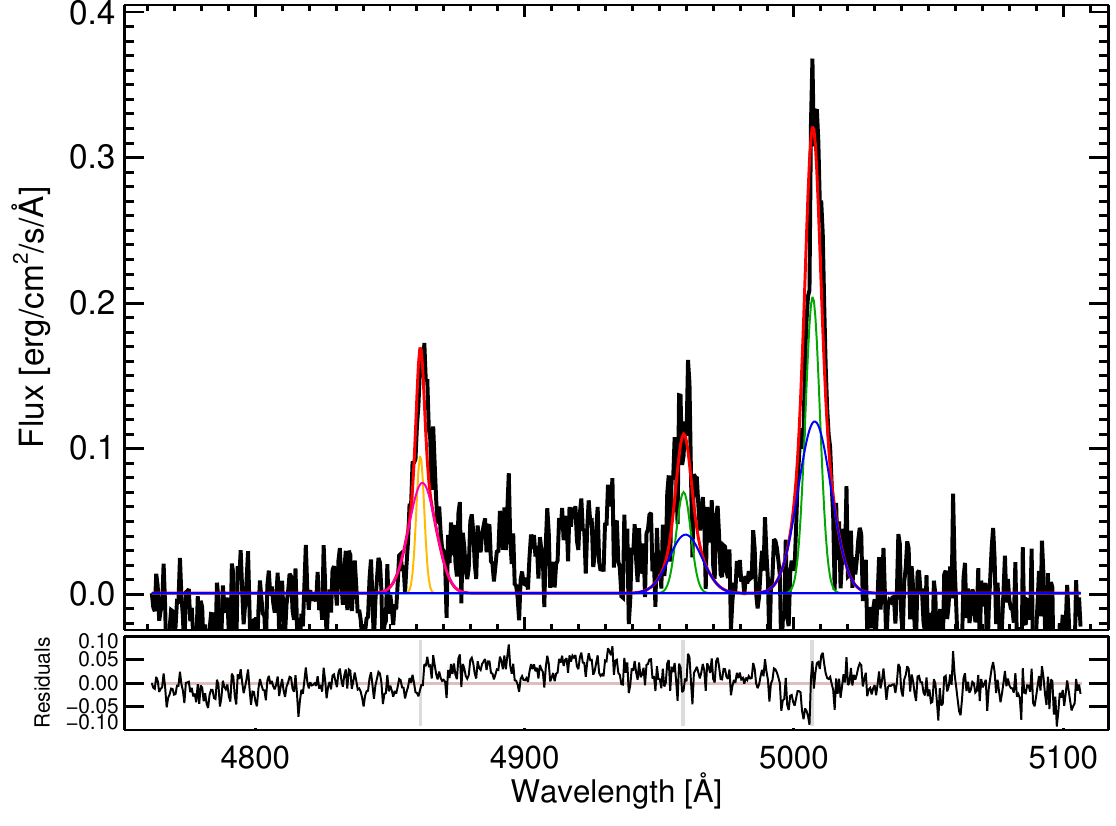}
\hspace{0.1cm} 
\includegraphics[trim = 5.55cm 13.25cm 5.25cm 6.3cm, clip=true, width=.445\textwidth]{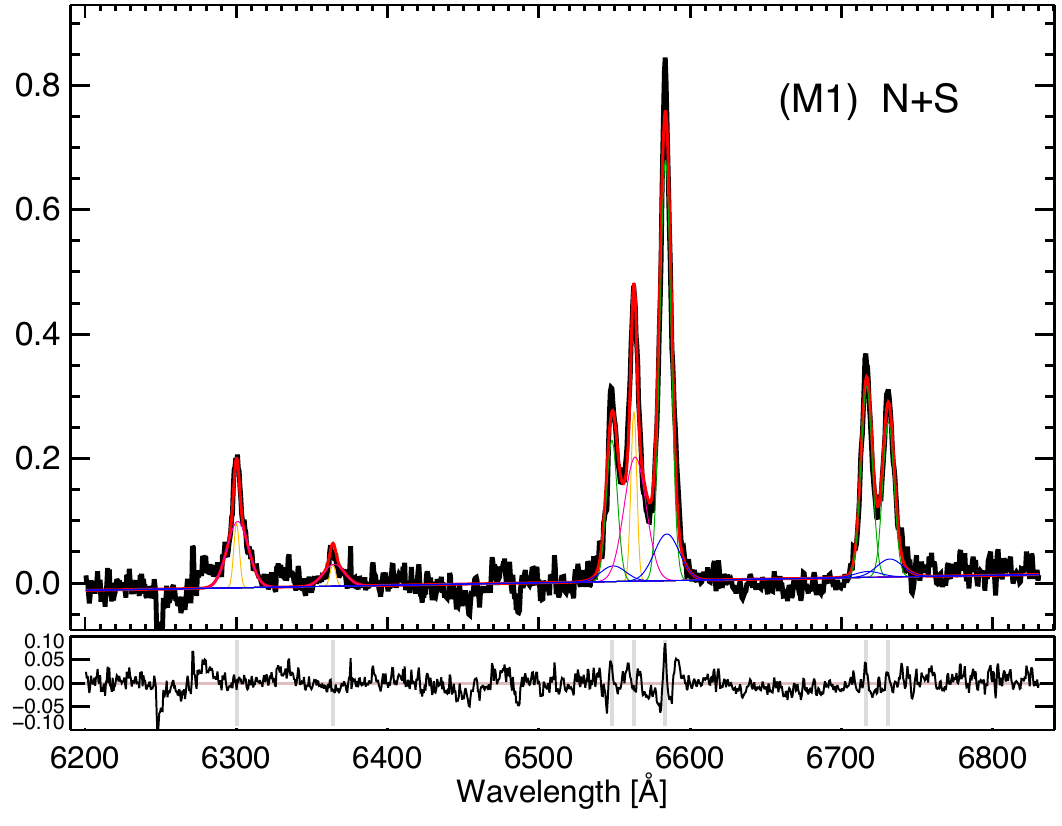}  
\caption{(General description as in Fig.\,\ref{Panel_NGC0266}.) NGC\,4036: the presence of wings in all emission lines suggests that a second component, whose inclusion produces a small improvement in the modelling of [O\,I] (Table Table\,\ref{T_rms}), is true and not an artifact. [S\,II] and [O\,I] line profiles are quite different. The second component is less evident in [S\,II] (wings are almost absent). We did not find evidence of  the broad component in our ground-based data, whereas it is  however rather faint in the Palomar spectrum (\textit{HFS97}). The evidence for such a broad component in H$\alpha$ seems clear in the \textit{HST}/STIS spectrum.} 
\label{Panel_NGC4036} 		 		 
\end{figure*}
\clearpage

\begin{figure*}
\vspace{-0.25cm} 
\includegraphics[trim = 1.10cm .85cm 11.0cm 17.75cm, clip=true, width=.40\textwidth]{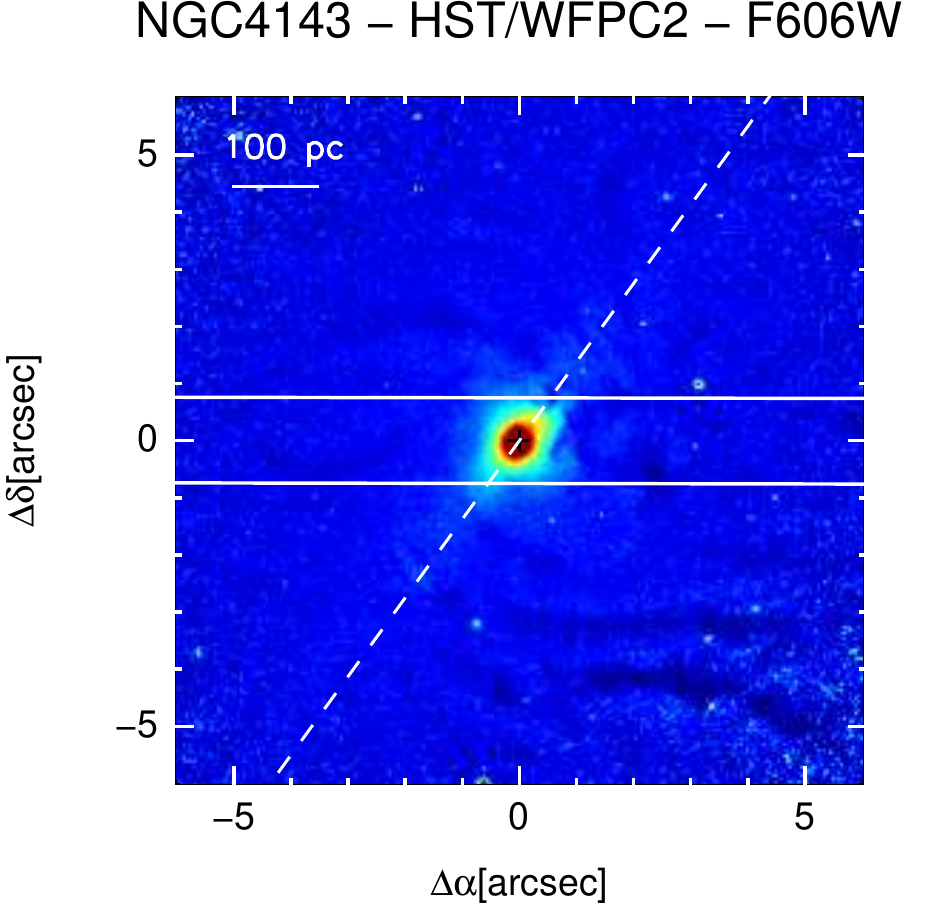} 
\hspace{-0.3cm} 
\includegraphics[trim = 4.5cm 13.cm 5.25cm 6.25cm, clip=true, width=.475\textwidth]{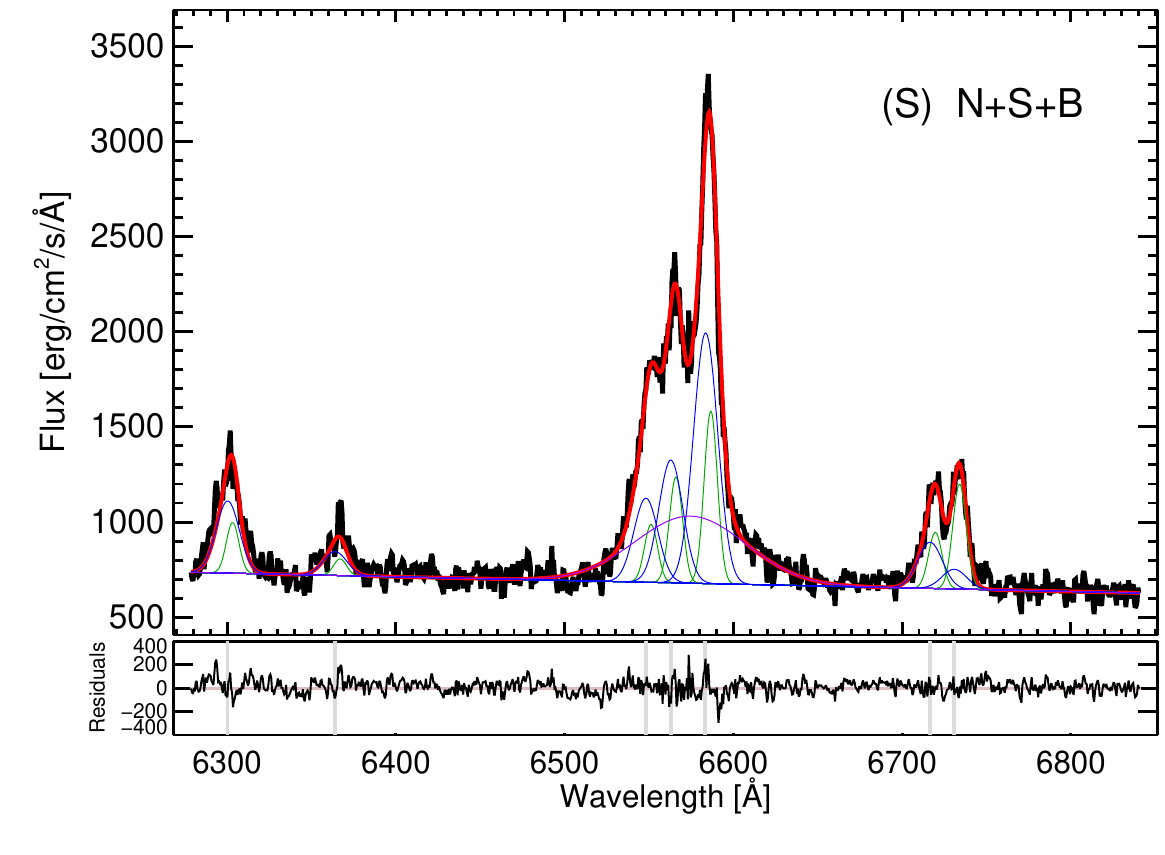}  \\
\vspace{-0.10cm}
\includegraphics[trim = 2.4cm 19.75cm 2.7cm 3.75cm, clip=true, width=.915\textwidth]{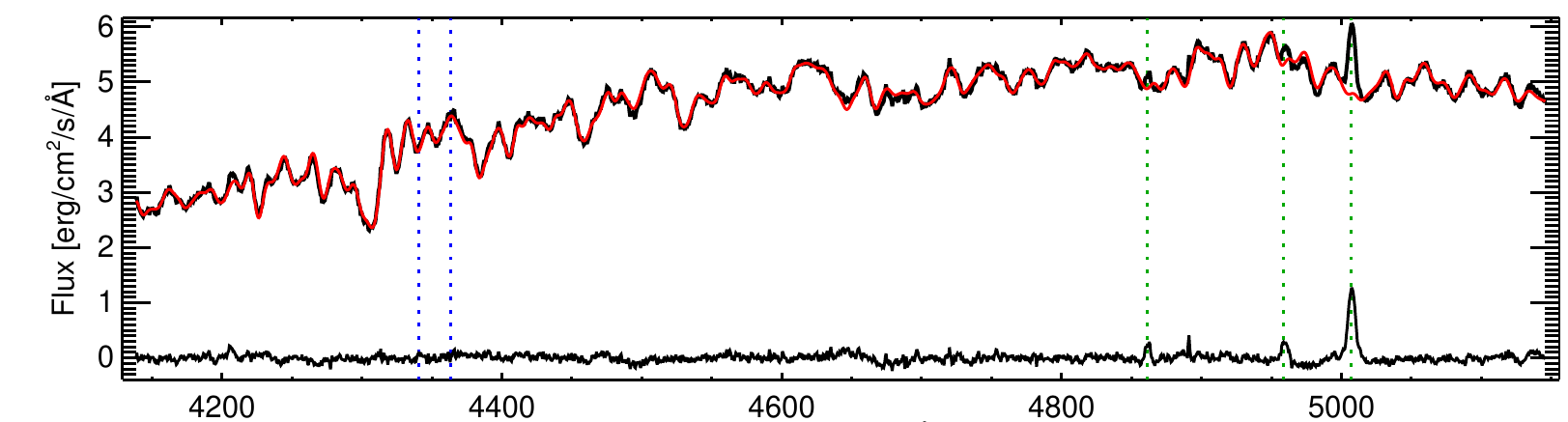} \\
\vspace{-0.10cm}
\includegraphics[trim = 2.4cm 18.75cm 2.7cm 3.75cm, clip=true, width=.91\textwidth]{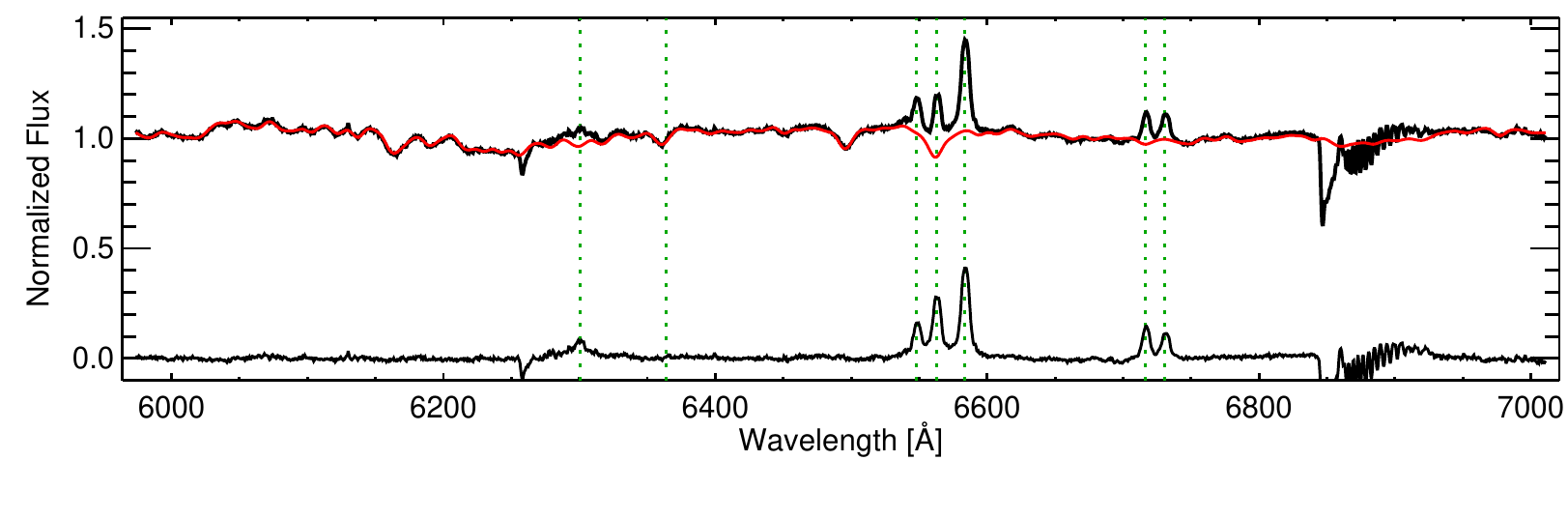} \\
\vspace{-0.45cm}
\includegraphics[trim = 5cm 13.25cm 5.25cm 6.3cm, clip=true, width=.465\textwidth]{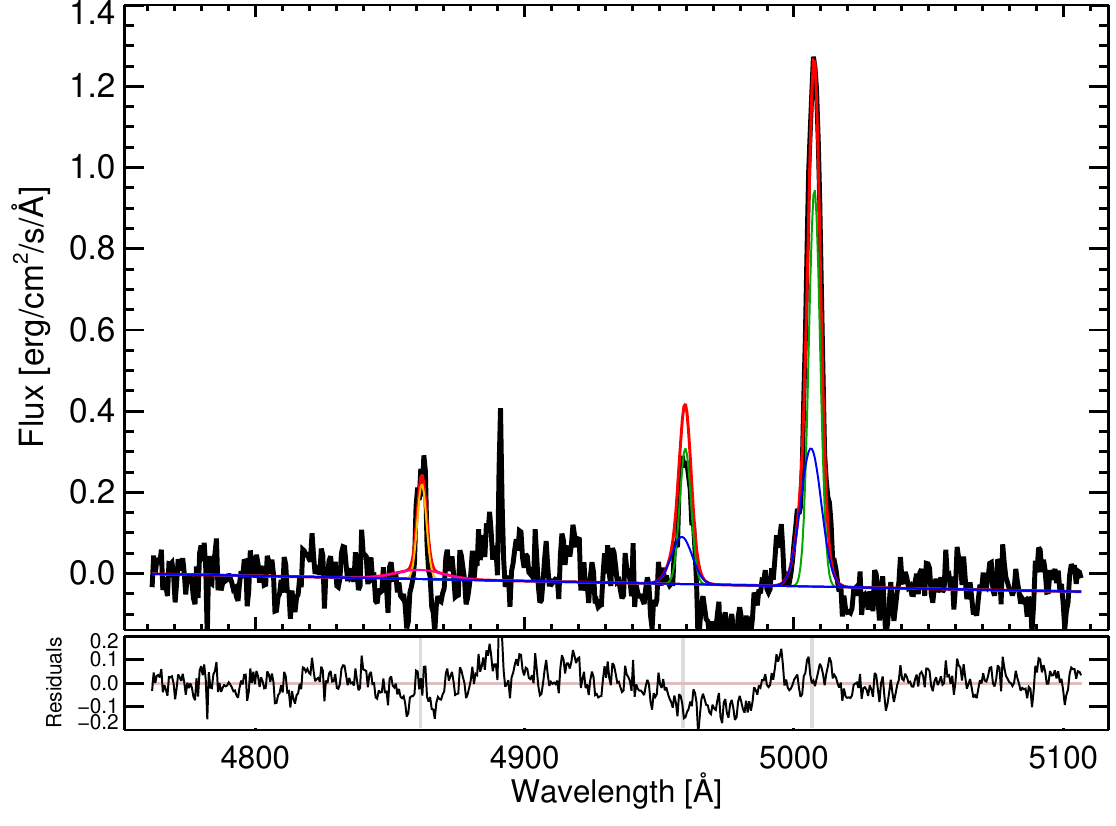}
\hspace{0.1cm} 
\includegraphics[trim = 5.55cm 13.25cm 5.25cm 6.3cm, clip=true, width=.445\textwidth]{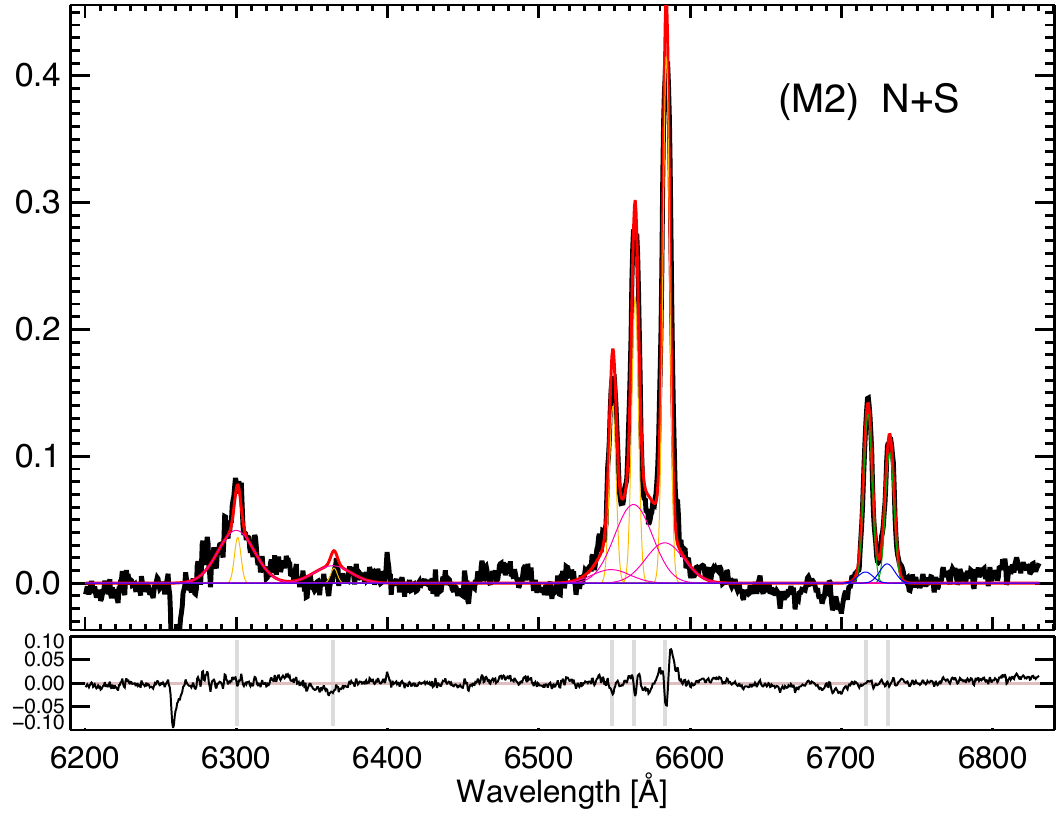}
\caption{(General description as in Fig.\,\ref{Panel_NGC0266}.) NGC\,4143: [S\,II] and [O\,I] line profiles are quite different with the latter having more pronounced  wings.  Similarly to the case of  NGC\,2682, redward to [O\,I] there is a dip. H$\beta$ is rather weak and symmetric. \textit{HFS97}  assumed   [S\,II] as template while we considered  [O\,I], this may be at the origin of the contradictory results  concerning the broad H$\alpha$ component (Table\,\ref{T_FWHM}).}
 \label{Panel_NGC4143} 		 		 
\end{figure*}
\clearpage
\begin{figure*}
\vspace{-0.25cm} 
\includegraphics[trim = 1.10cm .85cm 11.0cm 17.75cm, clip=true, width=.40\textwidth]{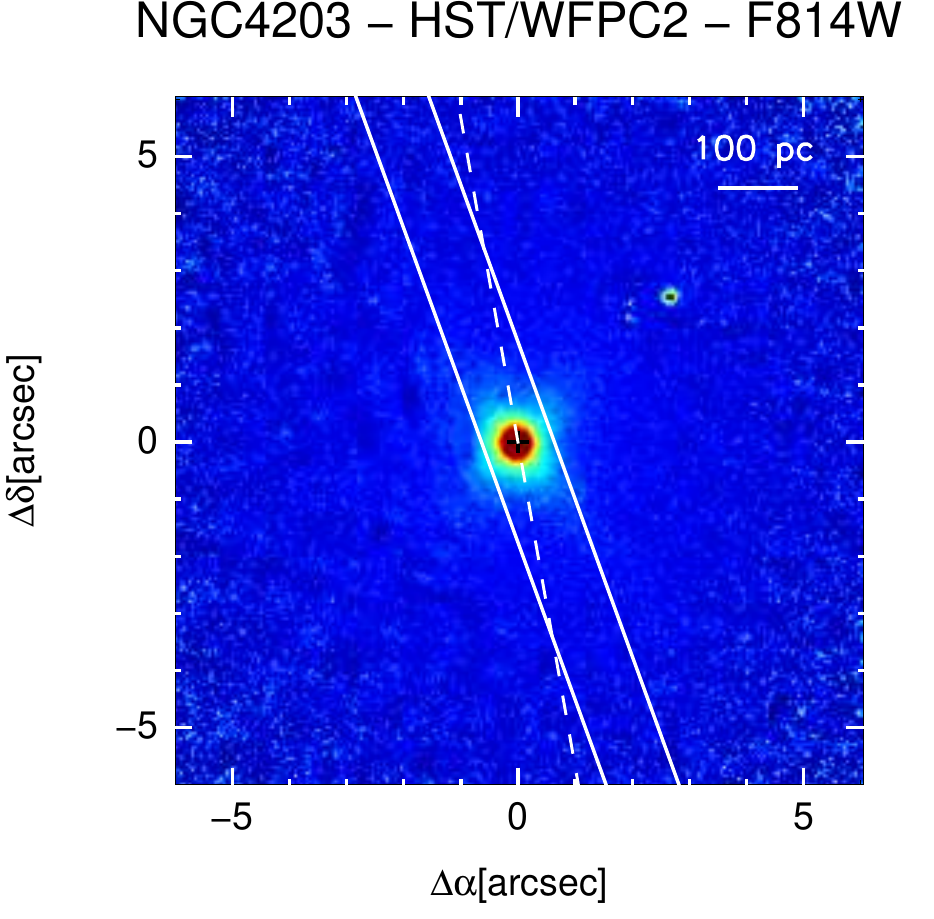} 
\hspace{-0.3cm} 
\includegraphics[trim = 4.5cm 13.cm 5.25cm 6.25cm, clip=true, width=.475\textwidth]{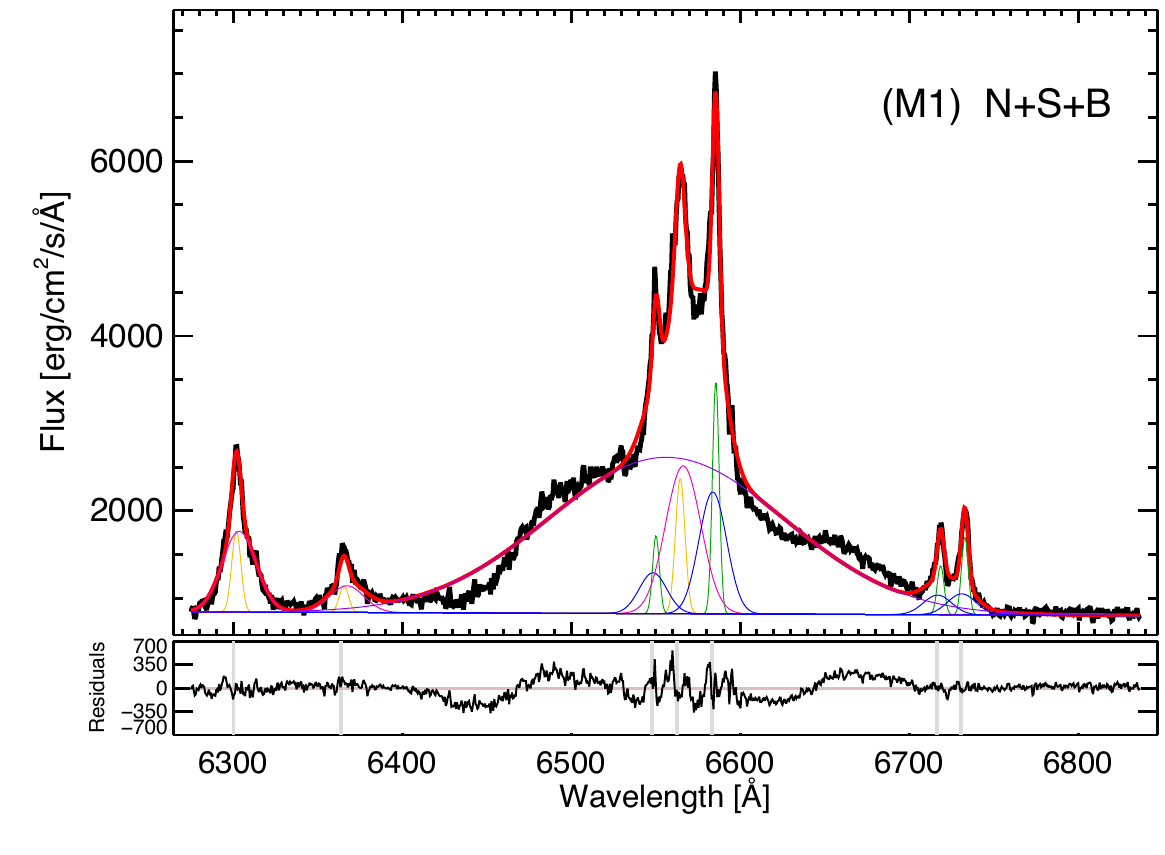}  \\
\vspace{-0.10cm}
\includegraphics[trim = 2.4cm 19.75cm 2.7cm 3.75cm, clip=true, width=.915\textwidth]{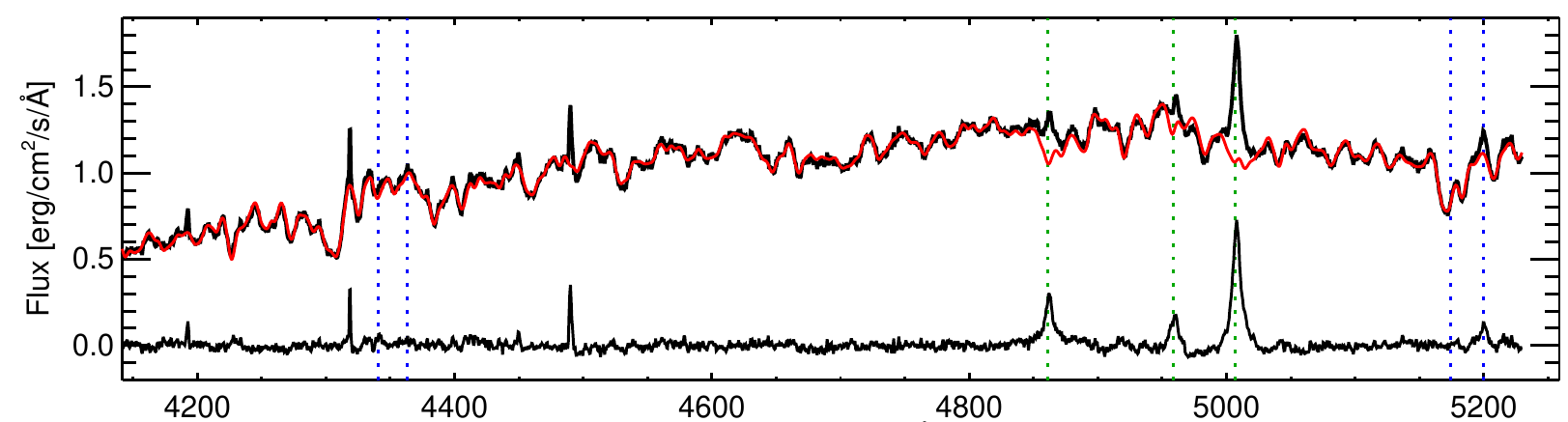} \\
\vspace{-0.10cm}
\includegraphics[trim = 2.4cm 18.75cm 2.7cm 3.75cm, clip=true, width=.91\textwidth]{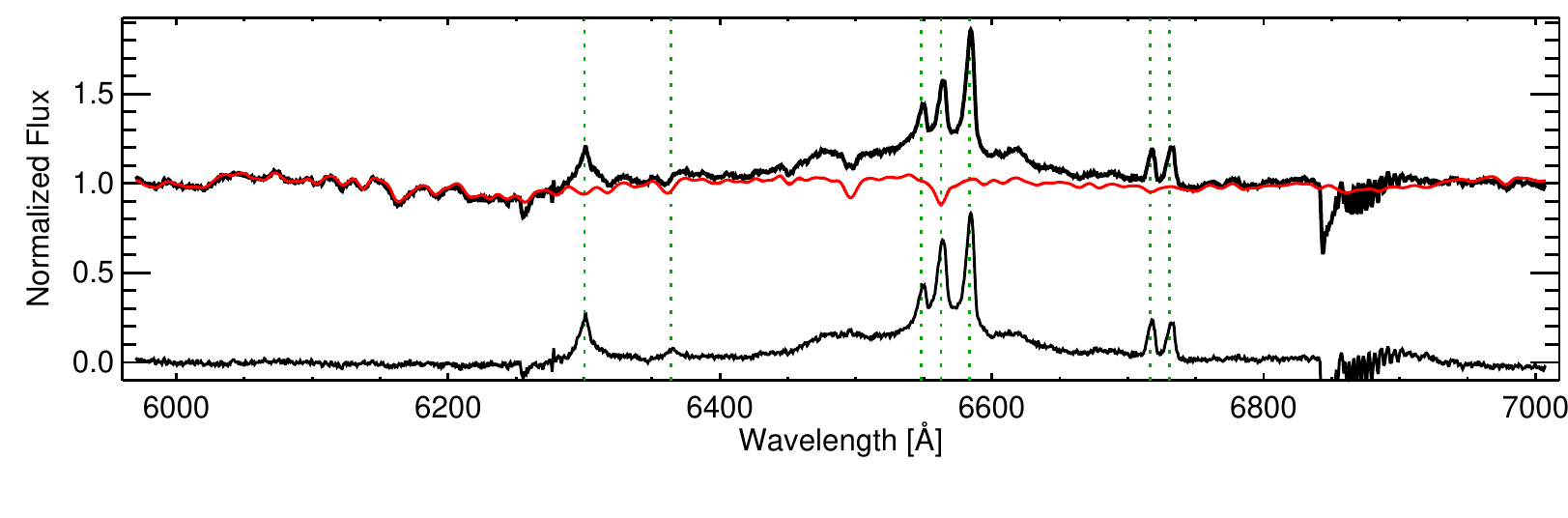}\\
\vspace{-0.45cm}
\includegraphics[trim = 5cm 13.25cm 5.25cm 6.3cm, clip=true, width=.465\textwidth]{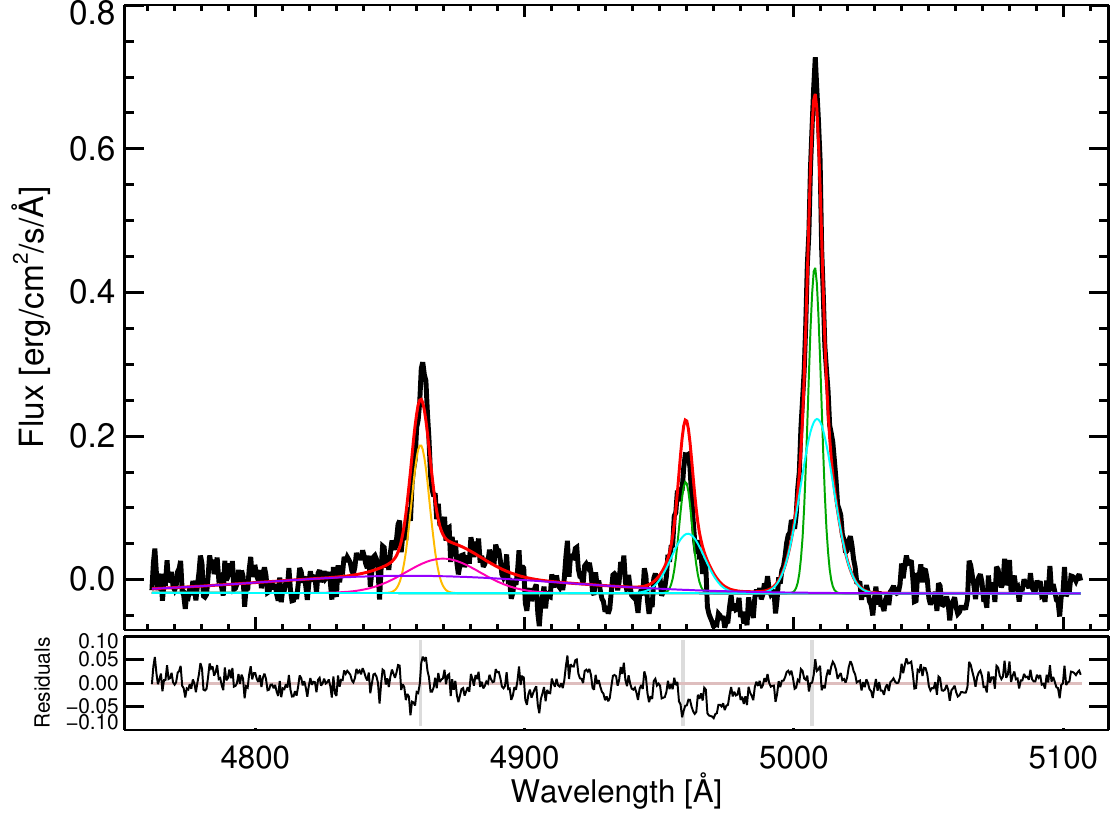}
\hspace{0.1cm} 
\includegraphics[trim = 5.55cm 13.25cm 5.25cm 6.3cm, clip=true, width=.445\textwidth]{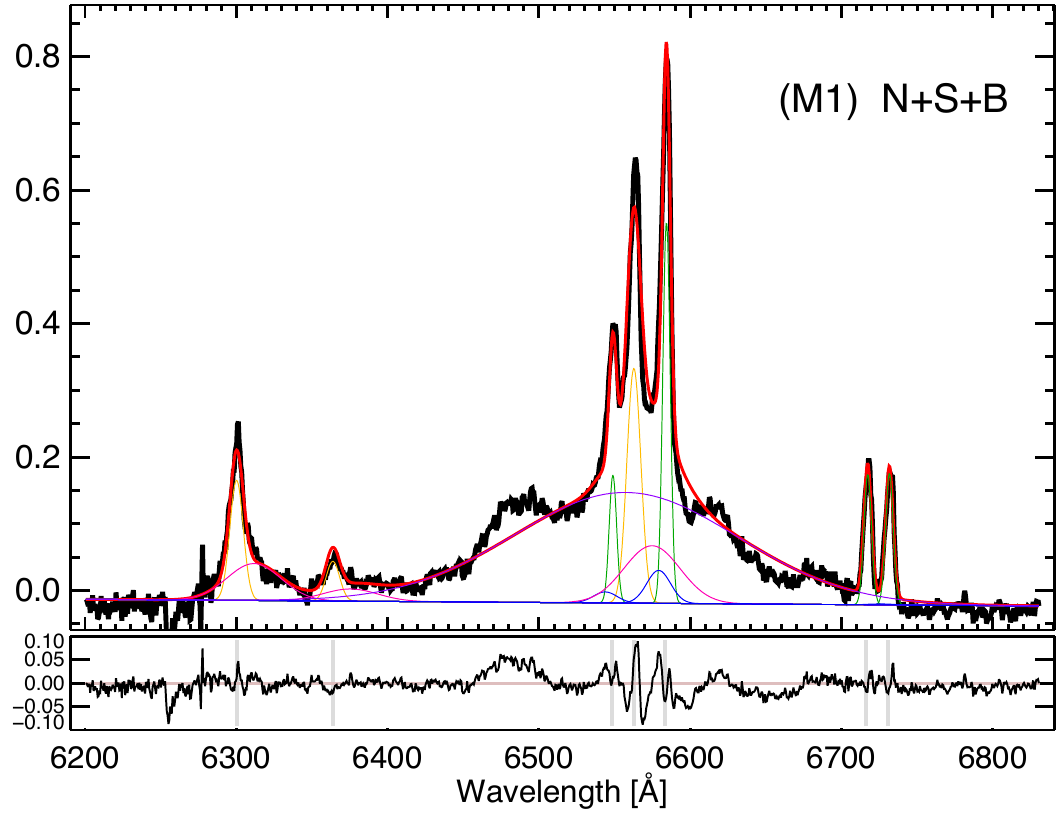}   
\caption{(General description as in Fig.\,\ref{Panel_NGC0266}.) NGC\,4203: already a  visual inspection indicates that H$\alpha$-[N\,II] lines seem quite different from those in the Palomar spectrum (\textit{HFS97}). A red wing is evident in the [O\,I] profiles but not in [S\,II] lines which are contaminated by the broad H$\alpha$ component. A simple Gaussian fitting cannot reproduce well the line profiles due to the strong effect of the  accretion disk  in this galaxy (see text in Appendix B).}
 \label{Panel_NGC4203} 		 		 
\end{figure*}
\clearpage

\begin{figure*}
\vspace{-0.25cm} 
\includegraphics[trim = 1.10cm .85cm 11.0cm 17.75cm, clip=true, width=.40\textwidth]{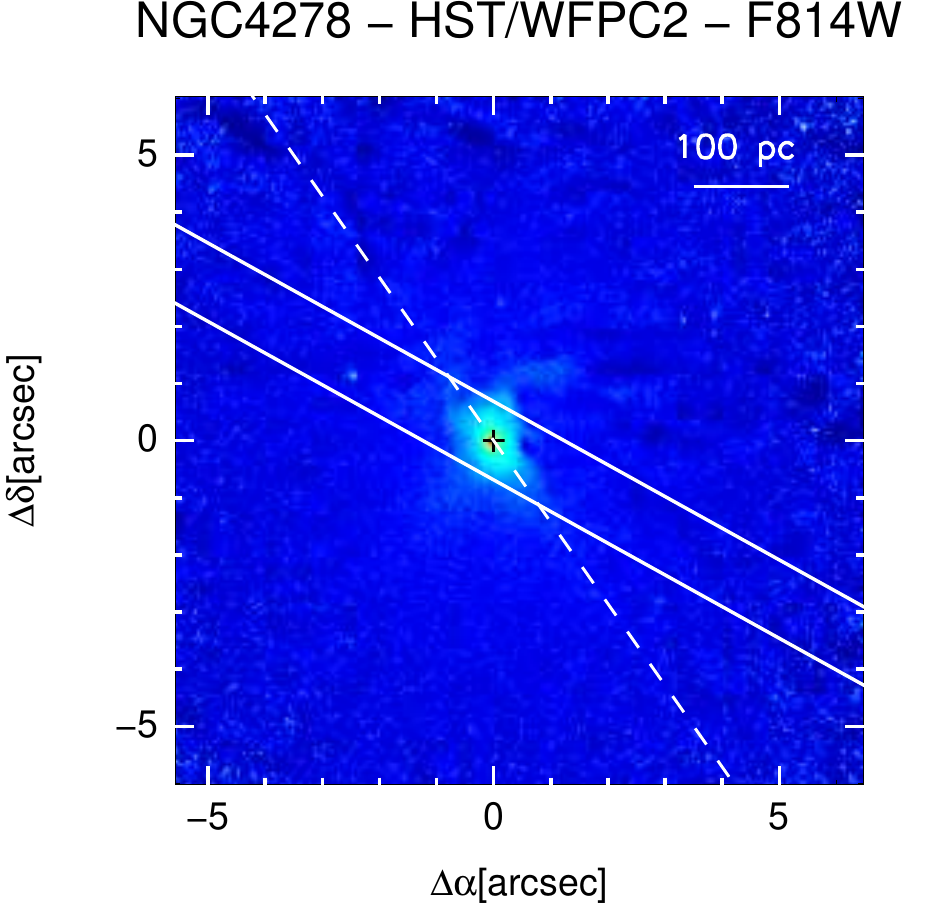} 
\hspace{-0.3cm} 
\includegraphics[trim = 4.5cm 13.cm 5.25cm 6.25cm, clip=true, width=.475\textwidth]{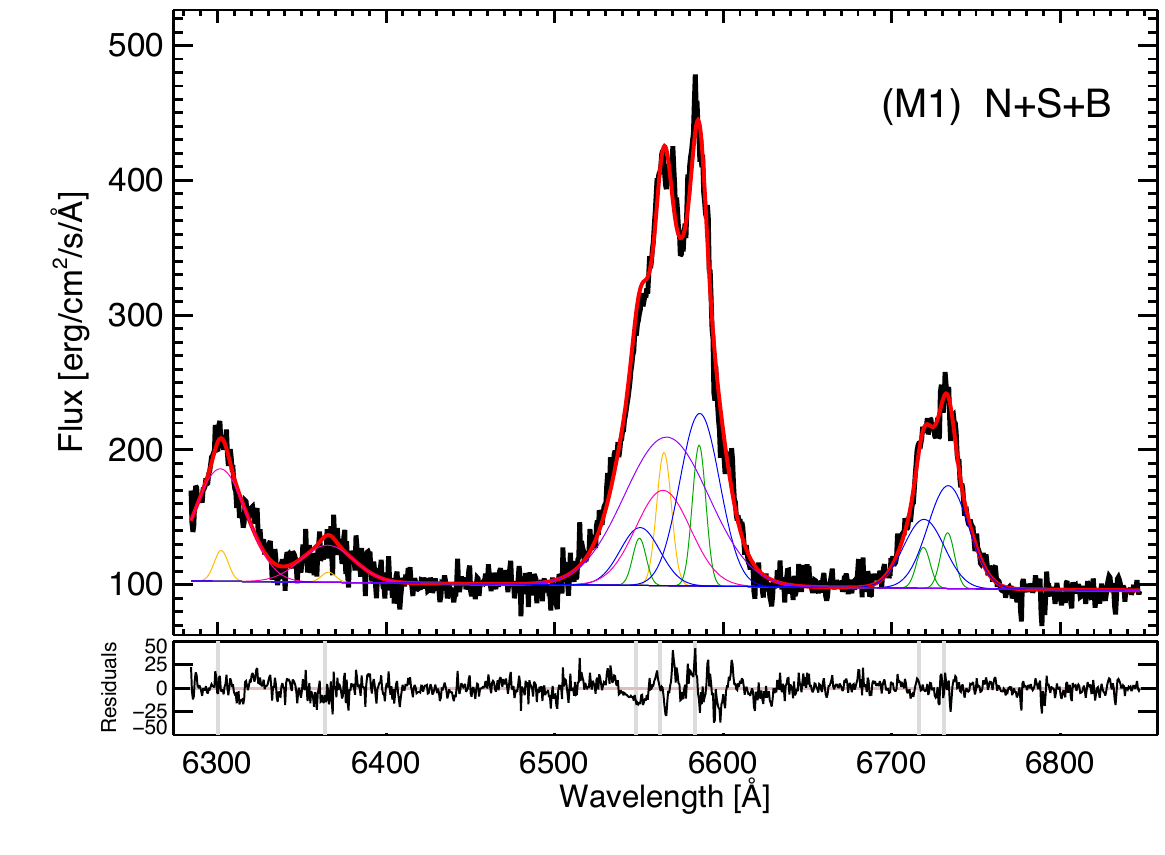}  \\
\vspace{-0.10cm}
\includegraphics[trim = 2.4cm 19.75cm 2.7cm 3.75cm, clip=true, width=.915\textwidth]{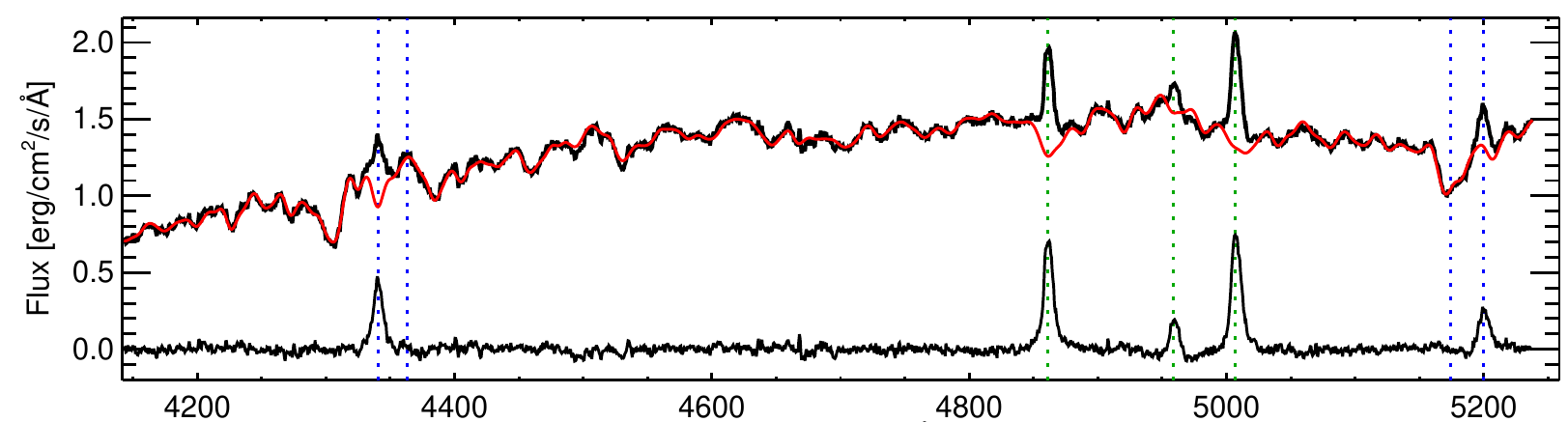}  \\
\vspace{-0.10cm}
\includegraphics[trim = 2.4cm 18.75cm 2.7cm 3.75cm, clip=true, width=.91\textwidth]{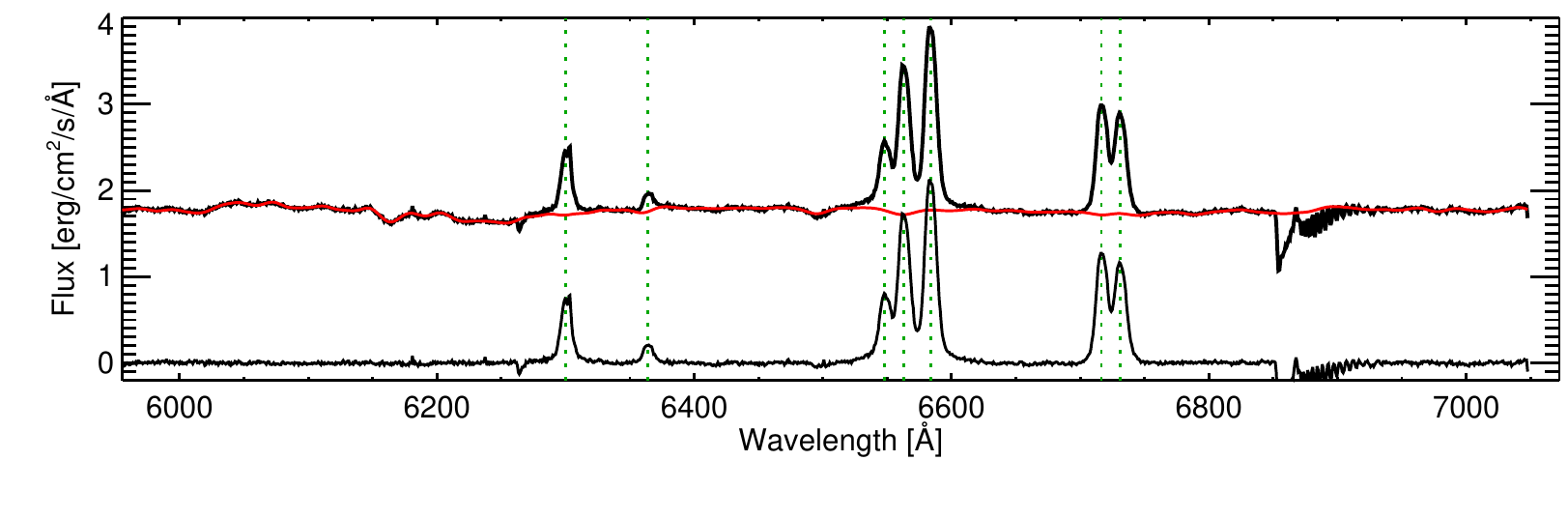} \\
\vspace{-0.45cm}
\includegraphics[trim = 5cm 13.25cm 5.25cm 6.3cm, clip=true, width=.465\textwidth]{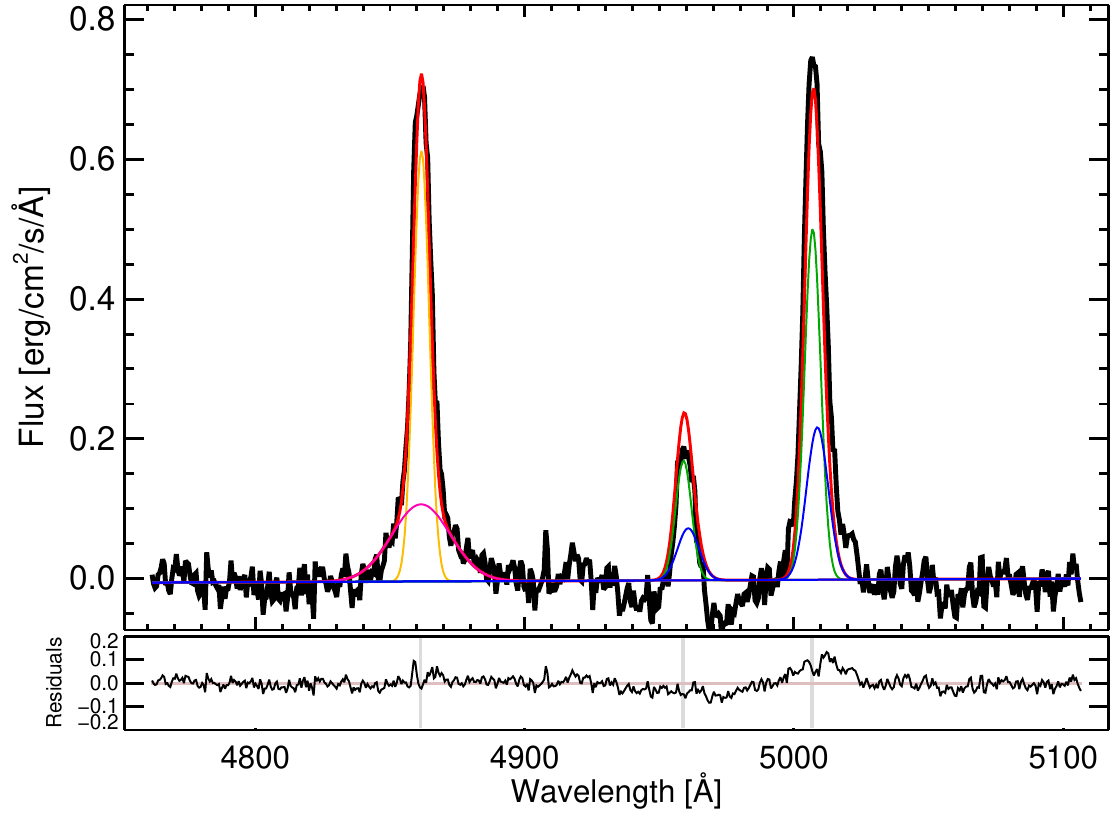}
\hspace{0.1cm} 
\includegraphics[trim = 5.55cm 13.25cm 5.25cm 6.3cm, clip=true, width=.445\textwidth]{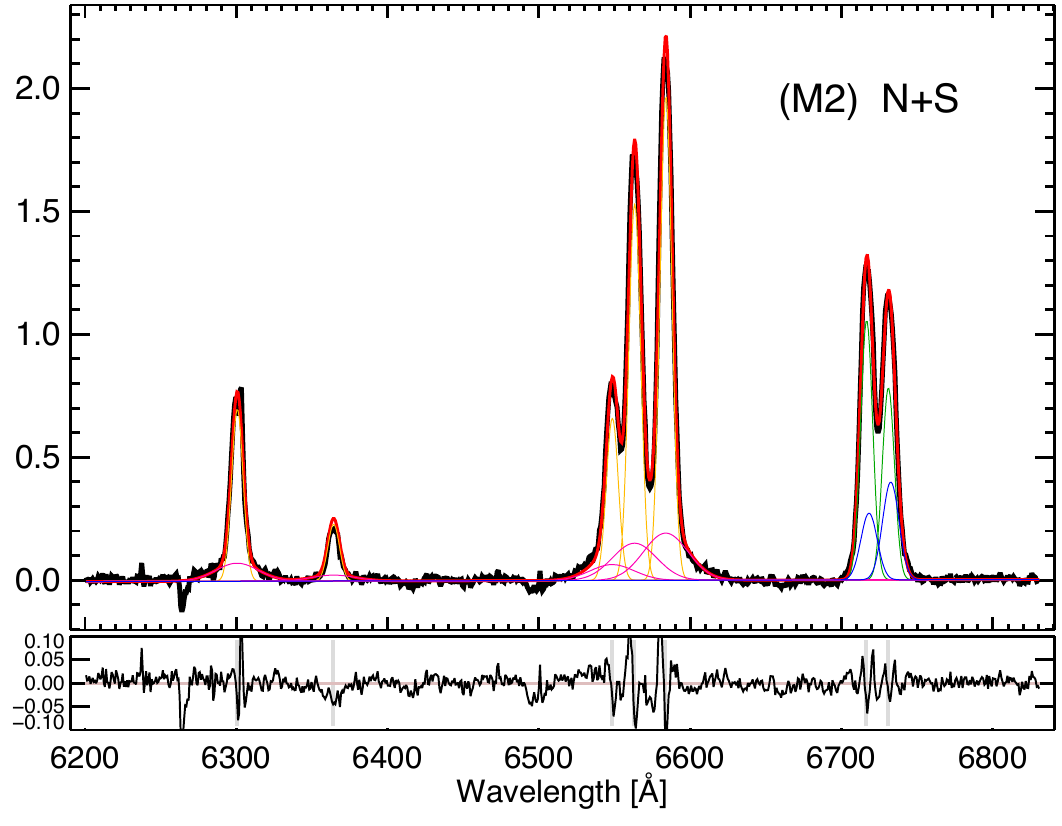}  
\caption{(General description as in Fig.\,\ref{Panel_NGC0266}.) NGC\,4278: the stellar H$\alpha$ absorption seems to be negligible. We did not find evidence for a broad component 
in our ground-based spectra, whereas it is relatively weak in the Palomar spectrum (\textit{HFS97}). 
} \label{Panel_NGC4278} 		 		 
\end{figure*}
\clearpage
\begin{figure*}
\vspace{-0.25cm} 
\includegraphics[trim = 1.10cm .85cm 11.0cm 17.75cm, clip=true, width=.40\textwidth]{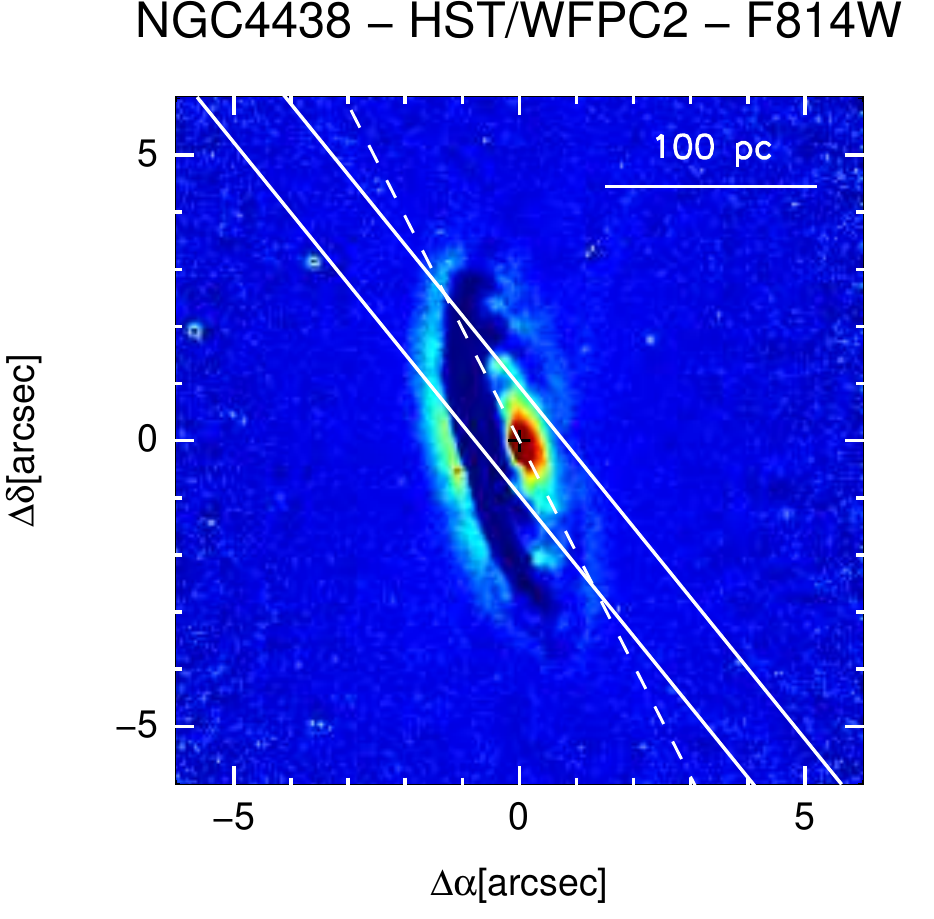}
\hspace{-0.3cm} 
\includegraphics[trim = 2.4cm 19.75cm 2.7cm 3.75cm, clip=true, width=.915\textwidth]{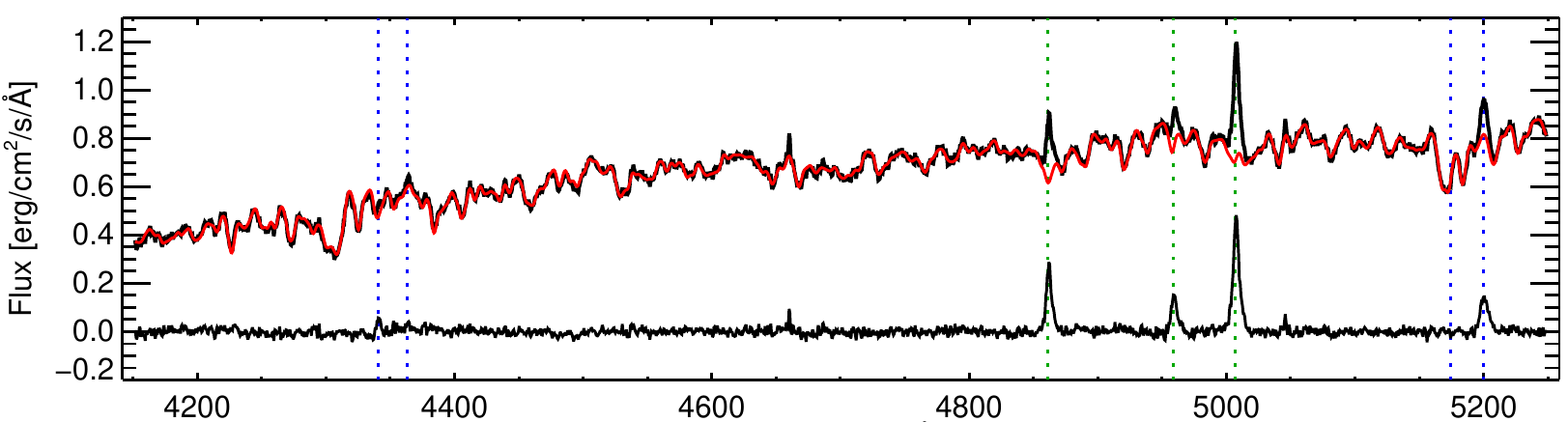}\\
\vspace{-0.10cm}
\includegraphics[trim = 2.4cm 18.75cm 2.7cm 3.75cm, clip=true, width=.91\textwidth]{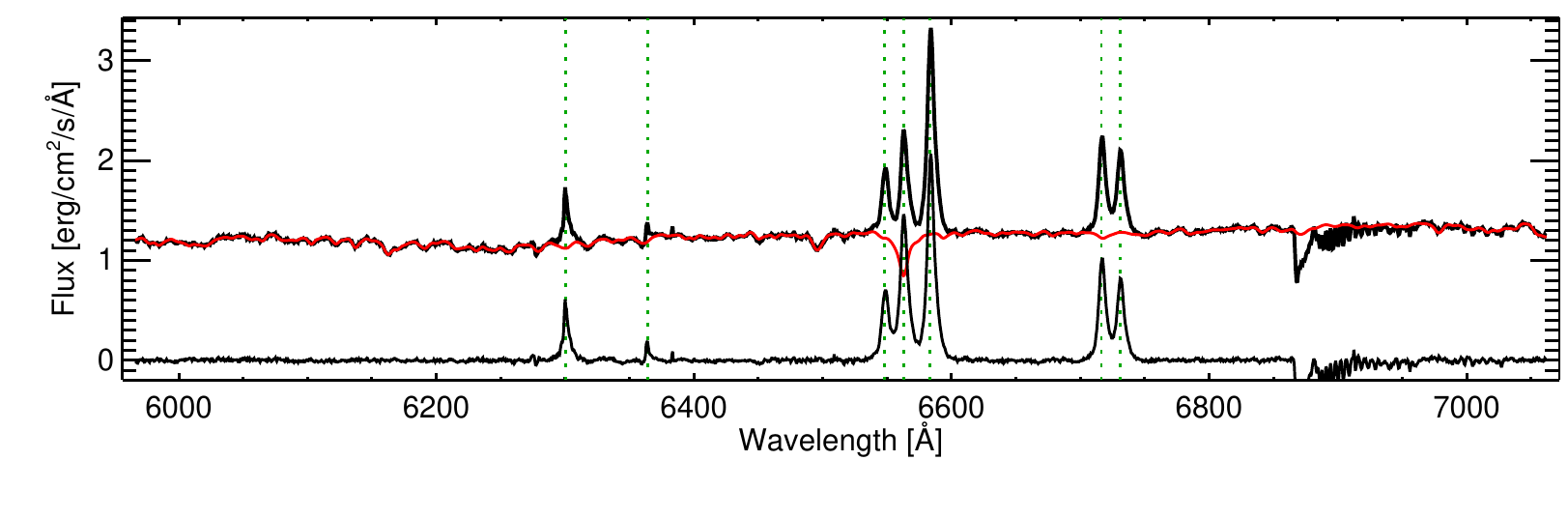}\\
\vspace{-0.45cm}
\includegraphics[trim = 5cm 13.25cm 5.25cm 6.3cm, clip=true, width=.465\textwidth]{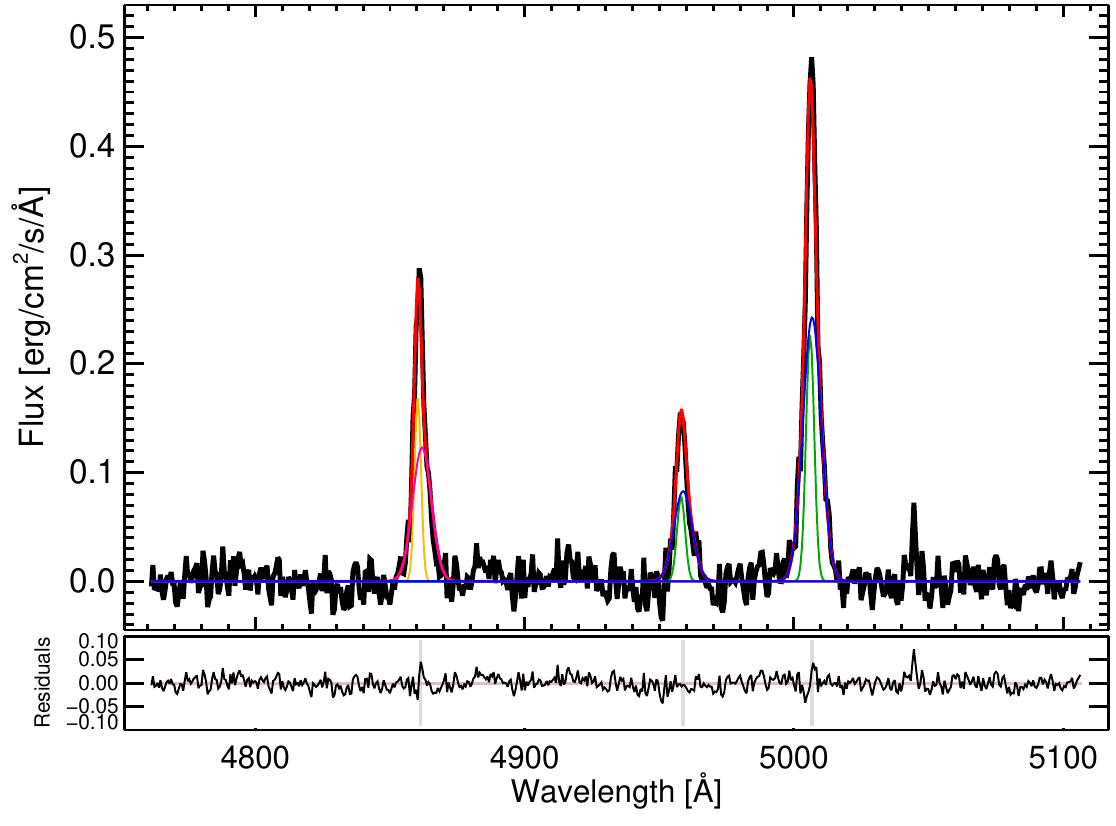}
\hspace{0.1cm} 
\includegraphics[trim = 5.55cm 13.25cm 5.25cm 6.3cm, clip=true, width=.445\textwidth]{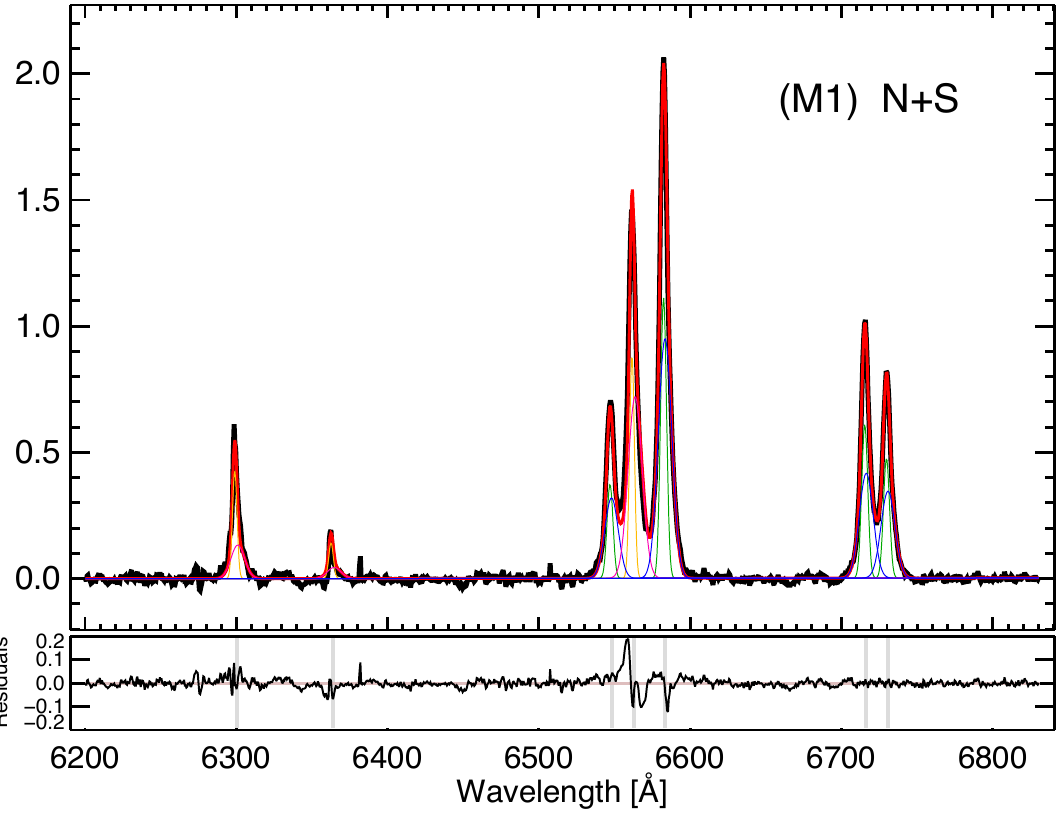} 
\caption{(General description as in Fig.\,\ref{Panel_NGC0266}.) NGC\,4438: all emission lines are relatively narrow. However,  blue and red wings are detected  in both [O\,I] and [S\,II] lines. The broad component is not required for modeling the H$\alpha$ emission  contrary to what was found by \textit{HFS97} in the analysis of the Palomar spectrum (note that their broad component was the weakest among their detections).}
 \label{Panel_NGC4438} 		 		 
\end{figure*}
\clearpage

\clearpage
\begin{figure*}
\vspace{-0.25cm} 
\includegraphics[trim = 1.10cm .85cm 11.0cm 17.75cm, clip=true, width=.40\textwidth]{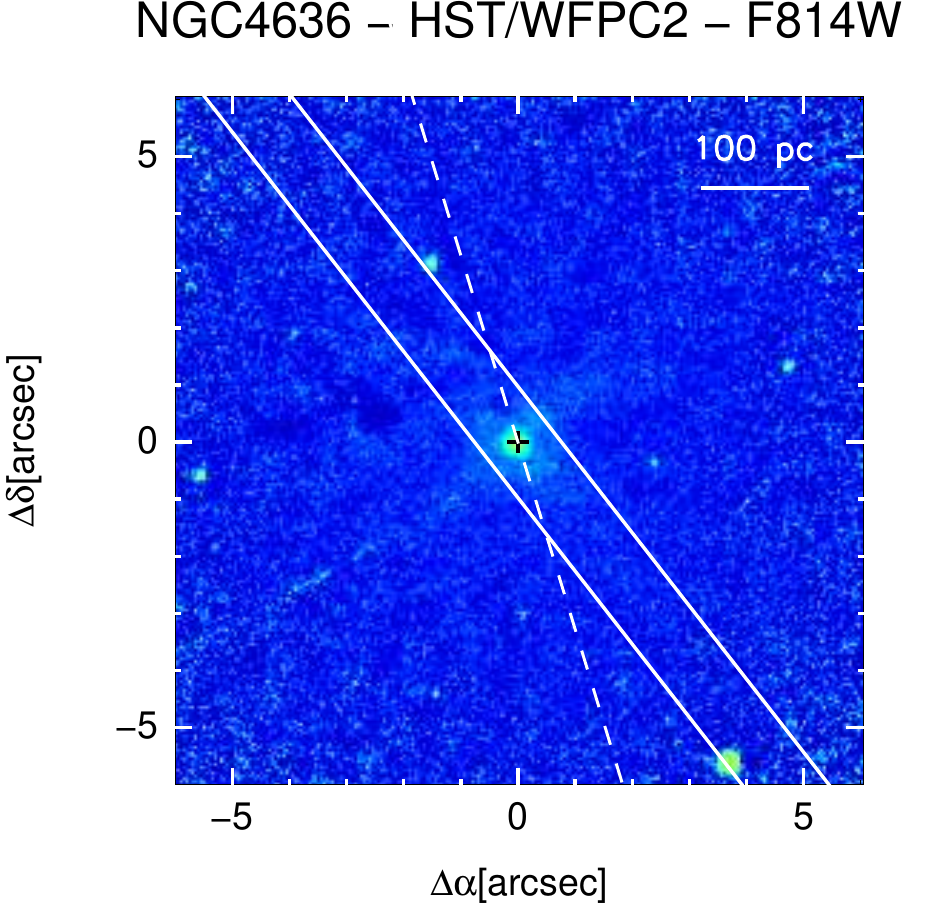}
\hspace{-0.3cm} 
\includegraphics[trim = 2.4cm 19.75cm 2.7cm 3.75cm, clip=true, width=.915\textwidth]{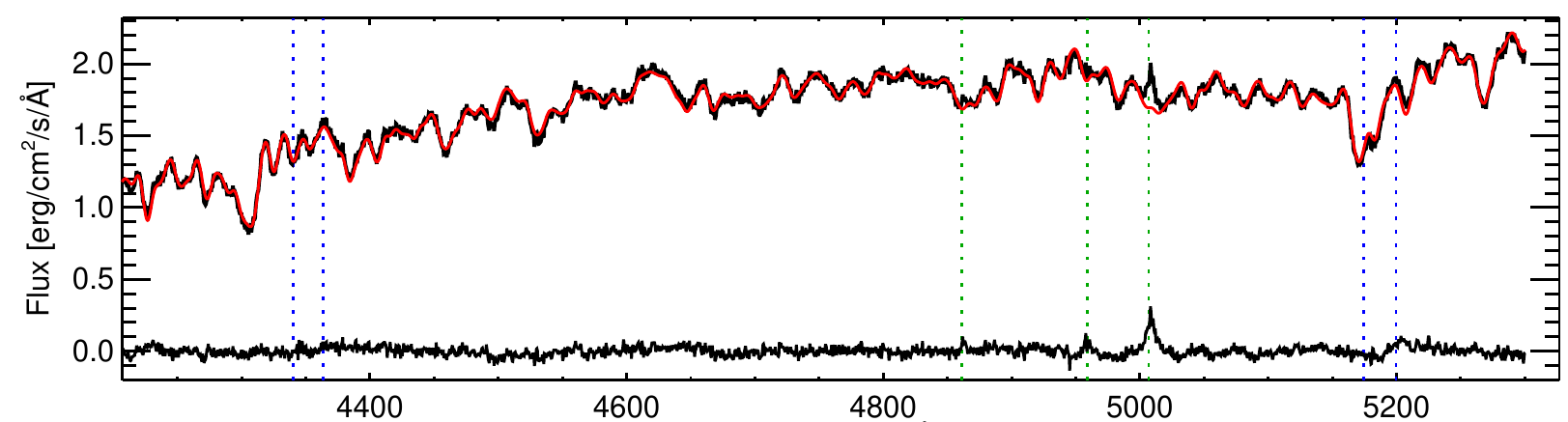} \\
\vspace{-0.10cm}
\includegraphics[trim = 2.4cm 18.75cm 2.7cm 3.75cm, clip=true, width=.91\textwidth]{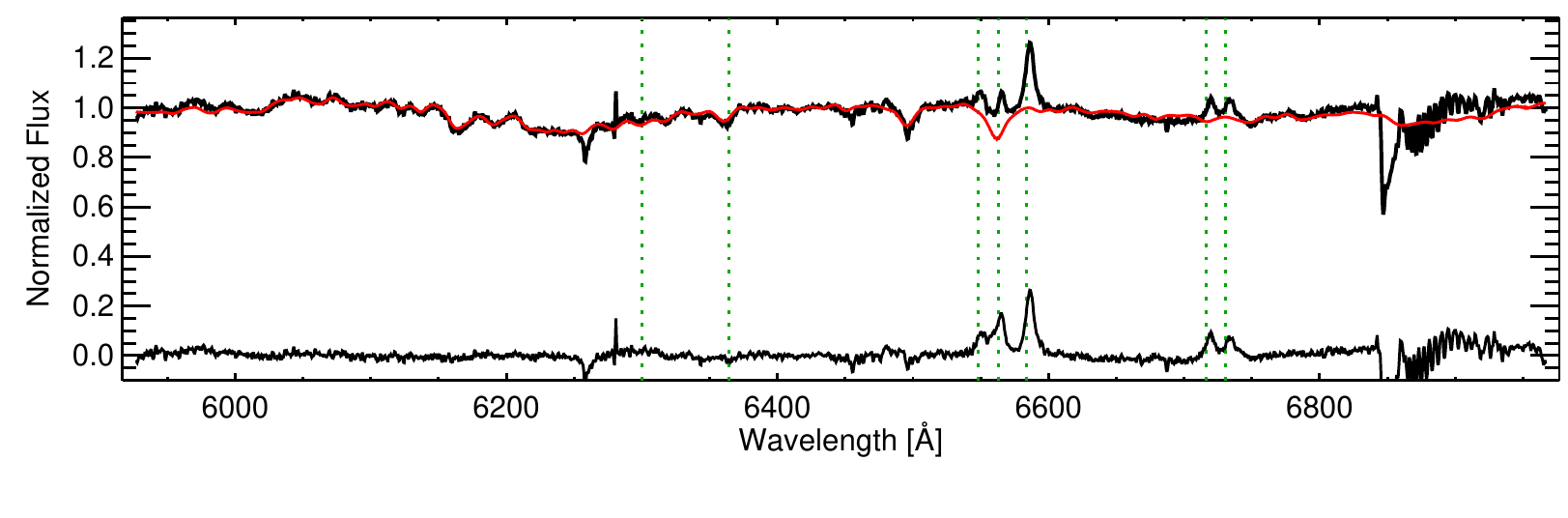} \\
\vspace{-0.45cm}
\includegraphics[trim = 5cm 13.25cm 5.25cm 6.3cm, clip=true, width=.465\textwidth]{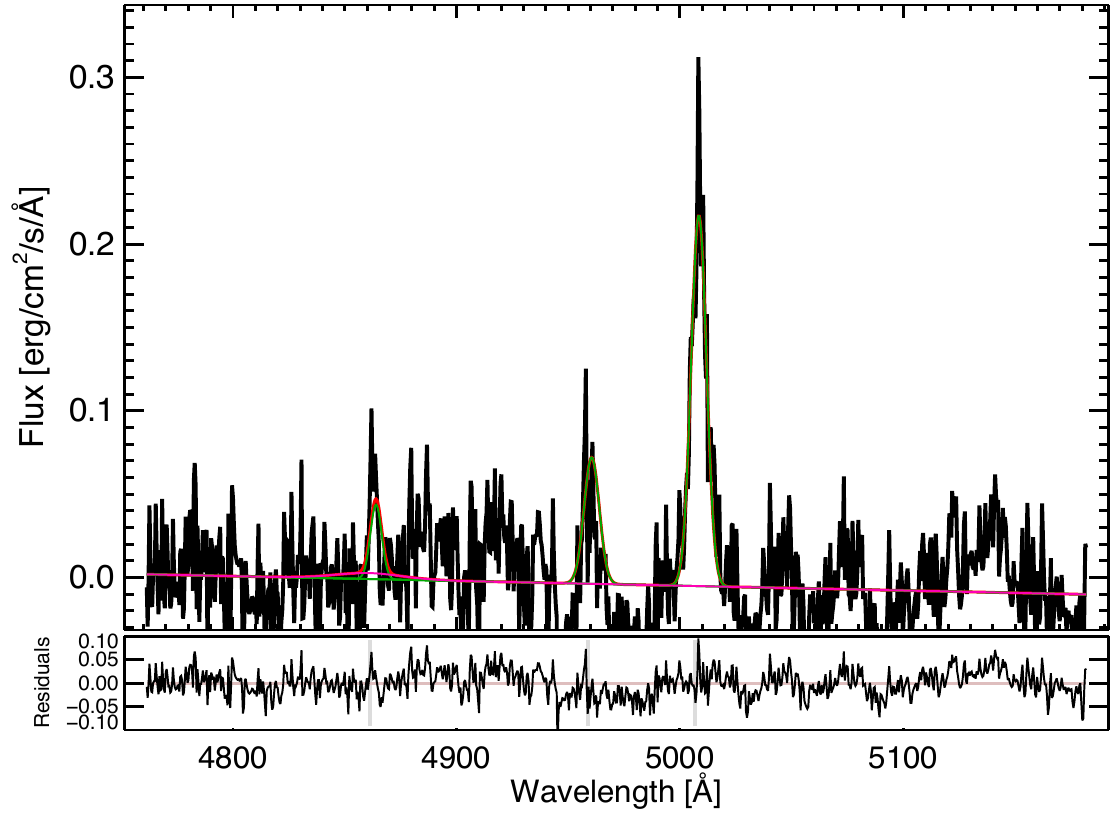}
\hspace{0.1cm} 
\includegraphics[trim = 5.55cm 13.25cm 5.25cm 6.3cm, clip=true, width=.445\textwidth]{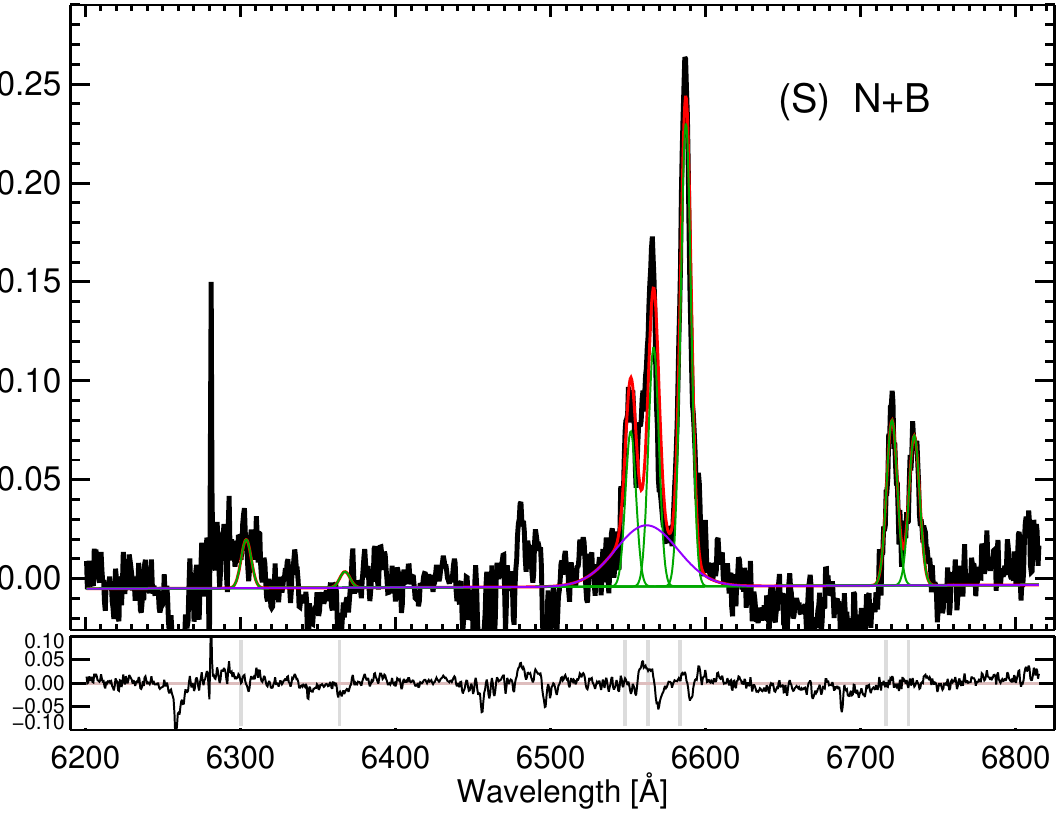} 
\caption{(General description as in Fig.\,\ref{Panel_NGC0266}.) NGC\,4636: [O\,I] lines are very weak even after starlight subtraction. [S\,II] is therefore more reliable  as a reference for modelling the narrow lines.}
 \label{Panel_NGC4636} 		 		 
\end{figure*}
\clearpage
\begin{figure*}
\vspace{-0.25cm} 
\includegraphics[trim = 1.10cm .85cm 11.0cm 17.75cm, clip=true, width=.40\textwidth]{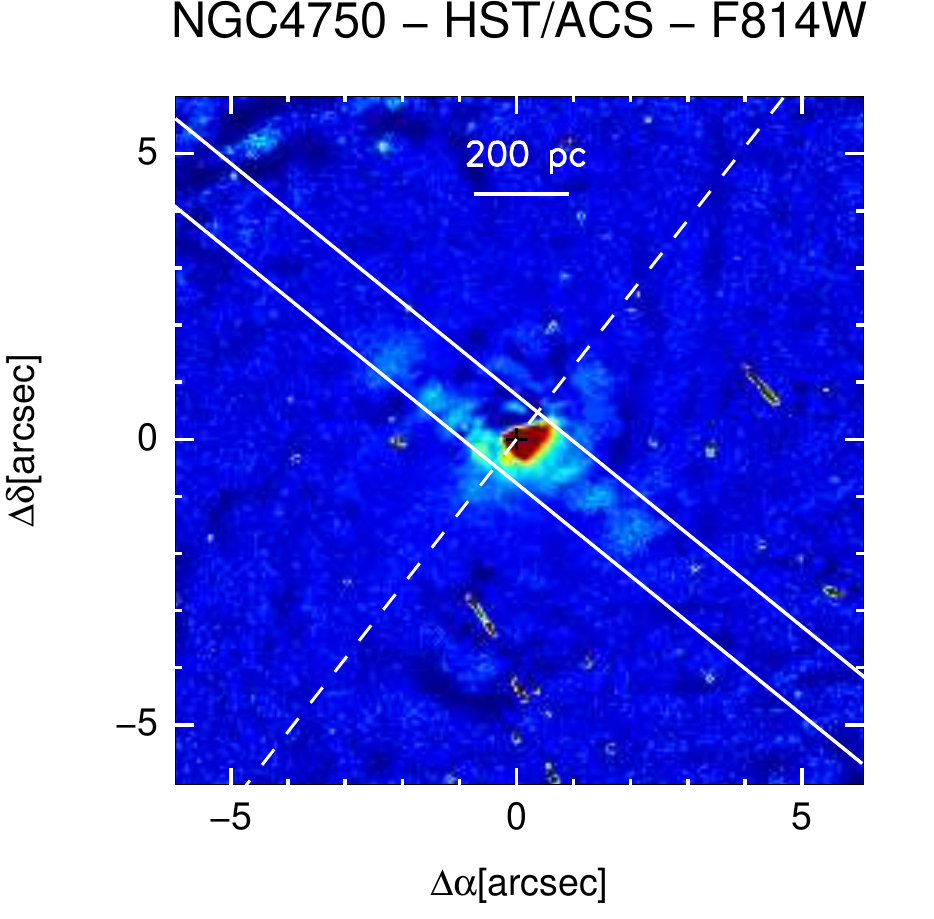} 
\hspace{-0.3cm} 
\includegraphics[trim = 2.4cm 19.75cm 2.7cm 3.75cm, clip=true, width=.915\textwidth]{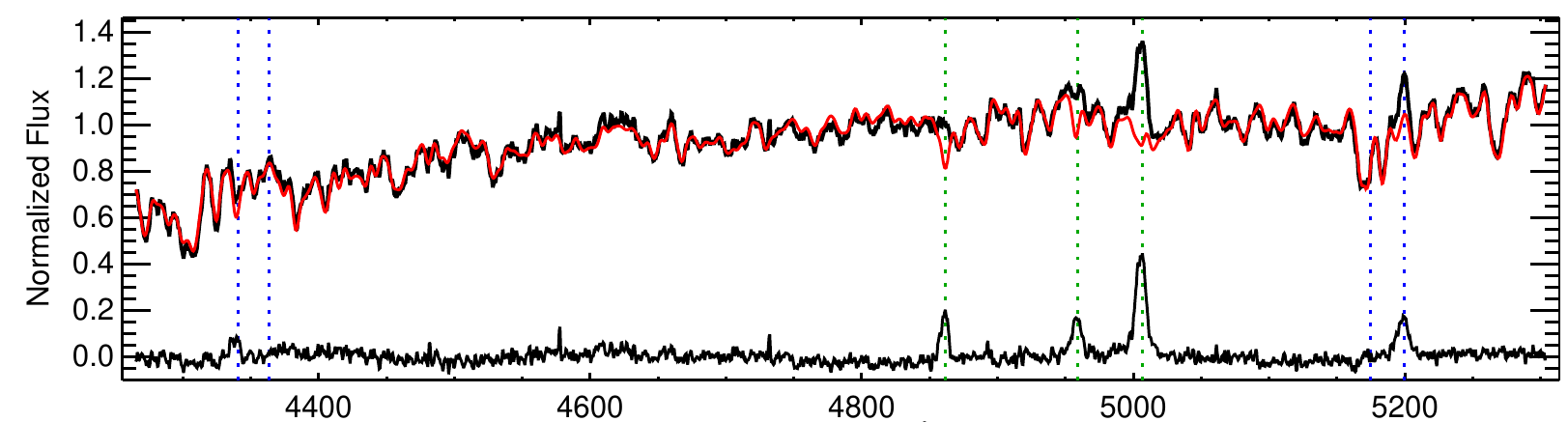} \\
\vspace{-0.10cm}
\includegraphics[trim = 2.4cm 18.75cm 2.7cm 3.75cm, clip=true, width=.91\textwidth]{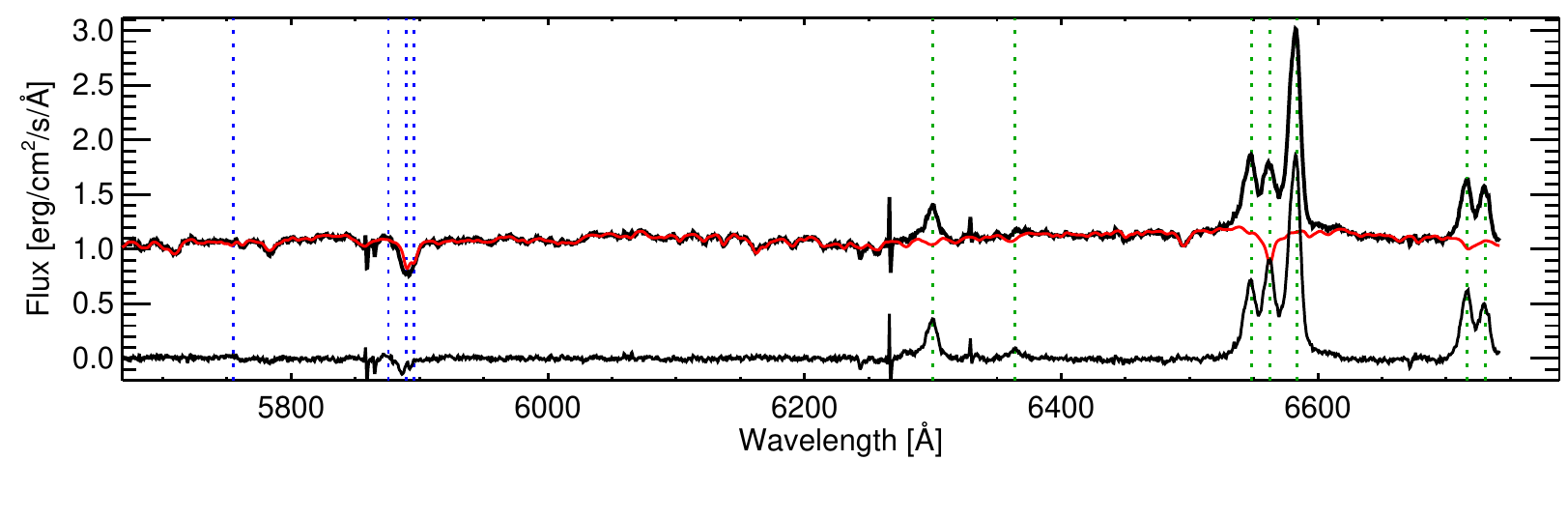} \\
\vspace{-0.45cm}
\includegraphics[trim = 5cm 13.25cm 5.25cm 6.3cm, clip=true, width=.465\textwidth]{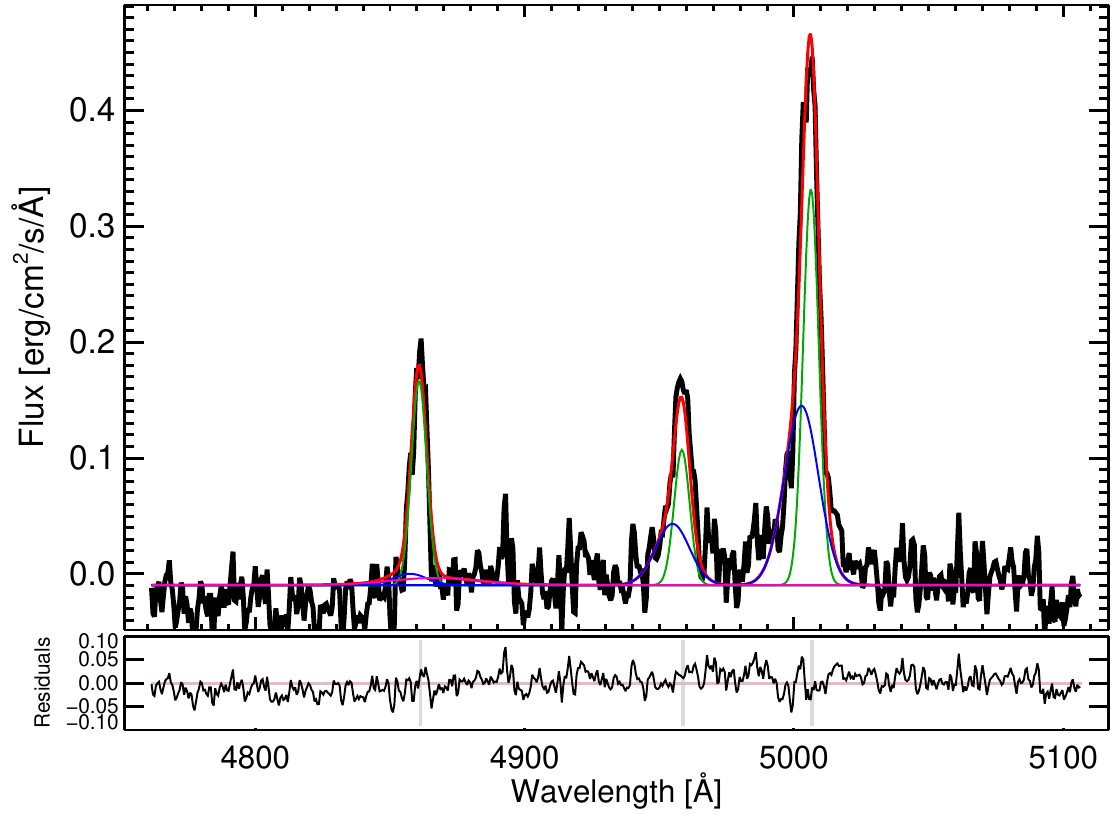}
\hspace{0.1cm} 
\includegraphics[trim = 5.55cm 13.25cm 5.25cm 6.3cm, clip=true, width=.445\textwidth]{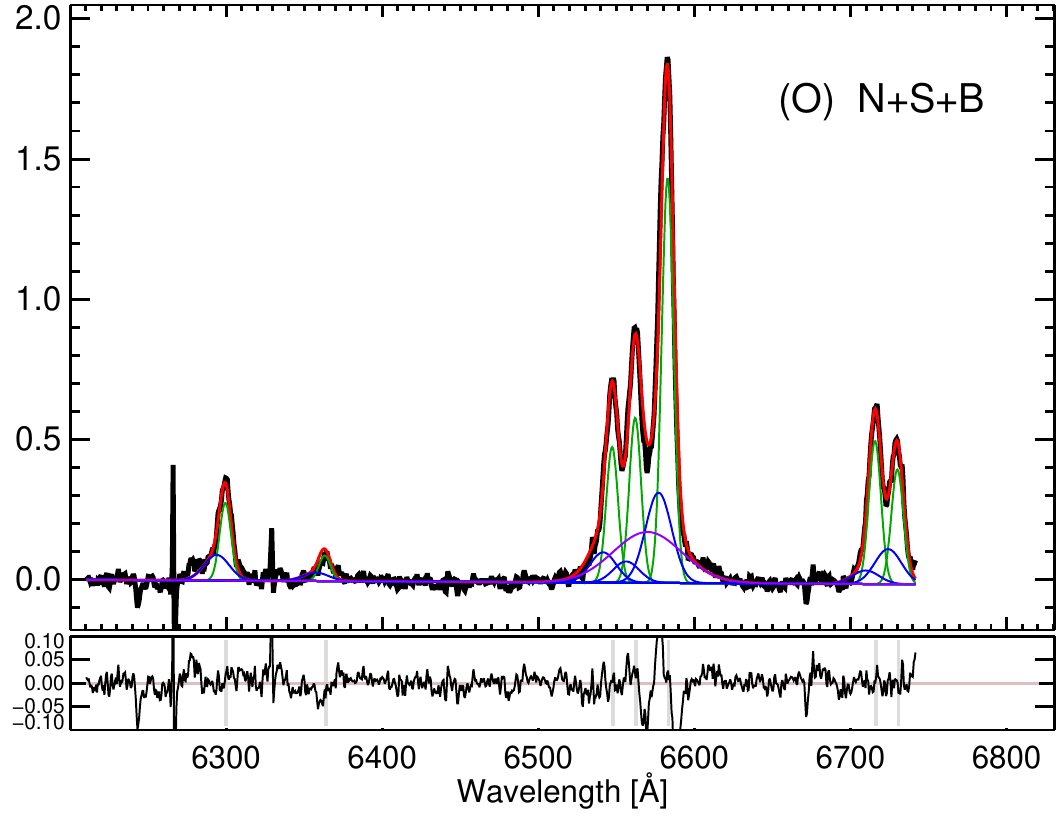}  
\caption{(General description as in Fig.\,\ref{Panel_NGC0266}.) NGC\,4750: some residuals from the sky subtraction affect the region of [O\,I] and that of [S\,II] is at the edge of the spectrum. In spite of this, we were able to  obtain a good fit for the forbidden lines with two components. We did not find a clear improvement of the standard deviation in H$\alpha$ for the ground-based data fitting when adding the broad component (Table\,\ref{T_rms}). However, the presence of a red wing near the base of [N\,II]$\lambda$6584 (absent in the other template lines) makes the broad component necessary to adequately model the H$\alpha$-[N\,II] complex. Thus a broad component is needed for a good fit, but its presence is not obvious and it is rather weak (in rather good agreement with \textit{HFS97}).}
 \label{Panel_NGC4750} 		 		 
\end{figure*}
\clearpage
\begin{figure*}
\vspace{-0.25cm} 
\includegraphics[trim = 1.10cm .85cm 11.0cm 17.75cm, clip=true, width=.40\textwidth]{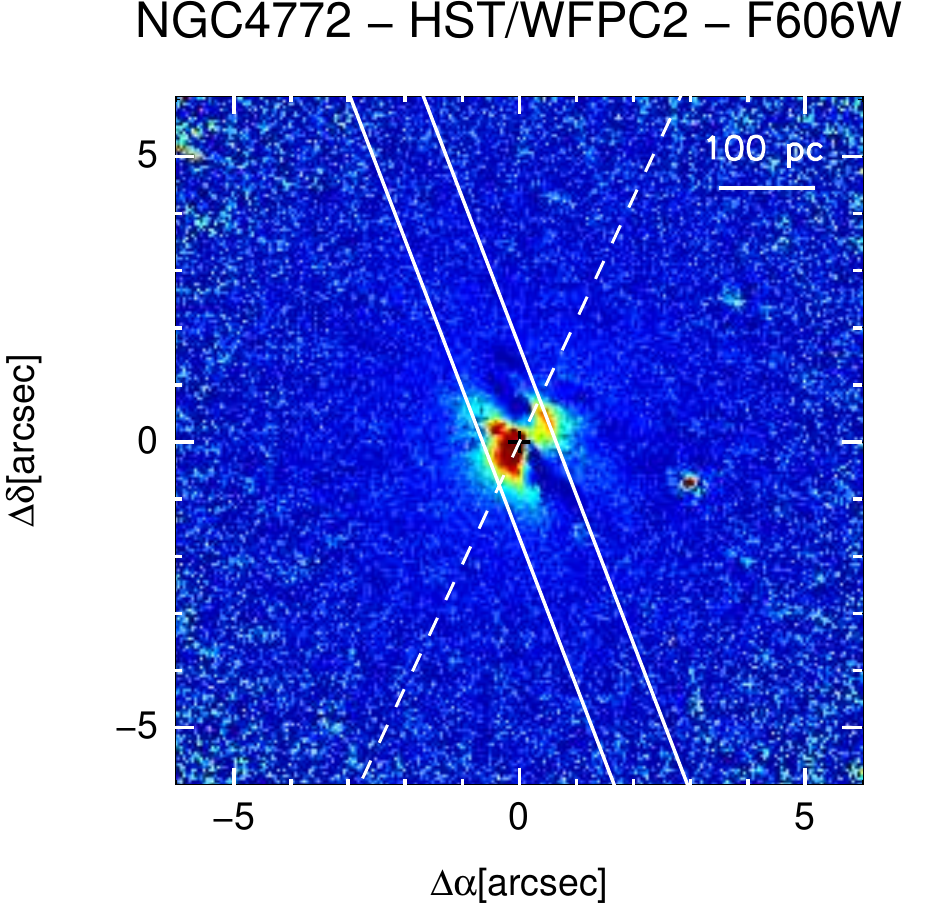}
\hspace{-0.3cm} 
\includegraphics[trim = 2.4cm 19.75cm 2.7cm 3.75cm, clip=true, width=.915\textwidth]{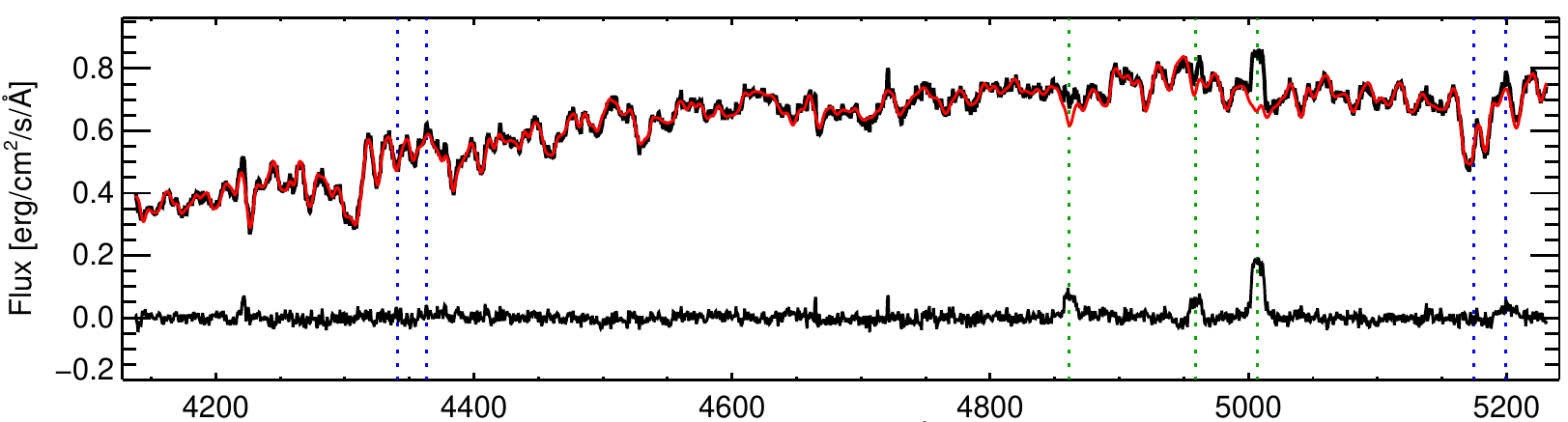} \\
\vspace{-0.10cm}
\includegraphics[trim = 2.4cm 18.75cm 2.7cm 3.75cm, clip=true, width=.91\textwidth]{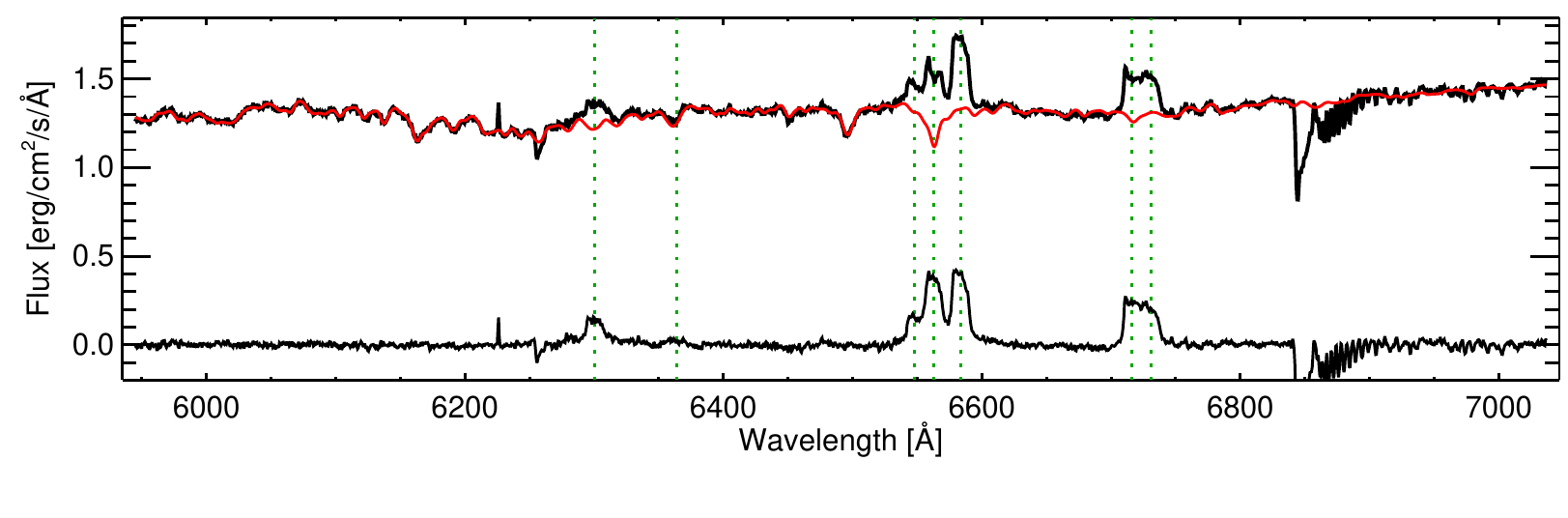} \\
\vspace{-0.45cm}
\includegraphics[trim = 4.9cm 13.25cm 5.25cm 6.3cm, clip=true, width=.4715\textwidth]{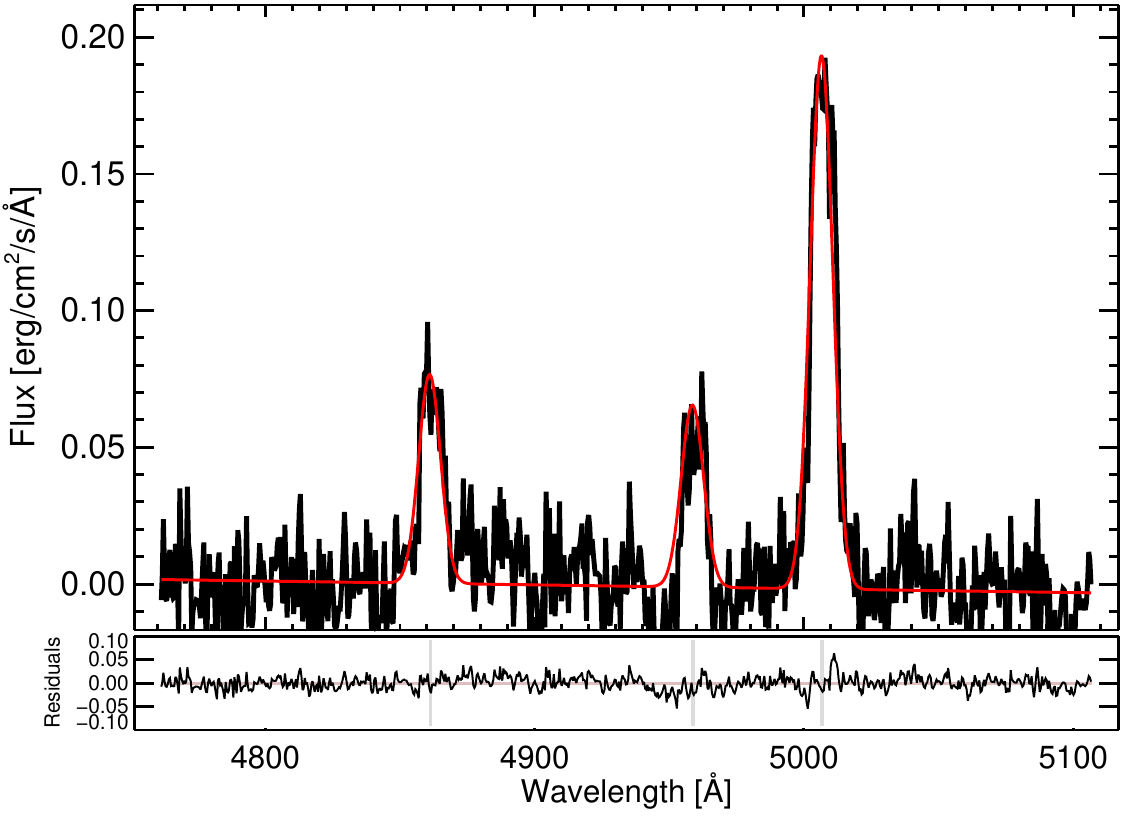}
\hspace{0.1cm} 
\includegraphics[trim = 5.55cm 13.25cm 5.25cm 6.3cm, clip=true, width=.445\textwidth]{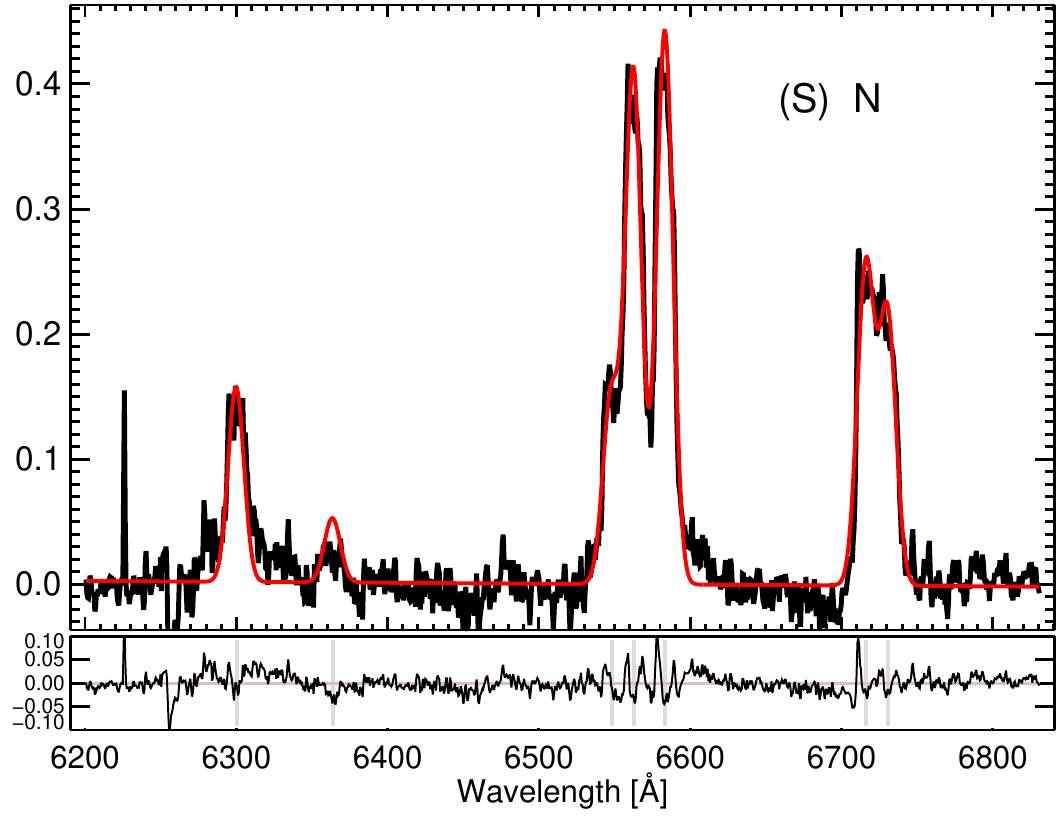}  
\caption{(General description as in Fig.\,\ref{Panel_NGC0266}.) NGC\,4772: all emission line profiles are quite complex. The velocity curve is very chaotic with multiple kinematic components  already  in the individual spectra of the nuclear aperture we considered. In the extracted nuclear spectrum, the presence of a second component is evident for [O\,I] but not for [S\,II]. We tested a two-components fit for the forbidden lines, without obtaining a good modelling. We hence preferred to fit a single Gaussian for the narrow lines, though it results to be rather broad. Higher spectral resolution and S/N are needed to confirm the absence of any second and/or BLR components.}
 \label{Panel_NGC4772} 		 		 
\end{figure*}
\clearpage

\begin{figure*}
\vspace{-0.25cm} 
\includegraphics[trim = 1.10cm .85cm 11.0cm 17.75cm, clip=true, width=.40\textwidth]{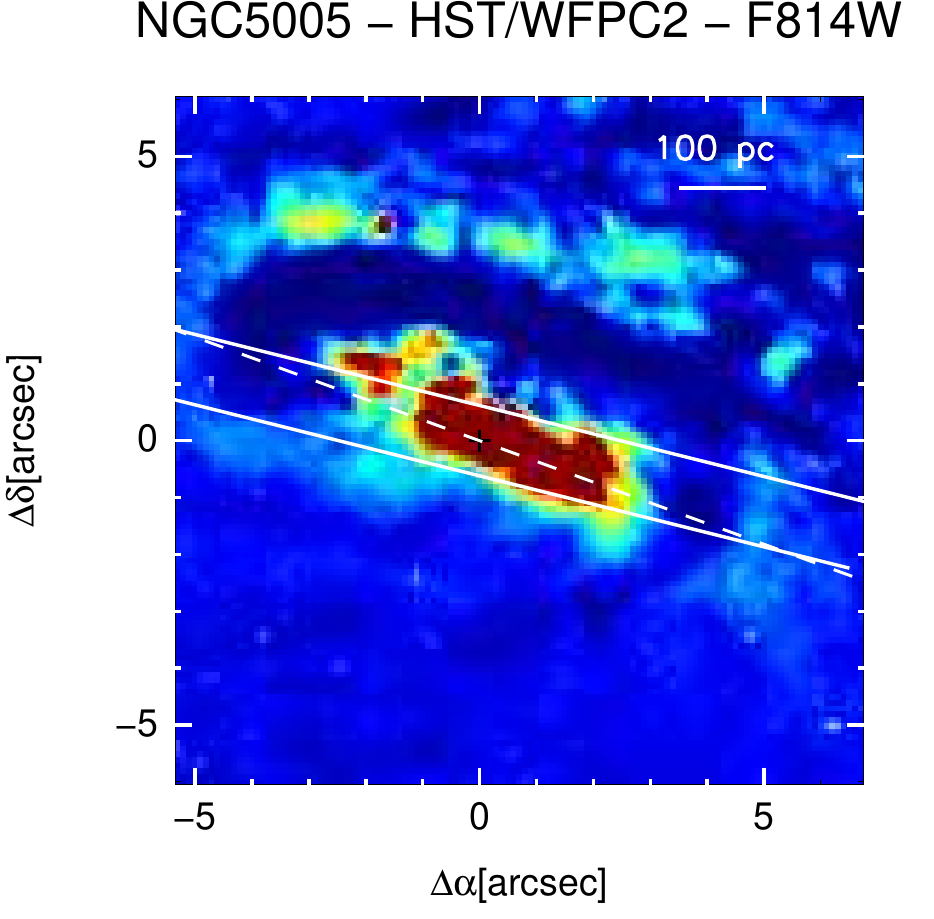} 
\hspace{-0.3cm} 
\includegraphics[trim = 4.5cm 13.cm 5.25cm 6.25cm, clip=true, width=.475\textwidth]{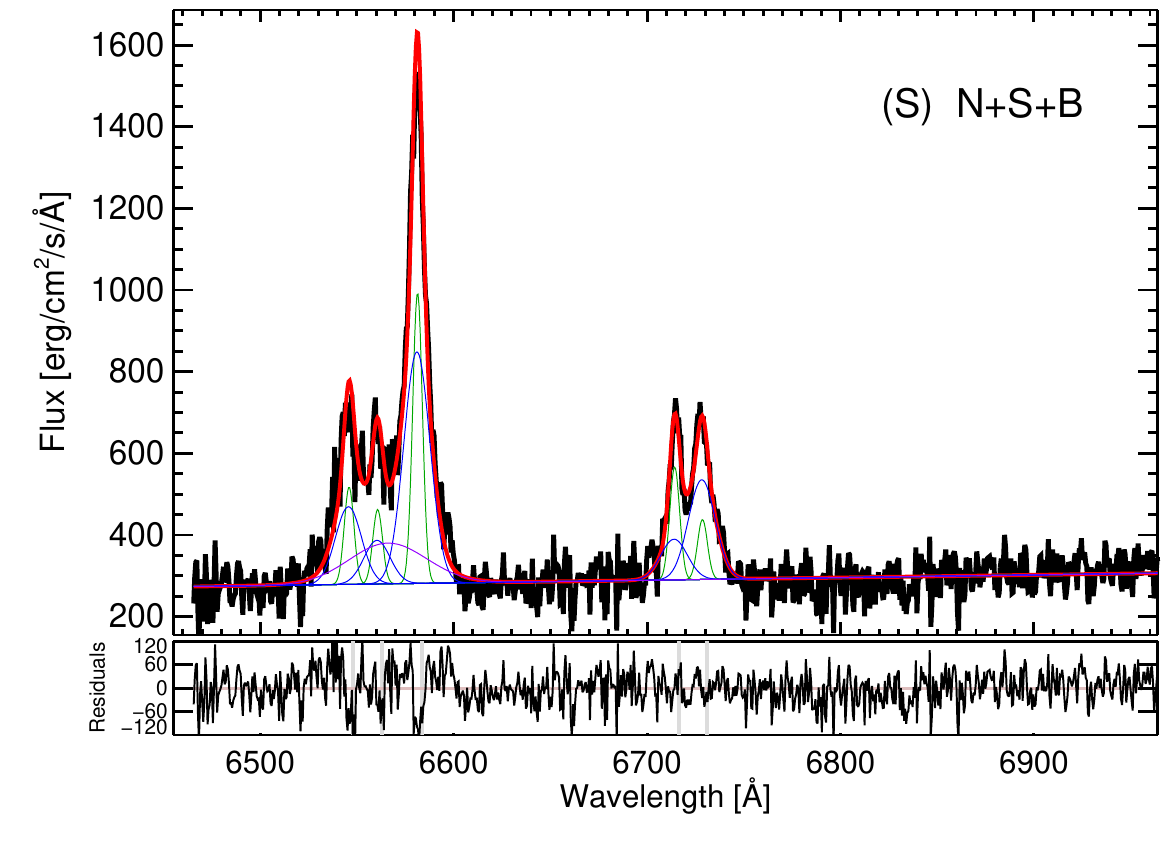}  \\
\vspace{-0.10cm}
\includegraphics[trim = 2.4cm 19.75cm 2.7cm 3.75cm, clip=true, width=.915\textwidth]{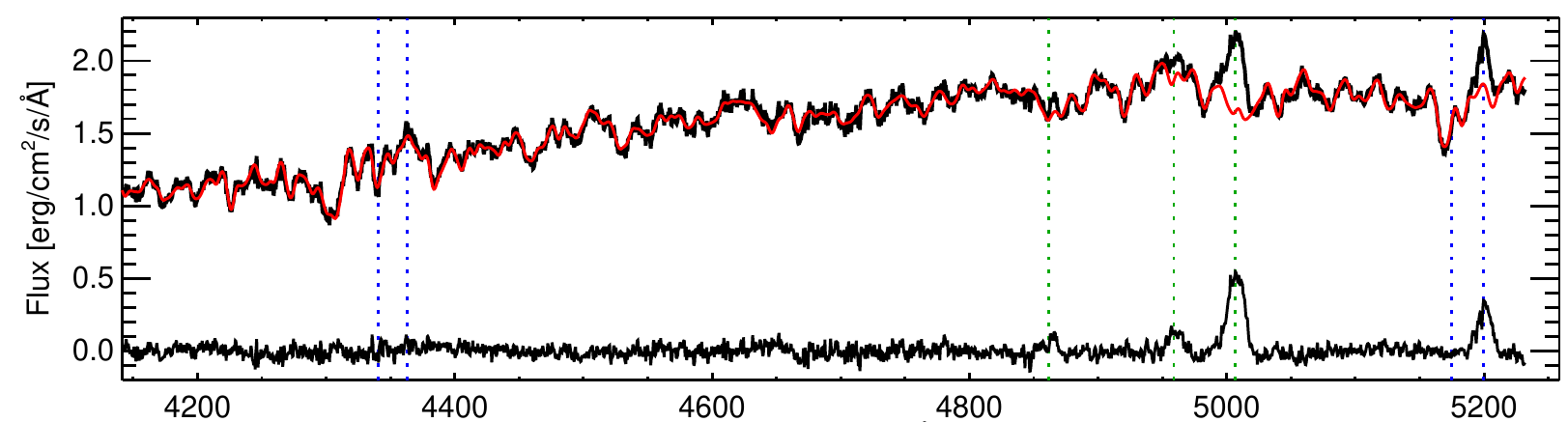} \\
\vspace{-0.10cm}
\includegraphics[trim = 2.4cm 18.75cm 2.7cm 3.75cm, clip=true, width=.91\textwidth]{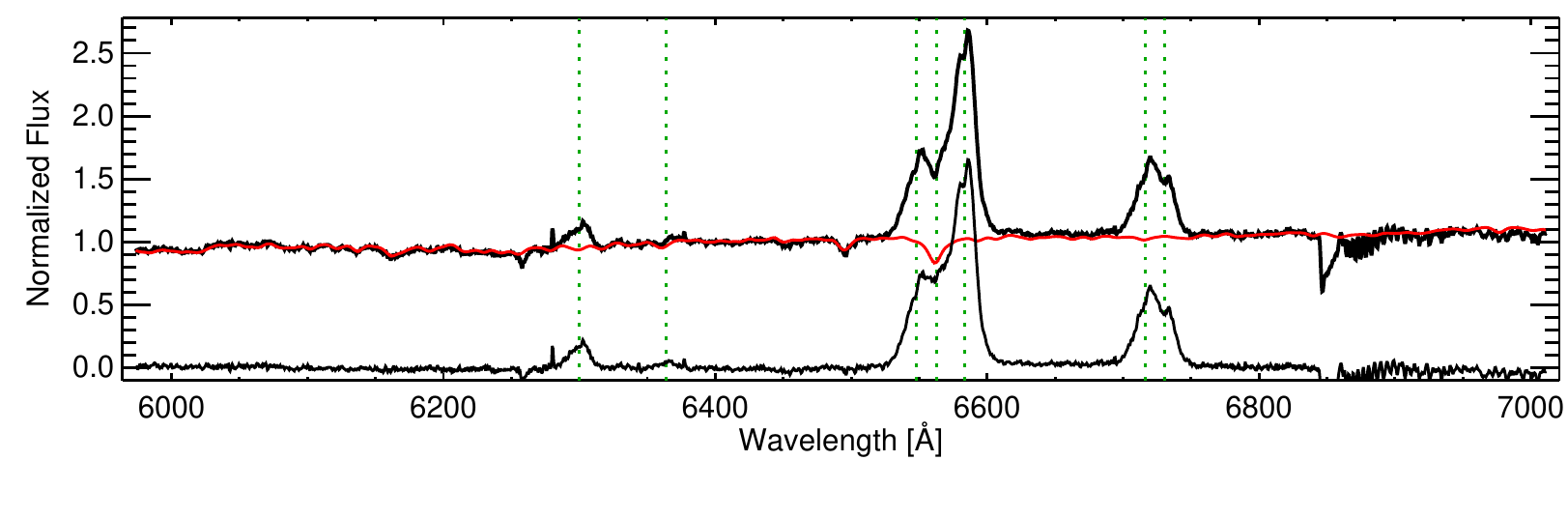} \\
\vspace{-0.45cm}
\includegraphics[trim = 5cm 13.25cm 5.25cm 6.3cm, clip=true, width=.465\textwidth]{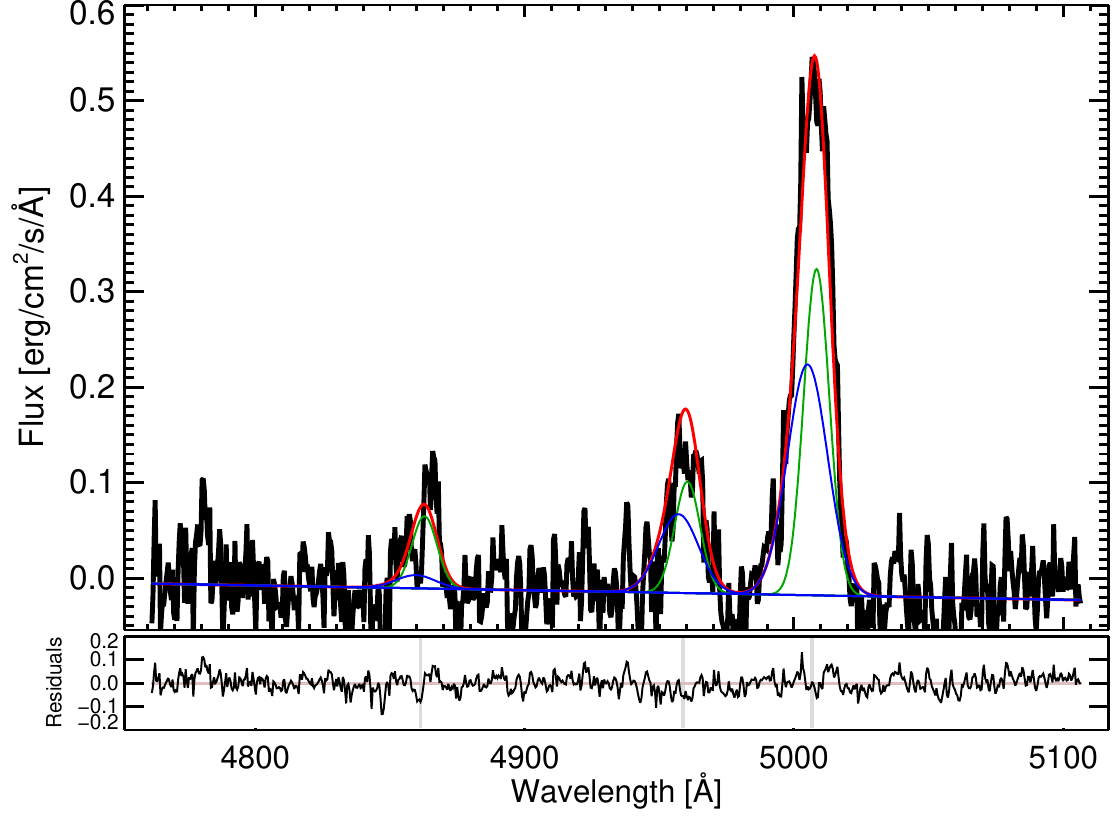}
\hspace{0.1cm} 
\includegraphics[trim = 5.55cm 13.25cm 5.25cm 6.3cm, clip=true, width=.445\textwidth]{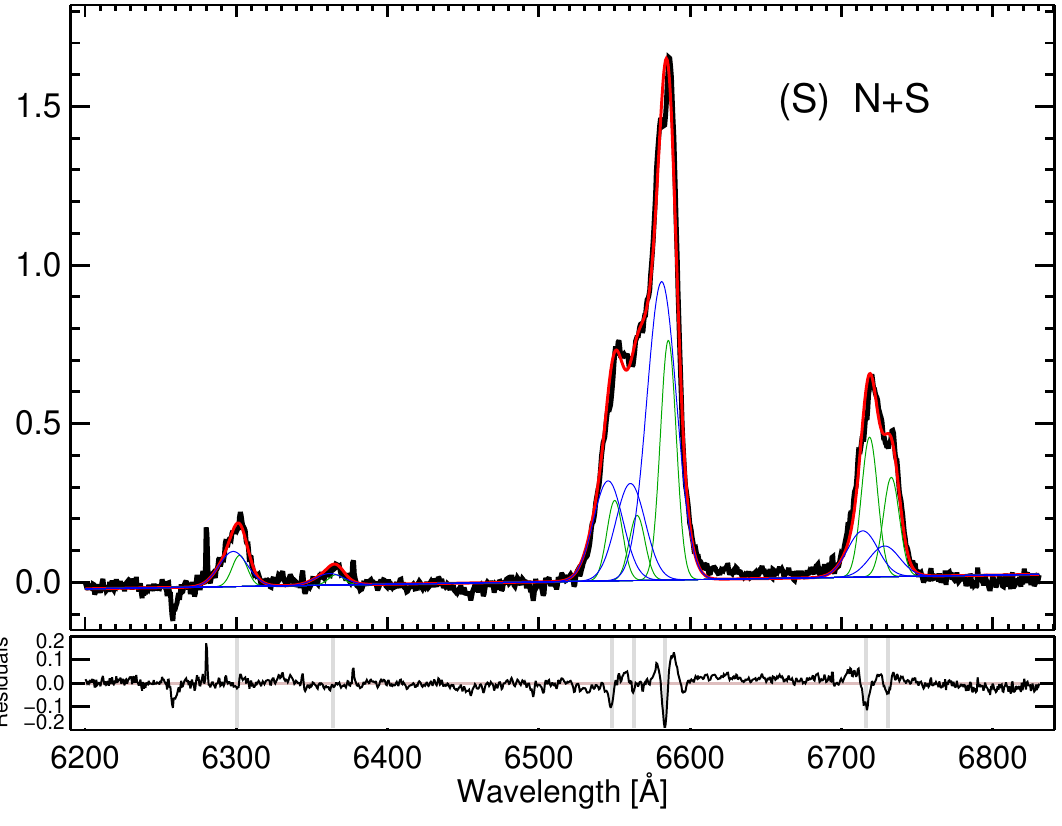}  
\caption{(General description as in Fig.\,\ref{Panel_NGC0266}.) NGC\,5005: both H$\alpha$-[N\,II] and [S\,II] lines are strongly blended. We carefully tested all models (Sect.\,\ref{Analysis_LF}), and finally selected that based on [S\,II]. Even if a single-component model of [O\,I] produces lower residuals than the double Gaussians one, its resulting width is unrealistically large, what made us conclude that a two-component model is better in this case.} \label{Panel_NGC5005} 		 		 
\end{figure*}
\clearpage

\begin{figure*}
\vspace{-0.25cm} 
\includegraphics[trim = 1.10cm .85cm 11.0cm 17.75cm, clip=true, width=.40\textwidth]{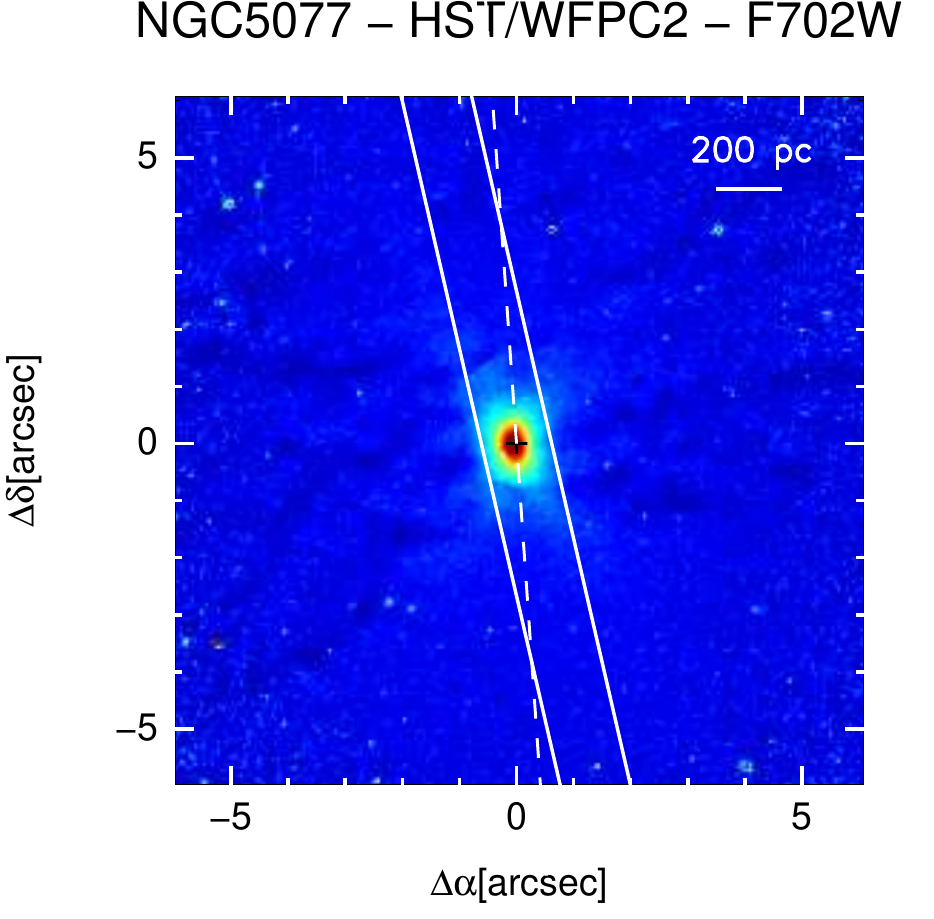} 
\hspace{-0.3cm} 
\includegraphics[trim = 4.5cm 13.cm 5.25cm 6.25cm, clip=true, width=.475\textwidth]{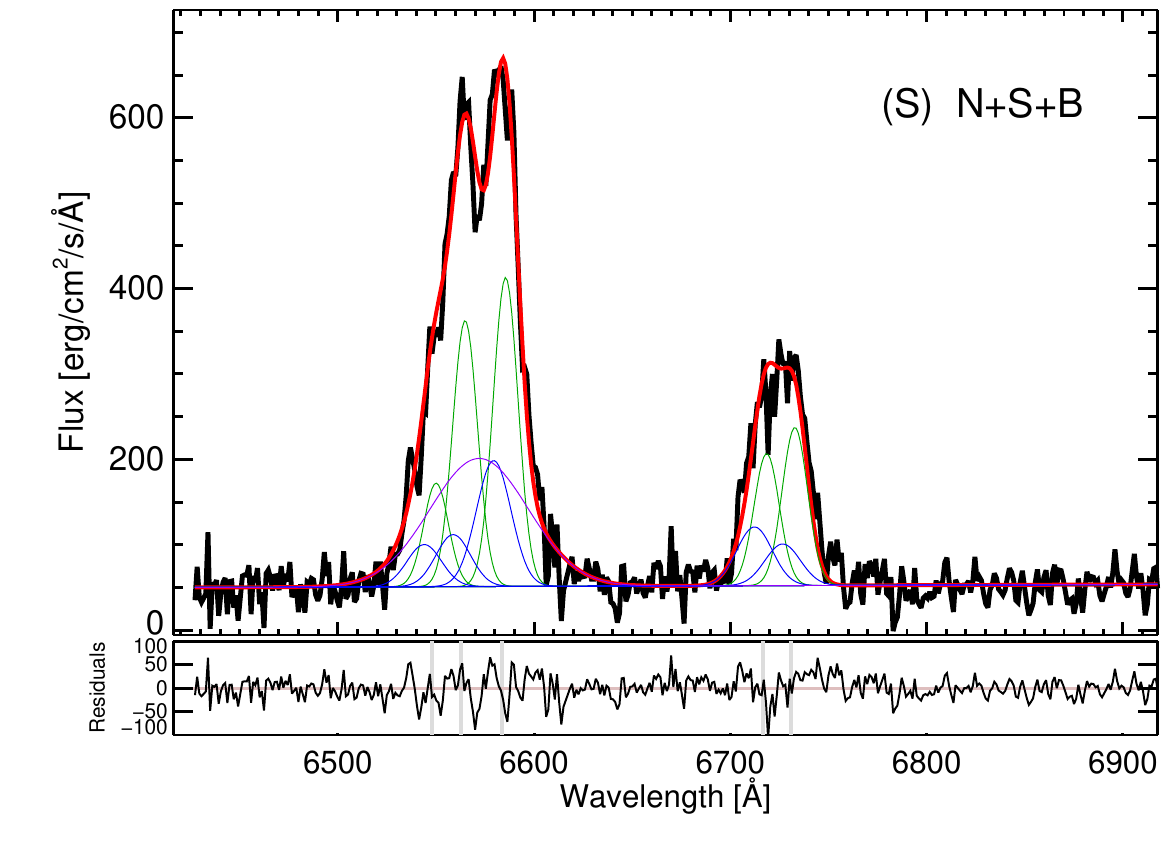}  \\
\vspace{-0.10cm}
\includegraphics[trim = 2.4cm 19.75cm 2.7cm 3.75cm, clip=true, width=.915\textwidth]{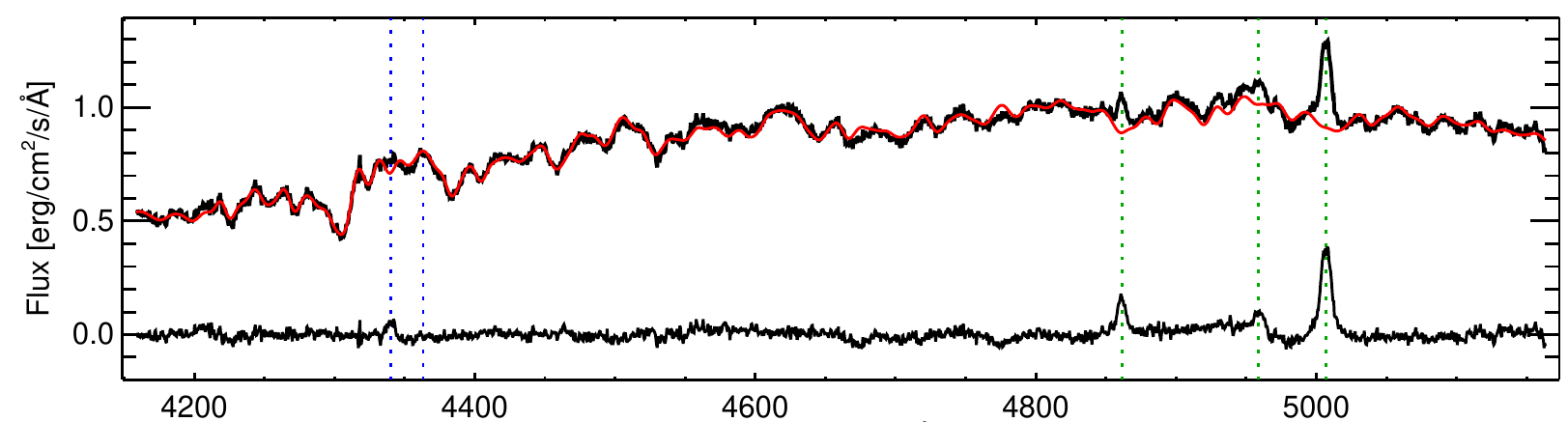}  \\
\vspace{-0.10cm}
\includegraphics[trim = 2.4cm 18.75cm 2.7cm 3.75cm, clip=true, width=.91\textwidth]{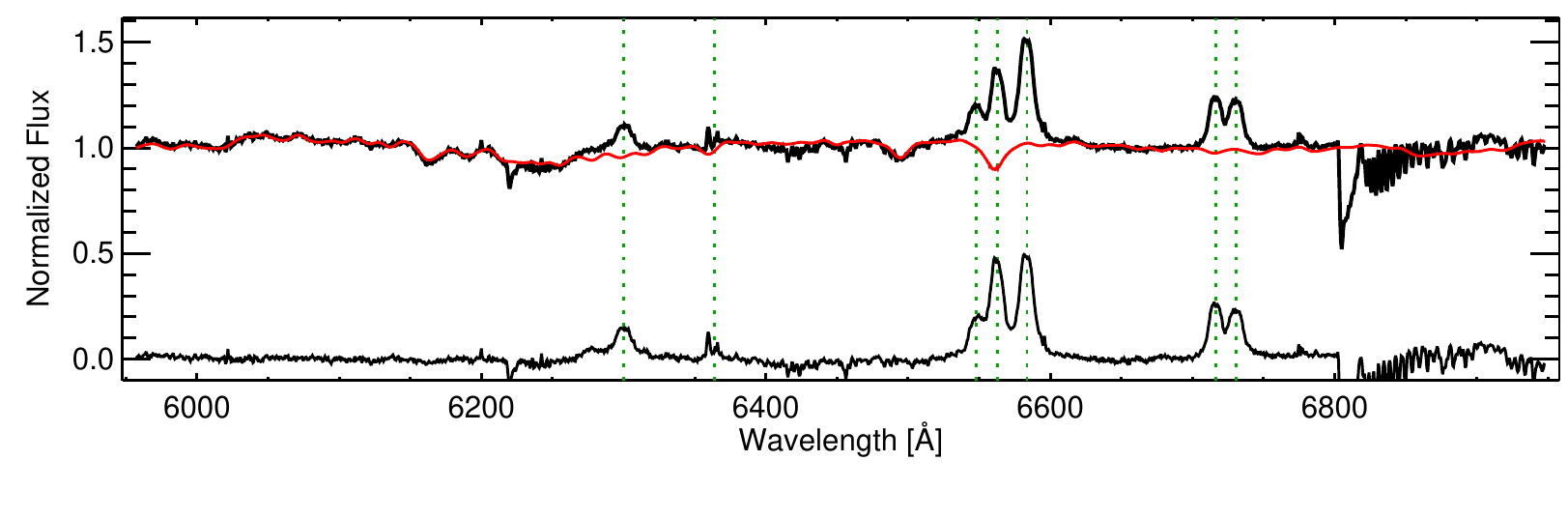} \\
\vspace{-0.45cm}
\includegraphics[trim = 5cm 13.25cm 5.25cm 6.3cm, clip=true, width=.465\textwidth]{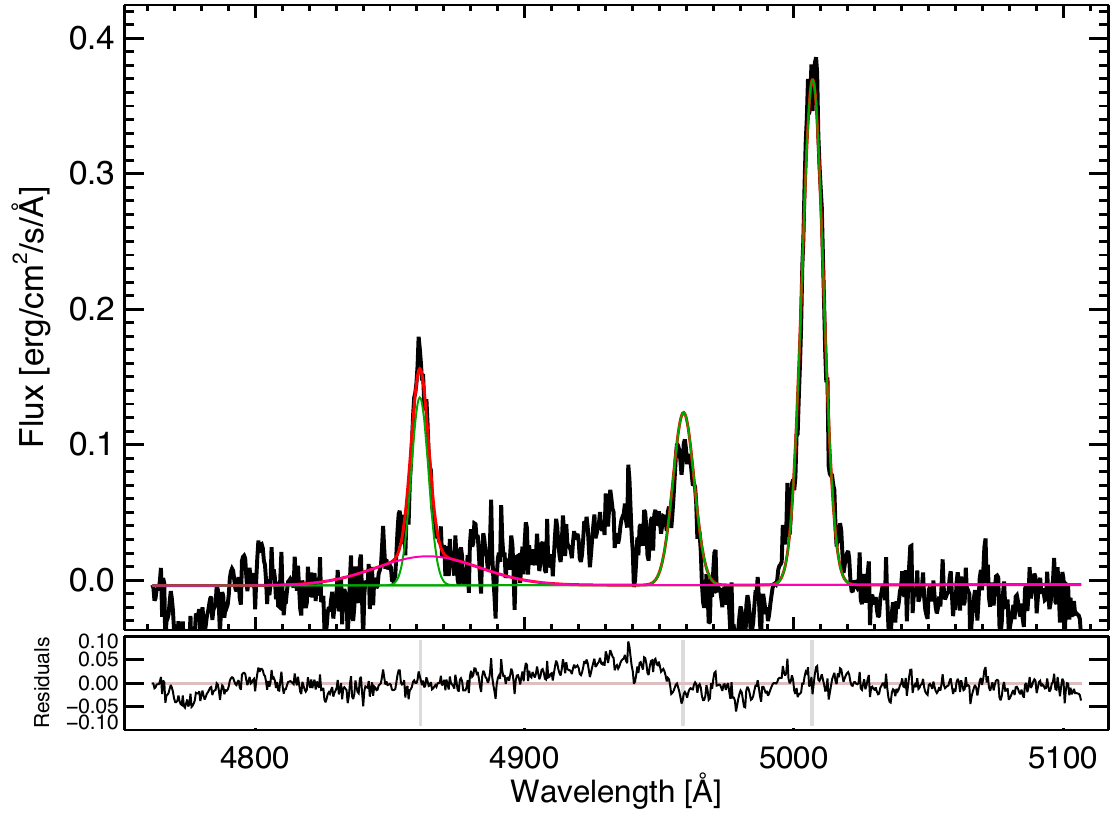}
\hspace{0.1cm} 
\includegraphics[trim = 5.55cm 13.25cm 5.25cm 6.3cm, clip=true, width=.445\textwidth]{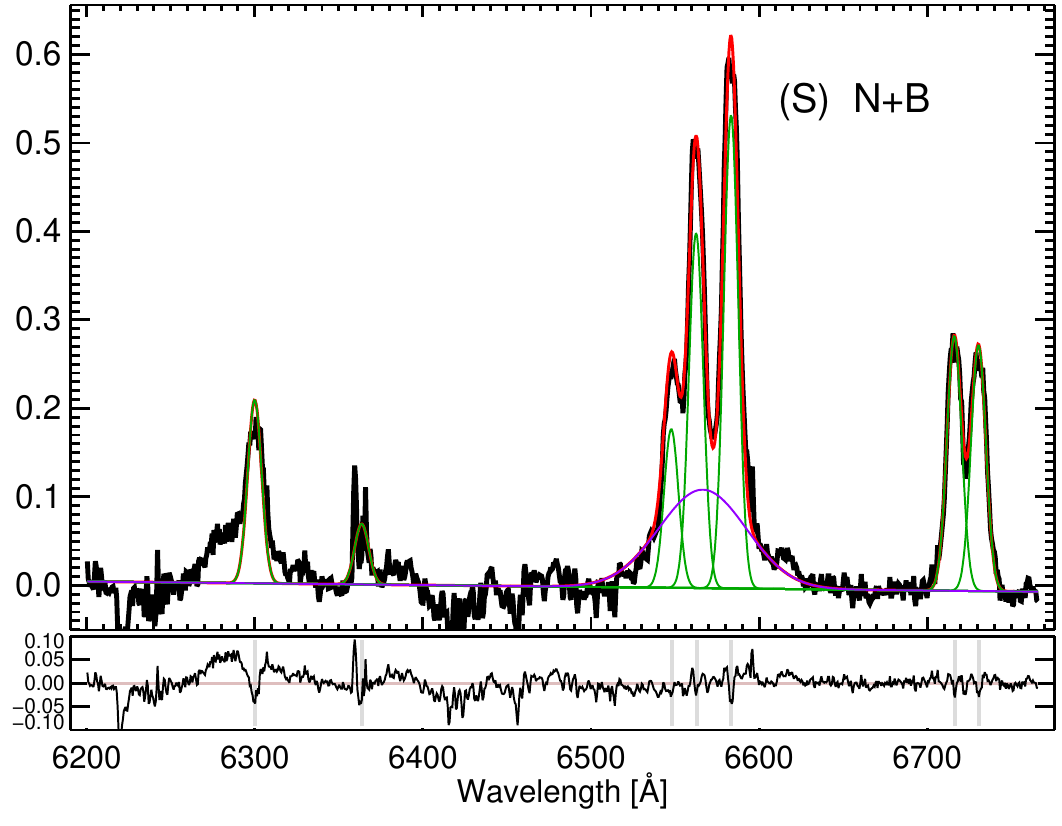} 
\caption{(General description as in Fig.\,\ref{Panel_NGC0266}.) NGC\,5077: the [S\,II] profile is well  fitted with one component, but  [O\,I] seems to have a strong, broad and blueshifted wing. A broad H$\alpha$ component is necessary for a satisfactory fit.  It is worth to note that the modelling proposed by \textit{HFS97} (which includes the broad component) depends on [S\,II] assumptions.} \label{Panel_NGC5077} 		 		 
\end{figure*}
\clearpage

\begin{figure*}
\includegraphics[trim = 2.cm 13.15cm 2.cm 1.5cm, clip=true, width=1.\textwidth]{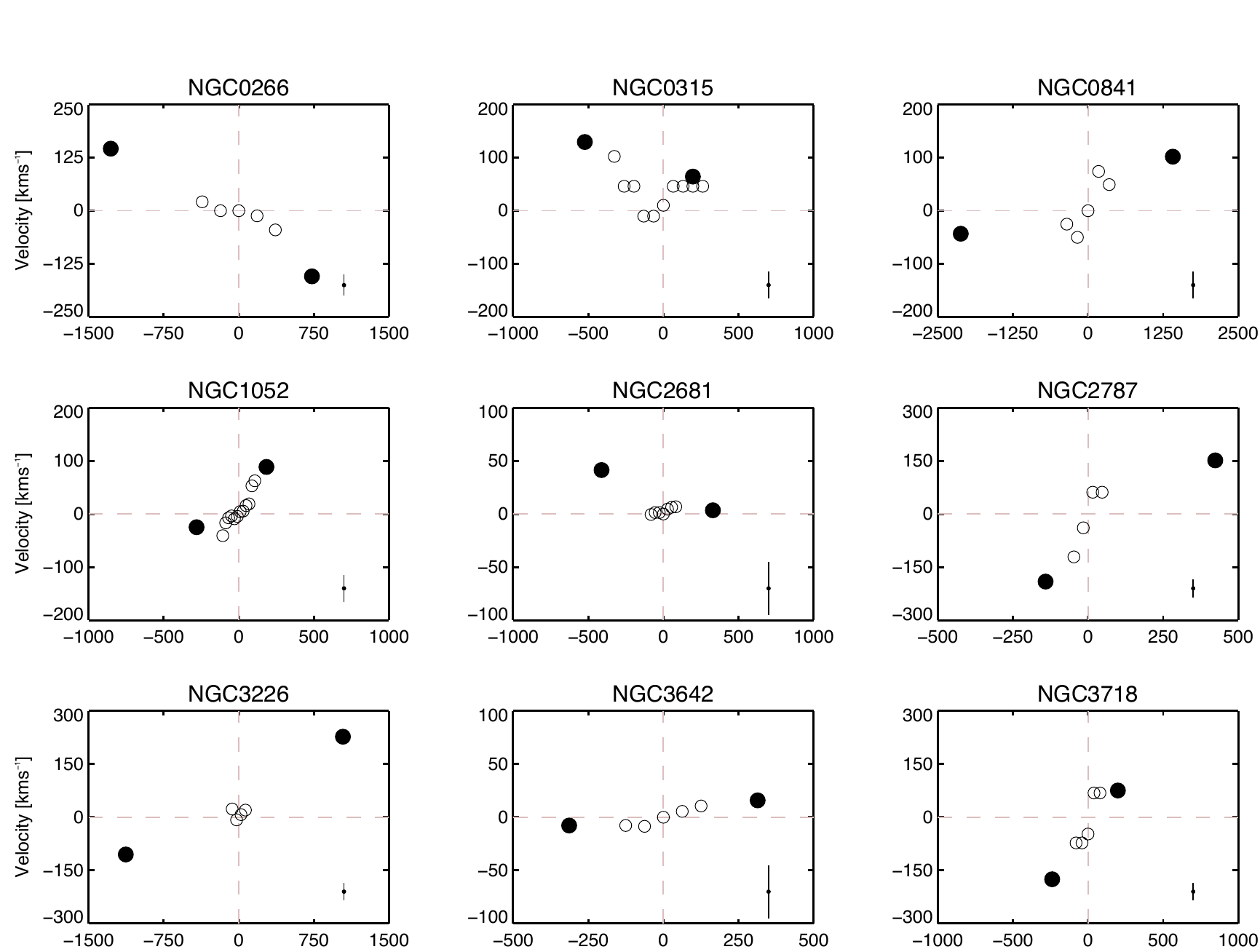} \\
\includegraphics[trim = 2.cm 12.60cm 2.cm 2.0cm, clip=true, width=1.\textwidth]{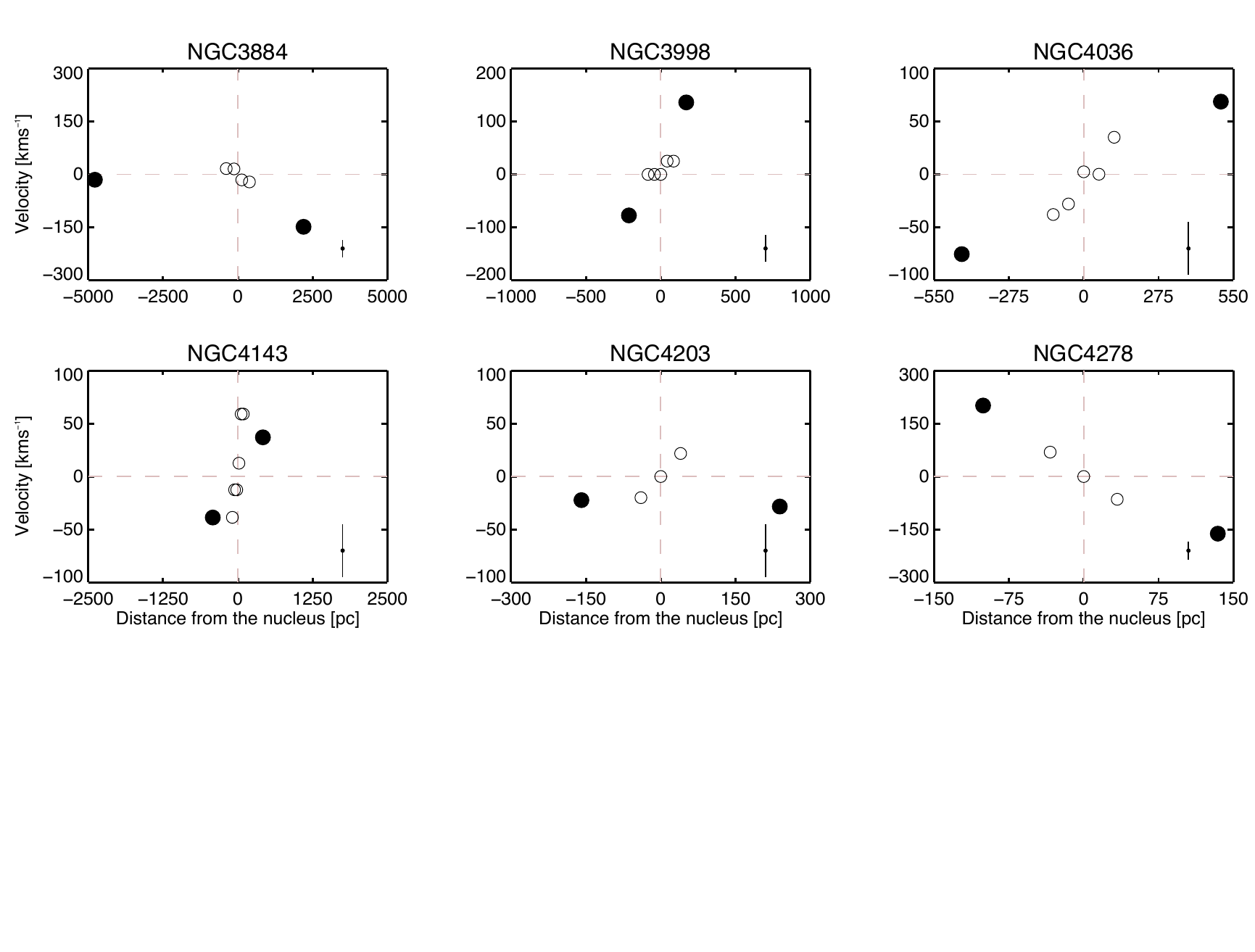}
 \captionsetup{labelformat=empty}{Fig.\,\ref{Panel_PVD}\,--\,continues.} 		 
\end{figure*}
\clearpage

\clearpage

\label{lastpage}
\end{document}